\documentclass{aa} 
\usepackage{graphicx}
\usepackage{natbib}
\usepackage{amssymb}
\usepackage{amsmath}
\usepackage{url}
\usepackage[section]{placeins}
\usepackage{multirow}
\usepackage{lscape}
\usepackage{txfonts}
\begin{document}
   \title{OH far-infrared emission from low- and intermediate-mass protostars surveyed with Herschel-PACS\thanks{\textit{Herschel} is an ESA space observatory with science instruments provided by European-led Principal Investigator consortia and with important participation from NASA.}}

\author{S.~F.~Wampfler \inst{\ref{inst1},\ref{inst2}} 
\and S.~Bruderer\inst{\ref{inst3}}
\and A.~Karska\inst{\ref{inst3}}
\and G.~J.~Herczeg\inst{\ref{inst4}}
\and E.~F.~van~Dishoeck\inst{\ref{inst3},\ref{inst5}}
\and L.~E.~Kristensen\inst{\ref{inst5}}
\and J.~R.~Goicoechea\inst{\ref{inst6}}
\and A.~O.~Benz\inst{\ref{inst1}}
\and S.~D.~Doty\inst{\ref{inst7}}
\and C.~M$^{\mathrm{c}}$Coey\inst{\ref{inst8}}
\and A.~Baudry\inst{\ref{inst9}}
\and T.~Giannini\inst{\ref{inst10}}
\and B.~Larsson\inst{\ref{inst11}}
}

\institute{
Institute for Astronomy, ETH Zurich, 8093 Zurich, Switzerland\label{inst1}
\and
Centre for Star and Planet Formation, Natural History Museum of Denmark, University of Copenhagen, \O{}ster Voldgade 5-7, DK-1350 K\o{}benhavn K, Denmark\label{inst2}
\and
Max Planck Institut f\"{u}r Extraterrestrische Physik, Giessenbachstrasse 1, 85748 Garching, Germany\label{inst3}
\and
Kavli Institute for Astronomy and Astrophysics at Peking University, China\label{inst4}
\and
Leiden Observatory, Leiden University, PO Box 9513, 2300 RA Leiden, The Netherlands\label{inst5}
\and
Centro de Astrobiolog\'{\i}a. CSIC-INTA. Carretera de Ajalvir, Km 4, Torrej\'{o}n de Ardoz. 28850, Madrid, Spain.\label{inst6}
\and
Department of Physics and Astronomy, Denison University, Granville, OH, 43023, USA\label{inst7}
\and
Department of Physics and Astronomy, University of Waterloo, Waterloo, Ontario, N2L 3G1, Canada\label{inst8}
\and
Universit\'{e} de Bordeaux, Laboratoire d'Astrophysique de Bordeaux, France; CNRS/INSU, UMR 5804, Floirac, France\label{inst9}
\and
INAF - Osservatorio Astronomico di Roma, 00040 Monte Porzio Catone, Italy\label{inst10}
\and
Department of Astronomy, Stockholm University, AlbaNova, 106 91 Stockholm, Sweden\label{inst11}
}

\date{accepted December 18, 2012} \titlerunning{OH survey with PACS}


\def\placefigureOHLevels{
\begin{figure*}
\sidecaption 
 \centering
 \resizebox{\hsize}{!}{\includegraphics[angle=0,bb=0 0 498 566]{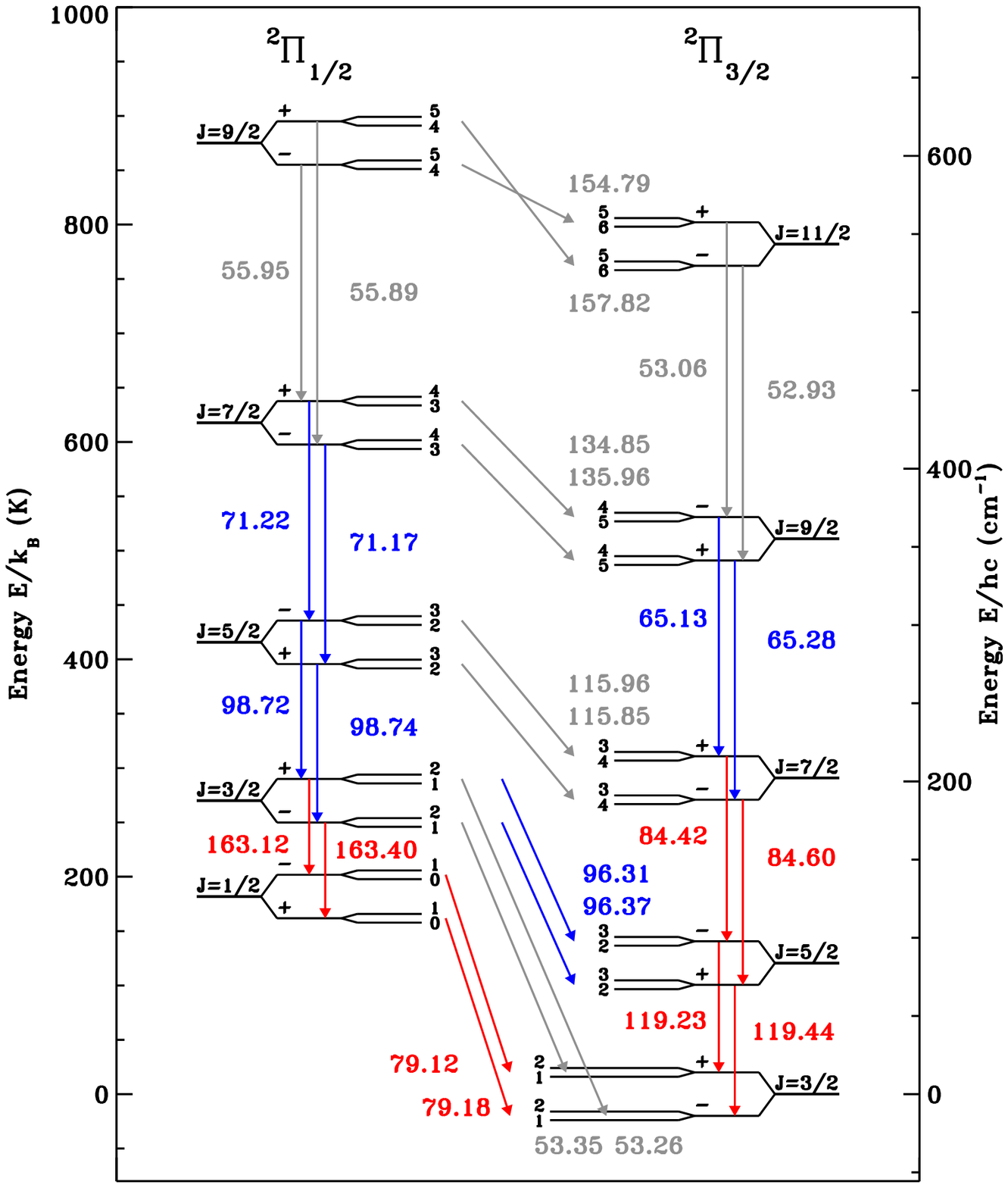}} 
 \caption{OH transitions accessible with Herschel/PACS. Wavelengths are given in units of microns. Transitions that were observed for the entire source sample are shown in red. Transitions targeted only in the fraction of sources observed in range scan mode are depicted in blue if detected from at least one source and in gray if undetected.}
 \label{fig:OH_levels}
\end{figure*}
}
 
\def\placefigureClass0Spectra{
\begin{figure*}
 \centering
 \resizebox{0.93\hsize}{!}{\includegraphics[angle=0,bb=0 0 510 680]{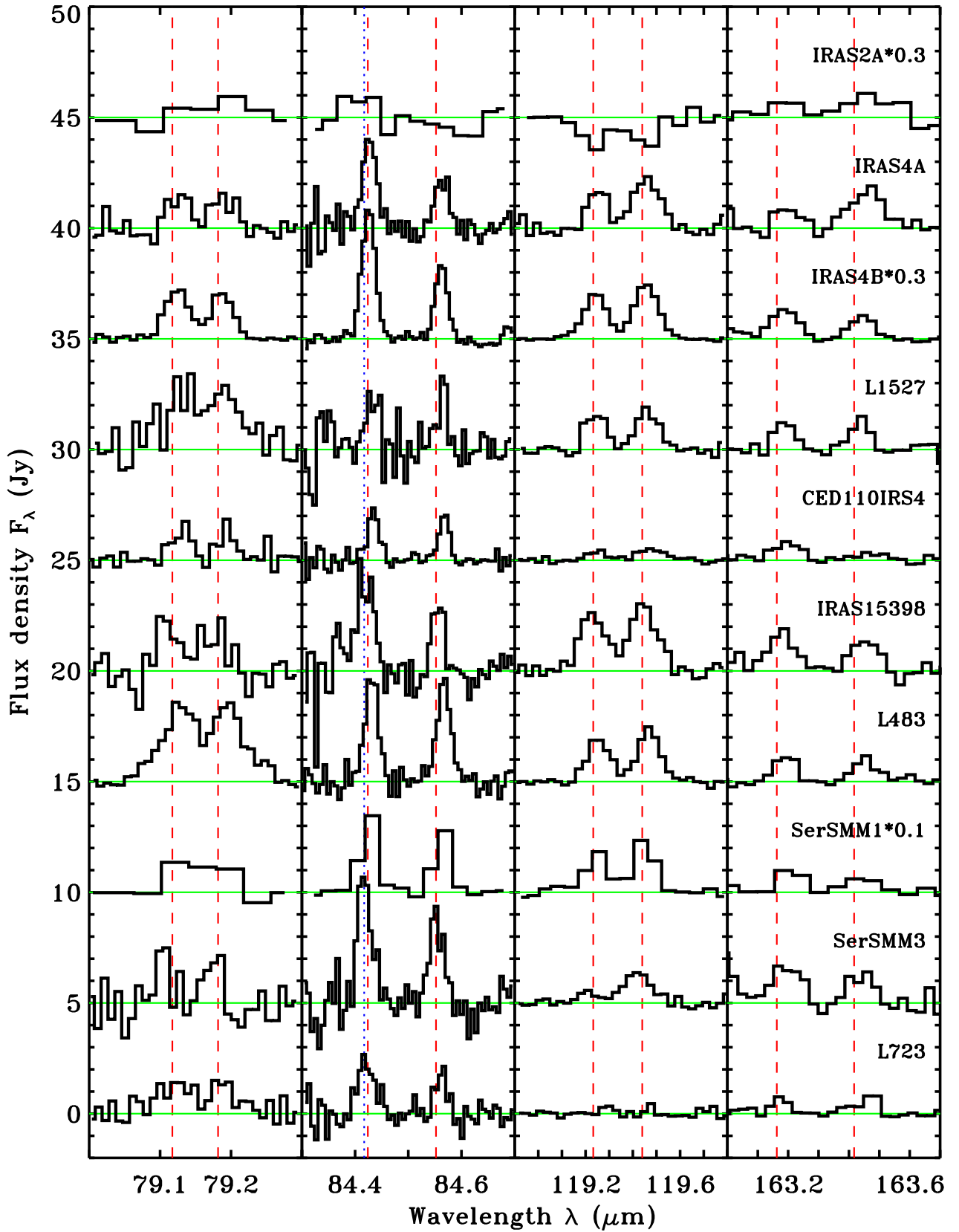}} 
 \caption{PACS line scan OH spectra of the low-mass class 0 young stellar objects in our sample. The x-axis is wavelengths in $\mu$m, the y-axis is continuum subtracted flux density in Jy (plus a constant offset). The red dashed lines indicate the rest frequencies of the OH transitions. The blue dotted line represents the rest frequency of CO(31-30), which is blended with OH at 84.42~$\mu$m. The sampling of range scans (Ser~SMM1 and NGC~1333~IRAS~2A) is different from line scans.}
 \label{fig:class0_spectra}
\end{figure*}
}

\def\placefigureClassISpectra{
\begin{figure*}
 \centering
 \resizebox{0.93\hsize}{!}{\includegraphics[angle=0,bb=0 0 510 566]{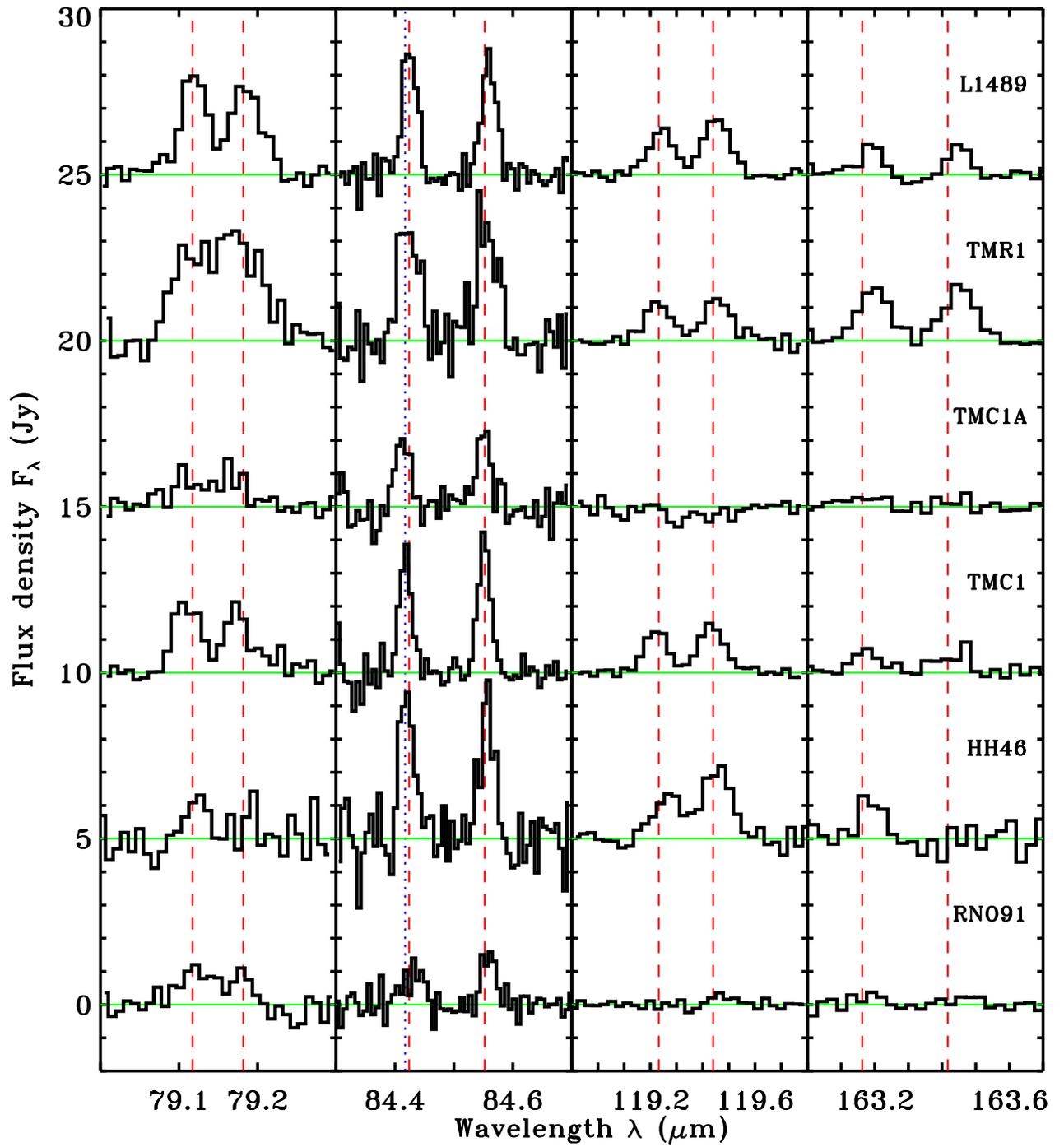}} 
 \caption{The same as Fig.~\ref{fig:class0_spectra} but for the low-mass class I YSOs.}
 \label{fig:classI_spectra}
\end{figure*}
}

\def\placefigureIMSpectra{
\begin{figure*}
 \centering
 \resizebox{0.93\hsize}{!}{\includegraphics[angle=0,bb=0 0 510 566]{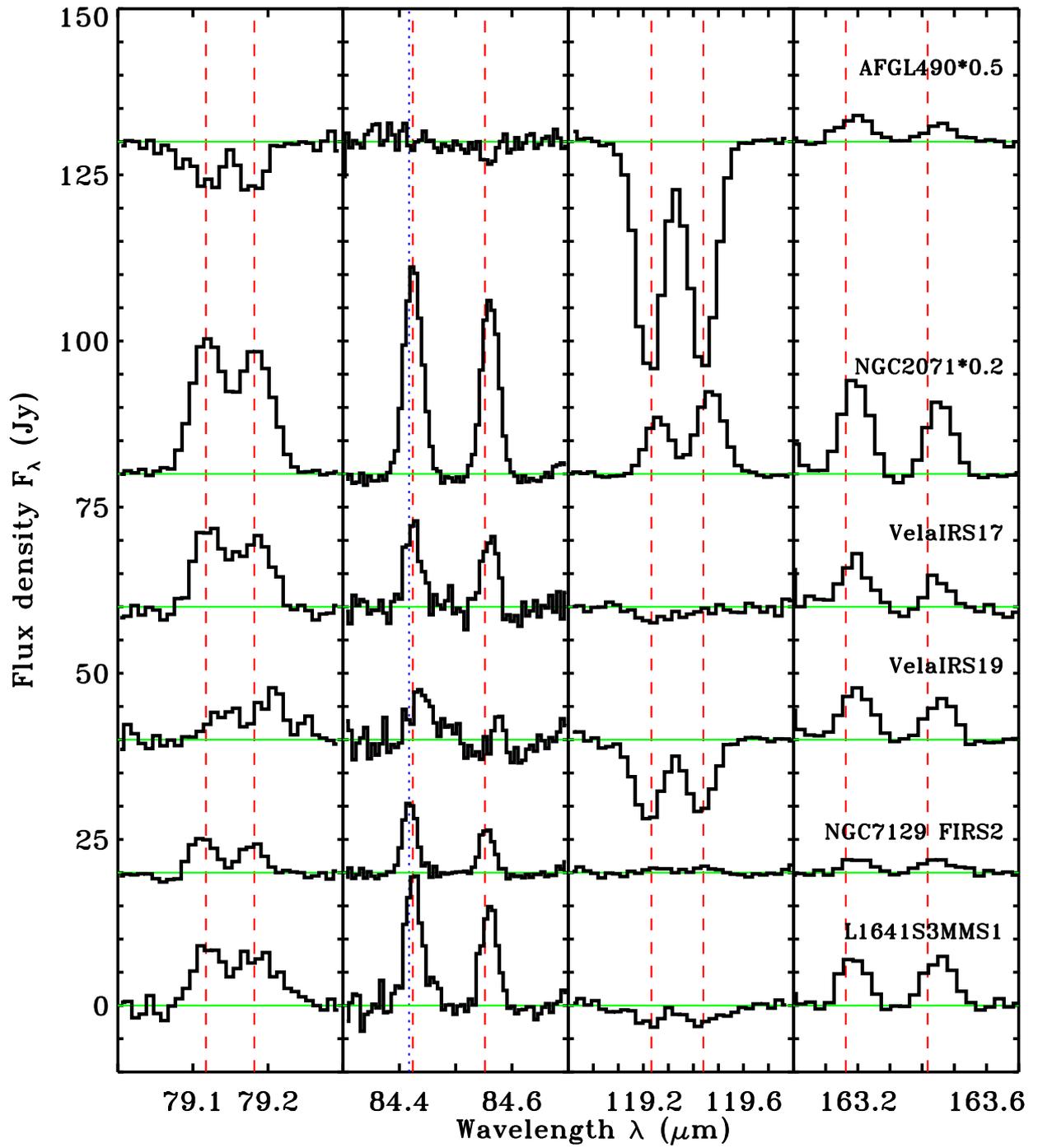}} 
 \caption{The same as Fig.~\ref{fig:class0_spectra} but for the intermediate-mass protostars.}
 \label{fig:im_spectra}
\end{figure*}
}

\def\placefigureLOHvsLbol{
\begin{figure}
 \centering
 \resizebox{1.00\hsize}{!}{\includegraphics[angle=0]{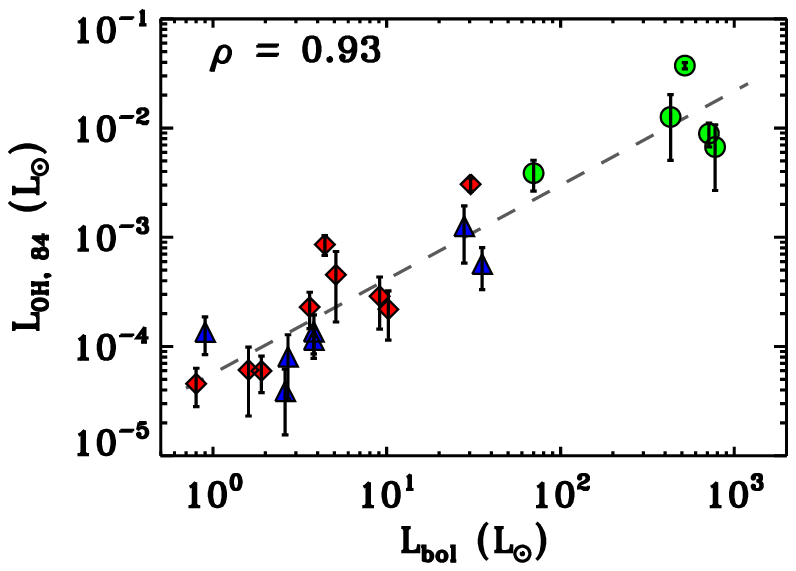}}
 \caption{OH luminosity of the 84~$\mu$m transition vs. bolometric luminosity of the source. Low-mass class 0 sources are shown as red diamonds, class I sources as blue triangles, and intermediate-mass protostars as green circles with $3~\sigma$ error bars from the flux determination (does not include the calibration uncertainty). The dashed line is a linear fit to the data.}
 \label{fig:LOH_Lbol}
\end{figure}
}

\def\placefigureLOHvsMenv{
\begin{figure}
 \centering
 \resizebox{1.00\hsize}{!}{\includegraphics[angle=0]{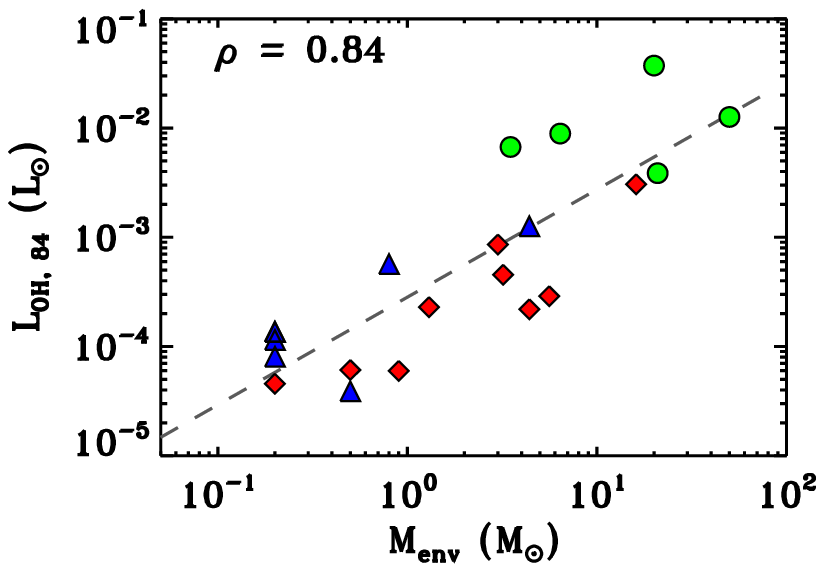}}
 \caption{OH 84~$\mu$m luminosity vs. envelope mass of the source.}
 \label{fig:LOH_Menv}
\end{figure}
}

\def\placefigureLbolvsMenv{
\begin{figure}
 \centering
 \resizebox{1.00\hsize}{!}{\includegraphics[angle=0]{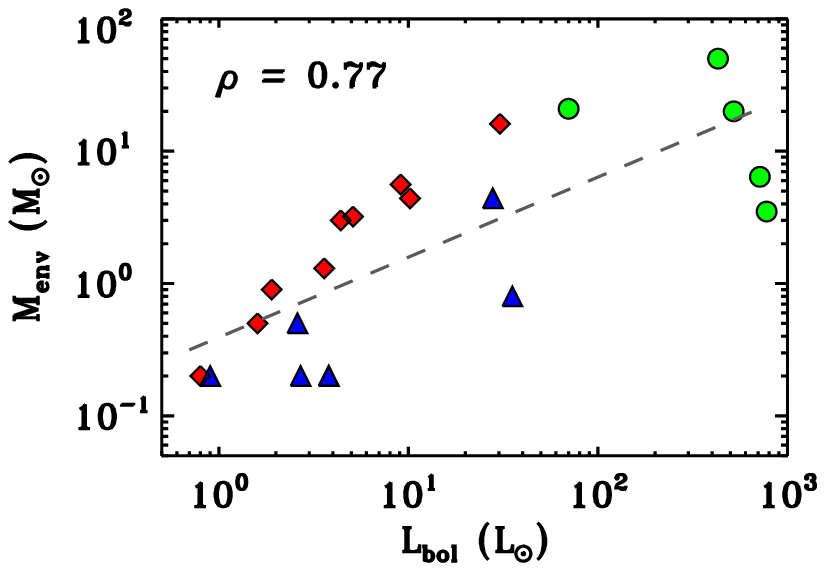}}
 \caption{Bolometric luminosity vs. envelope mass of the sources.}
 \label{fig:Lbol_Menv}
\end{figure}
}

\def\placefigureLOHvsTbol{
\begin{figure}
 \centering
 \resizebox{1.00\hsize}{!}{\includegraphics[angle=0]{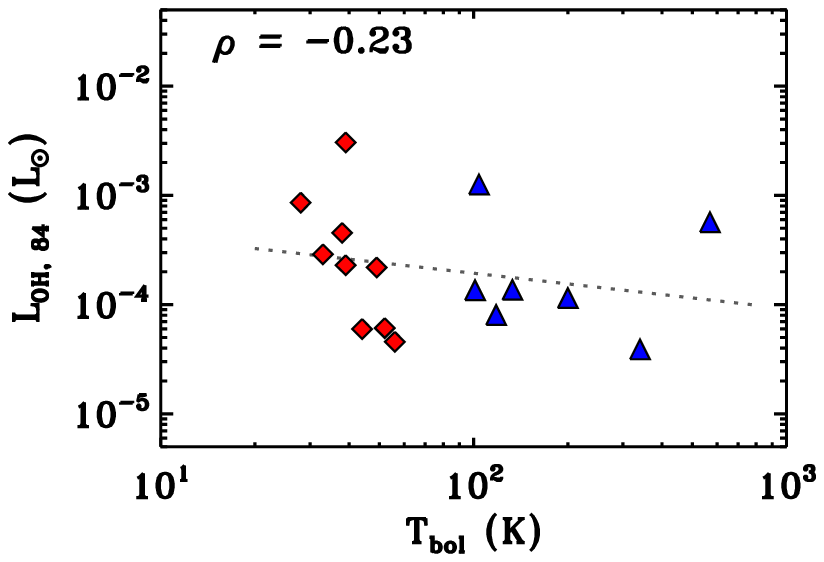}}
 \caption{OH 84~$\mu$m luminosity vs. bolometric temperature of the sources.}
 \label{fig:LOH_Tbol}
\end{figure}
}

\def\placefigureLOHvsEvo{
\begin{figure}
 \centering
 \resizebox{1.00\hsize}{!}{\includegraphics[angle=0]{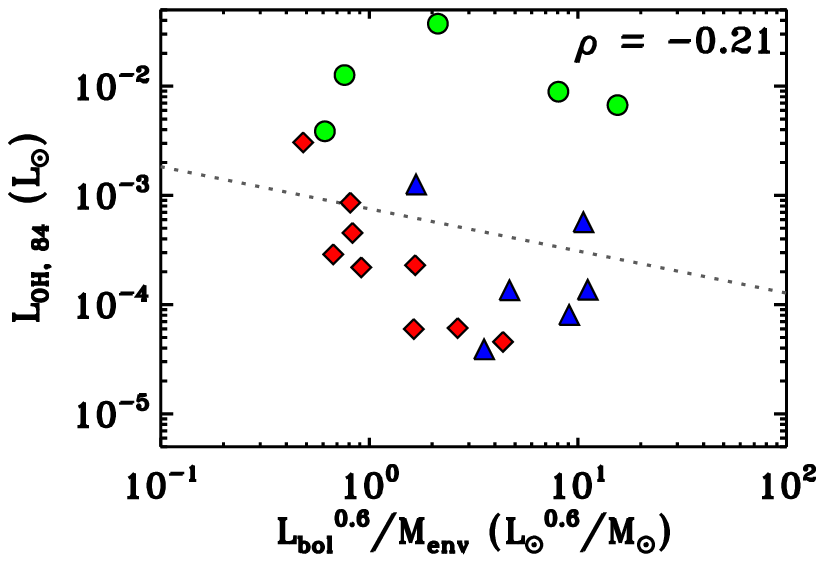}}
 \caption{OH 84~$\mu$m luminosity vs. luminosity to the 0.6 divided by the envelope mass of the sources (evolutionary tracer).}
 \label{fig:LOH_Evo}
\end{figure}
}

\def\placefigureLOHvsFCO{
\begin{figure}
 \centering
 \resizebox{1.00\hsize}{!}{\includegraphics[angle=0]{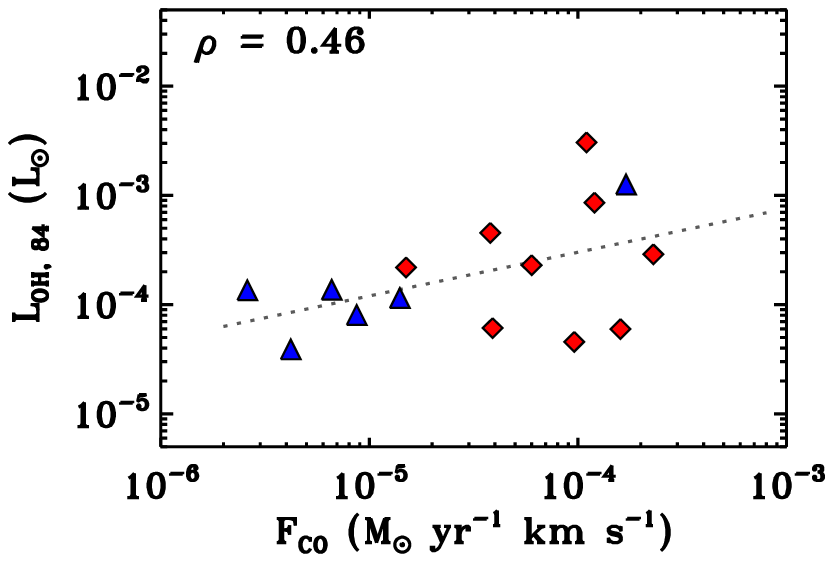}}
 \caption{OH 84~$\mu$m luminosity vs. outflow force.}
 \label{fig:LOH_fco}
\end{figure}
}

\def\placefigureLbolvsOHH2O89ratio{
\begin{figure}
 \centering
 \resizebox{1.00\hsize}{!}{\includegraphics[angle=0]{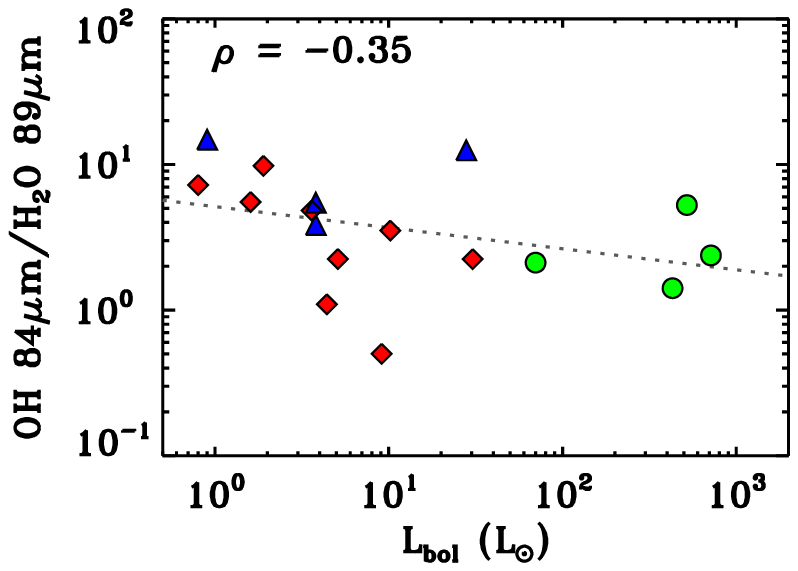}}
 \caption{OH $84~\mu\mathrm{m}$ flux divided by H$_2$O $89~\mu\mathrm{m}$ flux vs. bolometric luminosity of the sources. Low-mass class I sources have a higher ratio than class 0 sources on average.} 
 \label{fig:Lbol_OHH2O89ratio}
\end{figure}
}

\def\placefigureTbolvsOHH2O89ratio{
\begin{figure}
 \centering
 \resizebox{1.00\hsize}{!}{\includegraphics[angle=0]{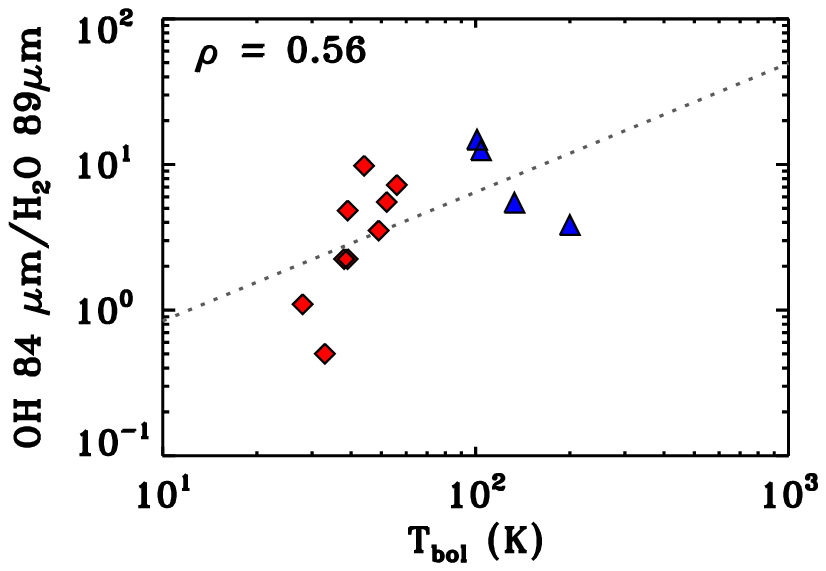}}
 \caption{OH $84~\mu\mathrm{m}$ flux divided by H$_2$O $89~\mu\mathrm{m}$ flux vs. bolometric temperature of the sources.}
 \label{fig:Tbol_OHH2O89ratio}
\end{figure}
}

\def\placefigureMenvvsOHH2O89ratio{
\begin{figure}
 \centering
 \resizebox{1.00\hsize}{!}{\includegraphics[angle=0]{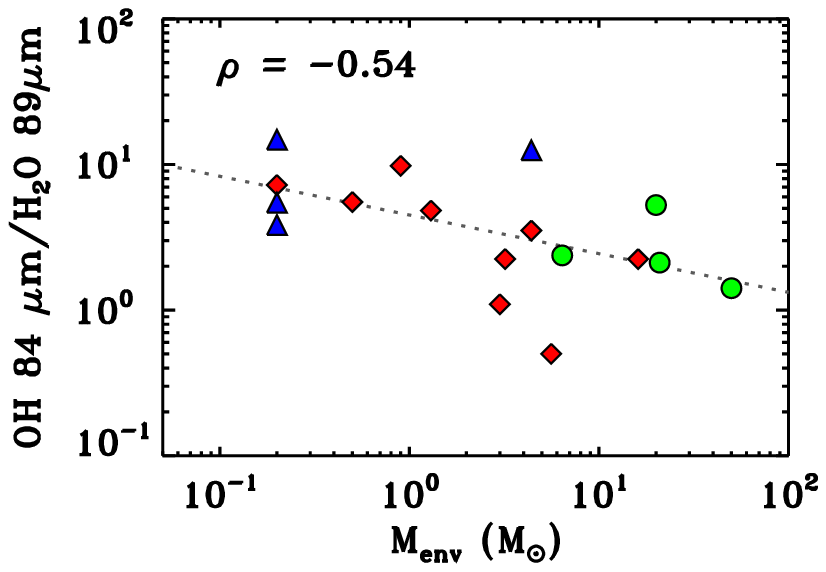}}
 \caption{OH $84~\mu\mathrm{m}$ flux divided by H$_2$O $89~\mu\mathrm{m}$ flux vs. envelope mass of the sources.}
 \label{fig:Menv_OHH2O89ratio}
\end{figure}
}

\def\placefigureEvovsOHH2O89ratio{
\begin{figure}
 \centering
 \resizebox{1.00\hsize}{!}{\includegraphics[angle=0]{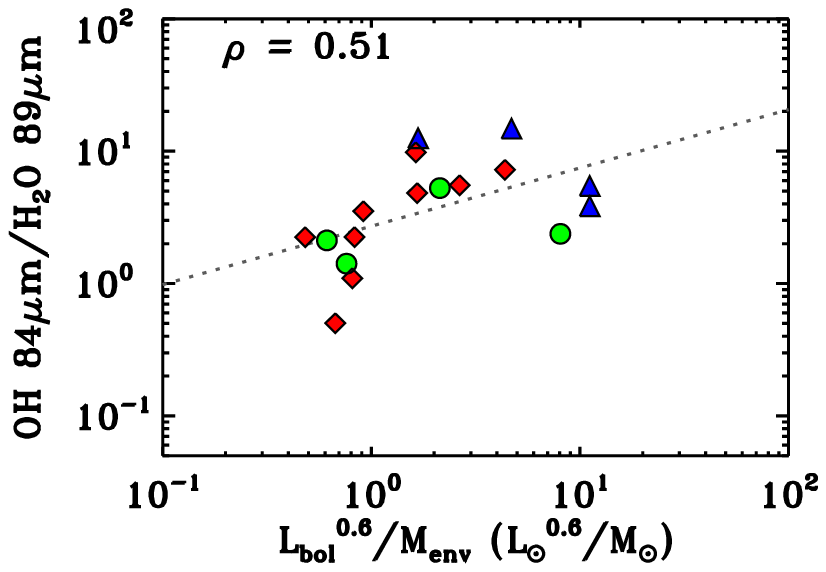}}
 \caption{OH $84~\mu\mathrm{m}$ flux divided by H$_2$O $89~\mu\mathrm{m}$ flux vs. evolutionary tracer.}
 \label{fig:Evo_OHH2O89ratio}
\end{figure}
}

\def\placefigureRotDiag{
\begin{figure}
 \centering
 \resizebox{1.00\hsize}{!}{\includegraphics[angle=0]{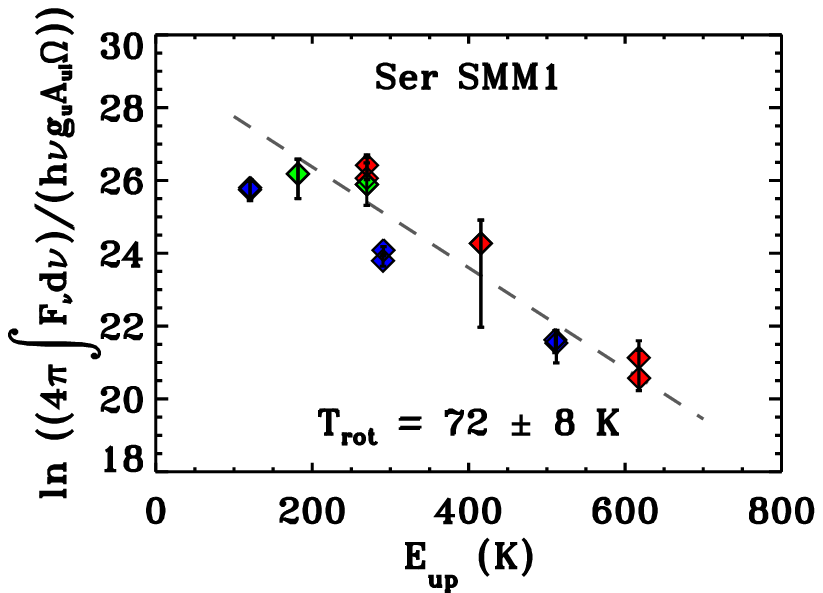}}
 \caption{Rotational diagram for the OH lines measured from Ser~SMM1. Transitions belonging to the ${}^2\Pi_{1/2}$ ladder are shown in red, those from the ${}^2\Pi_{3/2}$ ladder in blue, and cross-ladder transitions in green. The error bars indicate the $3~\sigma$ errors on the flux measurement (calibration errors are not included).}
 \label{fig:smm1_rotdiag}
\end{figure}
}

\def\placefigureLOHLOI{
\begin{figure}
 \centering
 \resizebox{1.00\hsize}{!}{\includegraphics[angle=0]{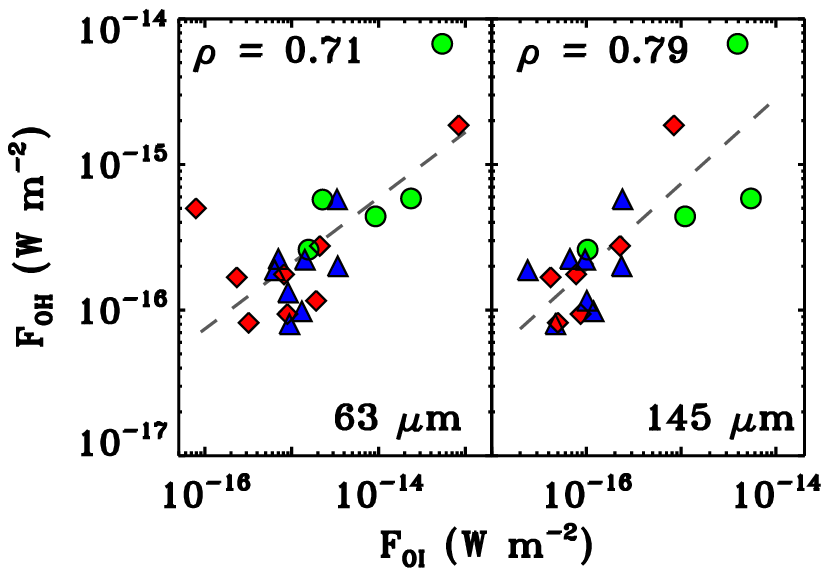}}
 \caption{Correlation of OH 84~$\mu$m and [OI] 63~$\mu$m and 145~$\mu$m fluxes.}
 \label{fig:LOH_LOI}
\end{figure}
}

\def\placefigureLOHH2O{
\begin{figure}
 \centering
 \resizebox{1.00\hsize}{!}{\includegraphics[angle=0]{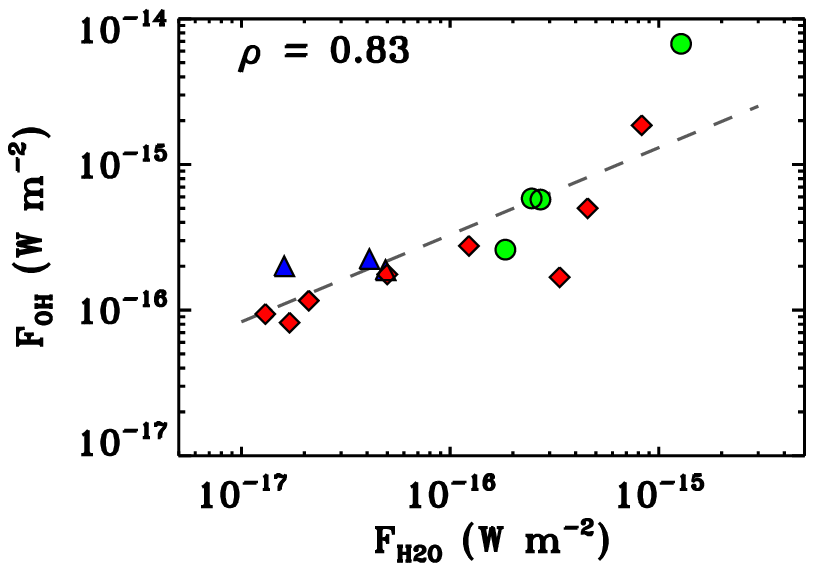}}
 \caption{Correlation of OH 84~$\mu$m and H$_2$O 89.99~$\mu$m fluxes.}
 \label{fig:LOH_H2O}
\end{figure}
}

\def\placefigureLOHrd100K{
\begin{figure}
 \centering
 \resizebox{1.00\hsize}{!}{\includegraphics[angle=0]{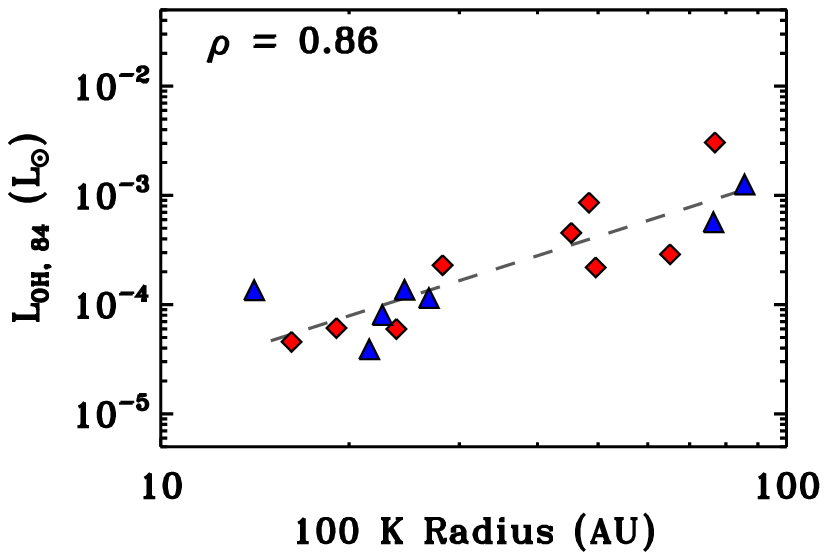}}
 \caption{Correlation of OH 84~$\mu$m luminosity with the 100 K radius of the spherical source models.}
 \label{fig:LOH_rd100K}
\end{figure}
}

\def\placefigureLOHn1000au{
\begin{figure}
 \centering
 \resizebox{1.00\hsize}{!}{\includegraphics[angle=0]{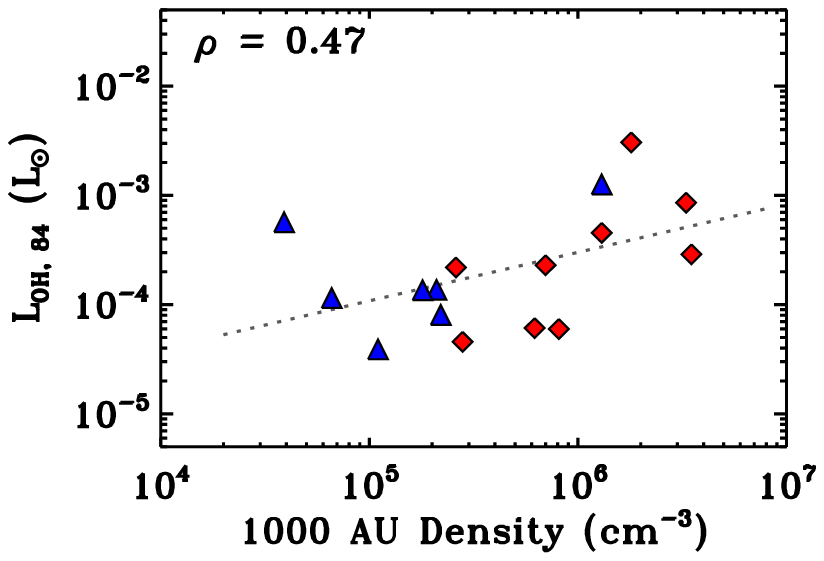}}
 \caption{Correlation of OH 84~$\mu$m luminosity with the H$_2$ density at 1000 AU from the spherical source models.}
 \label{fig:LOH_n1000au}
\end{figure}
}

\def\placefigureMapL1489{
\begin{figure}
 \centering
 \resizebox{0.90\hsize}{!}{\includegraphics[angle=90,bb=28 90 594 713]{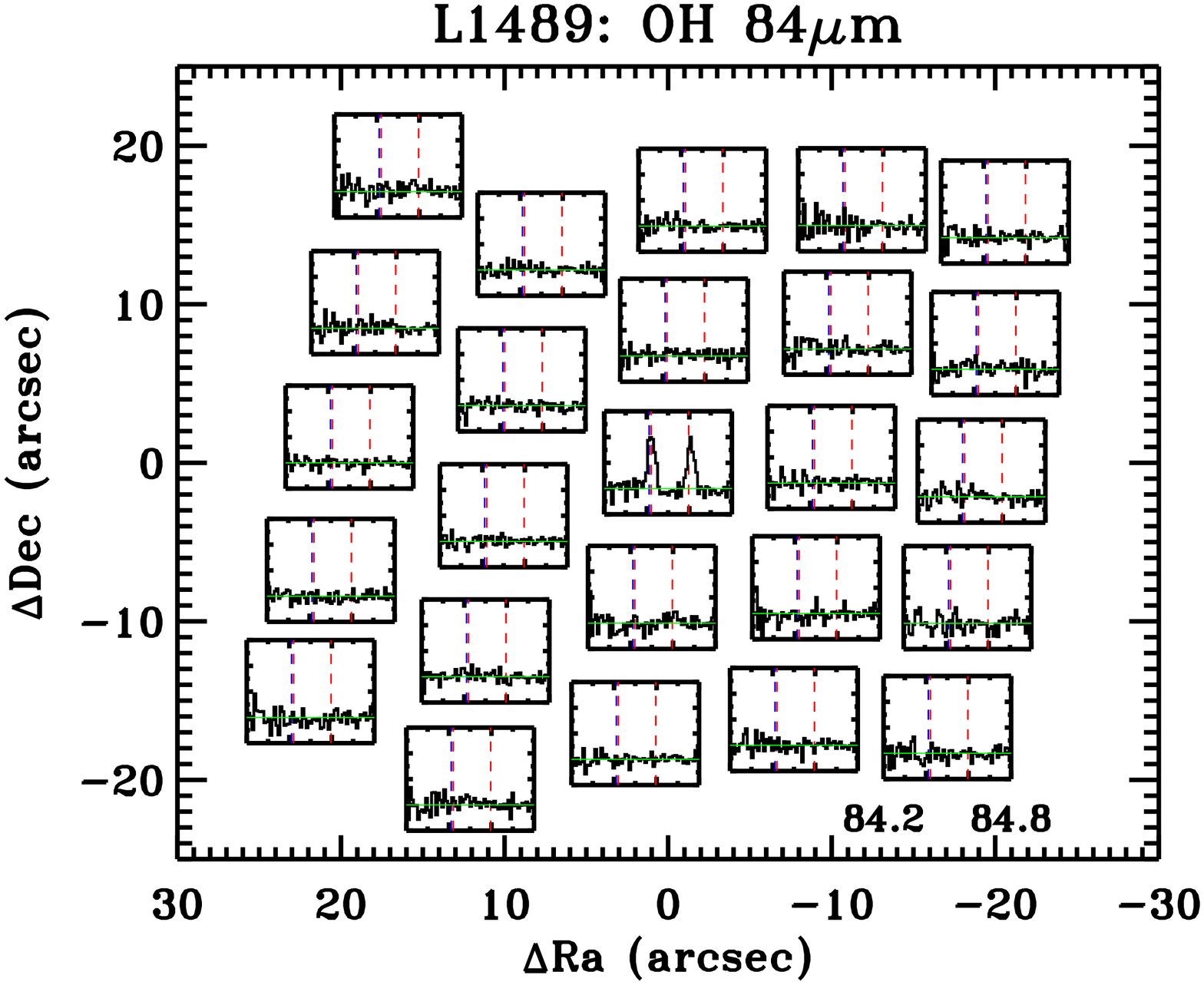}}
 \caption{Map of the compact OH $84~\mu\mathrm{m}$ emission from L~1489. The x-axis is wavelength in $\mu$m, the y-axis continuum subtracted and normalized flux density. All spaxels were normalized with respect to the central one. The red dashed lines indicate the rest frequencies of the OH transitions, the blue dashed line the CO(31-30).}
 \label{fig:map_l1489}
\end{figure}
}

\def\placefigureMapAFGL490{
\begin{figure}
 \centering
 \resizebox{0.90\hsize}{!}{\includegraphics[angle=0,bb=28 714 651 1280]{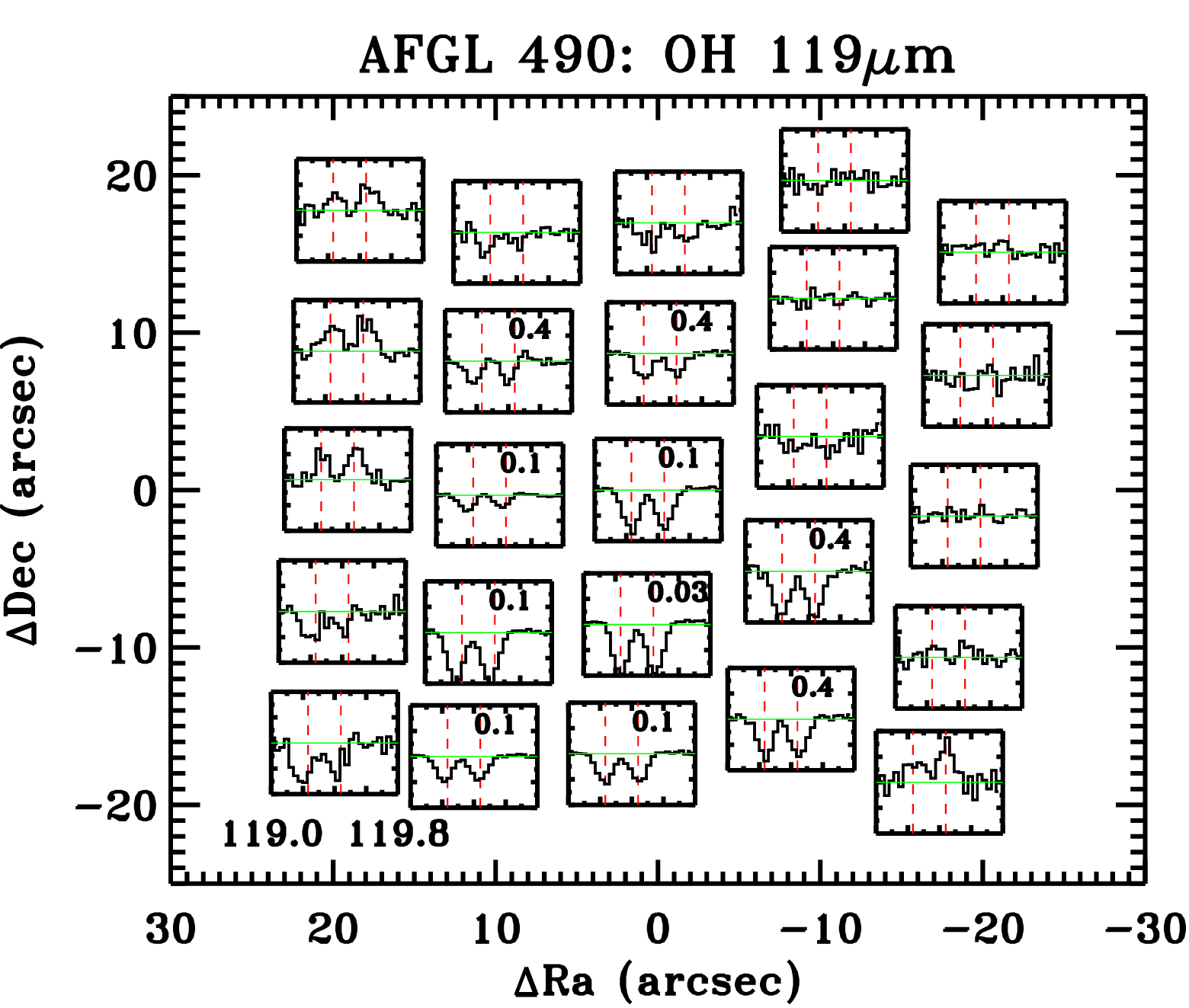}}
 \caption{Map of spatially extended OH $119~\mu\mathrm{m}$ absorption and emission from AFGL~490. The x-axis is wavelength in $\mu$m, the y-axis continuum subtracted and scaled flux density. The labels indicate the multiplicative scaling factors for each spaxel unless they are unity.}
 \label{fig:map_afgl490}
\end{figure}
}

\def\placefigureSpatOHOI{
\begin{figure}
 \centering
 \resizebox{0.90\hsize}{!}{\includegraphics[angle=0,bb=28 714 651 1280]{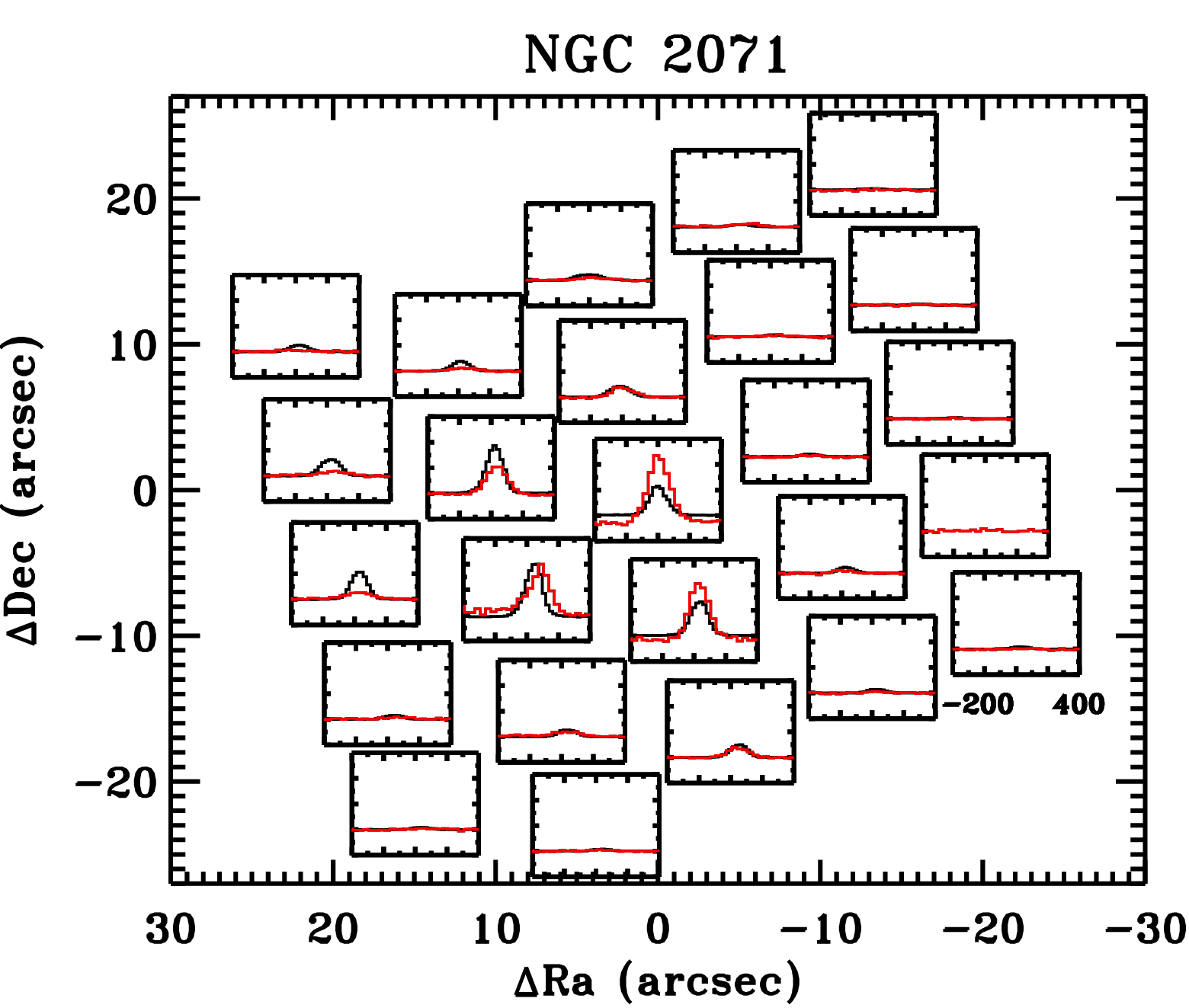}}
 \caption{Map of the OH $84~\mu\mathrm{m}$ (red) and [OI] $63~\mu\mathrm{m}$ emission (black) from NGC~2071, illustrating the similar spatial extent of the two transitions. The x-axis is velocity in km~s$^{-1}$, the y-axis continuum subtracted and normalized flux density. All spaxels were normalized with respect to the spaxel that contains the peak of the continuum emission at $63~\mu\mathrm{m}$. Note that the peak of the continuum emission at $84~\mu\mathrm{m}$ falls onto the central spaxel, although the two observations were carried out consecutively.}
 \label{fig:map_oh_oi_n2071}
\end{figure}
}

\def\placefigureModelTexTgas200KTdust100K{
\begin{figure*}
 \centering
 \resizebox{0.95\hsize}{!}{\includegraphics[angle=0,bb=0 0 498 374]{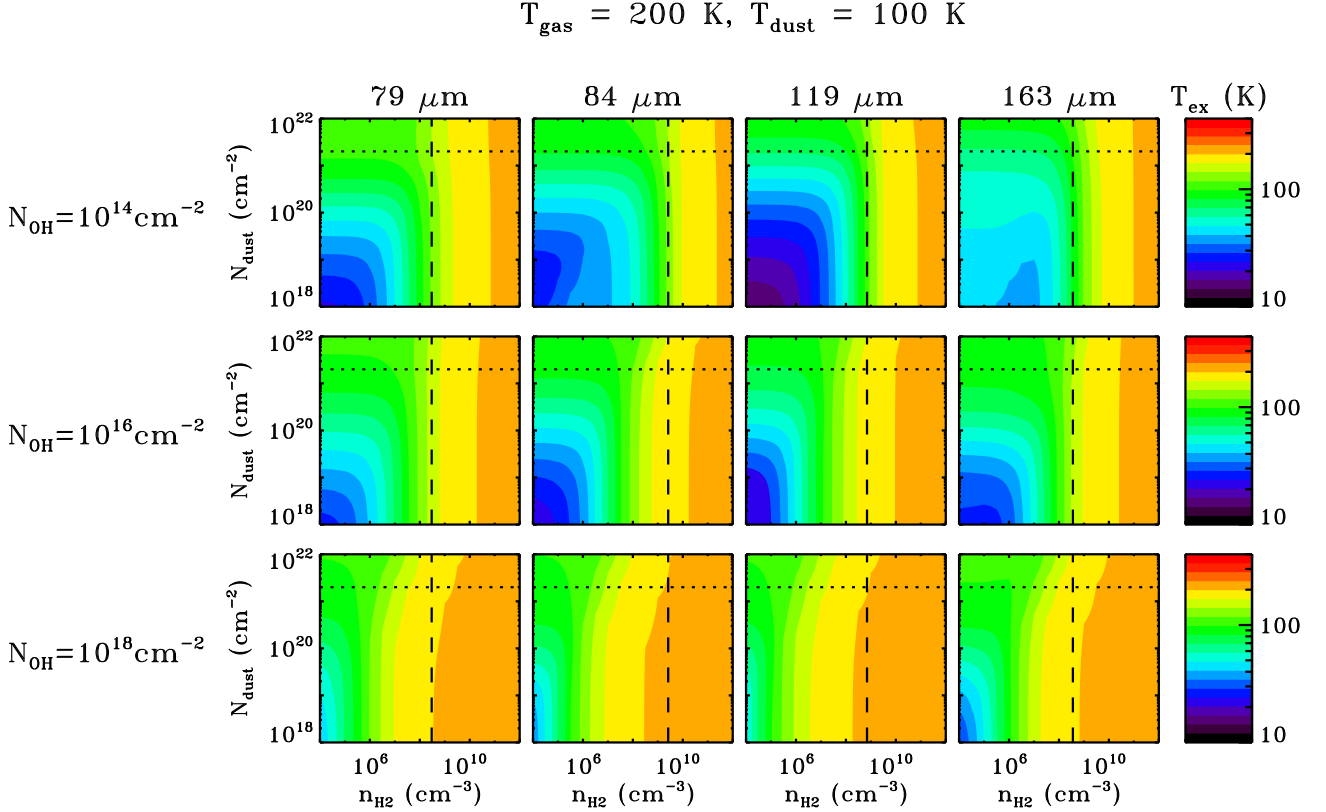}}
 \caption{Excitation temperature from slab models with $T_\mathrm{gas} = 200~\mathrm{K}$ and $T_\mathrm{dust} = 100~\mathrm{K}$ for the OH lines at 79, 84, 119, and $163~\mu\mathrm{m}$ for OH column densities of $10^{14}$, $10^{16}$, and $10^{18}~\mathrm{cm}^{-2}$ and varying H$_2$ density and dust column density. The dashed line indicates the critical density of the upper level of each transition, the dotted line marks where the dust becomes optically thick.
}
 \label{fig:model_tex_Tgas200K_Tdust100K}
\end{figure*}
}

\def\placefigureModelRatiosTgas200KTdust100K{
\begin{figure*}
 \centering
 \resizebox{1.00\hsize}{!}{\includegraphics[bb=0 0 680 374]{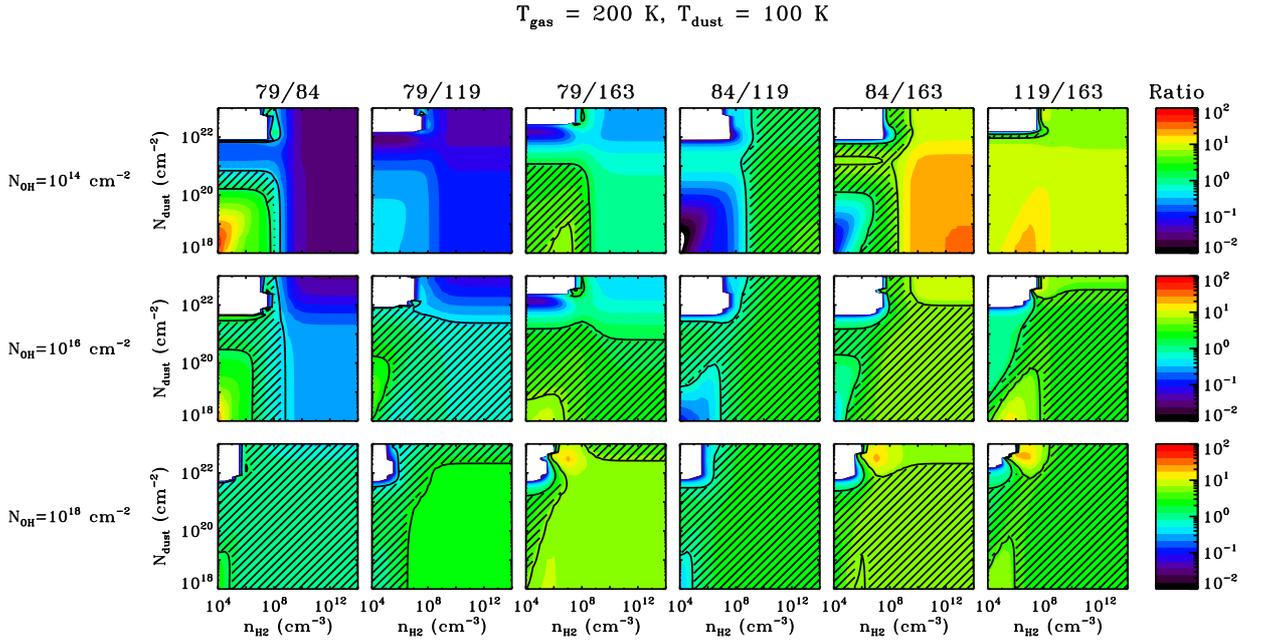}}
 \caption{Line flux ratios from slab models with $T_\mathrm{gas} = 200~\mathrm{K}$ and $T_\mathrm{dust} = 100~\mathrm{K}$ for the OH lines at 79, 84, 119, and $163~\mu\mathrm{m}$ for OH column densities of $10^{14}$, $10^{16}$, and $10^{18}~\mathrm{cm}^{-2}$ and varying H$_2$ density and dust column density. The shaded area marks the observed values. }
 \label{fig:model_ratios_Tgas200K_Tdust100K}
\end{figure*}
}

\def\placefigureRotDiagModel{
\begin{figure}
 \centering
 \resizebox{0.95\hsize}{!}{\includegraphics[angle=0,bb=0 0 249 175]{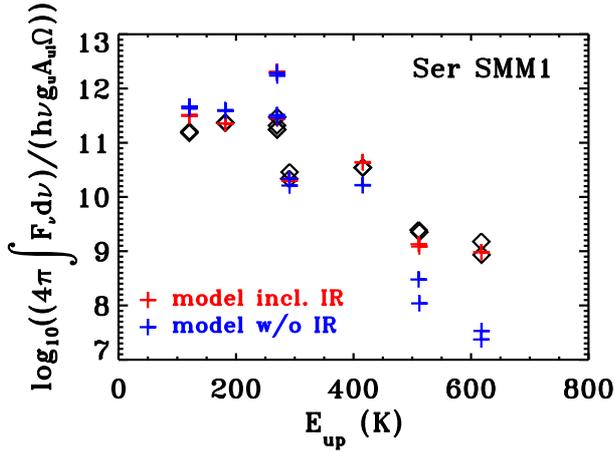}}
 \caption{Comparison of the rotational diagram for the OH lines measured from Ser~SMM1 (black diamonds) and a RADEX model with the source continuum included in the excitation (red plus signs) and a model without the continuum field of the source (blue plus signs). The model fluxes were scaled such that they match the observed $84.60~\mu\mathrm{m}$ emission. The model including the continuum field reproduces the higher excited lines better than the one without.}
 \label{fig:smm1_rotdiag_model}
\end{figure}
}

\def\placefigureLOH63continuum{
\begin{figure}
 \centering
 \resizebox{0.95\hsize}{!}{\includegraphics[angle=0]{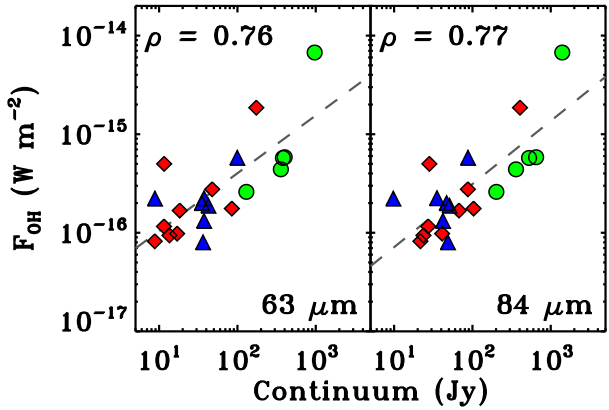}}
 \caption{Correlation of OH 84~$\mu$m flux with the continuum fluxes measured from the PACS data at $63~\mu\mathrm{m}$ and $84~\mu\mathrm{m}$.}
 \label{fig:LOH_63continuum}
\end{figure}
}

\def\placefigu2IRASAmaps{
\begin{landscape}
\begin{figure}
 \centering
  \begin{tabular}{c@{\extracolsep{40pt}}c}
   \resizebox{0.35\hsize}{!}{\includegraphics[angle=0,bb=28 714 651 1280]{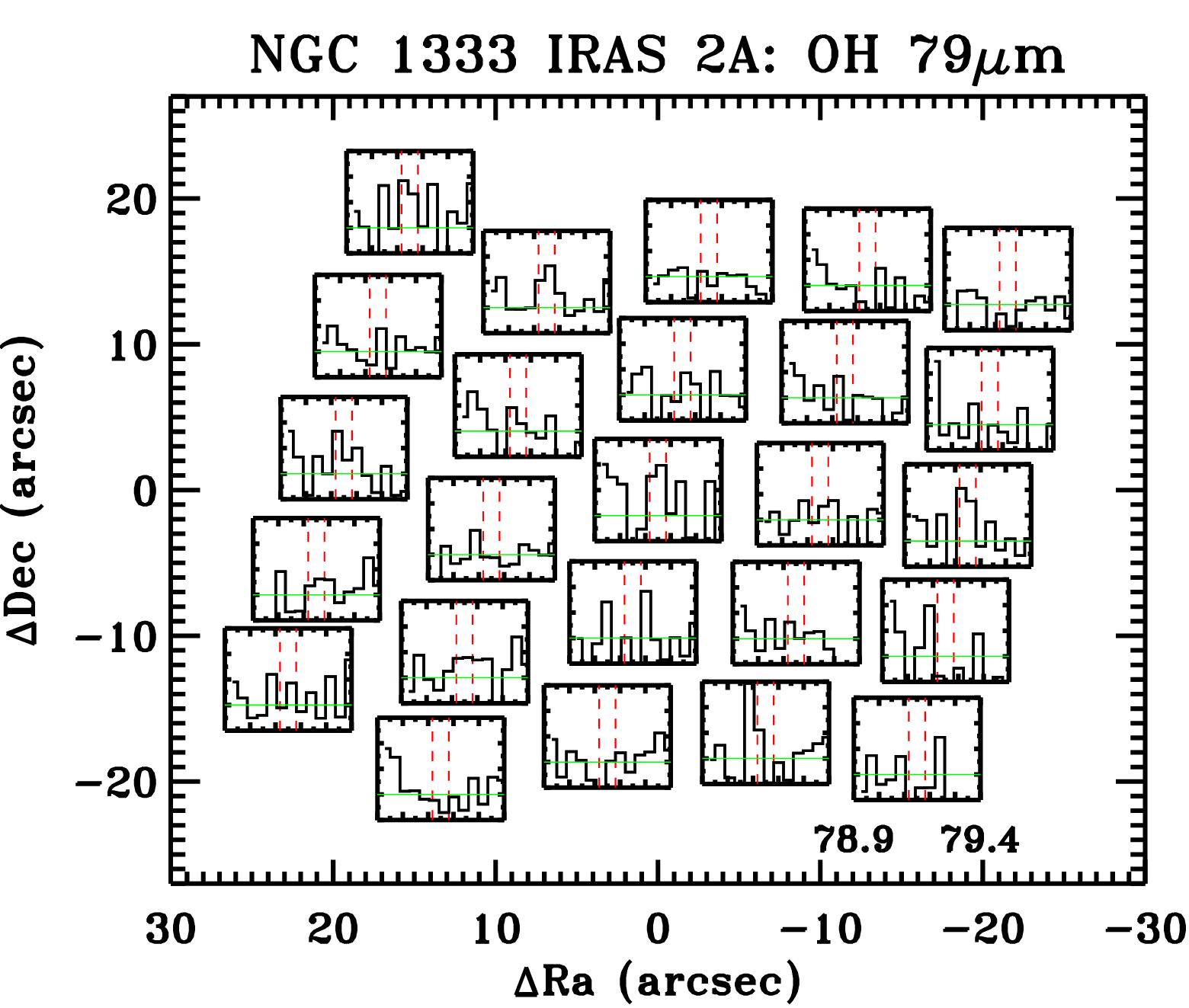}} &
   \resizebox{0.35\hsize}{!}{\includegraphics[angle=0,bb=28 714 651 1280]{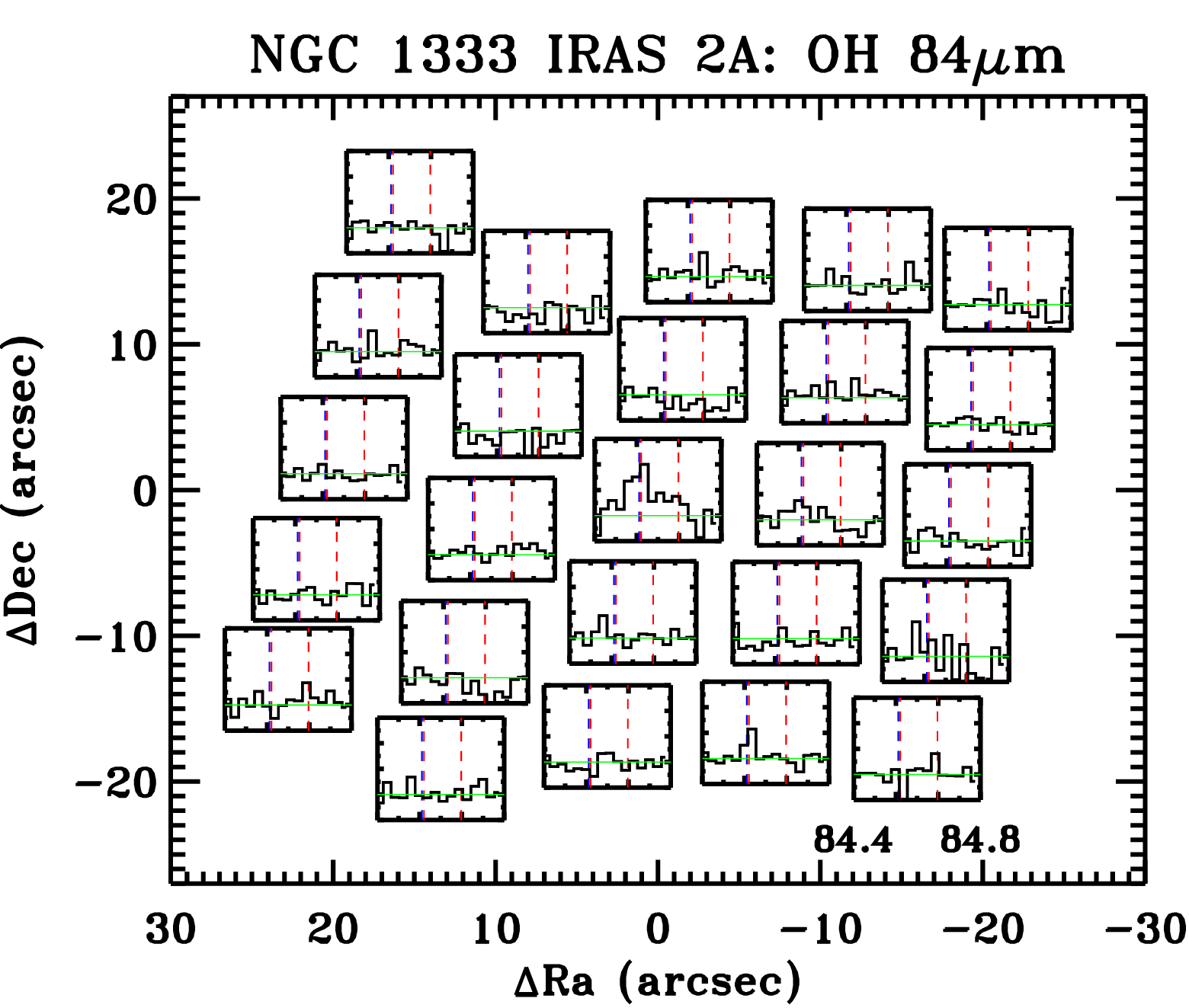}} \\
   & \\
   \resizebox{0.35\hsize}{!}{\includegraphics[angle=0,bb=28 714 651 1280]{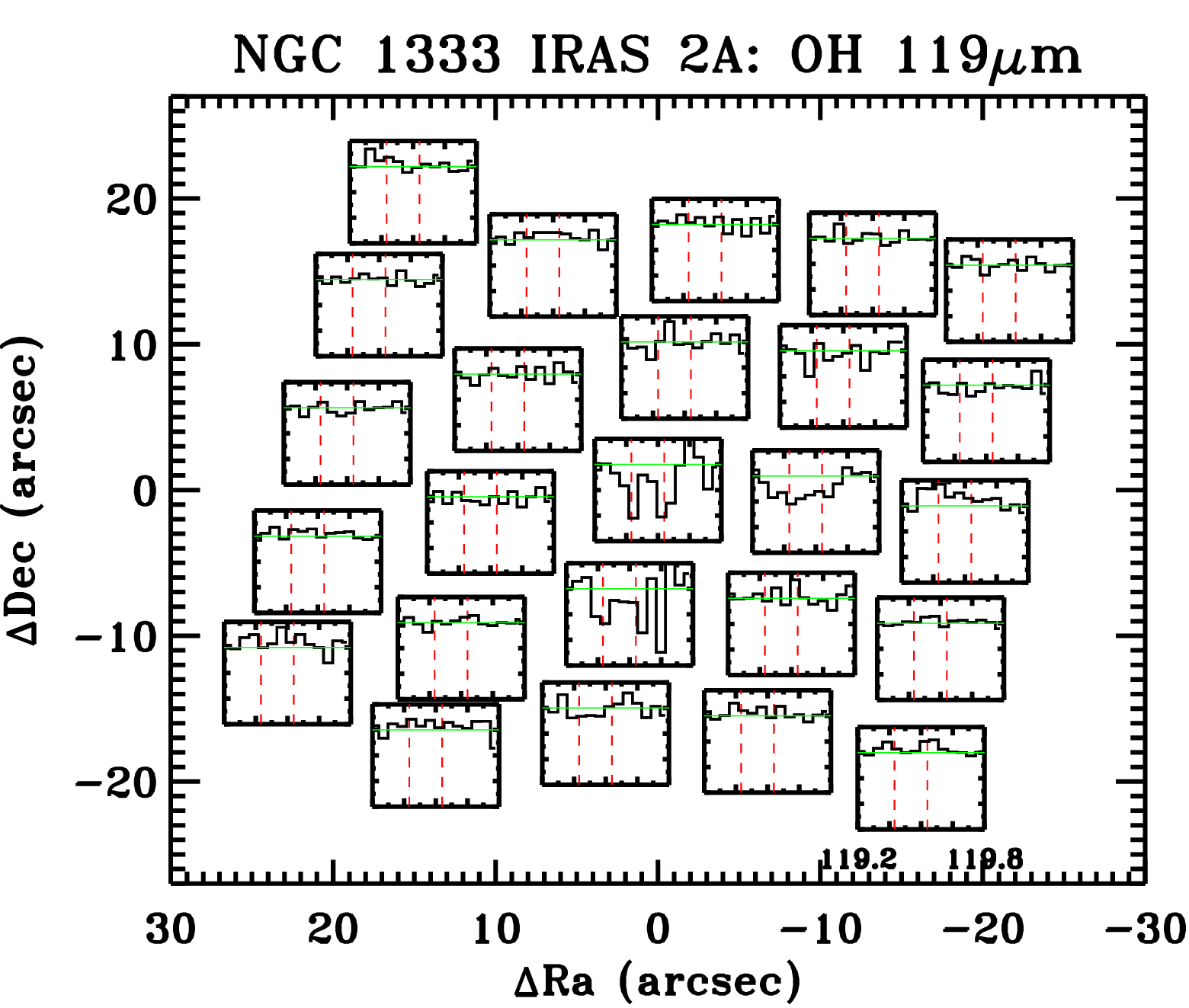}} &
   \resizebox{0.35\hsize}{!}{\includegraphics[angle=0,bb=28 714 651 1280]{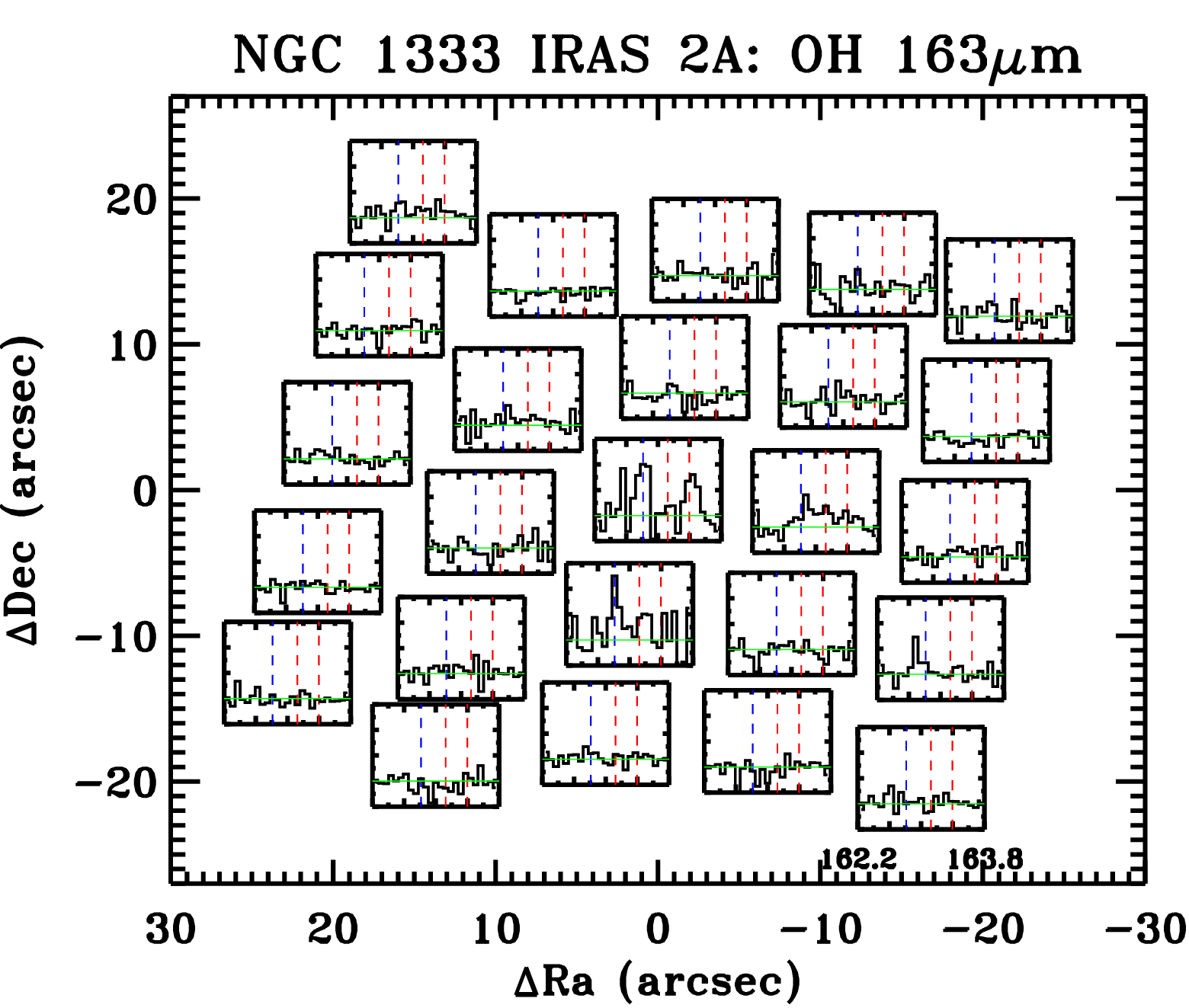}} \\
  \end{tabular}
  \caption{Maps of the OH emission from NGC~1333~IRAS~2A. The red and blue dashed lines indicate the rest wavelengths of the OH and CO transitions, respectively.}
 \label{fig:iras2amaps}
\end{figure}
\end{landscape}
}

\def\placefigIRAS4Amaps{
\begin{landscape}
\begin{figure}
 \centering
  \begin{tabular}{c@{\extracolsep{40pt}}c}
   \resizebox{0.35\hsize}{!}{\includegraphics[angle=0,bb=28 714 651 1280]{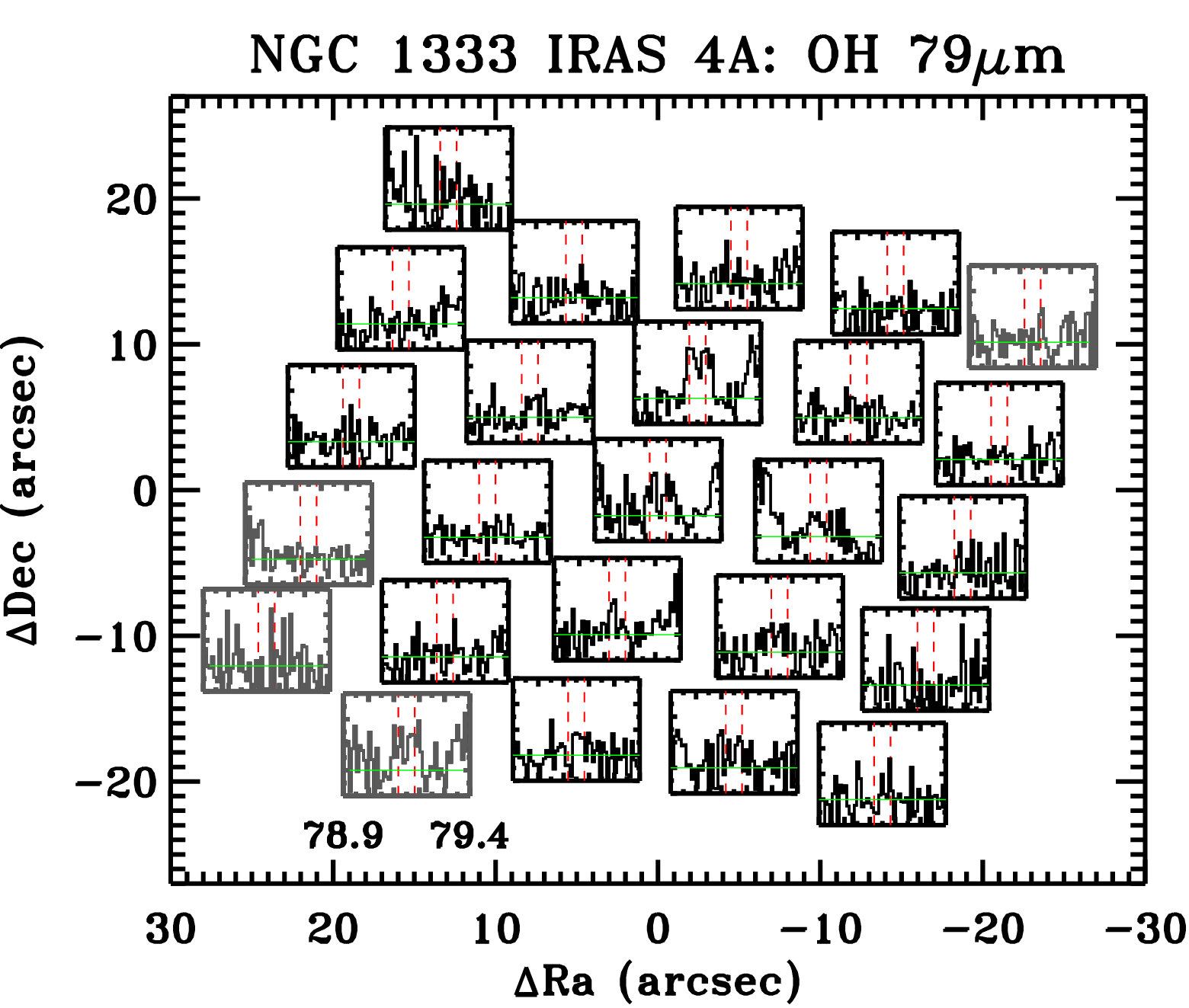}} &
   \resizebox{0.35\hsize}{!}{\includegraphics[angle=0,bb=28 714 651 1280]{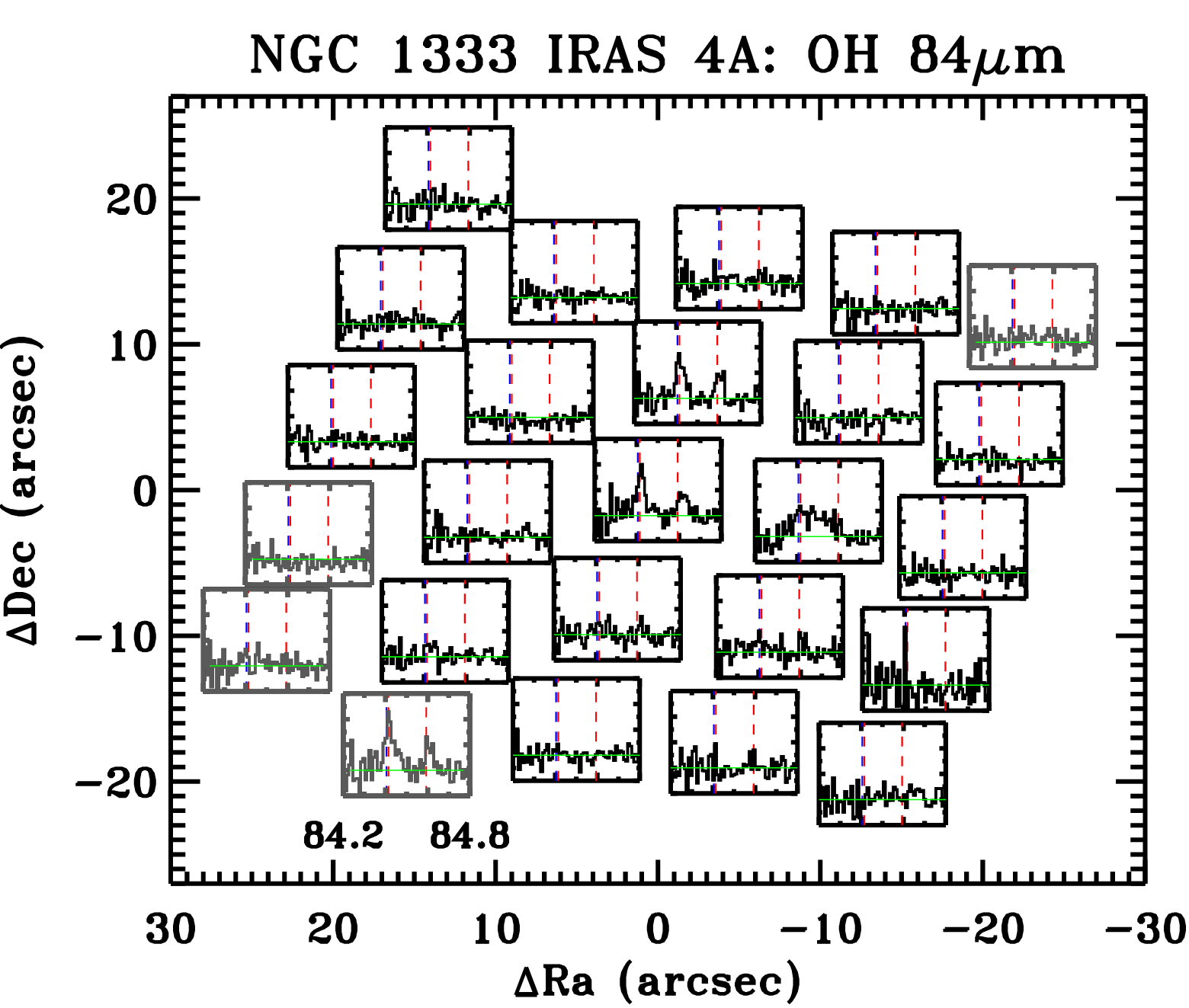}} \\
   & \\
   \resizebox{0.35\hsize}{!}{\includegraphics[angle=0,bb=28 714 651 1280]{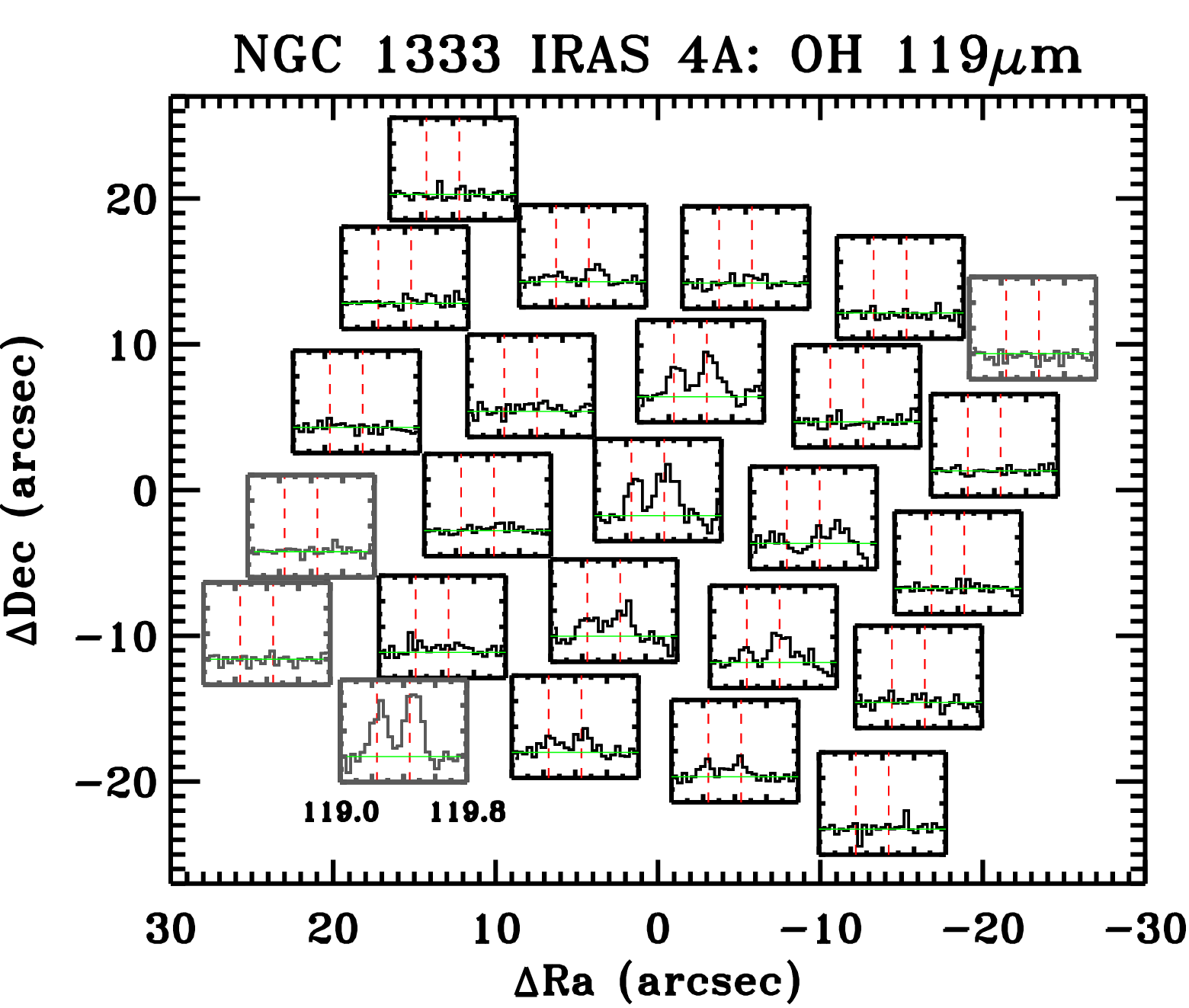}} &
   \resizebox{0.35\hsize}{!}{\includegraphics[angle=0,bb=28 714 651 1280]{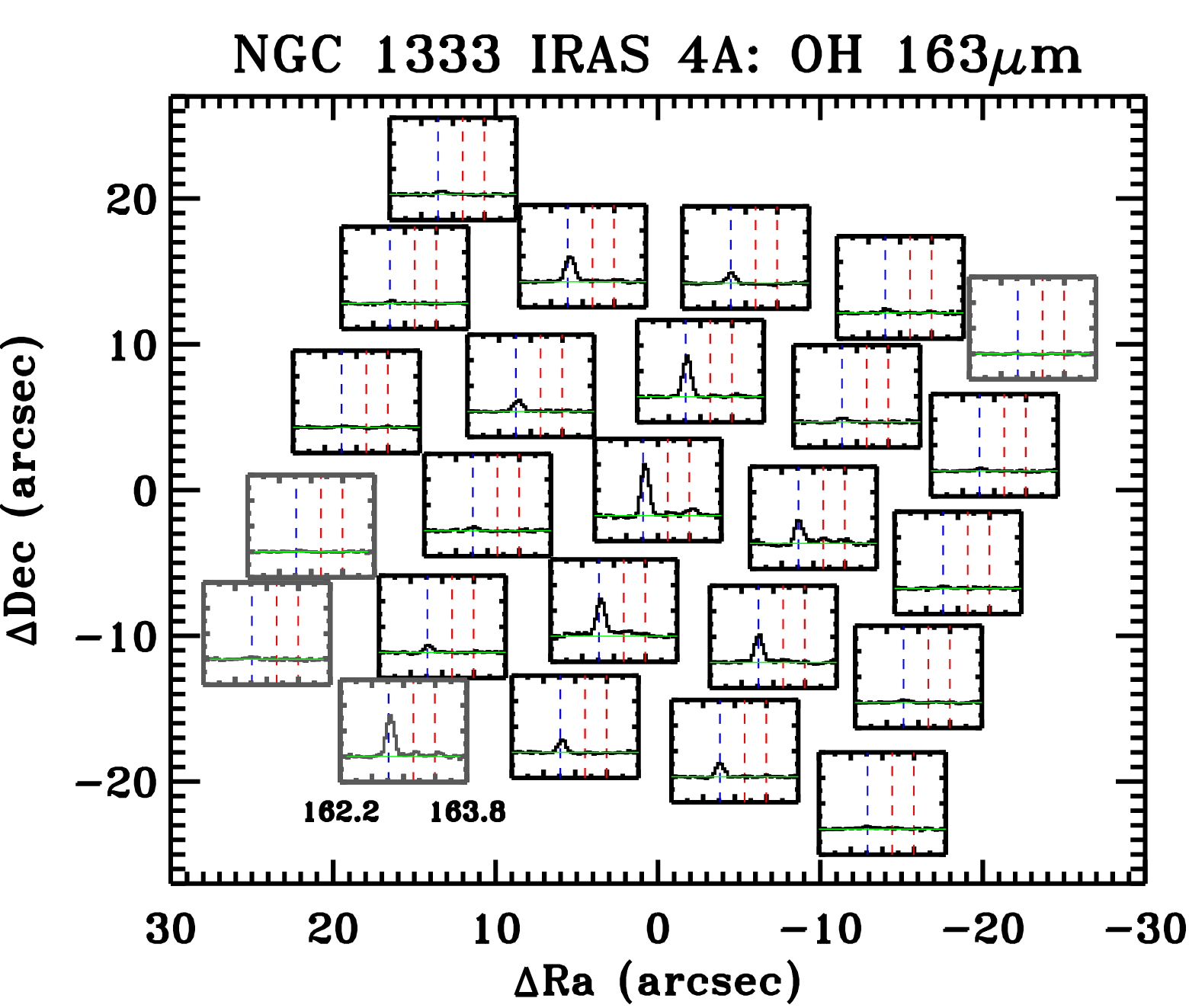}} \\
  \end{tabular}
  \caption{Maps of the OH emission from NGC~1333~IRAS~4A. The red and blue dashed lines indicate the rest wavelengths of the OH and CO transitions, respectively. Spaxels in gray were not included in the flux measurement.}
 \label{fig:iras4amaps}
\end{figure}
\end{landscape}
}

\def\placefigBIRAS4maps{
\begin{landscape}
\begin{figure}
 \centering
  \begin{tabular}{c@{\extracolsep{40pt}}c}
   \resizebox{0.35\hsize}{!}{\includegraphics[angle=0,bb=28 714 651 1280]{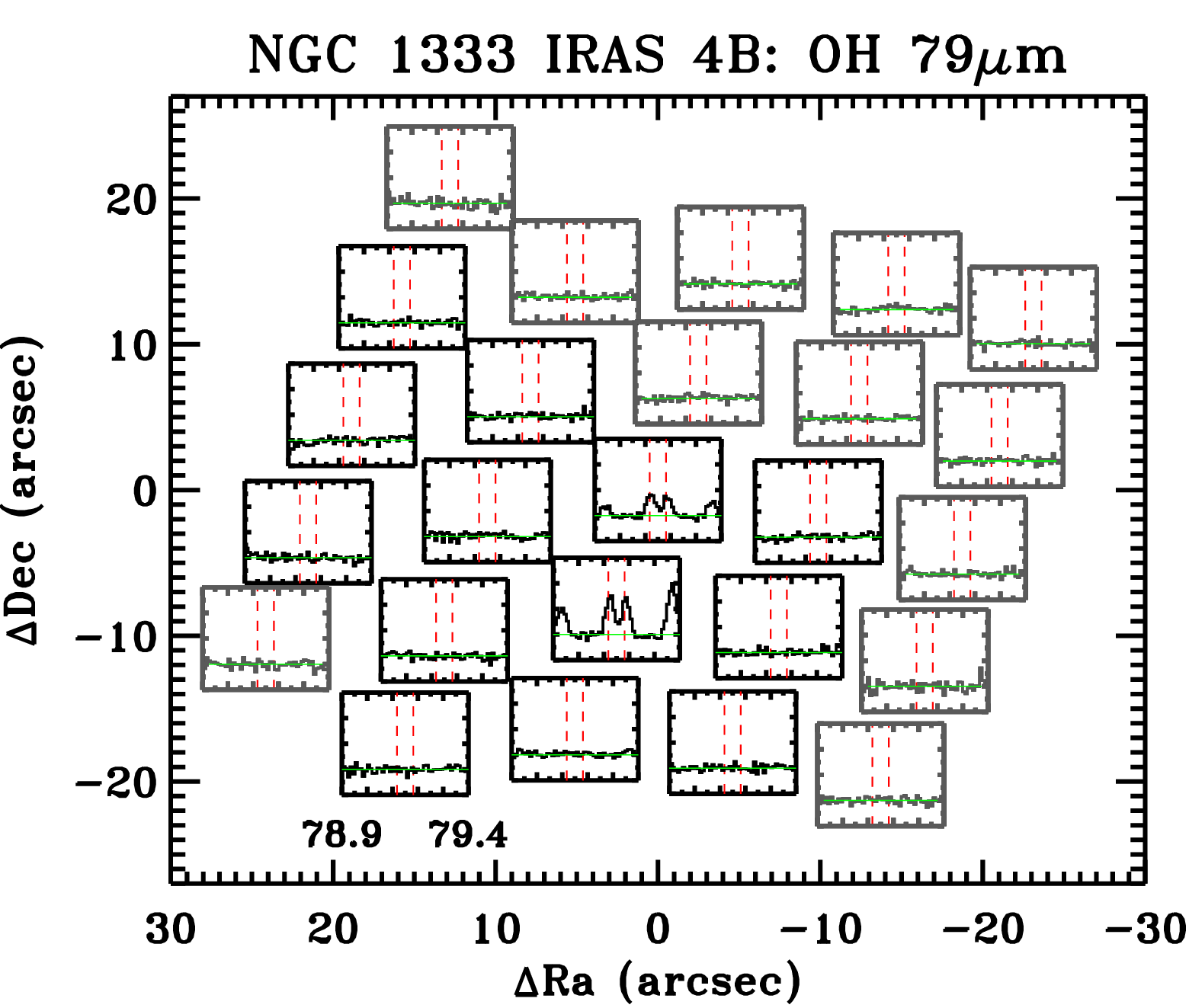}} &
   \resizebox{0.35\hsize}{!}{\includegraphics[angle=0,bb=28 714 651 1280]{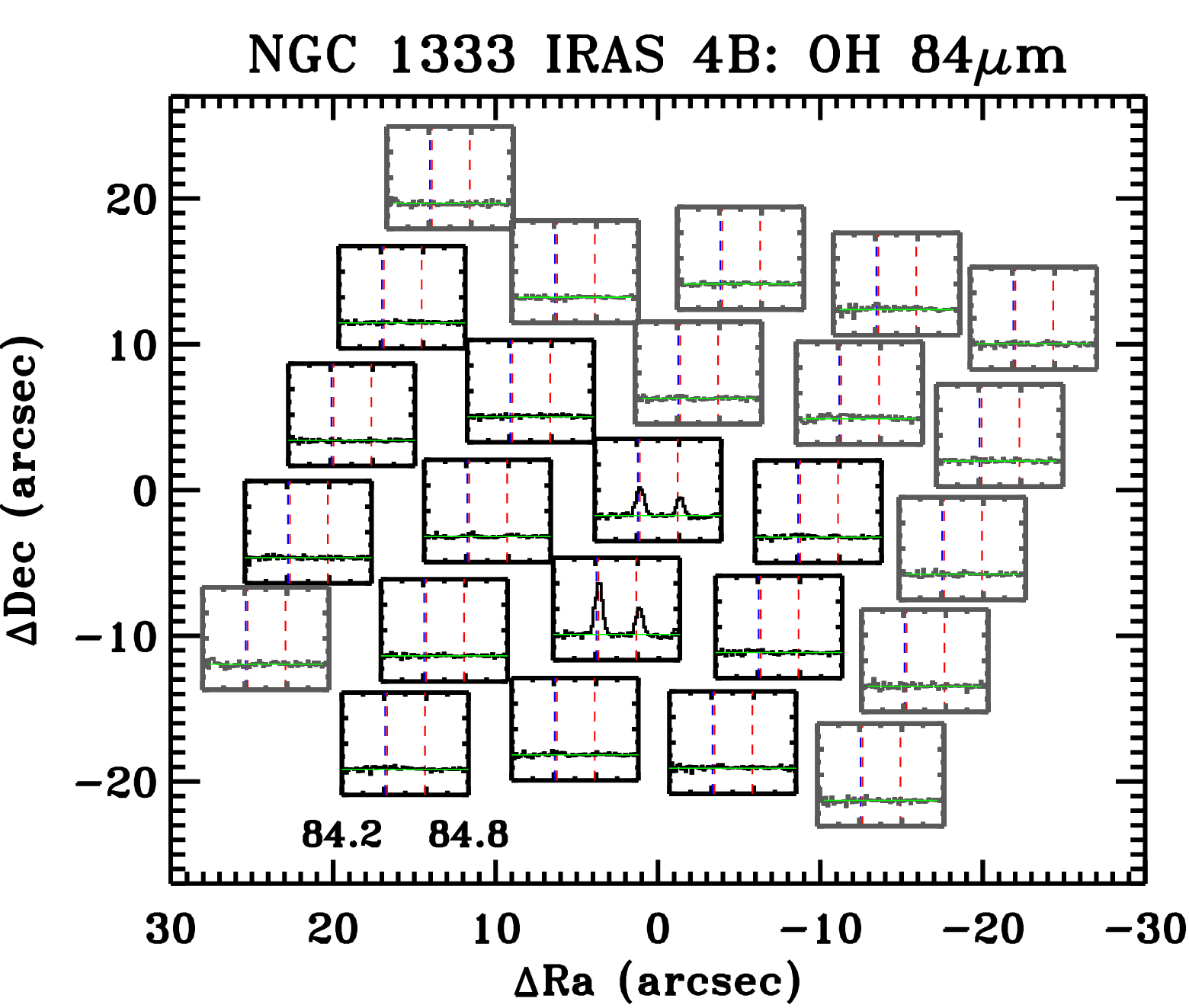}} \\
   & \\
   \resizebox{0.35\hsize}{!}{\includegraphics[angle=0,bb=28 714 651 1280]{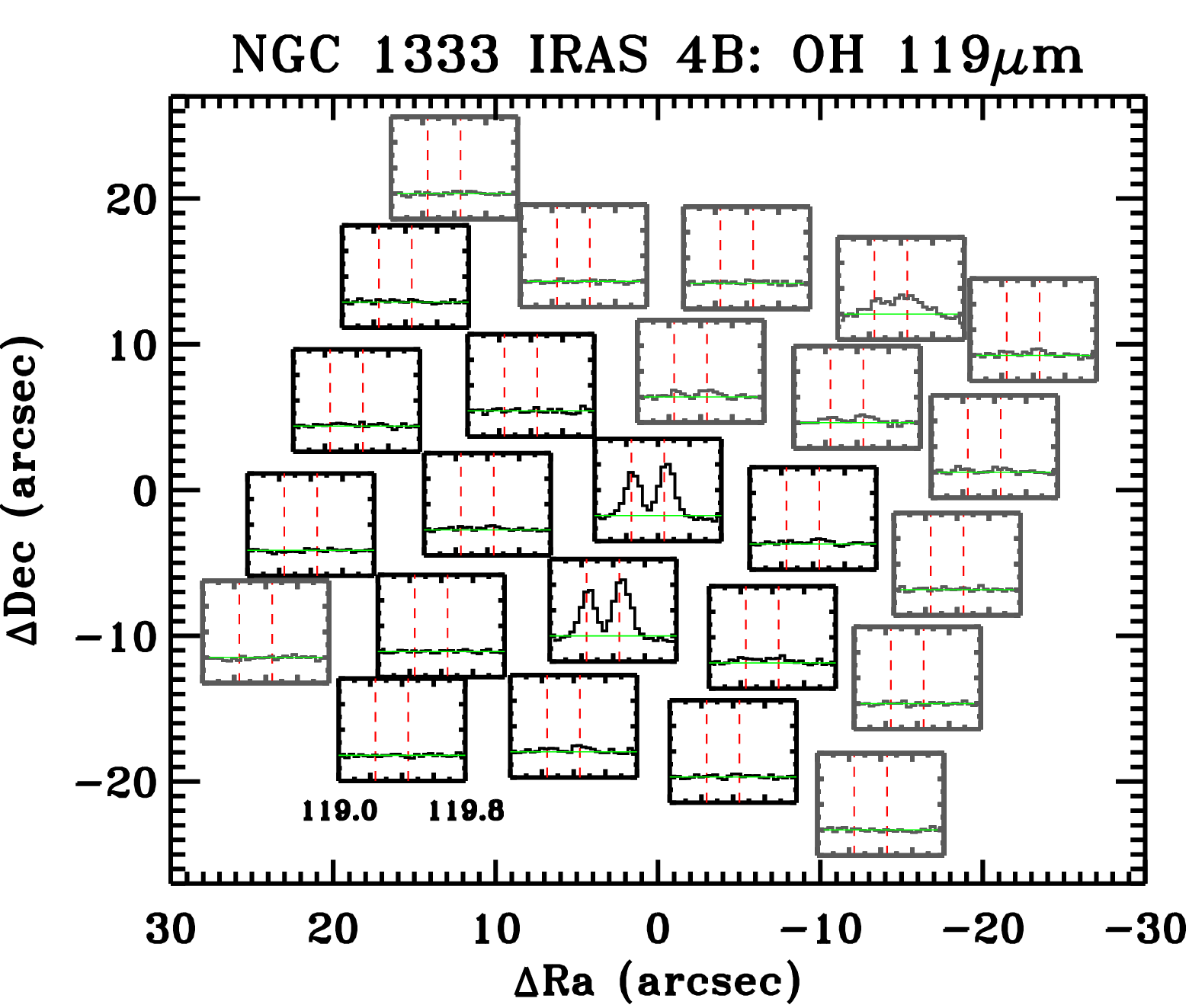}} &
   \resizebox{0.35\hsize}{!}{\includegraphics[angle=0,bb=28 714 651 1280]{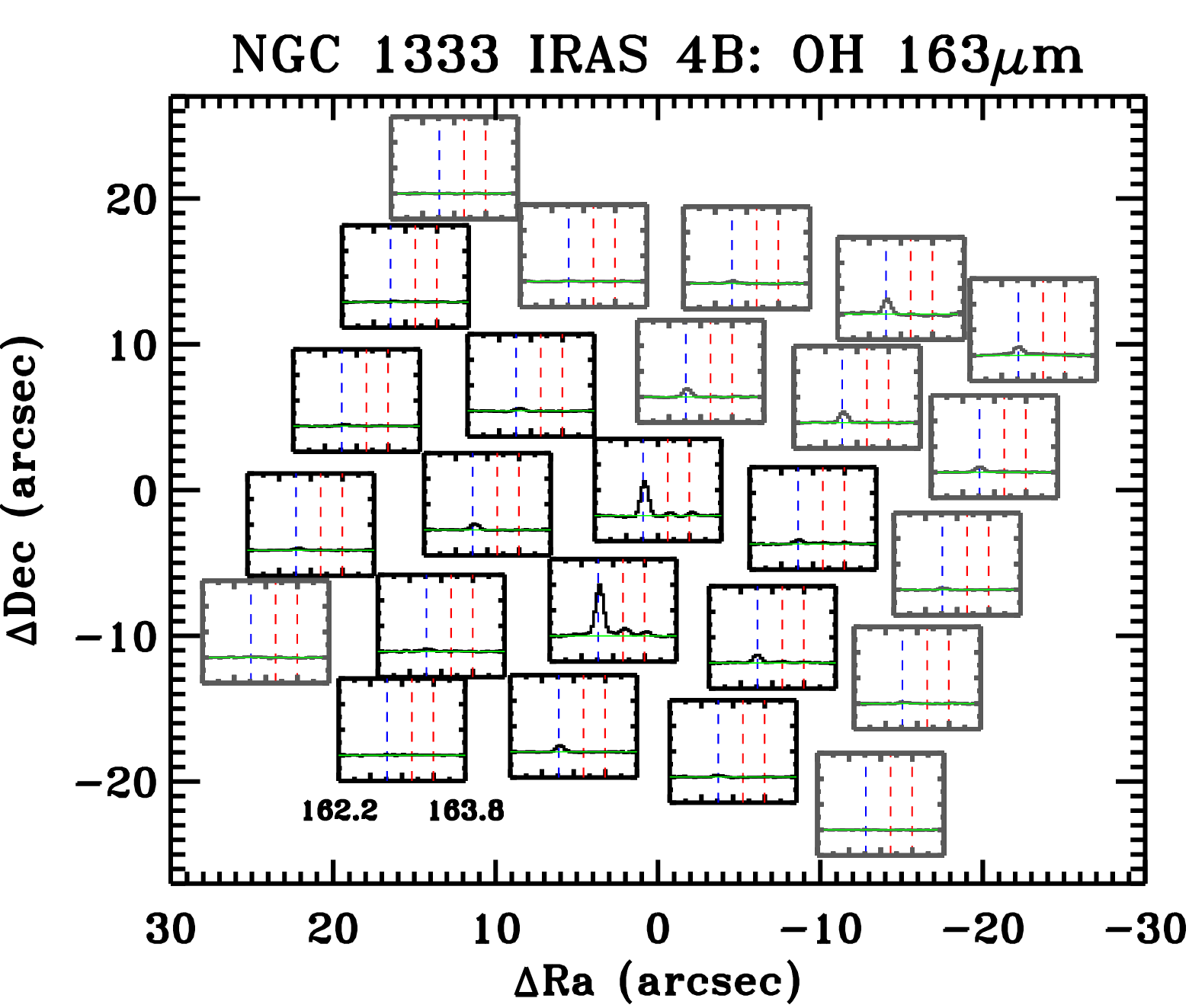}} \\
  \end{tabular}
  \caption{Maps of the OH emission from NGC~1333~IRAS~4B. The red and blue dashed lines indicate the rest wavelengths of the OH and CO transitions, respectively. Spaxels in gray were not included in the flux measurement.}
 \label{fig:iras4bmaps}
\end{figure}
\end{landscape}
}

\def\placefigL1527maps{
\begin{landscape}
\begin{figure}
 \centering
  \begin{tabular}{c@{\extracolsep{40pt}}c}
   \resizebox{0.35\hsize}{!}{\includegraphics[angle=0,bb=28 714 651 1280]{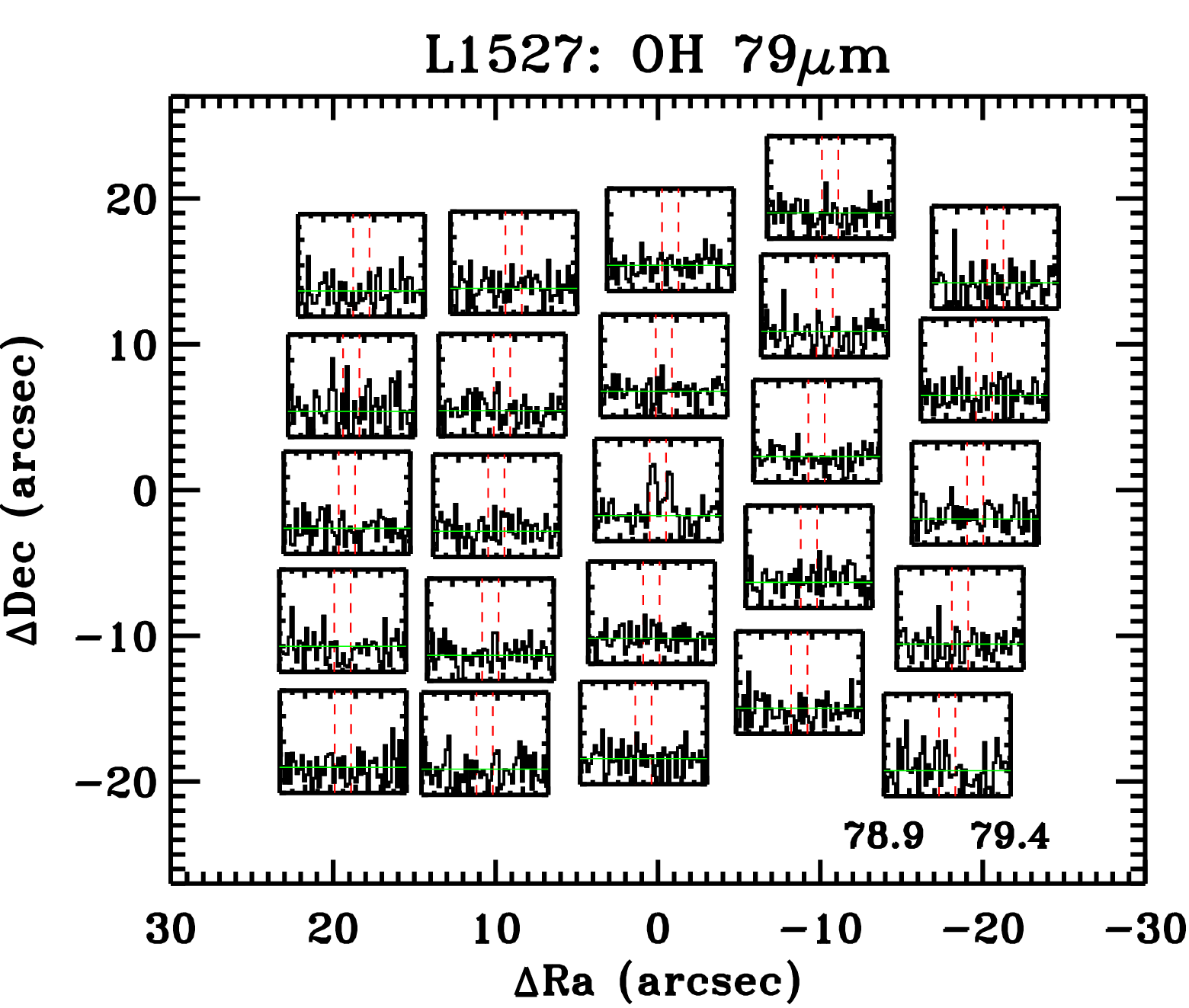}} &
   \resizebox{0.35\hsize}{!}{\includegraphics[angle=0,bb=28 714 651 1280]{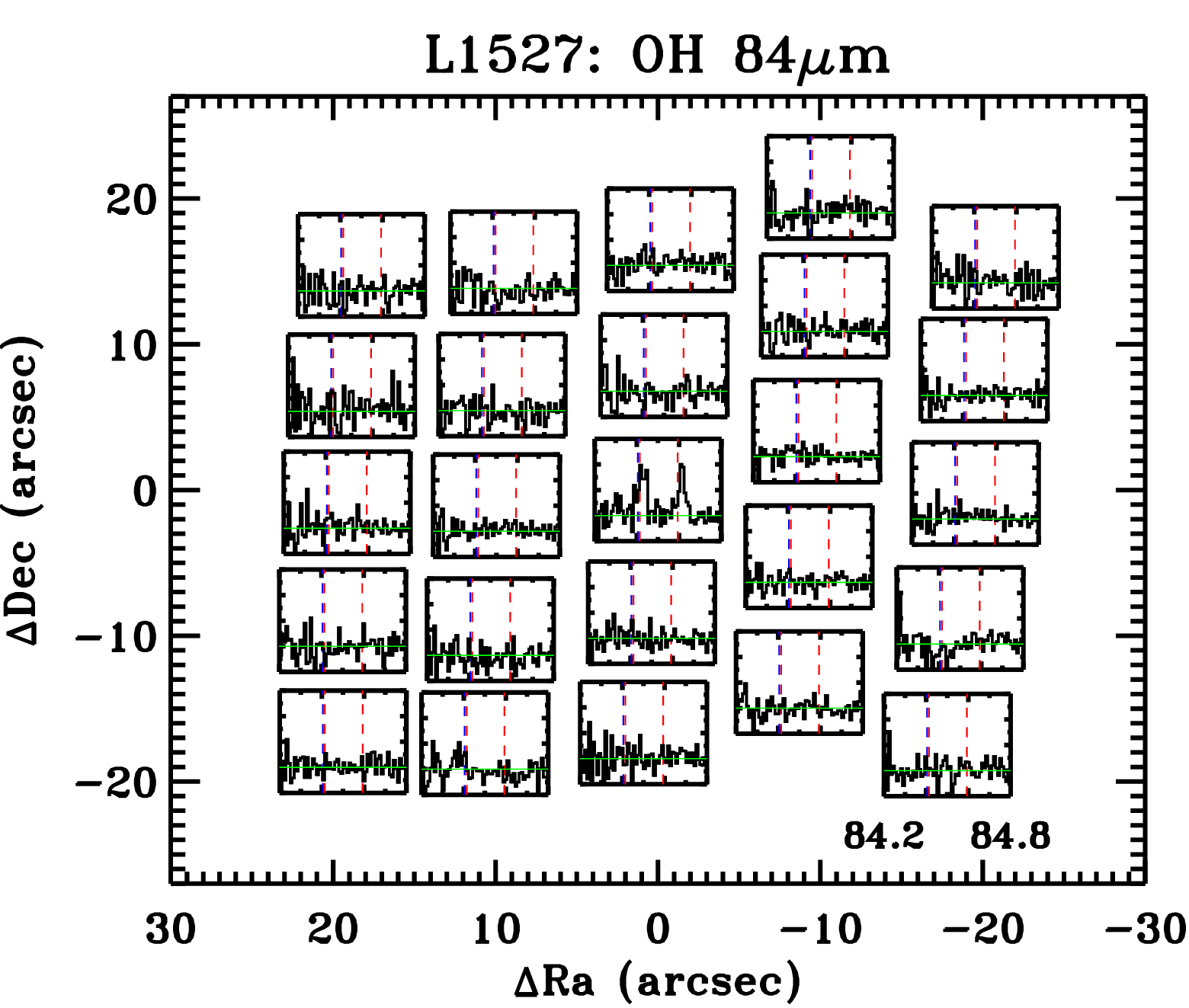}} \\
   & \\
   \resizebox{0.35\hsize}{!}{\includegraphics[angle=0,bb=28 714 651 1280]{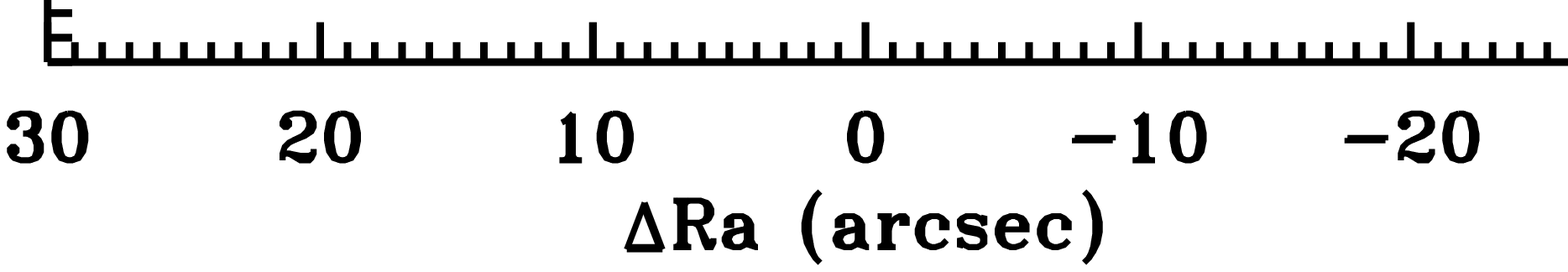}} &
   \resizebox{0.35\hsize}{!}{\includegraphics[angle=0,bb=28 714 651 1280]{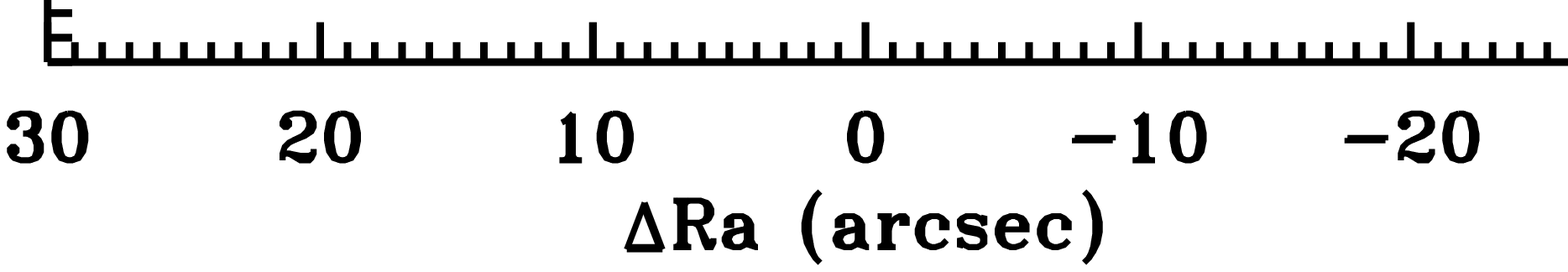}} \\
  \end{tabular}
  \caption{Maps of the OH emission from L~1527. The red and blue dashed lines indicate the rest wavelengths of the OH and CO transitions, respectively.}
 \label{fig:l1527maps}
\end{figure}
\end{landscape}
}

\def\placefigureced110irs4maps{
\begin{landscape}
\begin{figure}
 \centering
  \begin{tabular}{c@{\extracolsep{40pt}}c}
   \resizebox{0.35\hsize}{!}{\includegraphics[angle=0,bb=28 714 651 1280]{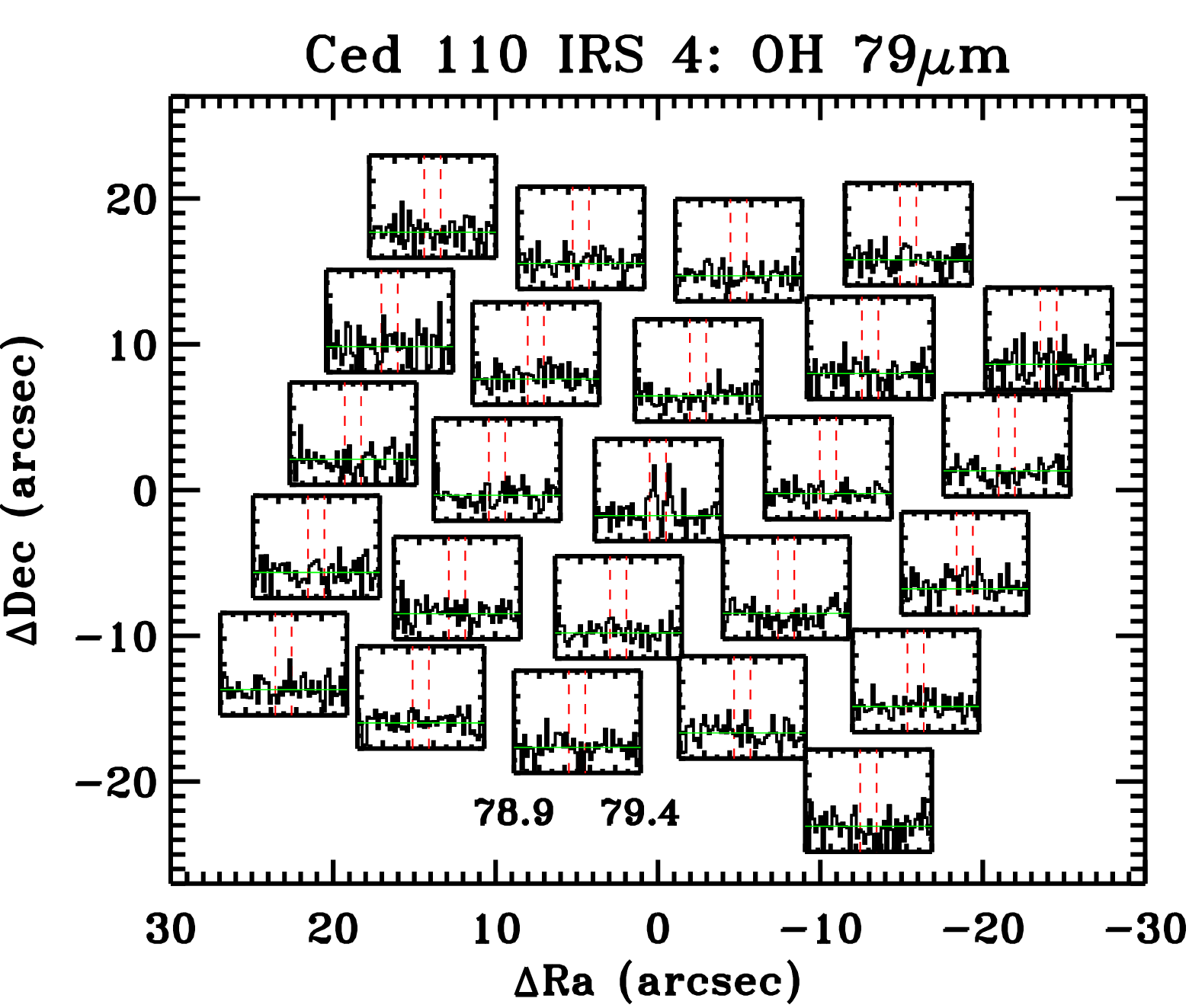}} &
   \resizebox{0.35\hsize}{!}{\includegraphics[angle=0,bb=28 714 651 1280]{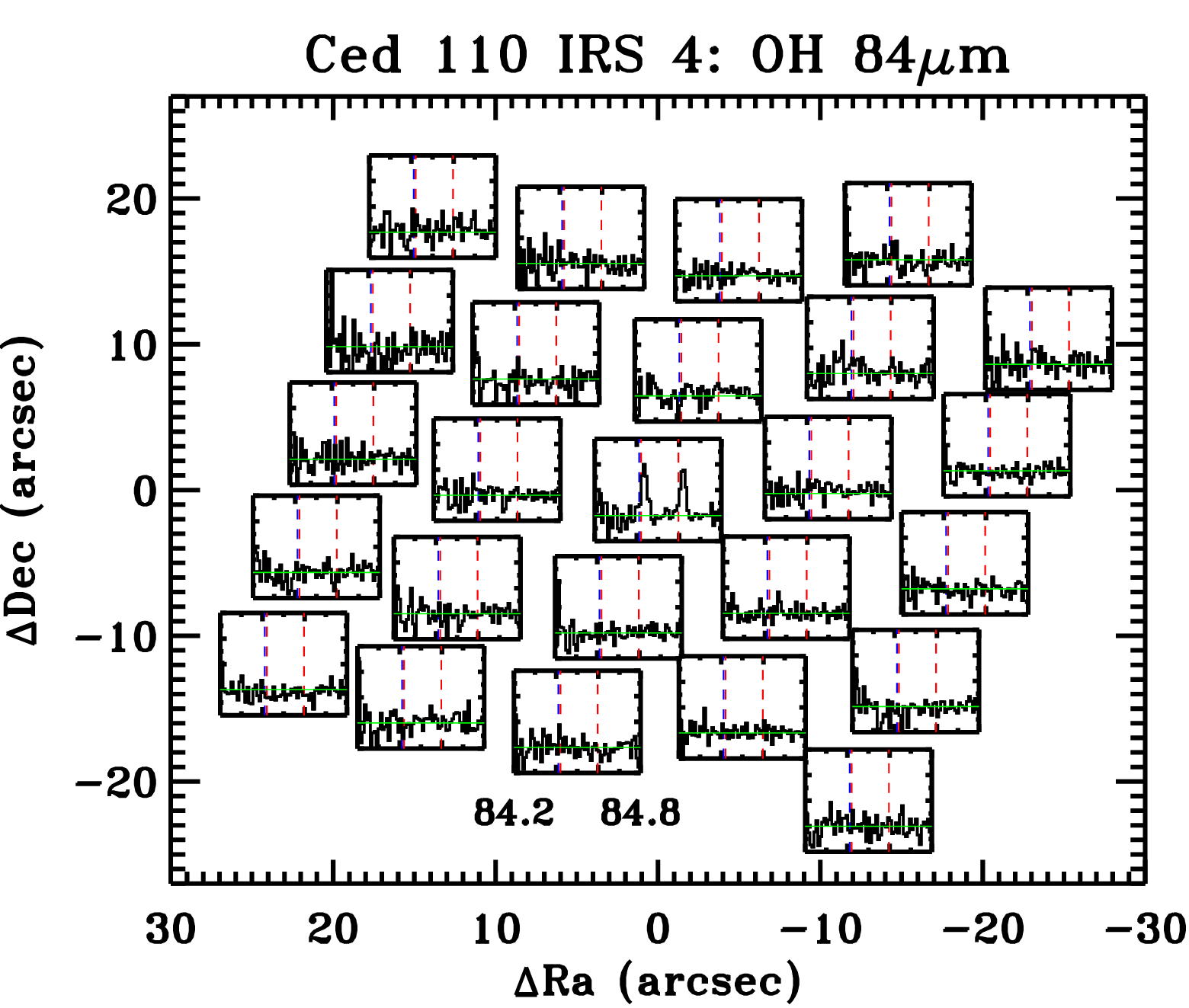}} \\
   & \\
   \resizebox{0.35\hsize}{!}{\includegraphics[angle=0,bb=28 714 651 1280]{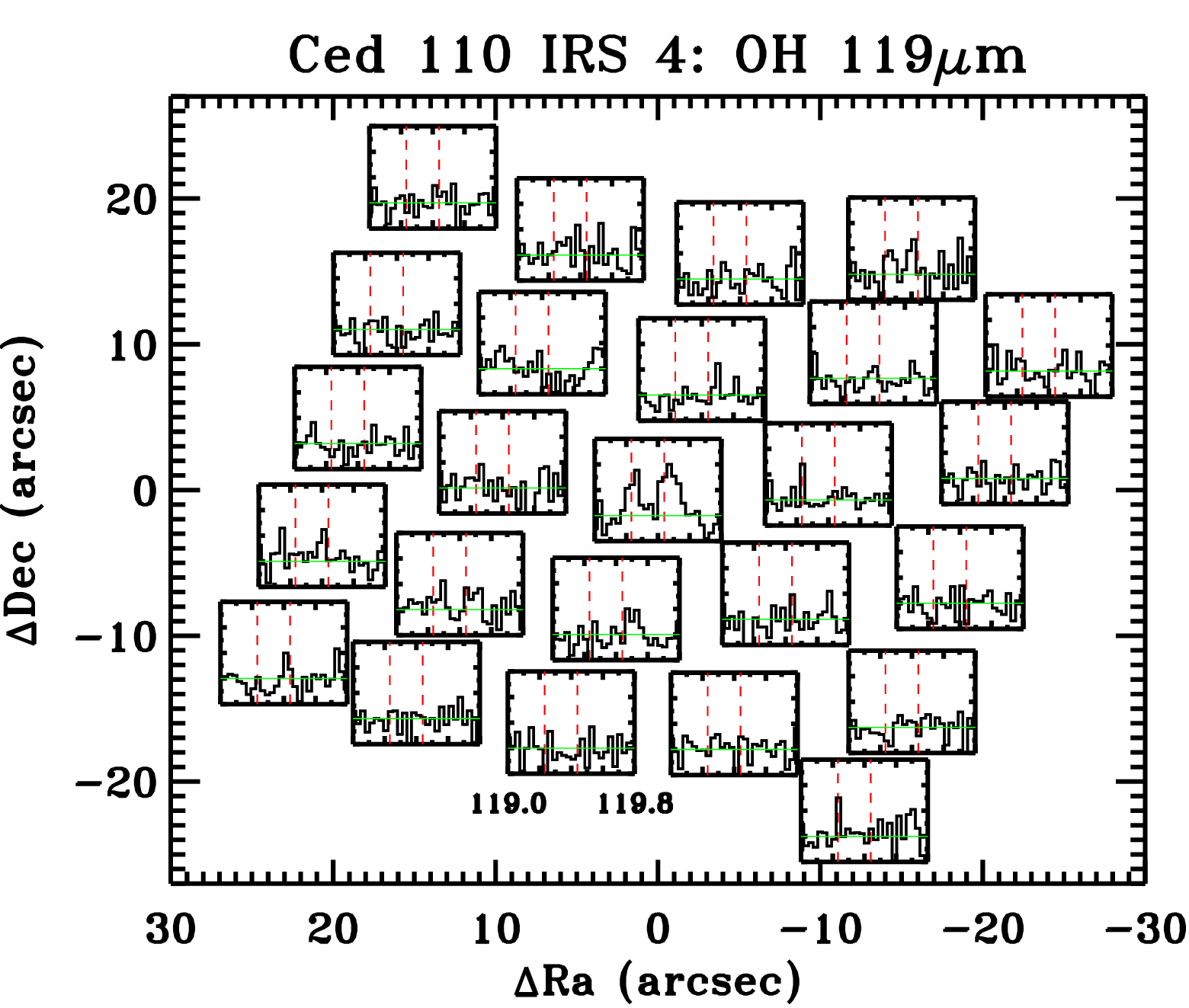}} &
   \resizebox{0.35\hsize}{!}{\includegraphics[angle=0,bb=28 714 651 1280]{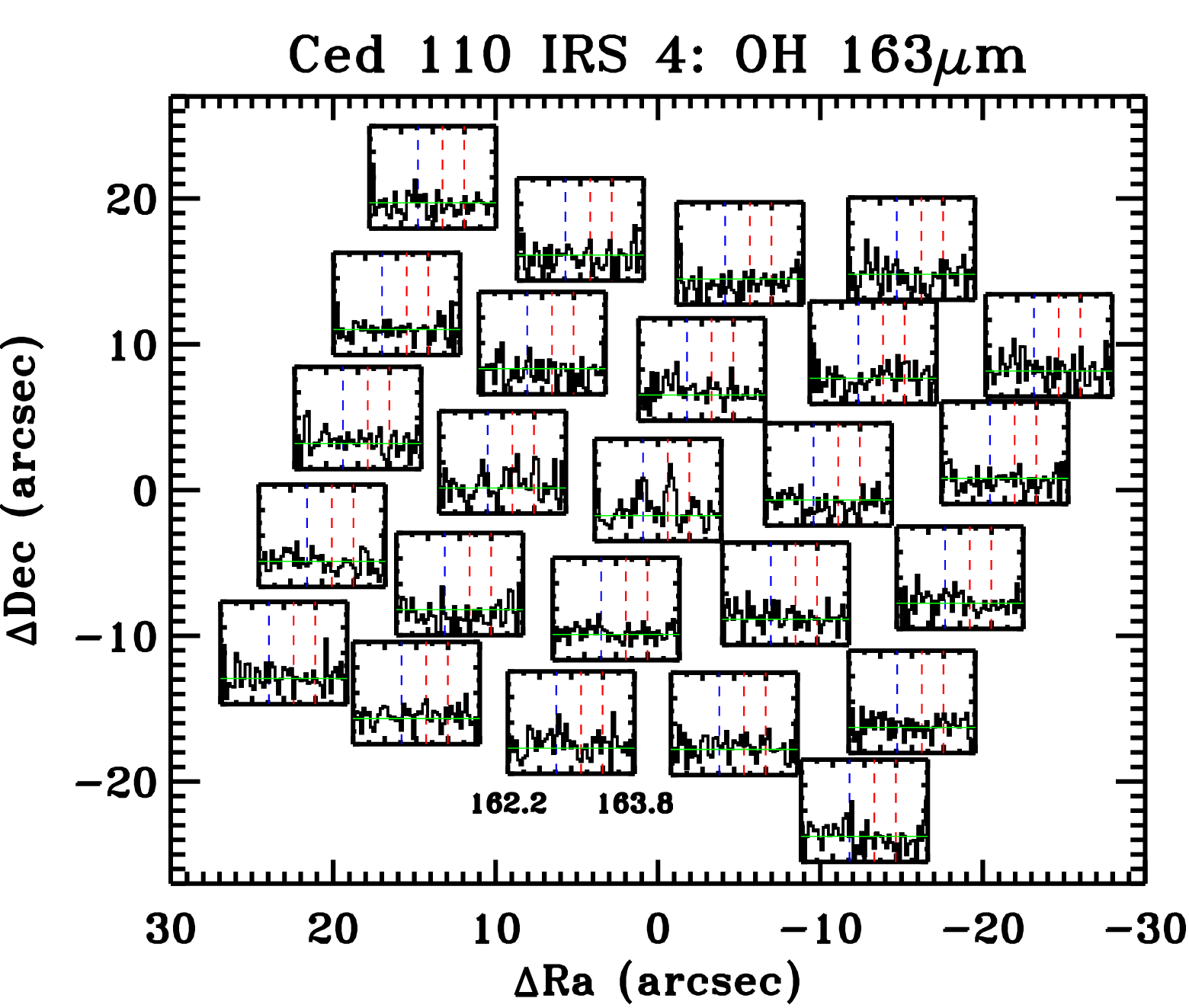}} \\
  \end{tabular}
  \caption{Maps of the OH emission from Ced~110~IRS~4. The red and blue dashed lines indicate the rest wavelengths of the OH and CO transitions, respectively.}
 \label{fig:ced110irs4maps}
\end{figure}
\end{landscape}
}

\def\placefigureiras15398maps{
\begin{landscape}
\begin{figure}
 \centering
  \begin{tabular}{c@{\extracolsep{40pt}}c}
   \resizebox{0.35\hsize}{!}{\includegraphics[angle=0,bb=28 714 651 1280]{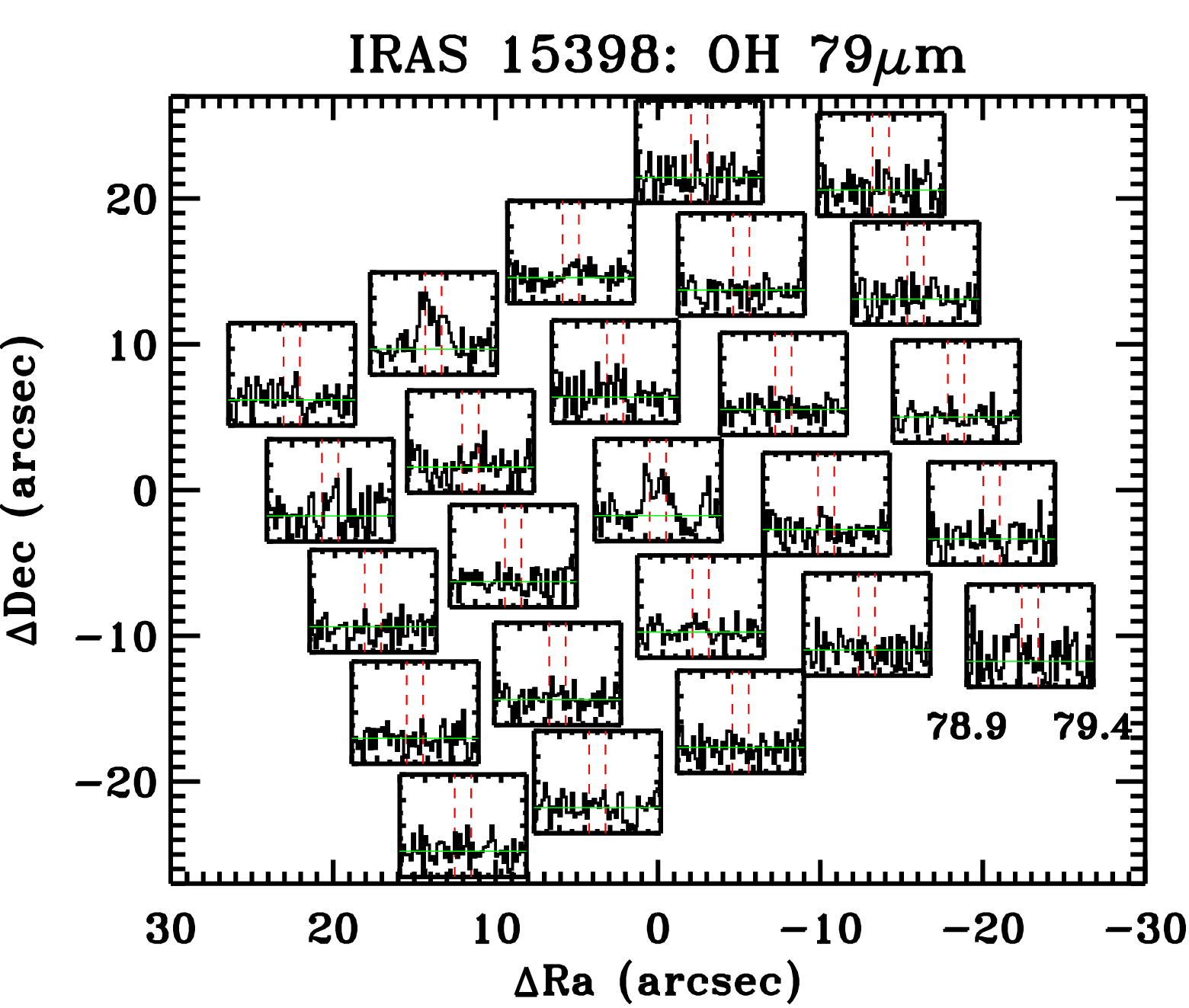}} &
   \resizebox{0.35\hsize}{!}{\includegraphics[angle=0,bb=28 714 651 1280]{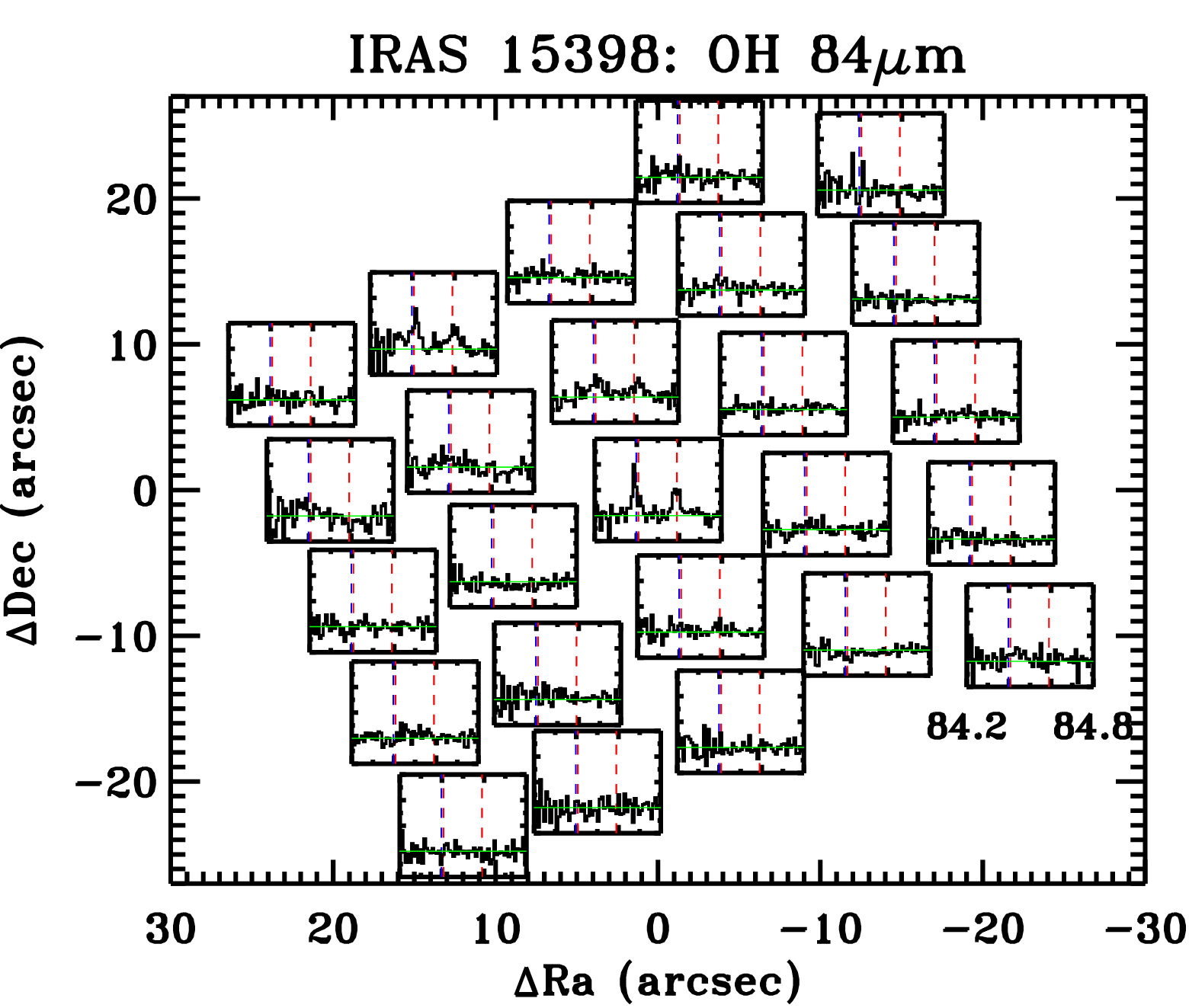}} \\
   & \\
   \resizebox{0.35\hsize}{!}{\includegraphics[angle=0,bb=28 714 651 1280]{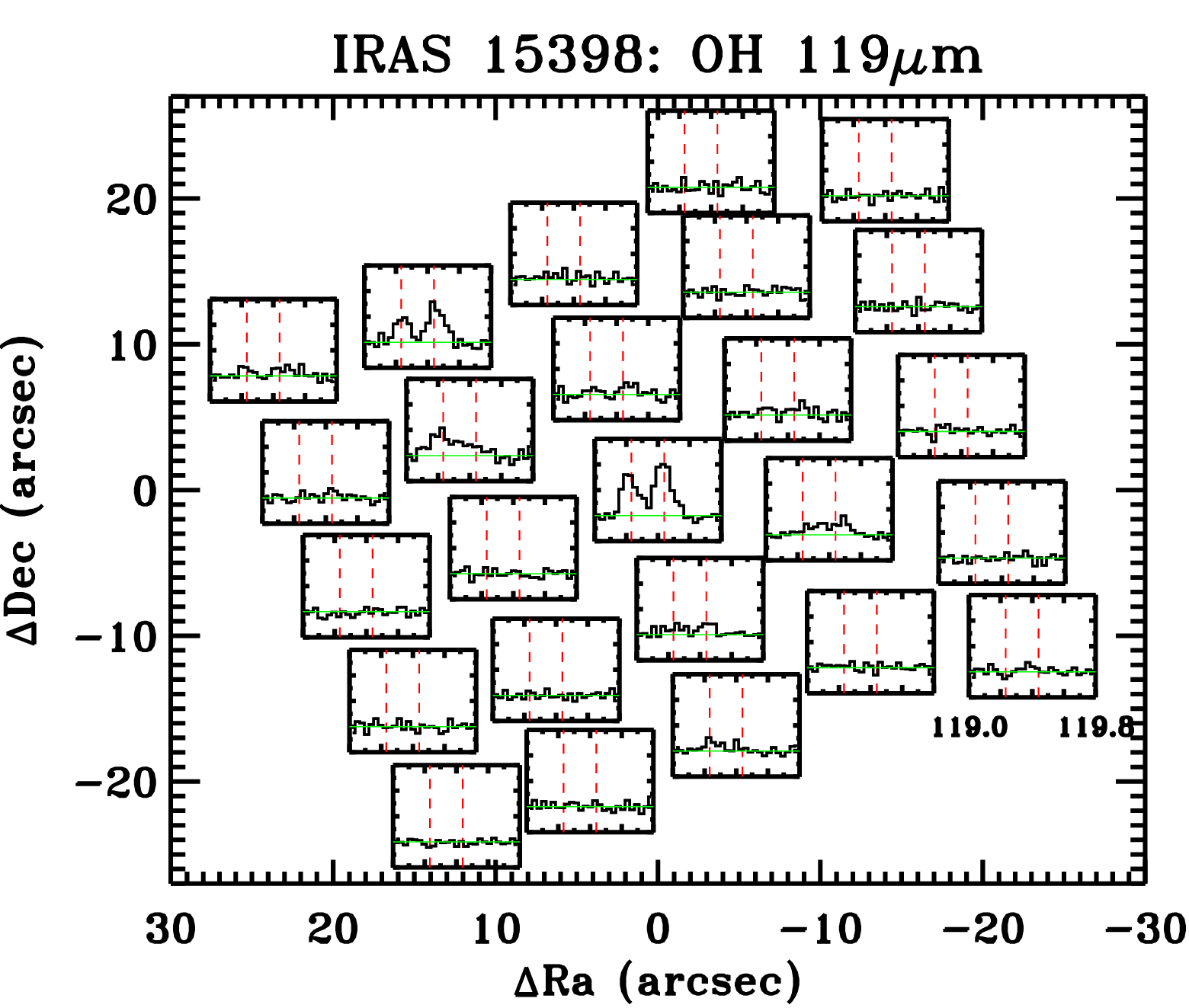}} &
   \resizebox{0.35\hsize}{!}{\includegraphics[angle=0,bb=28 714 651 1280]{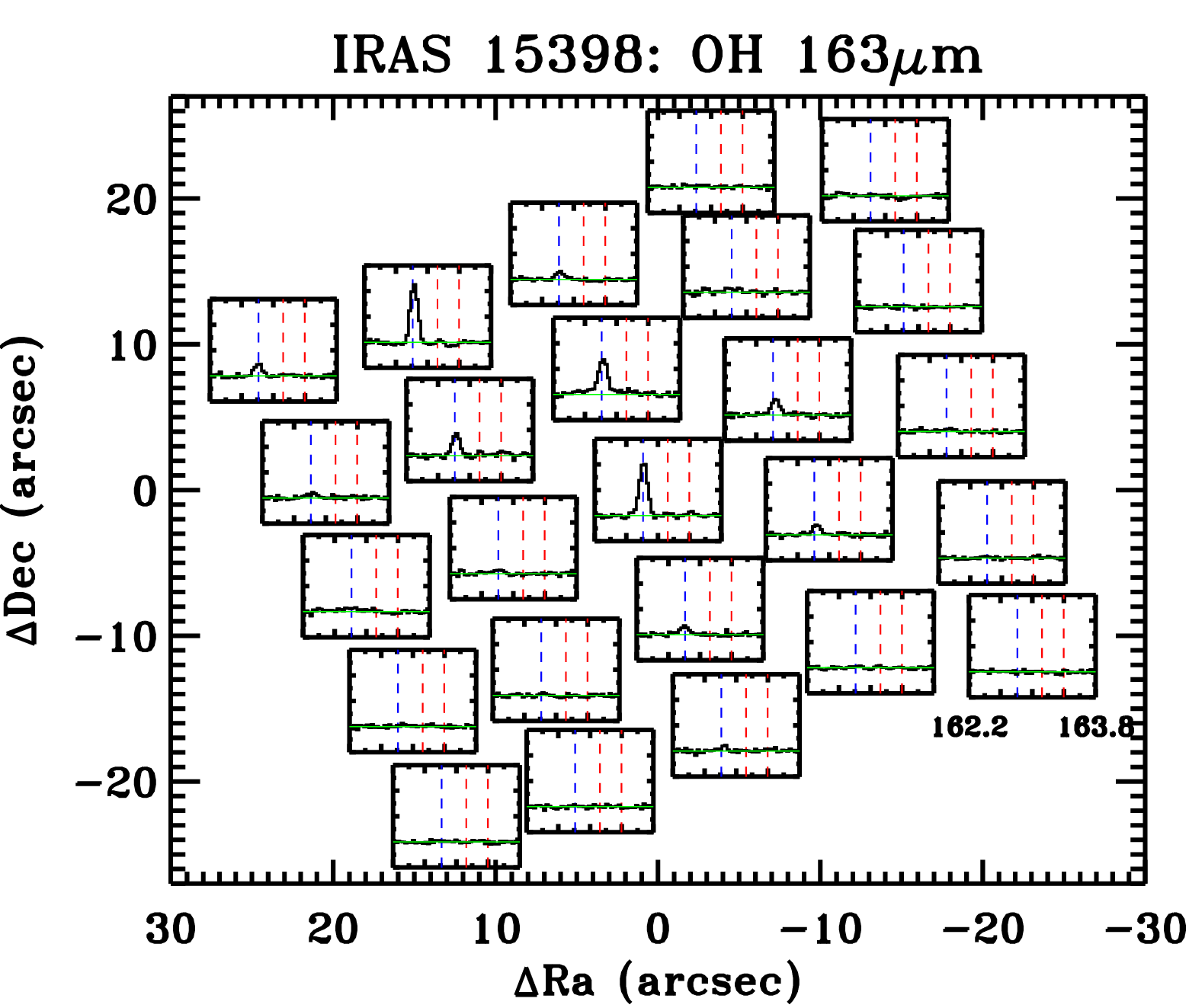}} \\
  \end{tabular}
  \caption{Maps of the OH emission from IRAS~15398. The red and blue dashed lines indicate the rest wavelengths of the OH and CO transitions, respectively.}
 \label{fig:iras15398maps}
\end{figure}
\end{landscape}
}

\def\placefigurel483maps{
\begin{landscape}
\begin{figure}
 \centering
  \begin{tabular}{c@{\extracolsep{40pt}}c}
   \resizebox{0.35\hsize}{!}{\includegraphics[angle=0,bb=28 714 651 1280]{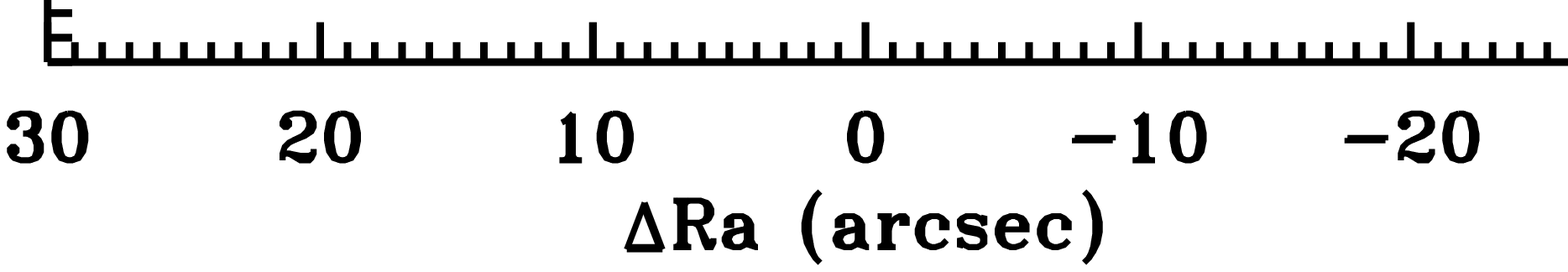}} &
   \resizebox{0.35\hsize}{!}{\includegraphics[angle=0,bb=28 714 651 1280]{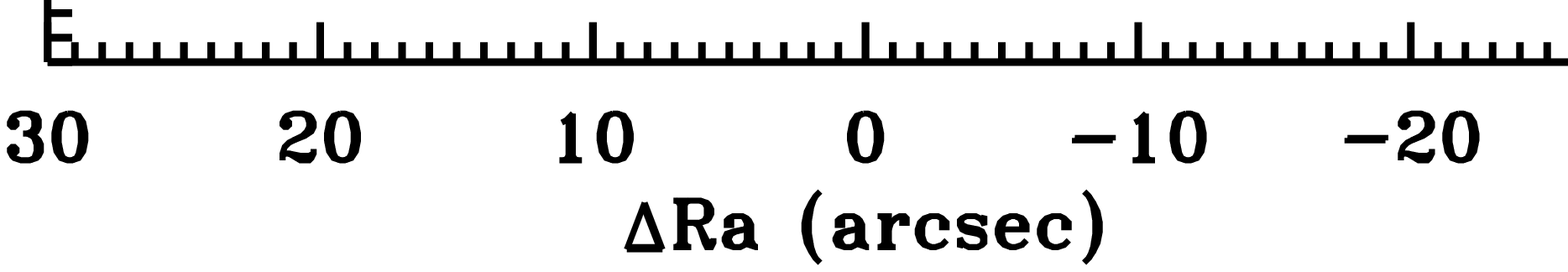}} \\
   & \\
   \resizebox{0.35\hsize}{!}{\includegraphics[angle=0,bb=28 714 651 1280]{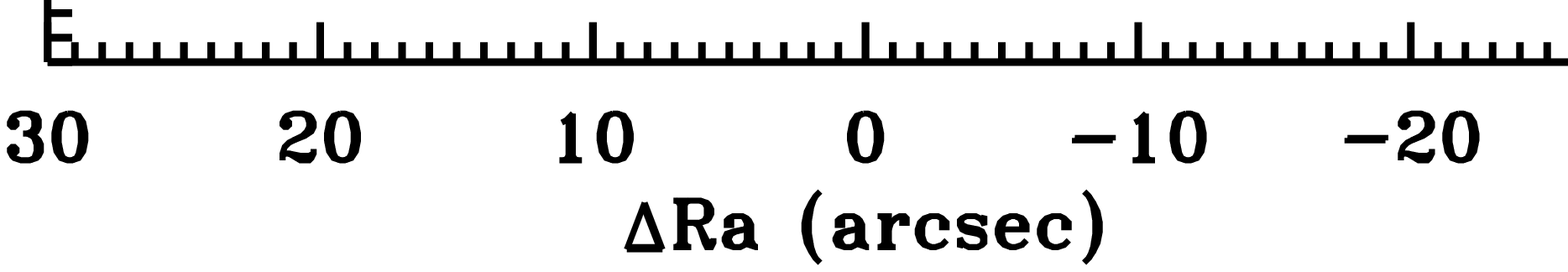}} &
   \resizebox{0.35\hsize}{!}{\includegraphics[angle=0,bb=28 714 651 1280]{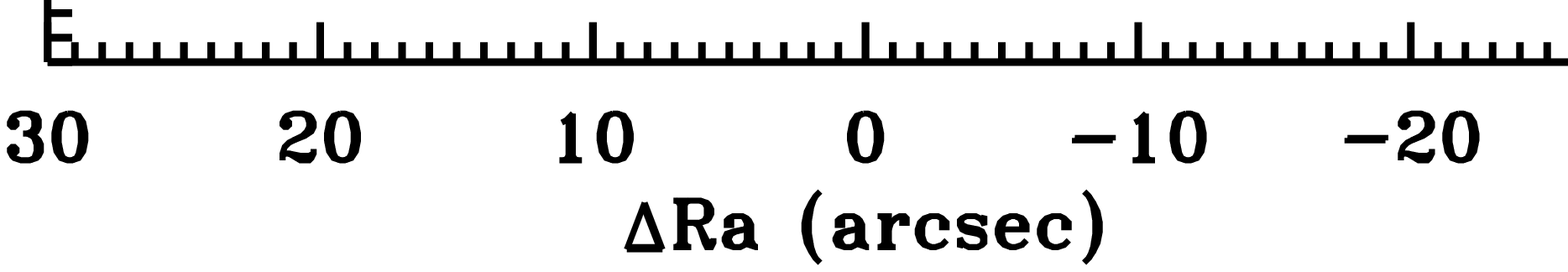}} \\
  \end{tabular}
  \caption{Maps of the OH emission from L~483. The red and blue dashed lines indicate the rest wavelengths of the OH and CO transitions, respectively.}
 \label{fig:l483maps}
\end{figure}
\end{landscape}
}

\def\placefiguresMM1maps{
\begin{landscape}
\begin{figure}
 \centering
  \begin{tabular}{c@{\extracolsep{40pt}}c}
   \resizebox{0.35\hsize}{!}{\includegraphics[angle=0,bb=28 714 651 1280]{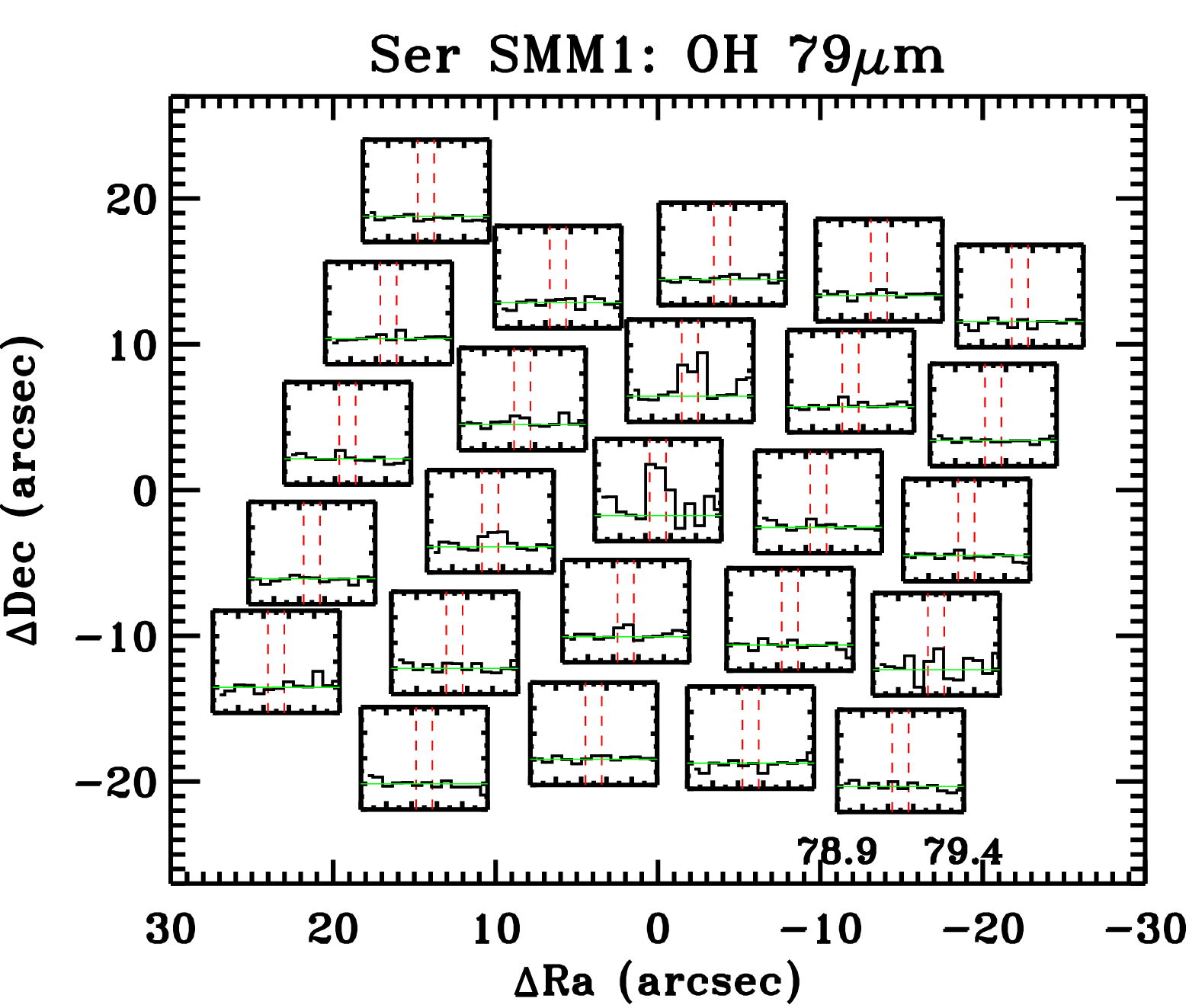}} &
   \resizebox{0.35\hsize}{!}{\includegraphics[angle=0,bb=28 714 651 1280]{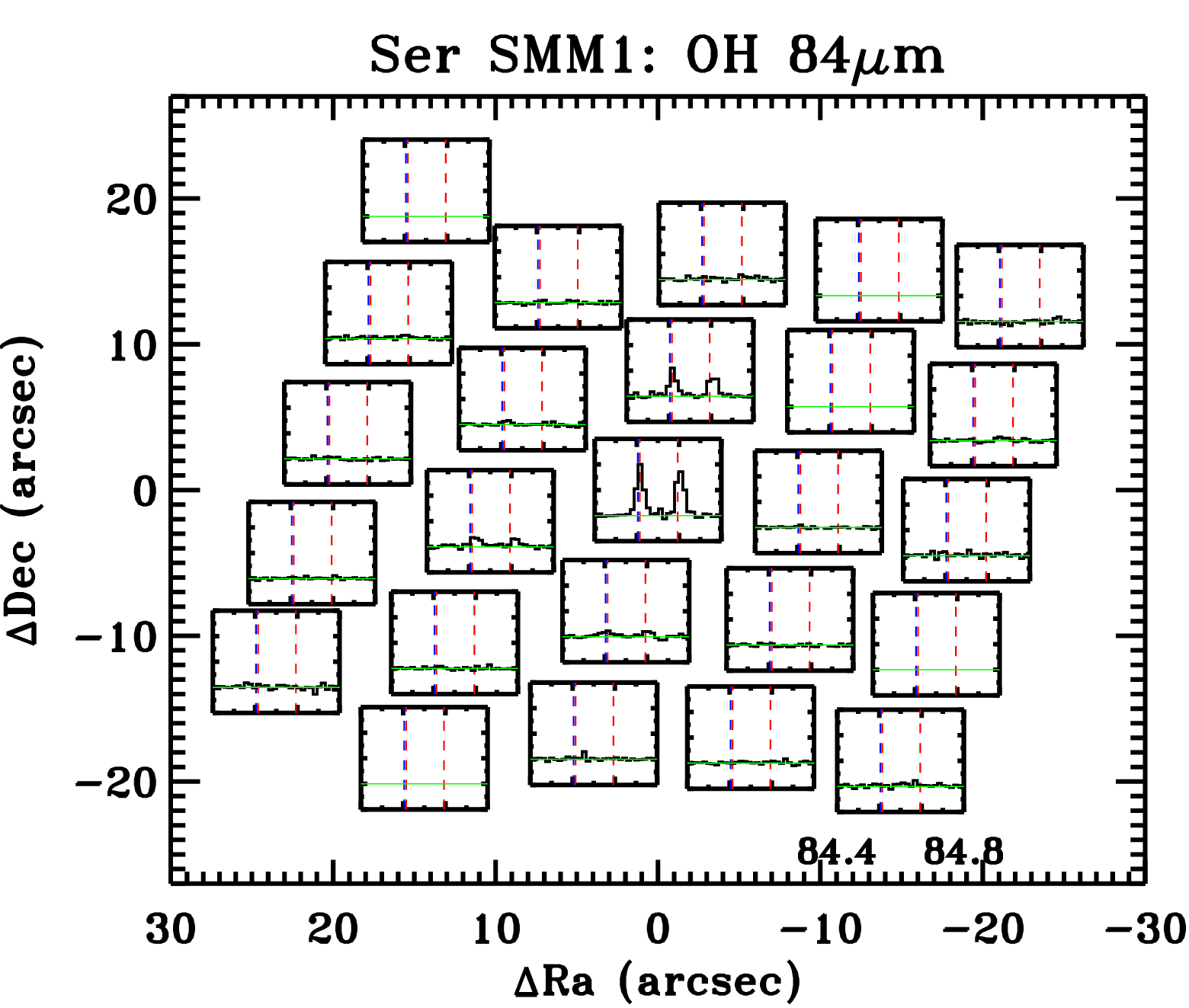}} \\
   & \\
   \resizebox{0.35\hsize}{!}{\includegraphics[angle=0,bb=28 714 651 1280]{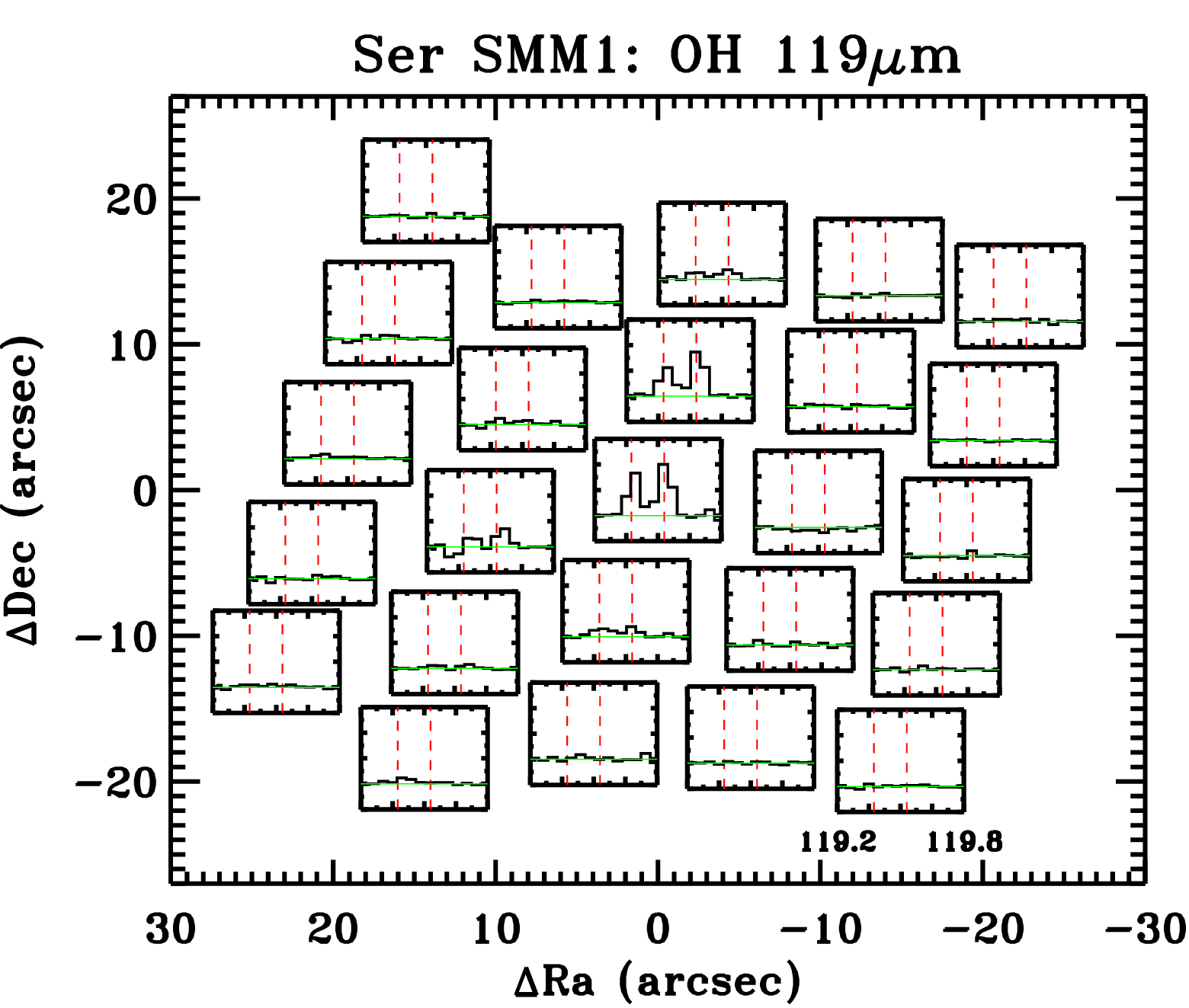}} &
   \resizebox{0.35\hsize}{!}{\includegraphics[angle=0,bb=28 714 651 1280]{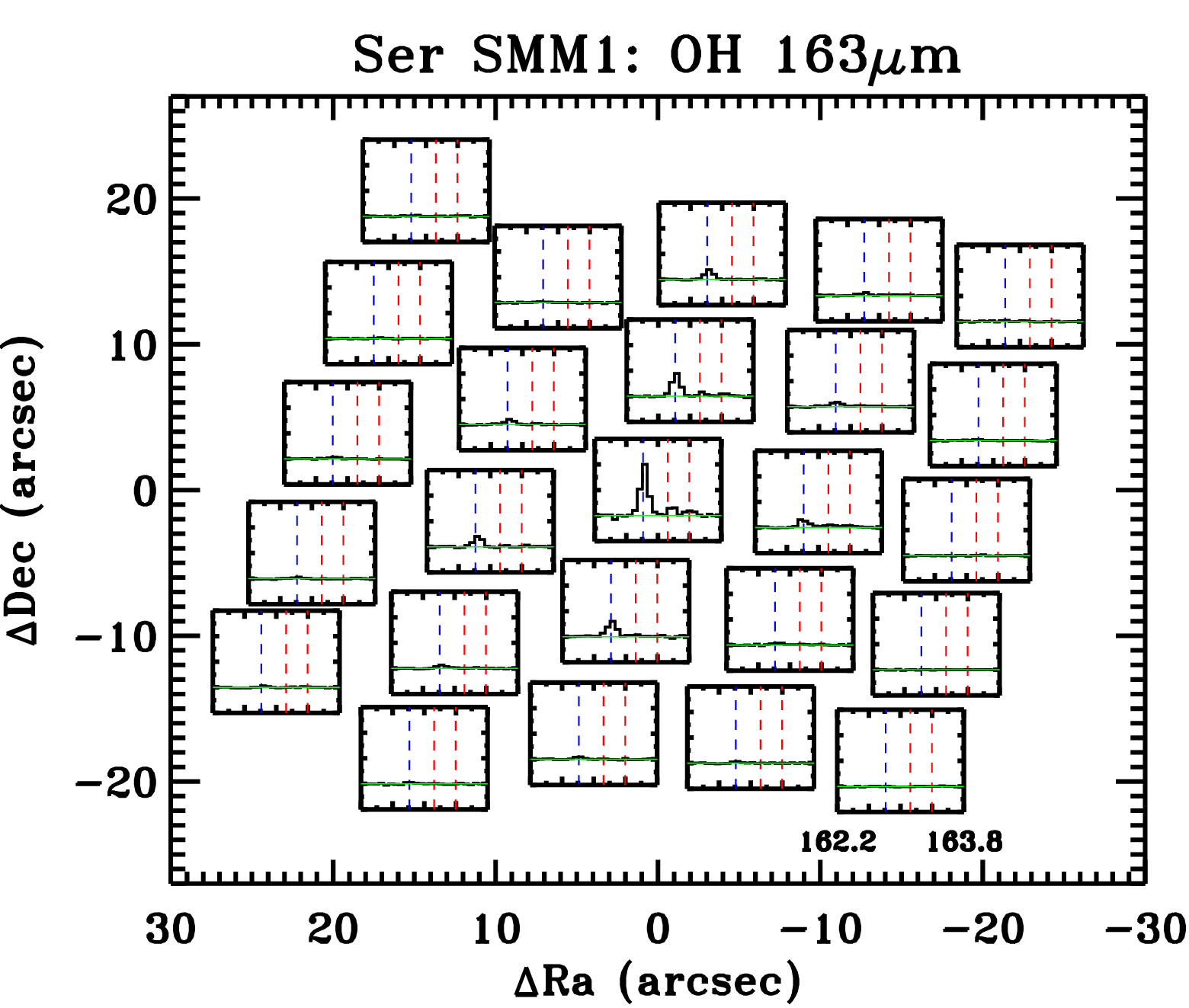}} \\
  \end{tabular}
  \caption{Maps of the OH emission from Ser~SMM1. The red and blue dashed lines indicate the rest wavelengths of the OH and CO transitions, respectively. See also \citet{Goicoechea12}.}
 \label{fig:smm1maps}
\end{figure}
\end{landscape}
}

\def\placefigureSMM3maps{
\begin{landscape}
\begin{figure}
 \centering
  \begin{tabular}{c@{\extracolsep{40pt}}c}
   \resizebox{0.35\hsize}{!}{\includegraphics[angle=0,bb=28 714 651 1280]{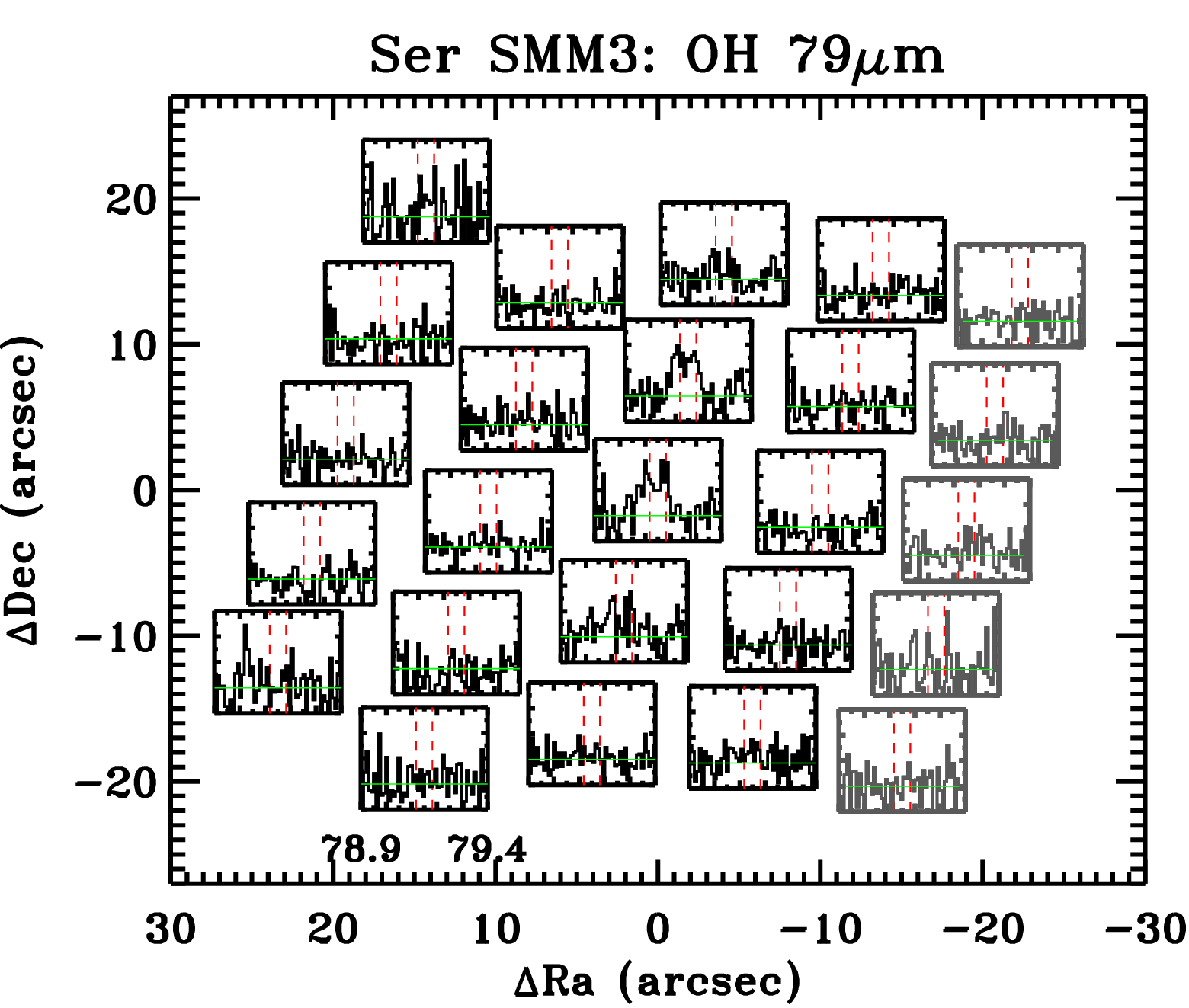}} &
   \resizebox{0.35\hsize}{!}{\includegraphics[angle=0,bb=28 714 651 1280]{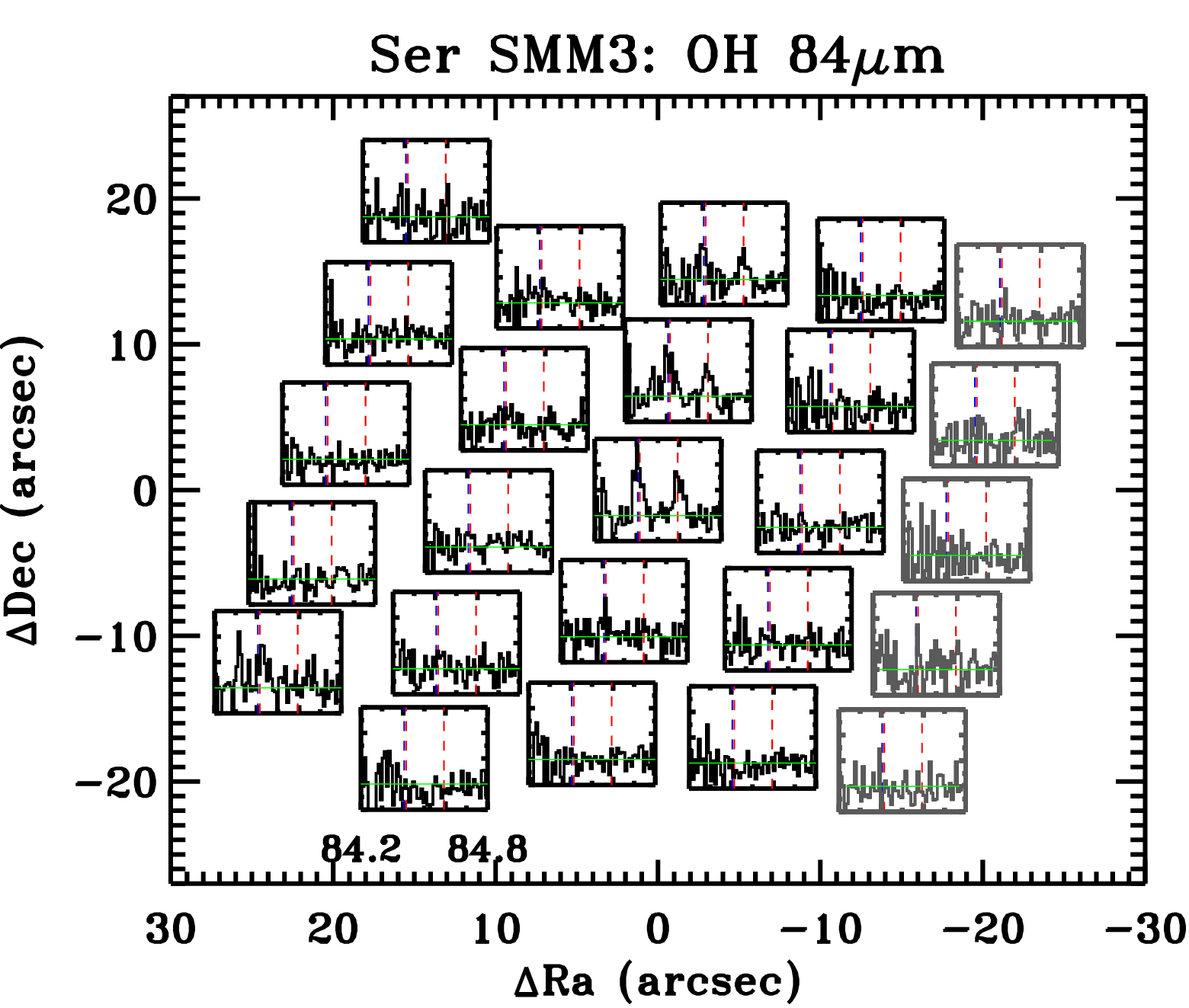}} \\
   & \\
   \resizebox{0.35\hsize}{!}{\includegraphics[angle=0,bb=28 714 651 1280]{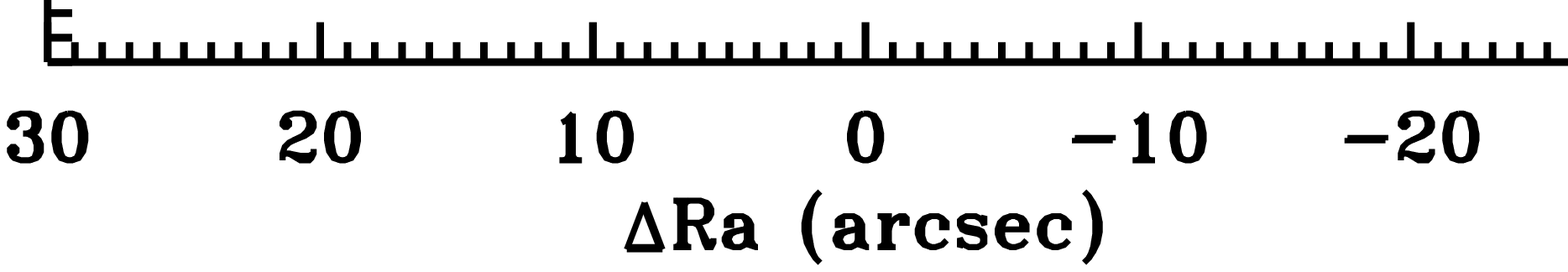}} &
   \resizebox{0.35\hsize}{!}{\includegraphics[angle=0,bb=28 714 651 1280]{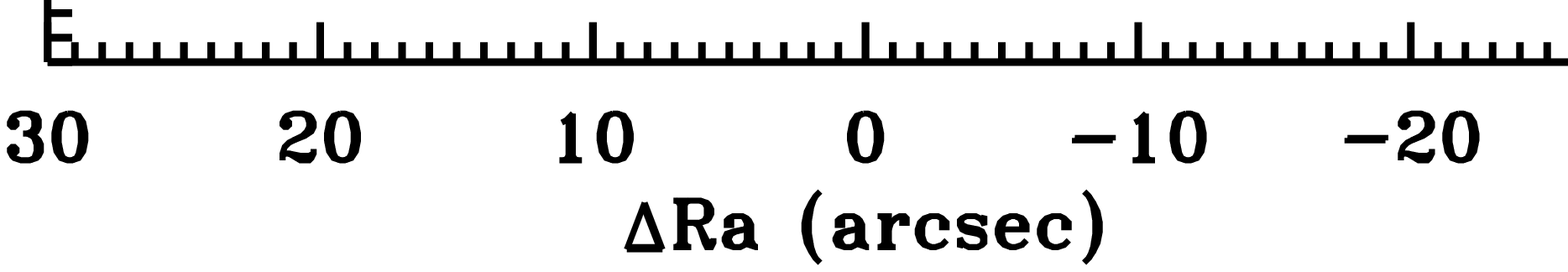}} \\
  \end{tabular}
  \caption{Maps of the OH emission from Ser~SMM3. The red and blue dashed lines indicate the rest wavelengths of the OH and CO transitions, respectively. Spaxels in gray were not included in the flux measurement.}
 \label{fig:smm3maps}
\end{figure}
\end{landscape}
}

\def\placefigure723Lmaps{
\begin{landscape}
\begin{figure}
 \centering
  \begin{tabular}{c@{\extracolsep{40pt}}c}
   \resizebox{0.35\hsize}{!}{\includegraphics[angle=0,bb=28 714 651 1280]{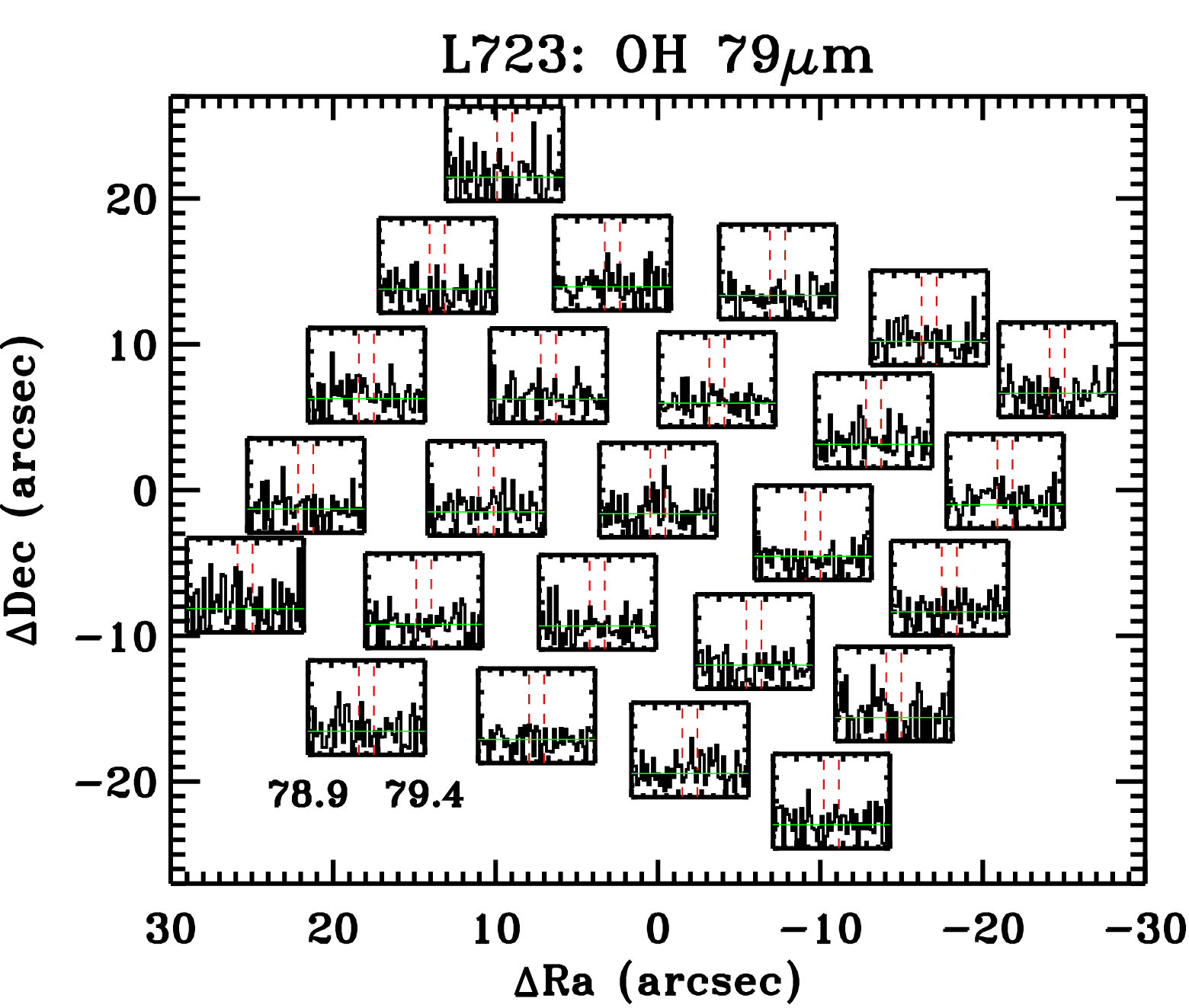}} &
   \resizebox{0.35\hsize}{!}{\includegraphics[angle=0,bb=28 714 651 1280]{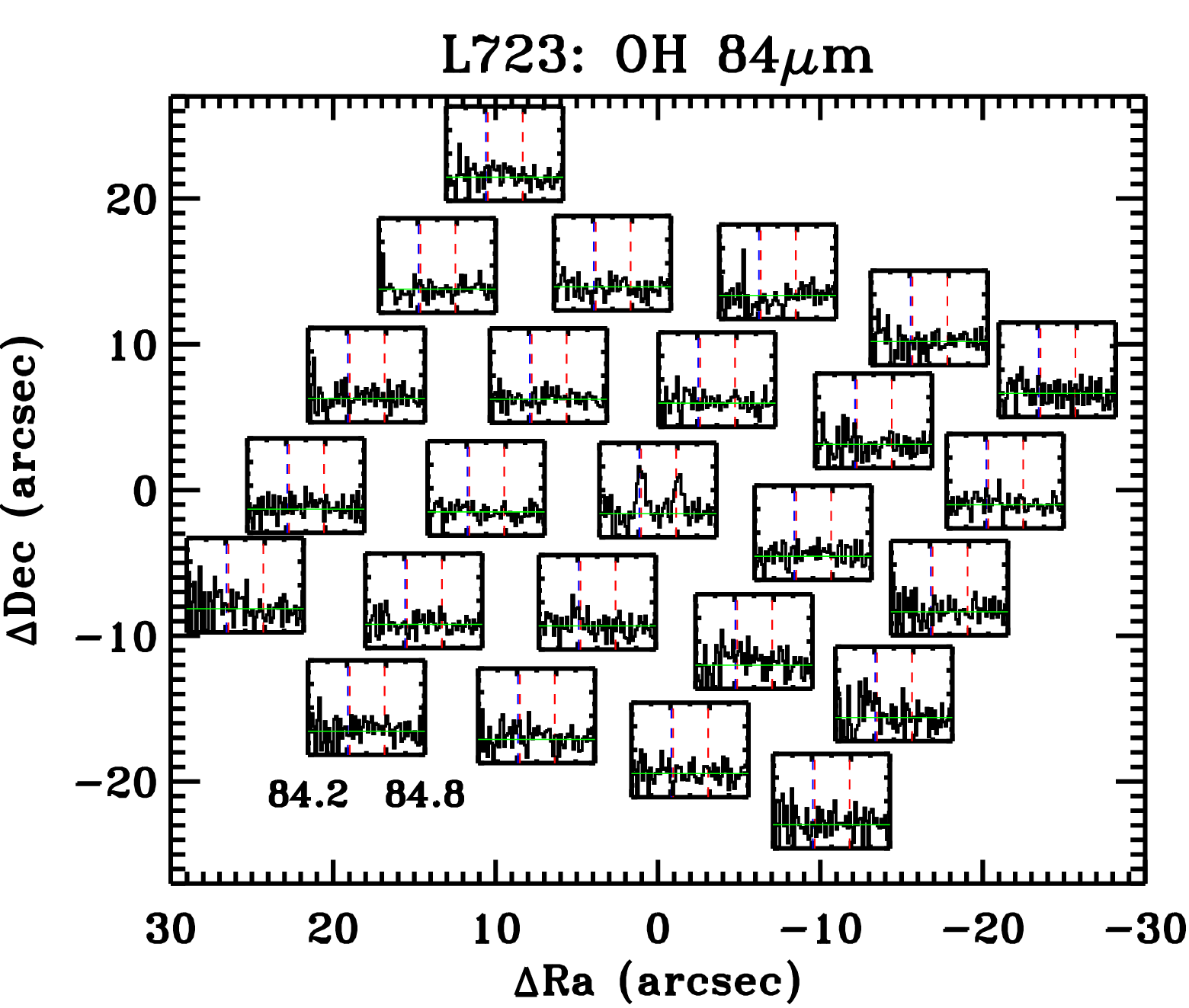}} \\
   & \\
   \resizebox{0.35\hsize}{!}{\includegraphics[angle=0,bb=28 714 651 1280]{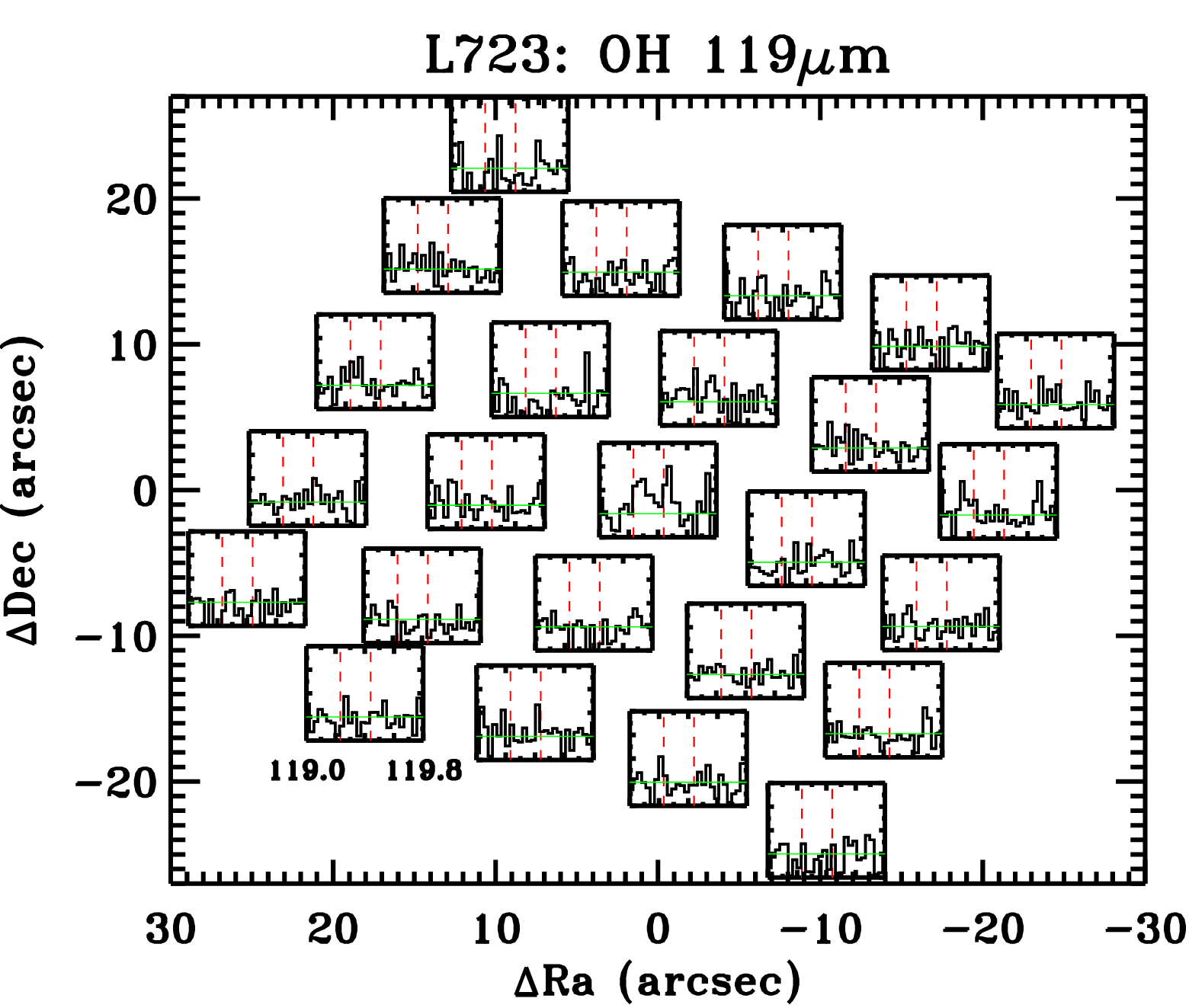}} &
   \resizebox{0.35\hsize}{!}{\includegraphics[angle=0,bb=28 714 651 1280]{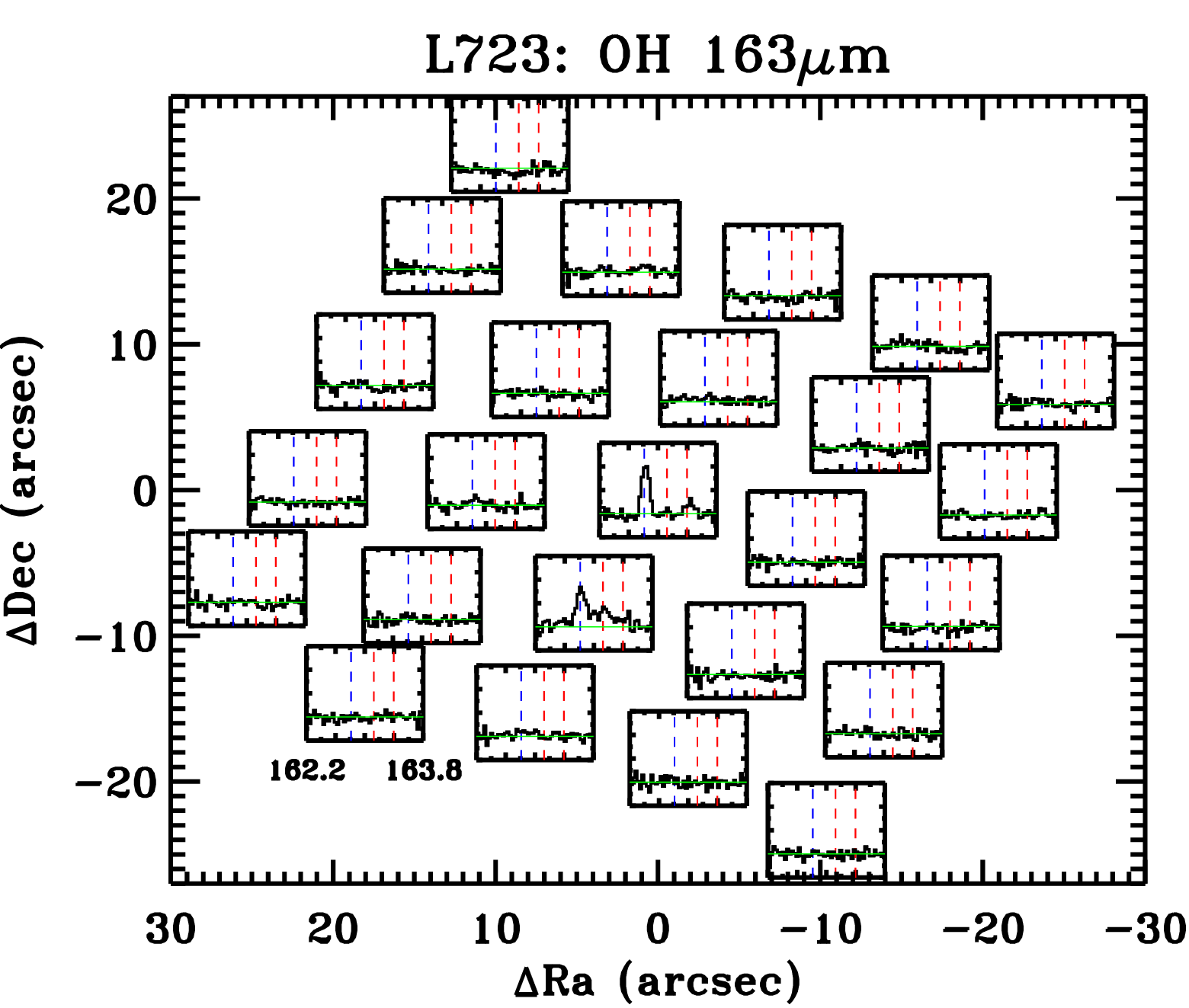}} \\
  \end{tabular}
  \caption{Maps of the OH emission from L~723. The red and blue dashed lines indicate the rest wavelengths of the OH and CO transitions, respectively.}
 \label{fig:l723maps}
\end{figure}
\end{landscape}
}

\def\placefigureL1489maps{
\begin{landscape}
\begin{figure}
 \centering
  \begin{tabular}{c@{\extracolsep{40pt}}c}
   \resizebox{0.35\hsize}{!}{\includegraphics[angle=0,bb=28 714 651 1280]{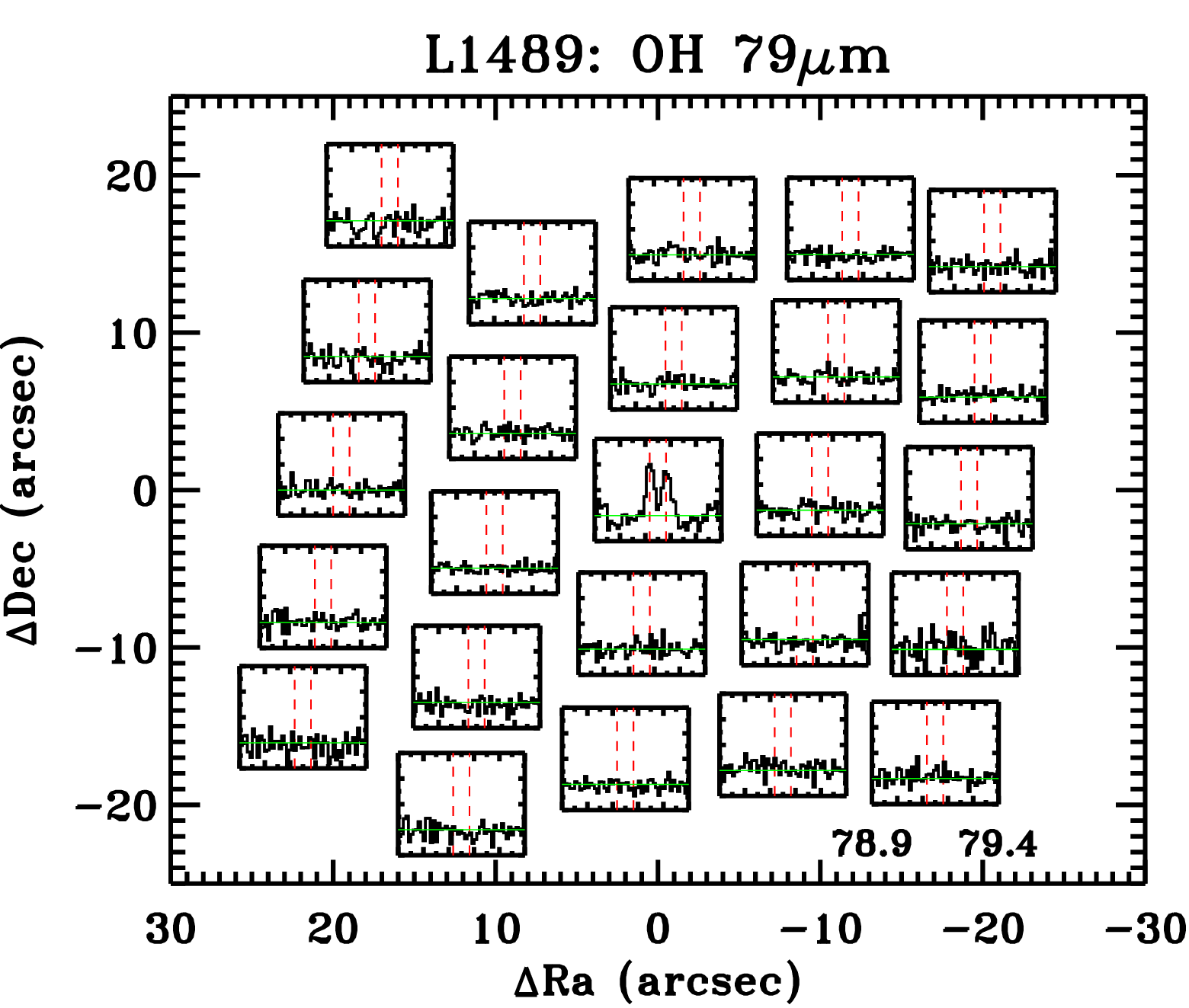}} &
   \resizebox{0.35\hsize}{!}{\includegraphics[angle=90,bb=28 90 594 713]{./ra_dec_map_l1489_84.eps}} \\
   & \\
   \resizebox{0.35\hsize}{!}{\includegraphics[angle=0,bb=28 714 651 1280]{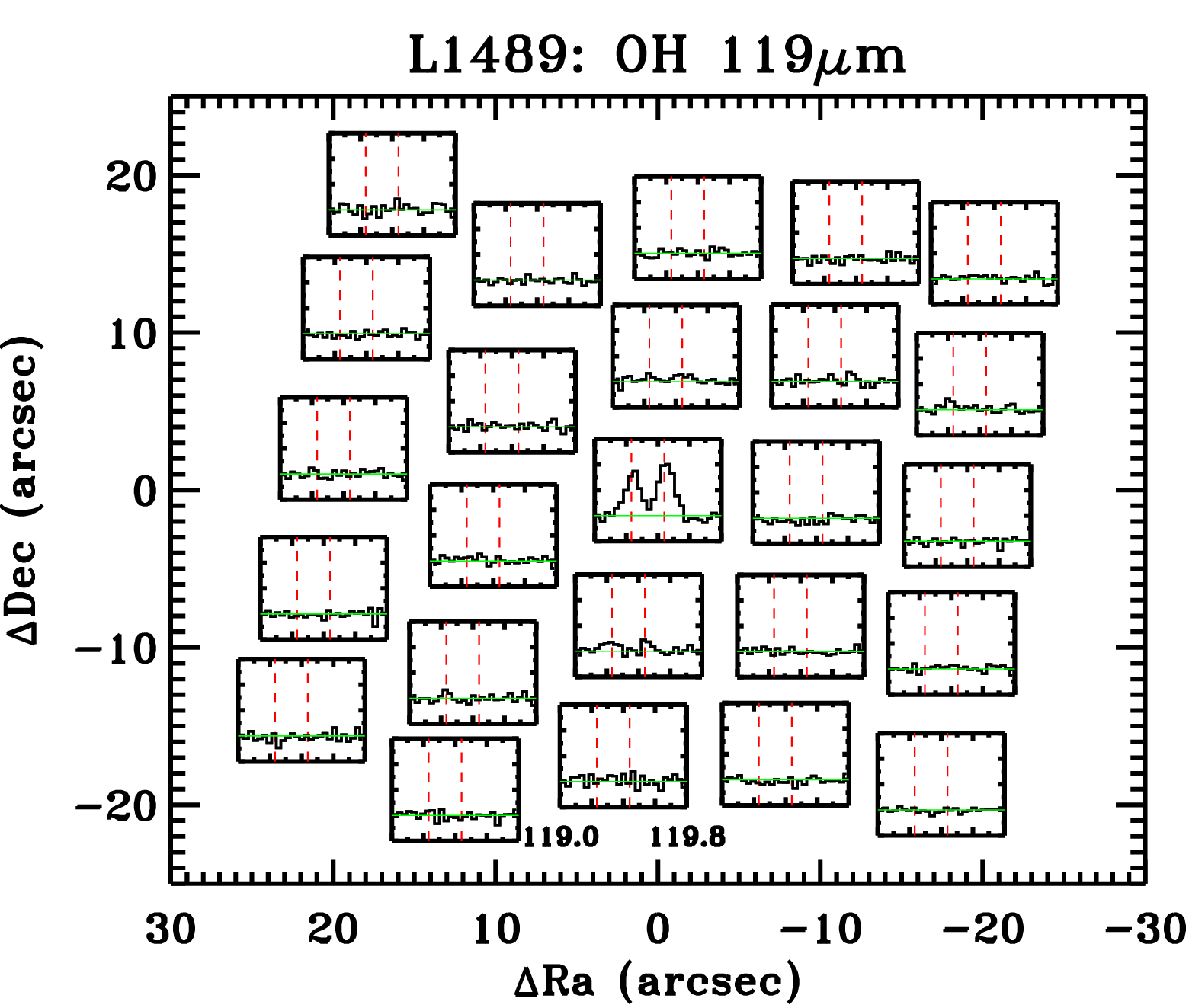}} &
   \resizebox{0.35\hsize}{!}{\includegraphics[angle=0,bb=28 714 651 1280]{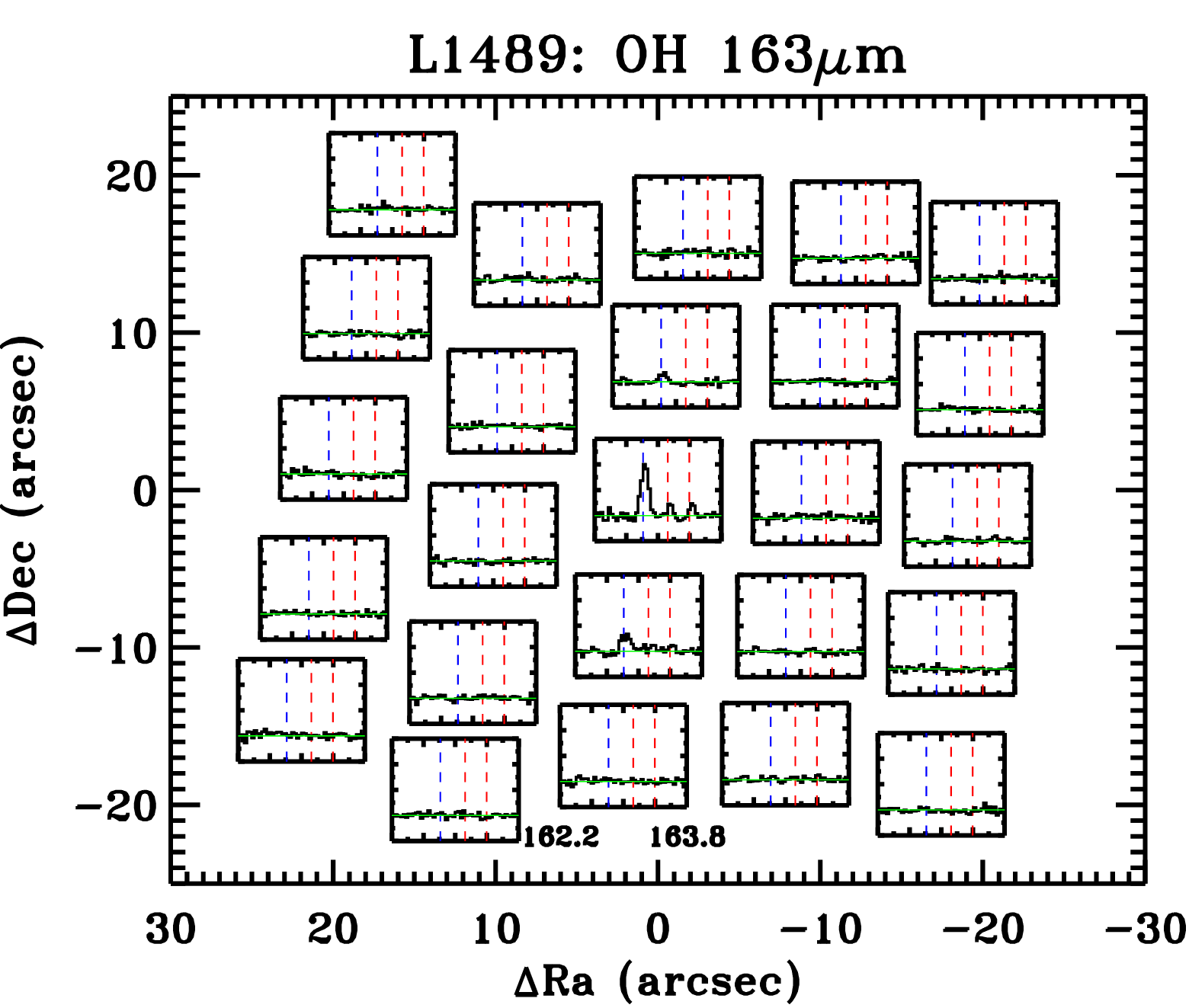}} \\
  \end{tabular}
  \caption{Maps of the OH emission from L~1489. The red and blue dashed lines indicate the rest wavelengths of the OH and CO transitions, respectively.}
 \label{fig:l1489maps}
\end{figure}
\end{landscape}
}

\def\placefigureTMR1maps{
\begin{landscape}
\begin{figure}
 \centering
  \begin{tabular}{c@{\extracolsep{40pt}}c}
   \resizebox{0.35\hsize}{!}{\includegraphics[angle=0,bb=28 714 651 1280]{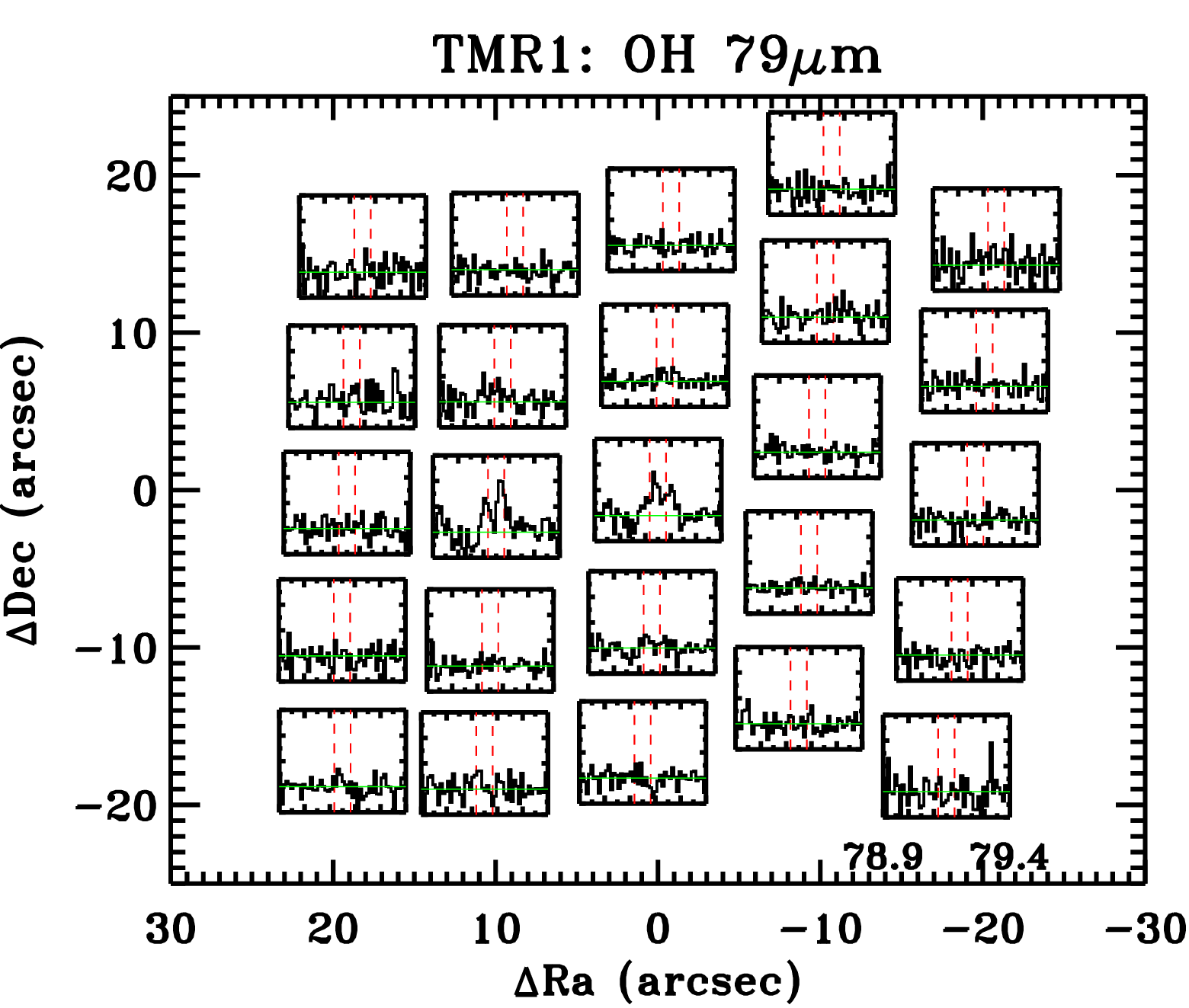}} &
   \resizebox{0.35\hsize}{!}{\includegraphics[angle=0,bb=28 714 651 1280]{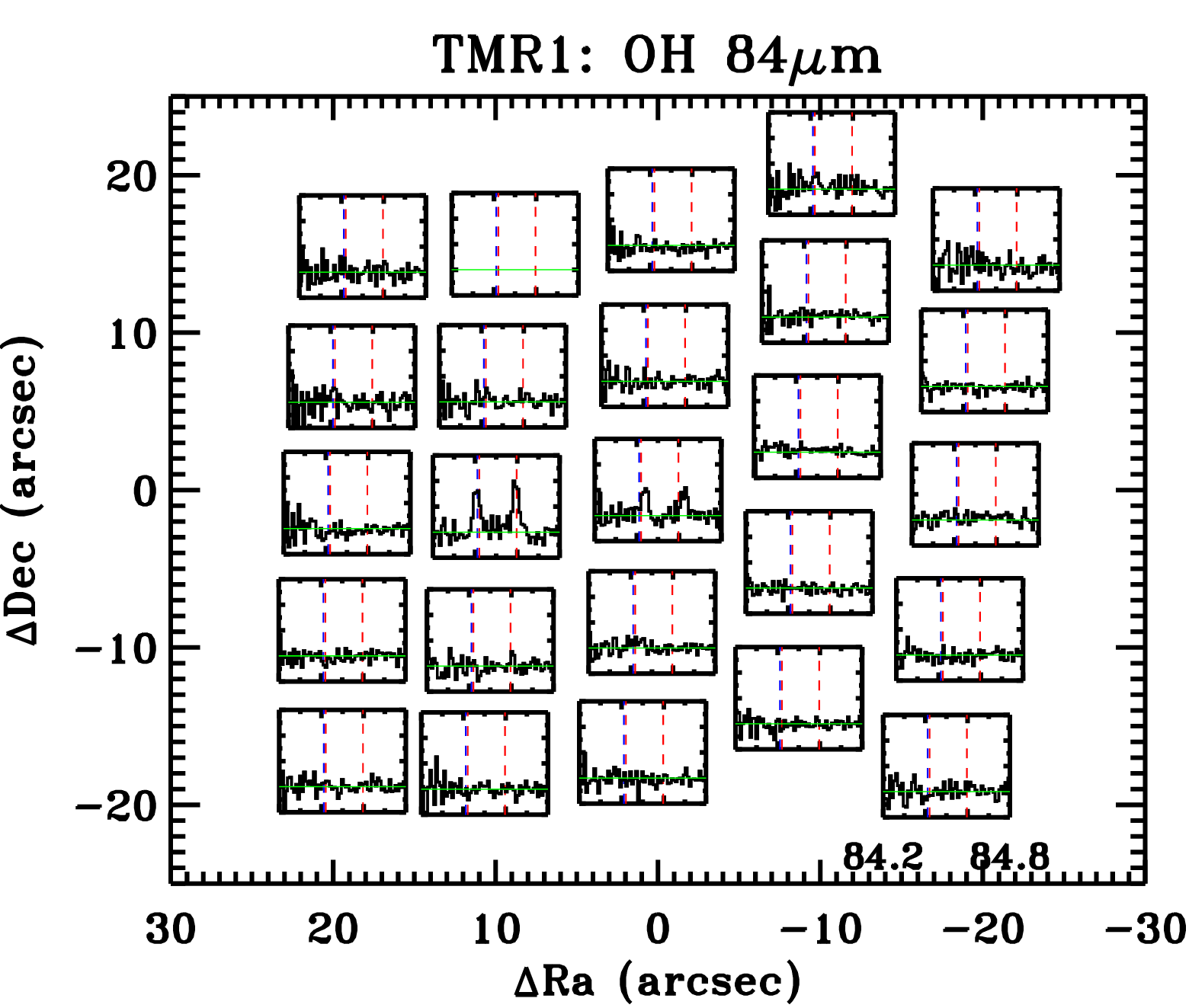}} \\
   & \\
   \resizebox{0.35\hsize}{!}{\includegraphics[angle=0,bb=28 714 651 1280]{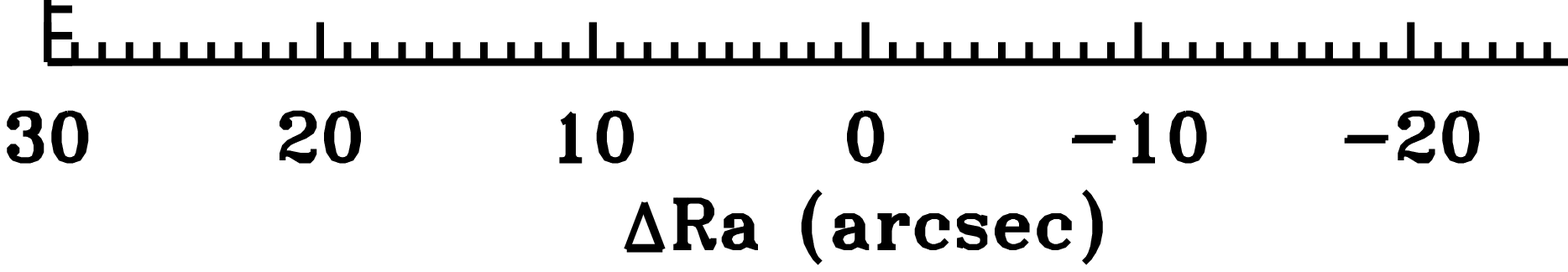}} &
   \resizebox{0.35\hsize}{!}{\includegraphics[angle=0,bb=28 714 651 1280]{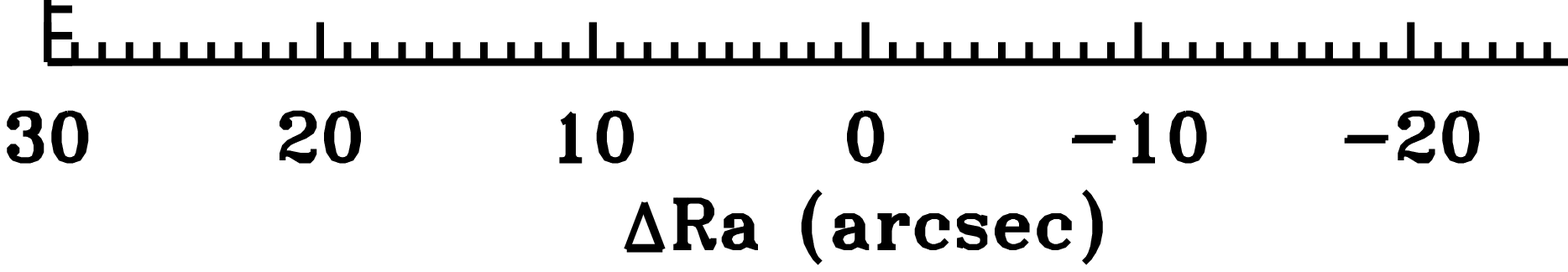}} \\
  \end{tabular}
  \caption{Maps of the OH emission from TMR~1. The red and blue dashed lines indicate the rest wavelengths of the OH and CO transitions, respectively.}
 \label{fig:tmr1maps}
\end{figure}
\end{landscape}
}

\def\placefigureATMC1maps{
\begin{landscape}
\begin{figure}
 \centering
  \begin{tabular}{c@{\extracolsep{40pt}}c}
   \resizebox{0.35\hsize}{!}{\includegraphics[angle=0,bb=28 714 651 1280]{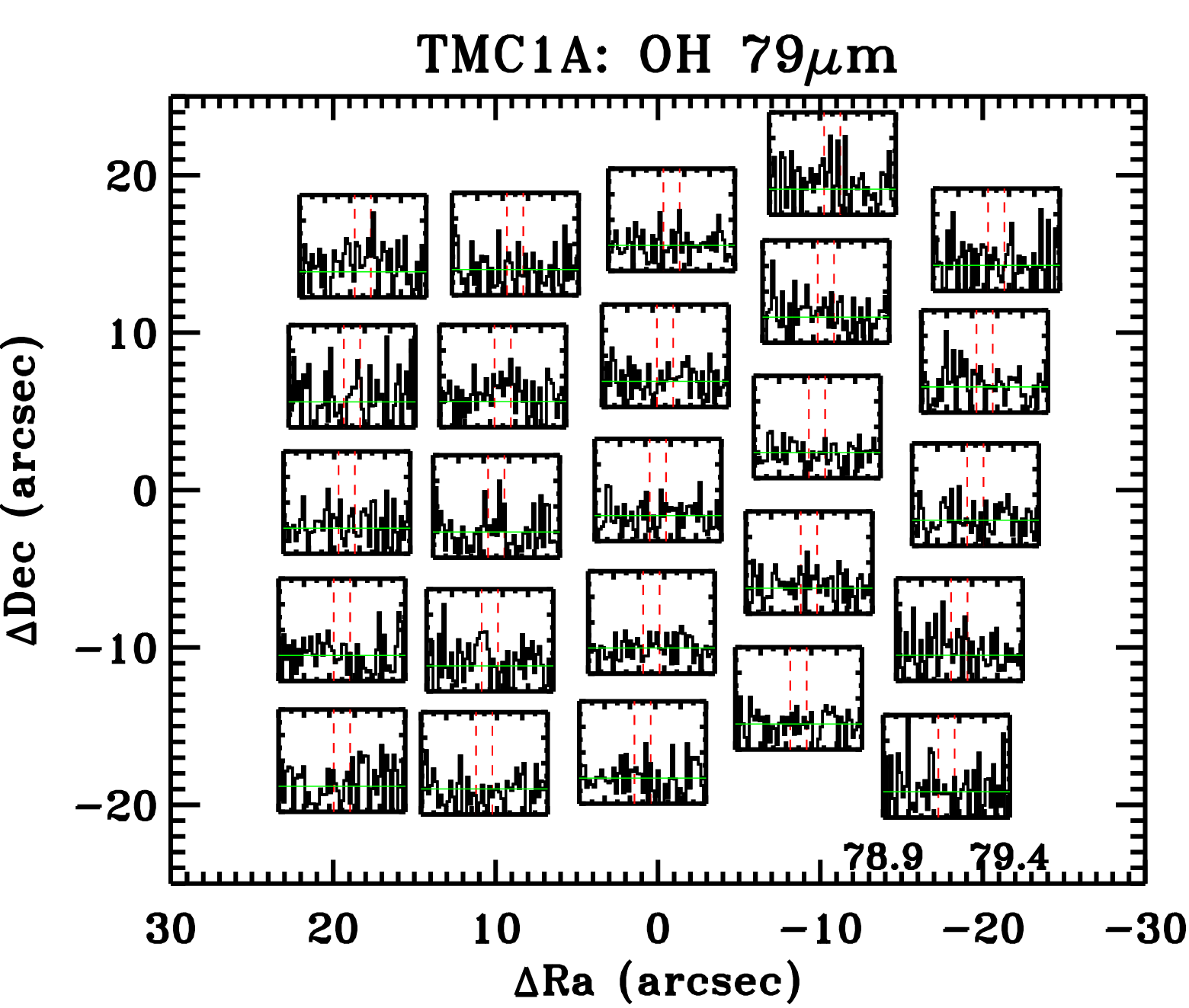}} &
   \resizebox{0.35\hsize}{!}{\includegraphics[angle=0,bb=28 714 651 1280]{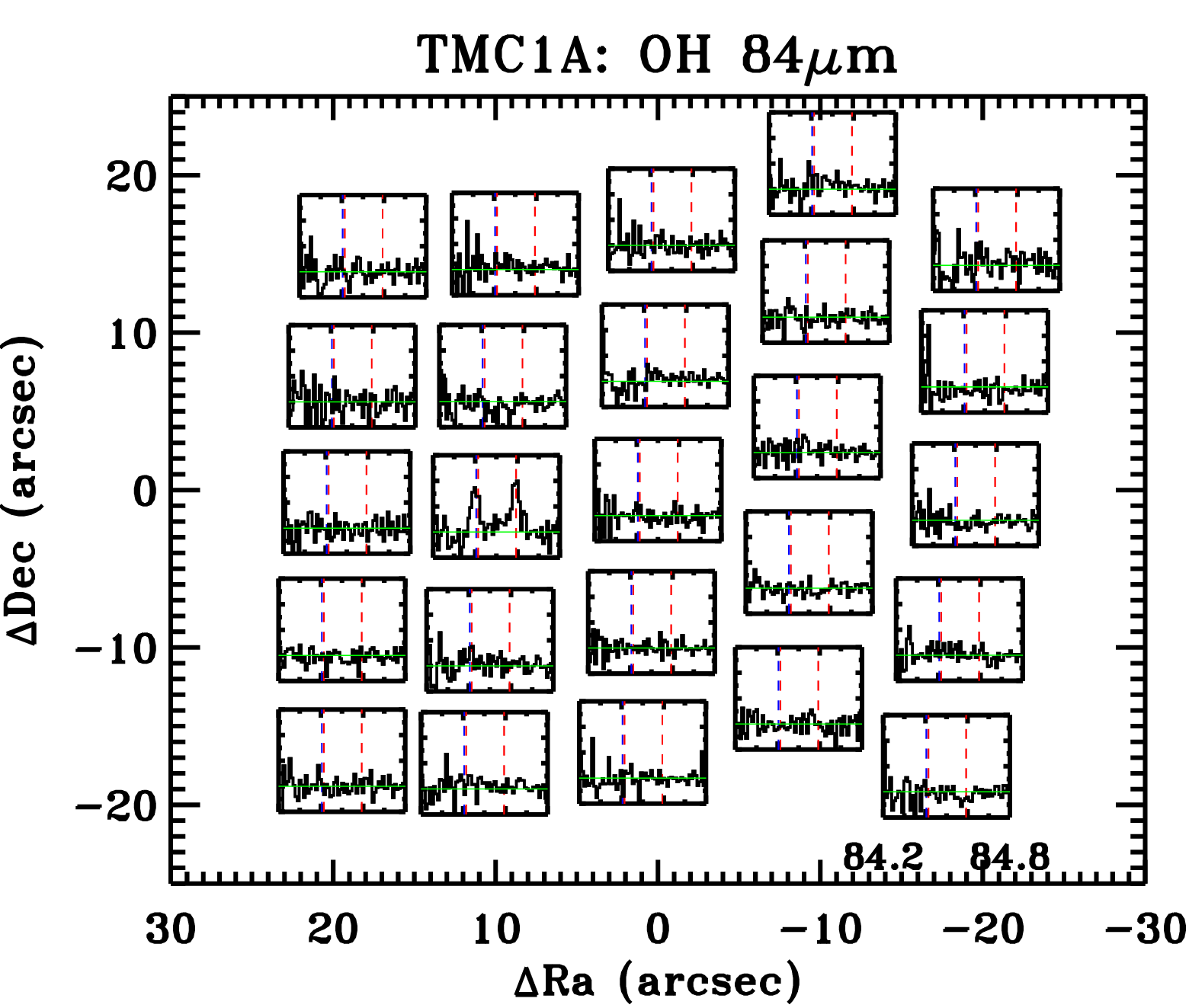}} \\
   & \\
   \resizebox{0.35\hsize}{!}{\includegraphics[angle=0,bb=28 714 651 1280]{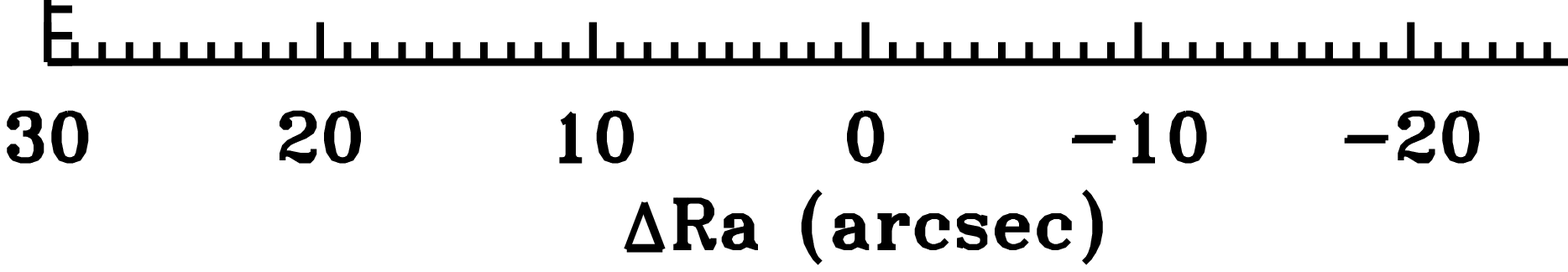}} &
   \resizebox{0.35\hsize}{!}{\includegraphics[angle=0,bb=28 714 651 1280]{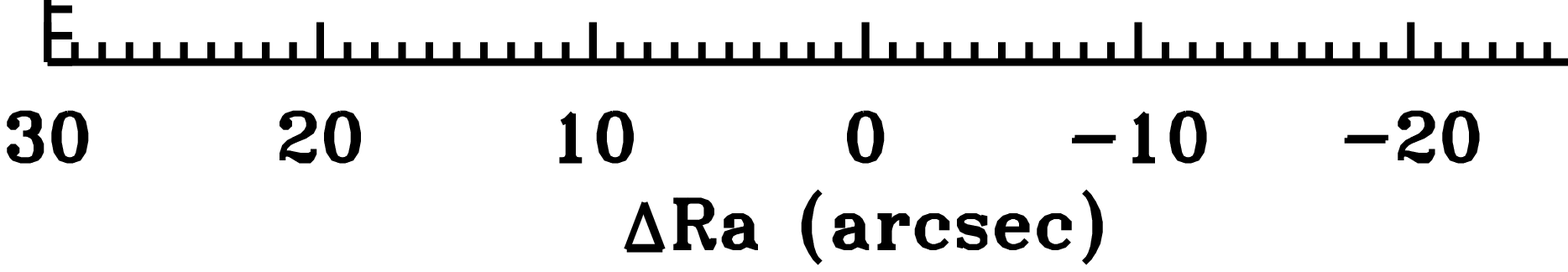}} \\
  \end{tabular}
  \caption{Maps of the OH emission from TMC~1A. The red and blue dashed lines indicate the rest wavelengths of the OH and CO transitions, respectively.}
 \label{fig:tmc1amaps}
\end{figure}
\end{landscape}
}

\def\placefigureTMC1maps{
\begin{landscape}
\begin{figure}
 \centering
  \begin{tabular}{c@{\extracolsep{40pt}}c}
   \resizebox{0.35\hsize}{!}{\includegraphics[angle=0,bb=28 714 651 1280]{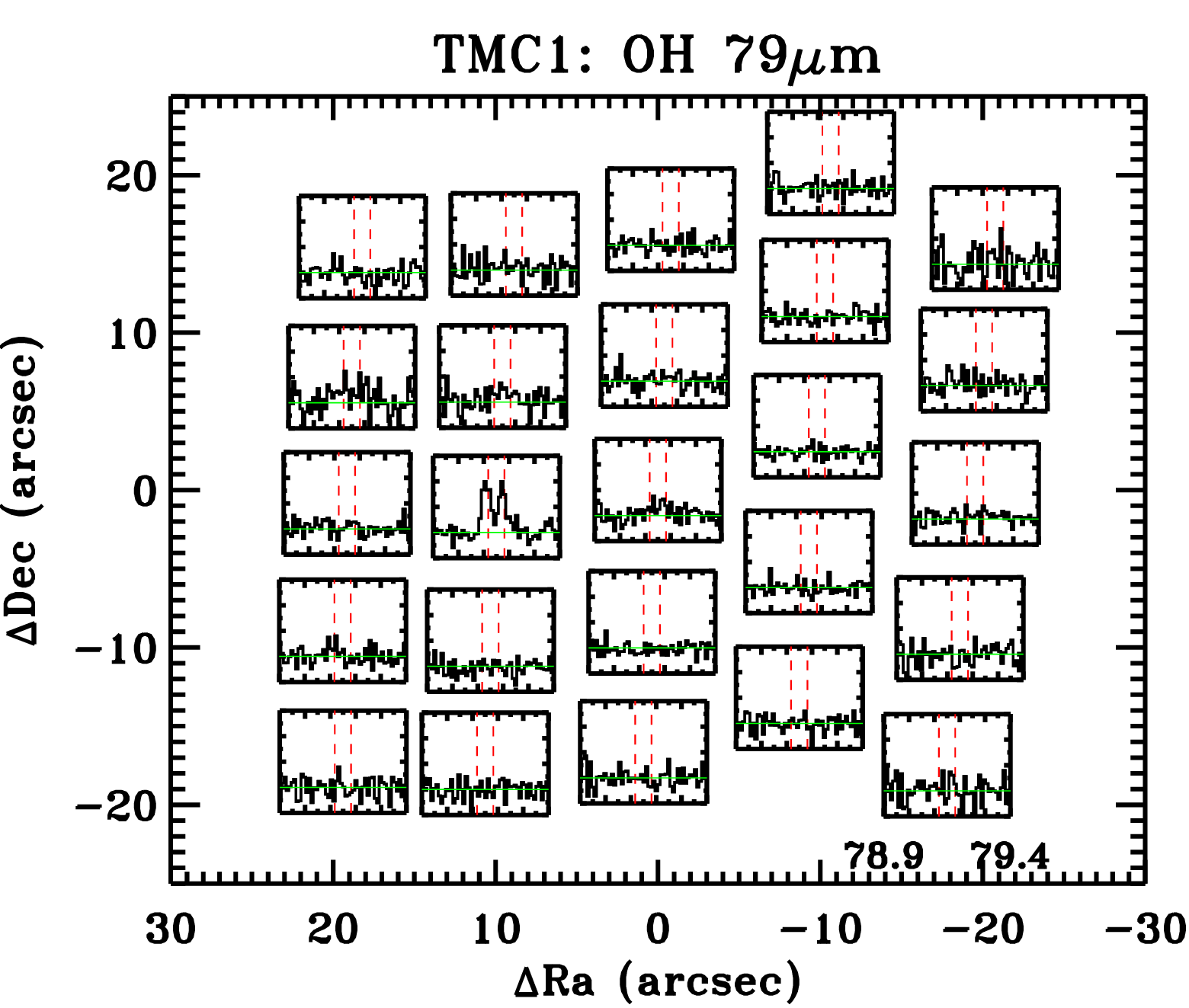}} &
   \resizebox{0.35\hsize}{!}{\includegraphics[angle=0,bb=28 714 651 1280]{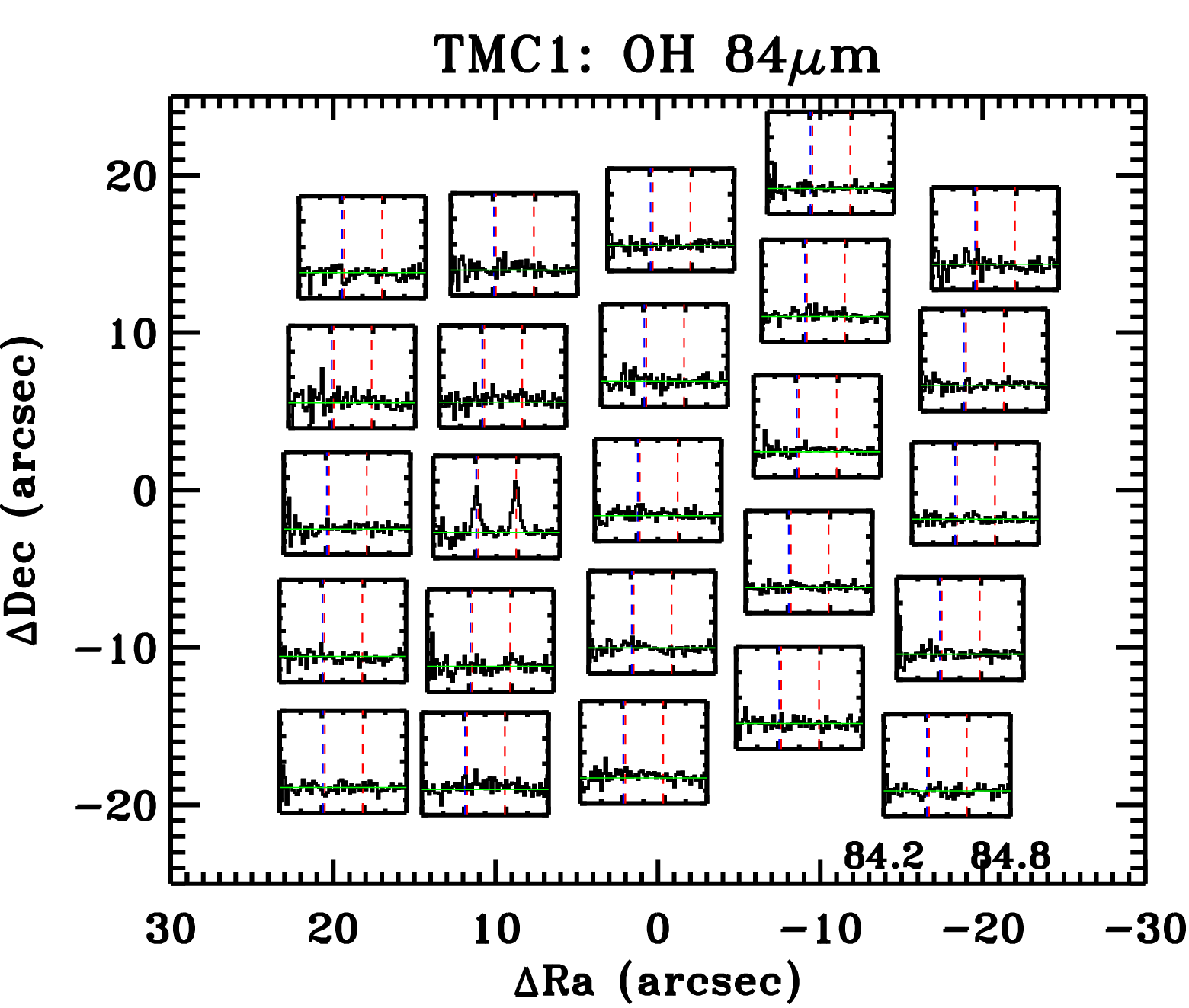}} \\
   & \\
   \resizebox{0.35\hsize}{!}{\includegraphics[angle=0,bb=28 714 651 1280]{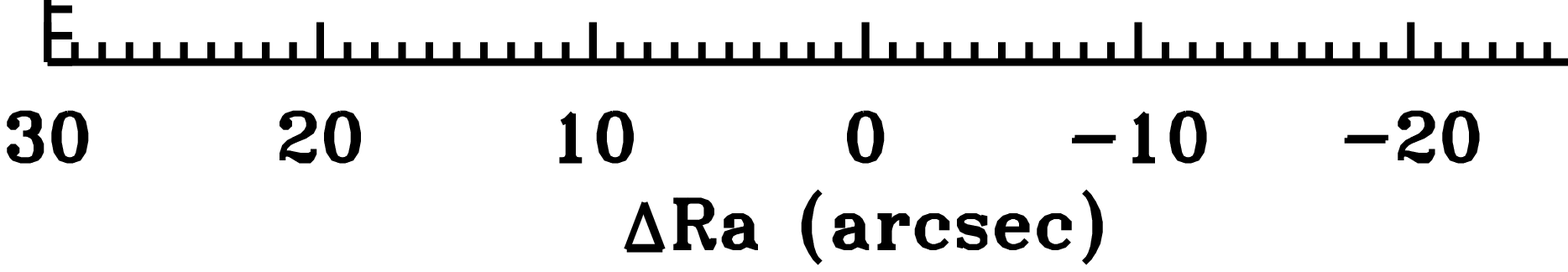}} &
   \resizebox{0.35\hsize}{!}{\includegraphics[angle=0,bb=28 714 651 1280]{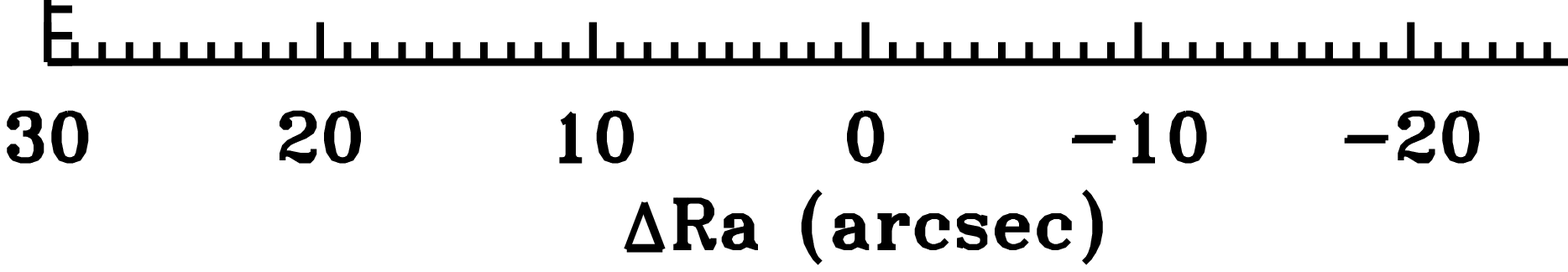}} \\
  \end{tabular}
  \caption{Maps of the OH emission from TMC~1. The red and blue dashed lines indicate the rest wavelengths of the OH and CO transitions, respectively.}
 \label{fig:tmc1maps}
\end{figure}
\end{landscape}
}

\def\placefigureHH46maps{
\begin{landscape}
\begin{figure}
 \centering
  \begin{tabular}{c@{\extracolsep{40pt}}c}
   \resizebox{0.35\hsize}{!}{\includegraphics[angle=0,bb=28 714 651 1280]{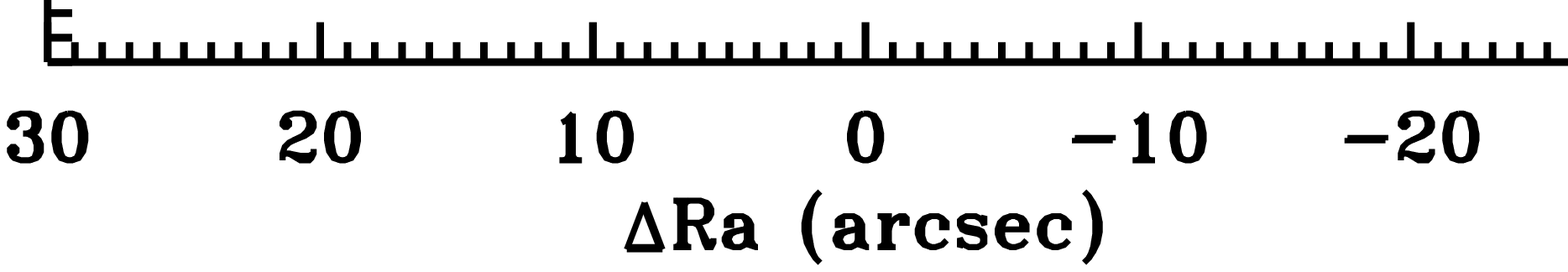}} &
   \resizebox{0.35\hsize}{!}{\includegraphics[angle=0,bb=28 714 651 1280]{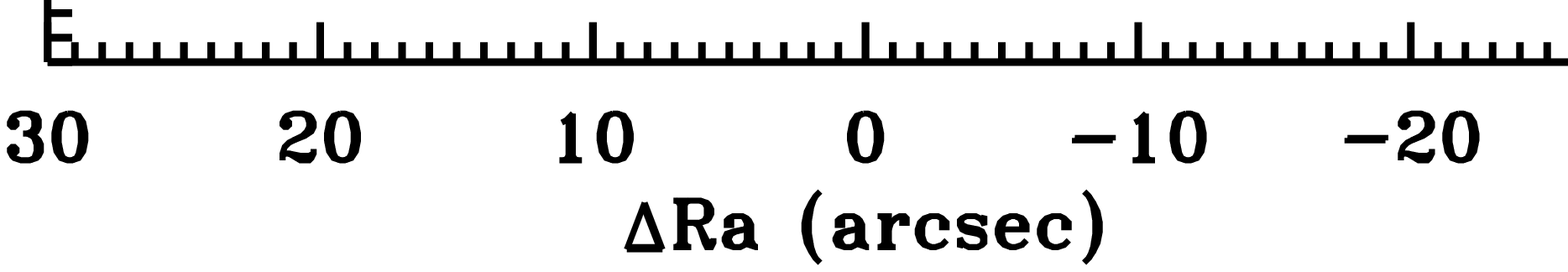}} \\
   & \\
   \resizebox{0.35\hsize}{!}{\includegraphics[angle=0,bb=28 714 651 1280]{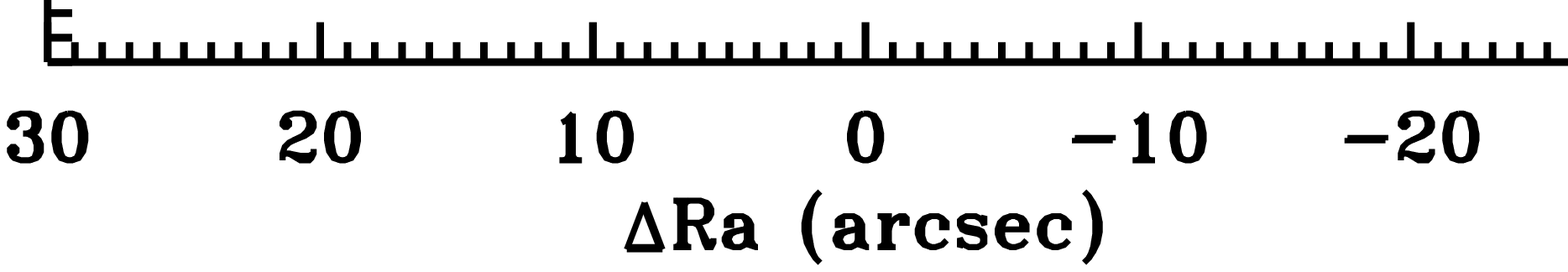}} &
   \resizebox{0.35\hsize}{!}{\includegraphics[angle=0,bb=28 714 651 1280]{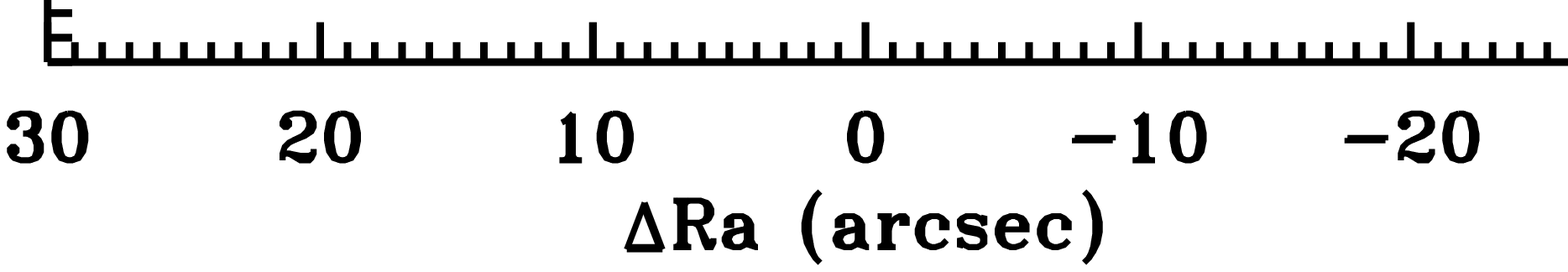}} \\
  \end{tabular}
  \caption{Maps of the OH emission from HH~46. The red and blue dashed lines indicate the rest wavelengths of the OH and CO transitions, respectively.}
 \label{fig:hh46maps}
\end{figure}
\end{landscape}
}

\def\placefigureRNO91maps{
\begin{landscape}
\begin{figure}
 \centering
  \begin{tabular}{c@{\extracolsep{40pt}}c}
   \resizebox{0.35\hsize}{!}{\includegraphics[angle=0,bb=28 714 651 1280]{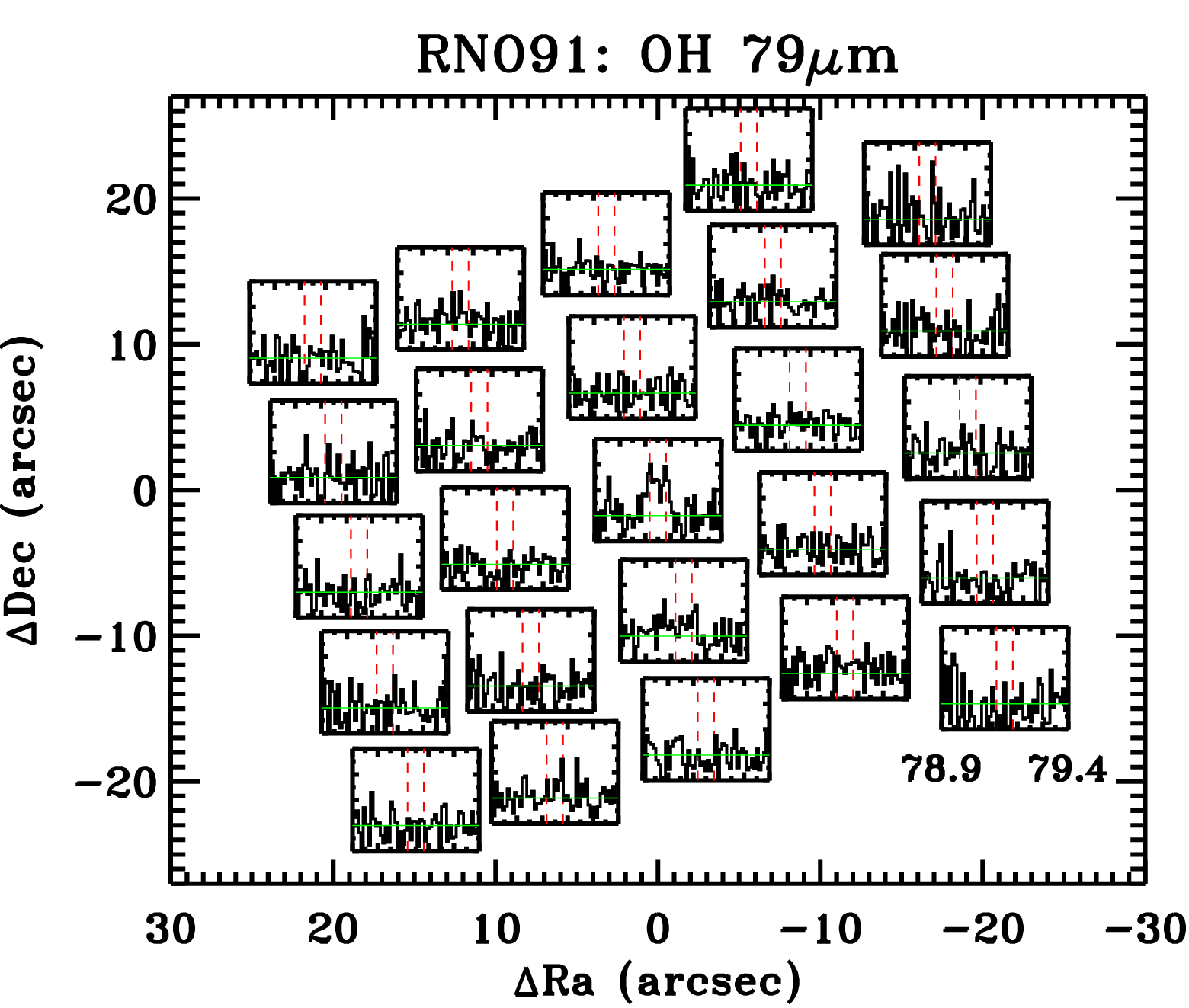}} &
   \resizebox{0.35\hsize}{!}{\includegraphics[angle=0,bb=28 714 651 1280]{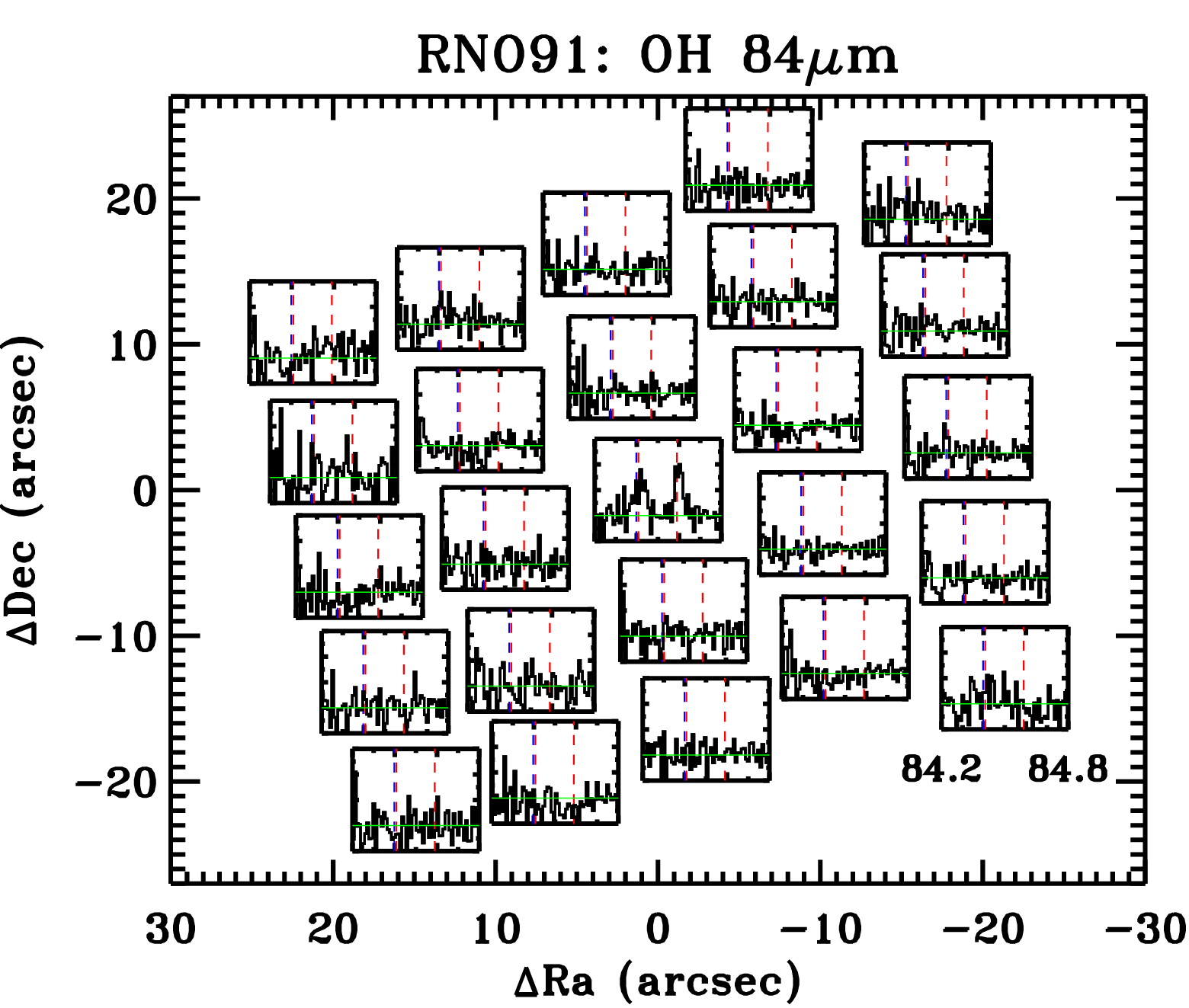}} \\
   & \\
   \resizebox{0.35\hsize}{!}{\includegraphics[angle=0,bb=28 714 651 1280]{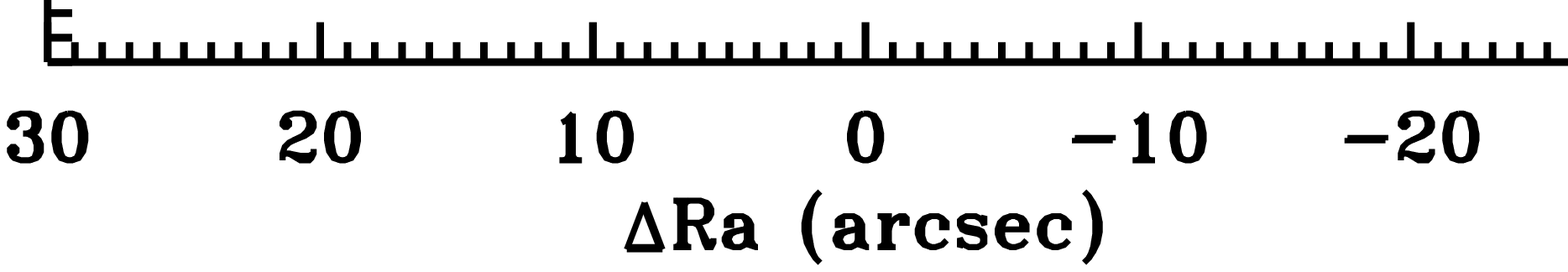}} &
   \resizebox{0.35\hsize}{!}{\includegraphics[angle=0,bb=28 714 651 1280]{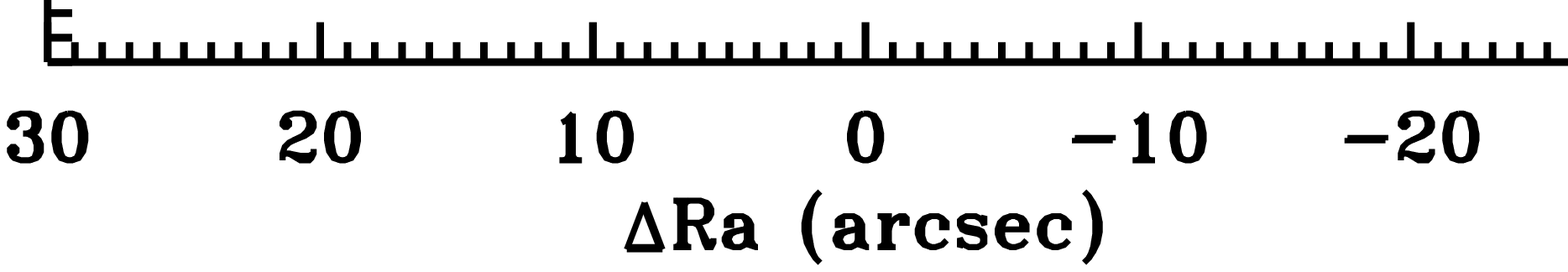}} \\
  \end{tabular}
  \caption{Maps of the OH emission from RNO~91. The red and blue dashed lines indicate the rest wavelengths of the OH and CO transitions, respectively.}
 \label{fig:rno91maps}
\end{figure}
\end{landscape}
}

\def\placefigureAFGL490maps{
\begin{landscape}
\begin{figure}
 \centering
  \begin{tabular}{c@{\extracolsep{40pt}}c}
   \resizebox{0.35\hsize}{!}{\includegraphics[angle=0,bb=28 714 651 1280]{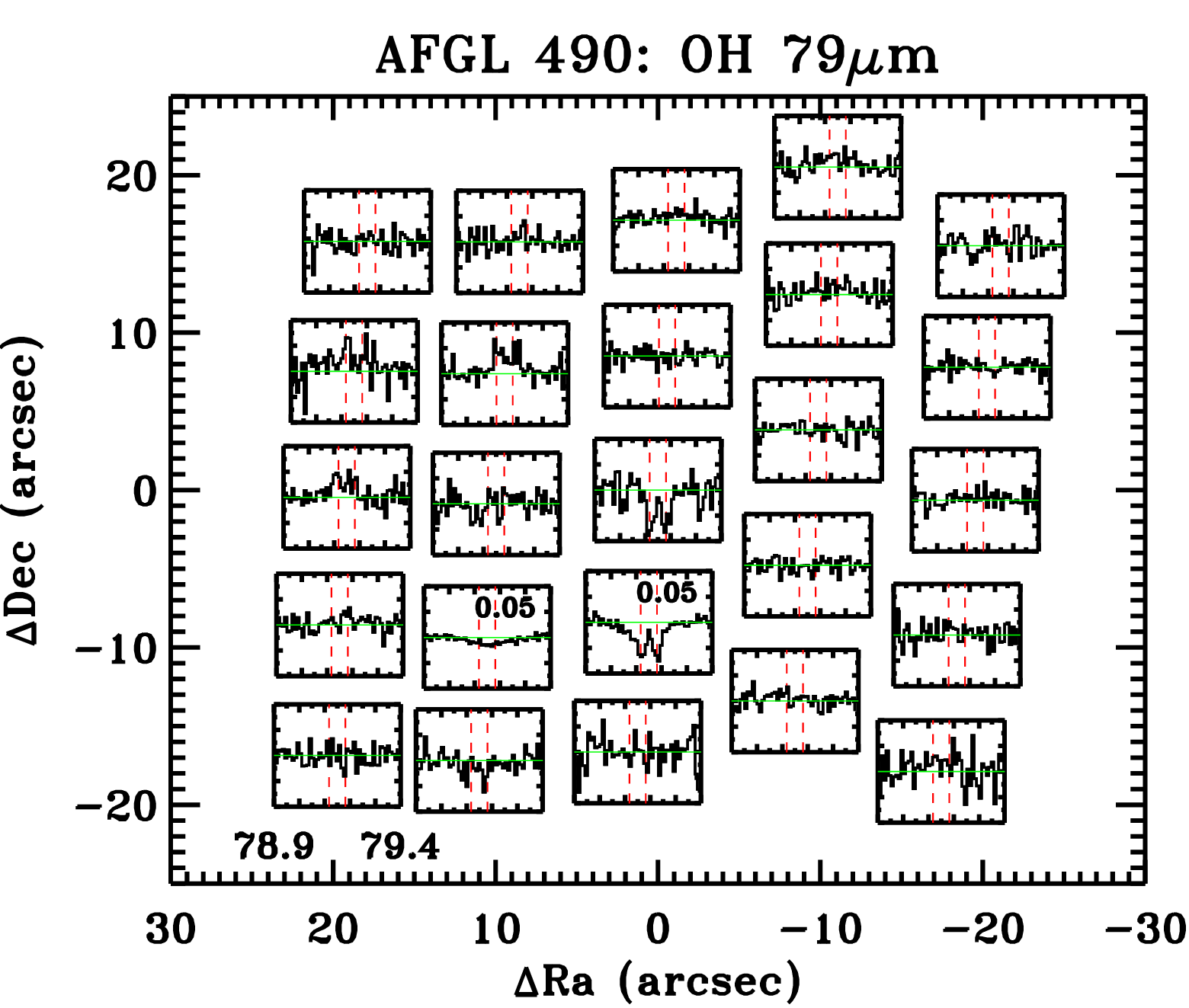}} &
   \resizebox{0.35\hsize}{!}{\includegraphics[angle=0,bb=28 714 651 1280]{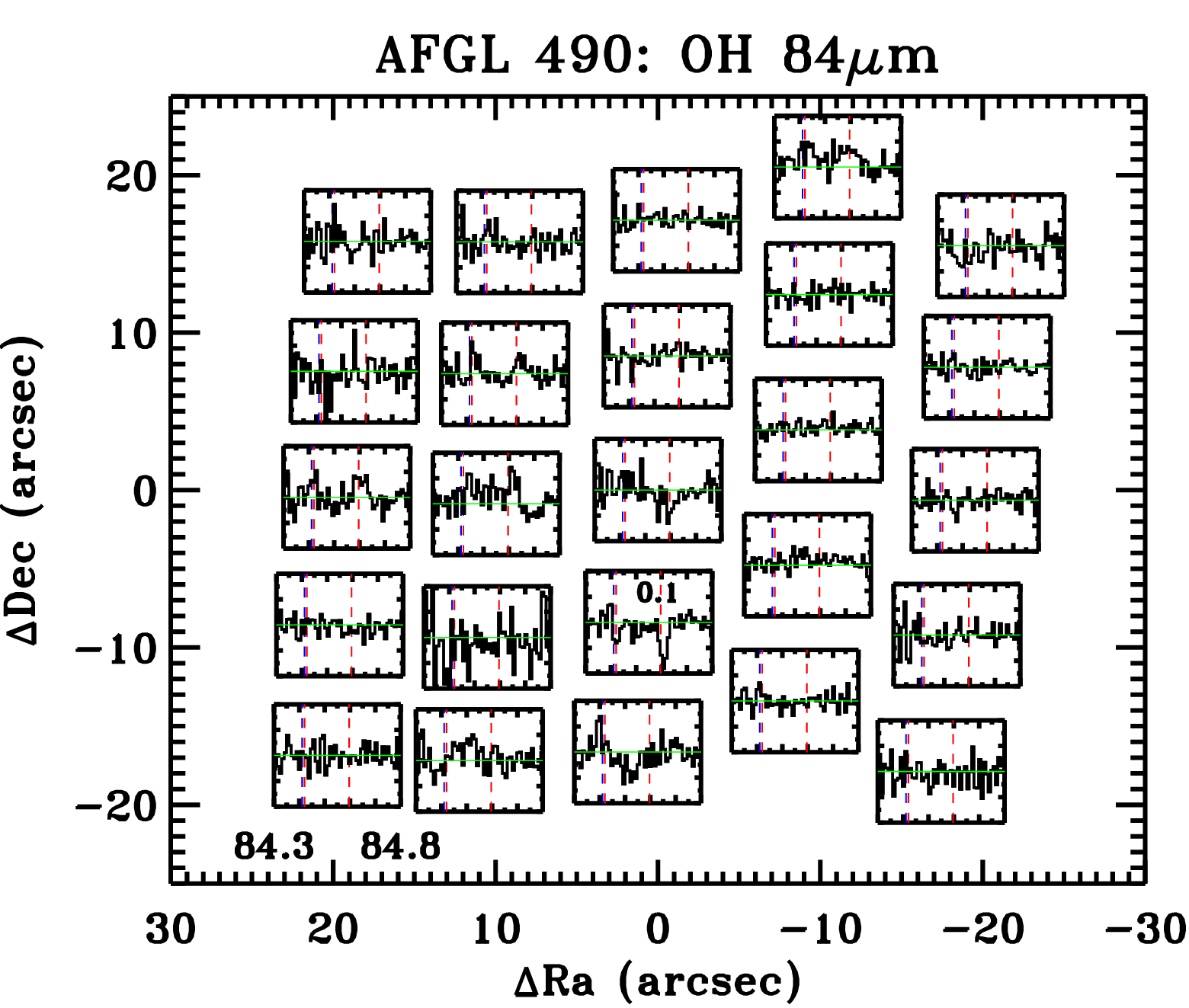}} \\
   & \\
   \resizebox{0.35\hsize}{!}{\includegraphics[angle=0,bb=28 714 651 1280]{./ra_dec_map_afgl490_119.eps}} &
   \resizebox{0.35\hsize}{!}{\includegraphics[angle=0,bb=28 714 651 1280]{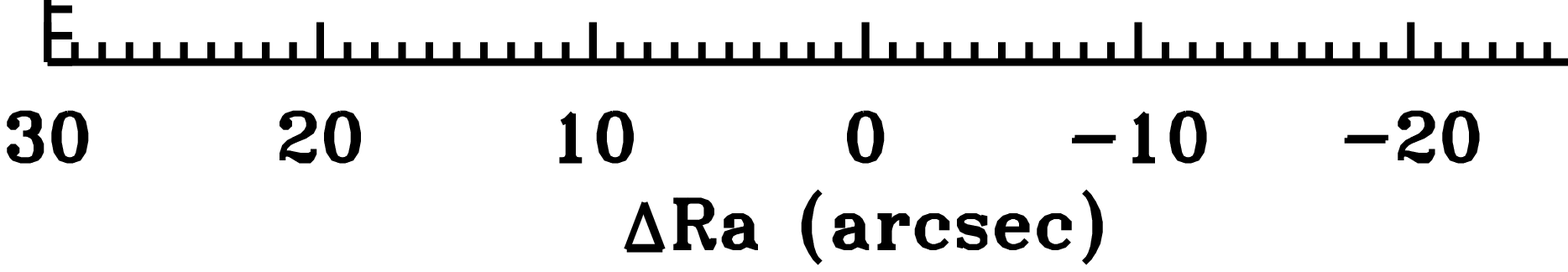}} \\
  \end{tabular}
  \caption{Maps of the OH emission from AFGL~490. The red and blue dashed lines indicate the rest wavelengths of the OH and CO transitions, respectively.}
 \label{fig:afgl490maps}
\end{figure}
\end{landscape}
}

\def\placefigureN2071maps{
\begin{landscape}
\begin{figure}
 \centering
  \begin{tabular}{c@{\extracolsep{40pt}}c}
   \resizebox{0.35\hsize}{!}{\includegraphics[angle=0,bb=28 714 651 1280]{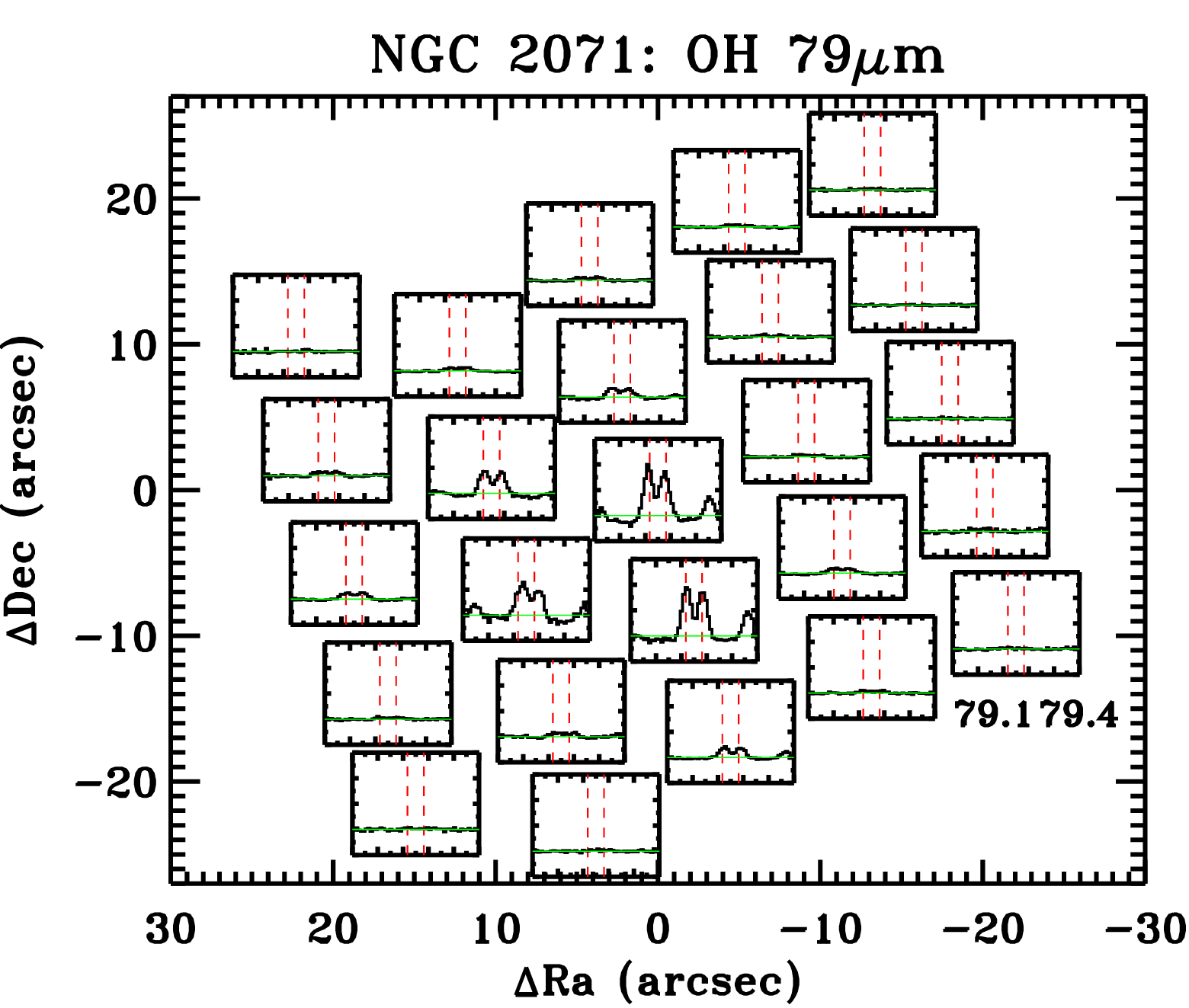}} &
   \resizebox{0.35\hsize}{!}{\includegraphics[angle=0,bb=28 714 651 1280]{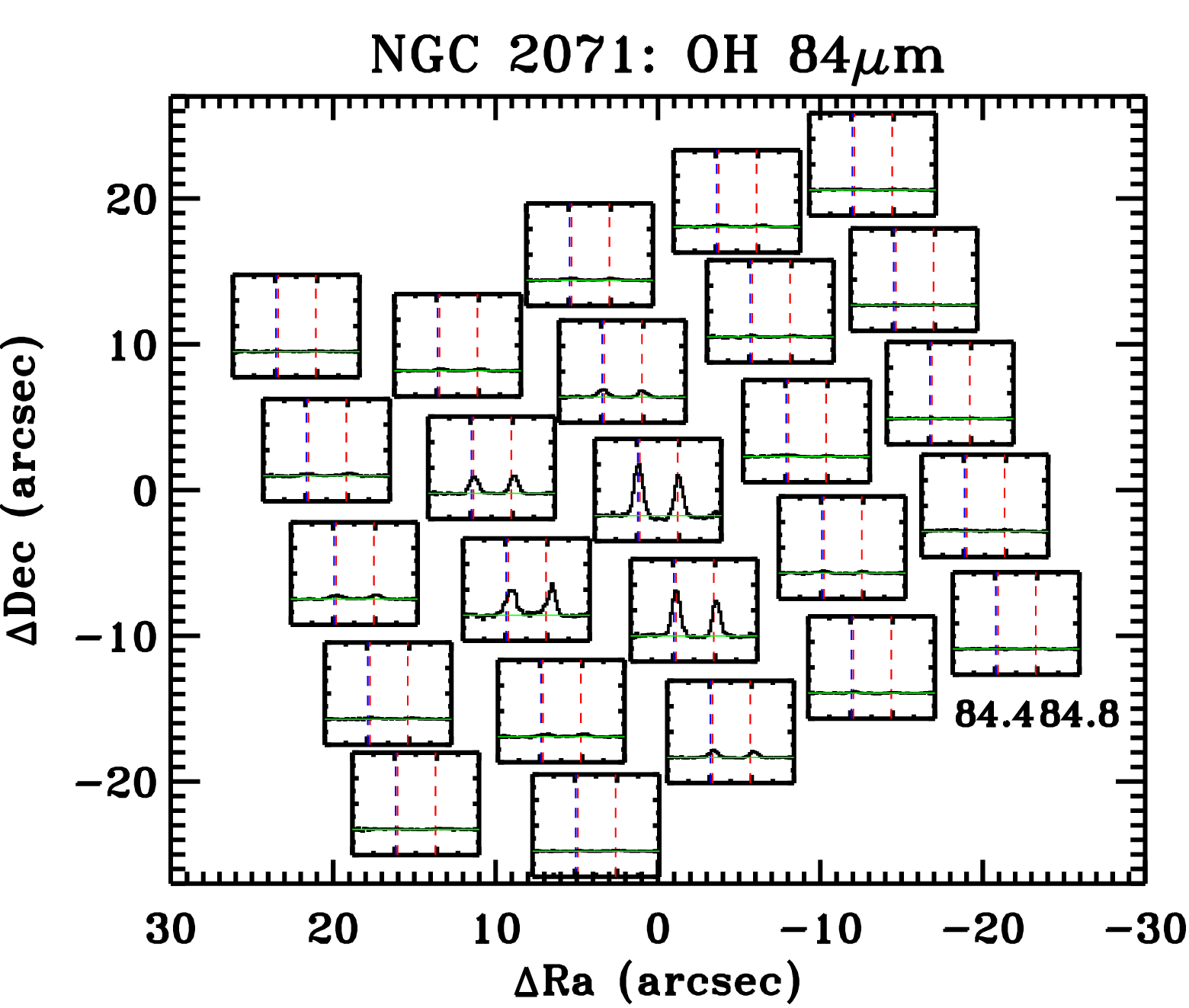}} \\
   & \\
   \resizebox{0.35\hsize}{!}{\includegraphics[angle=0,bb=28 714 651 1280]{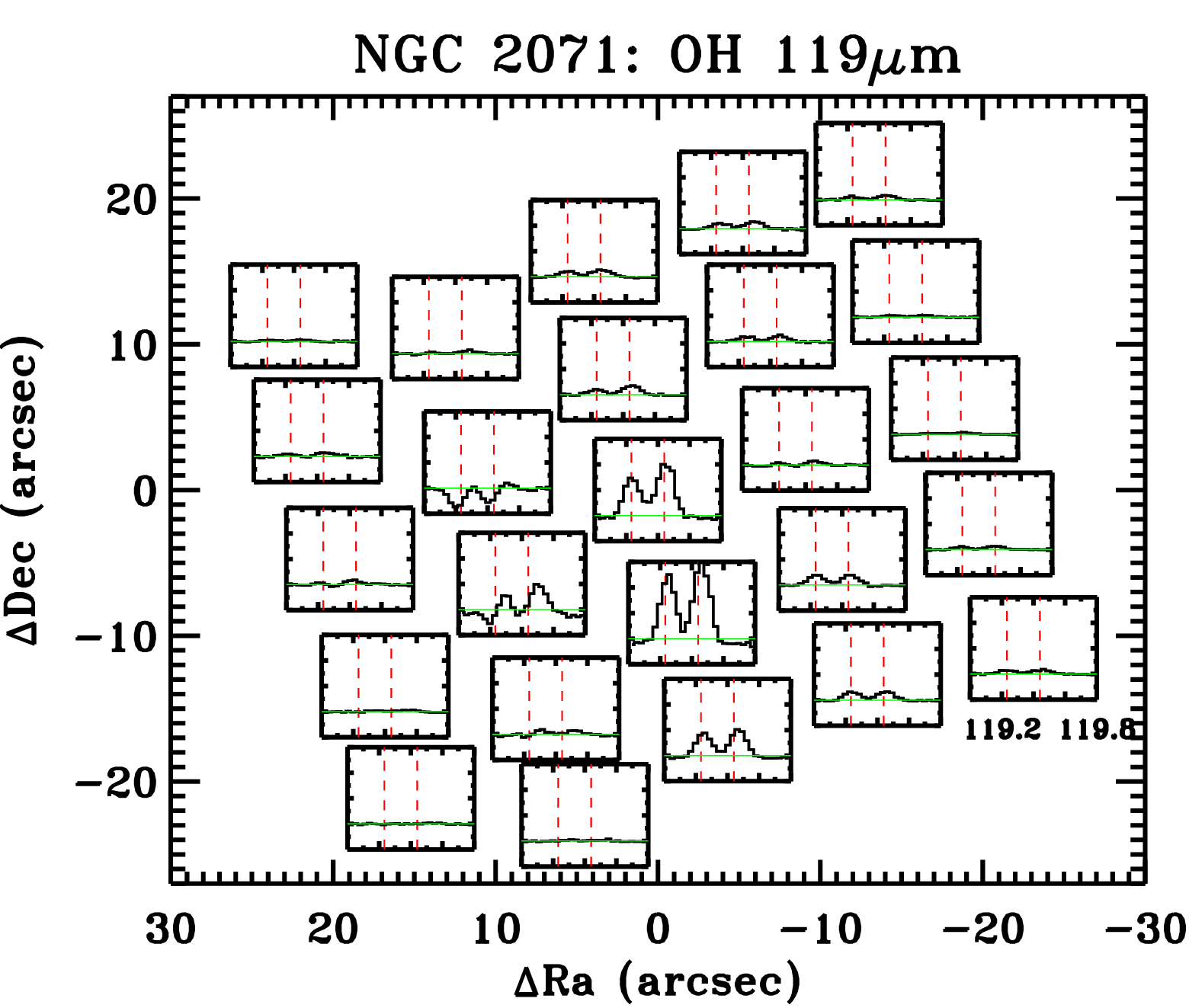}} &
   \resizebox{0.35\hsize}{!}{\includegraphics[angle=0,bb=28 714 651 1280]{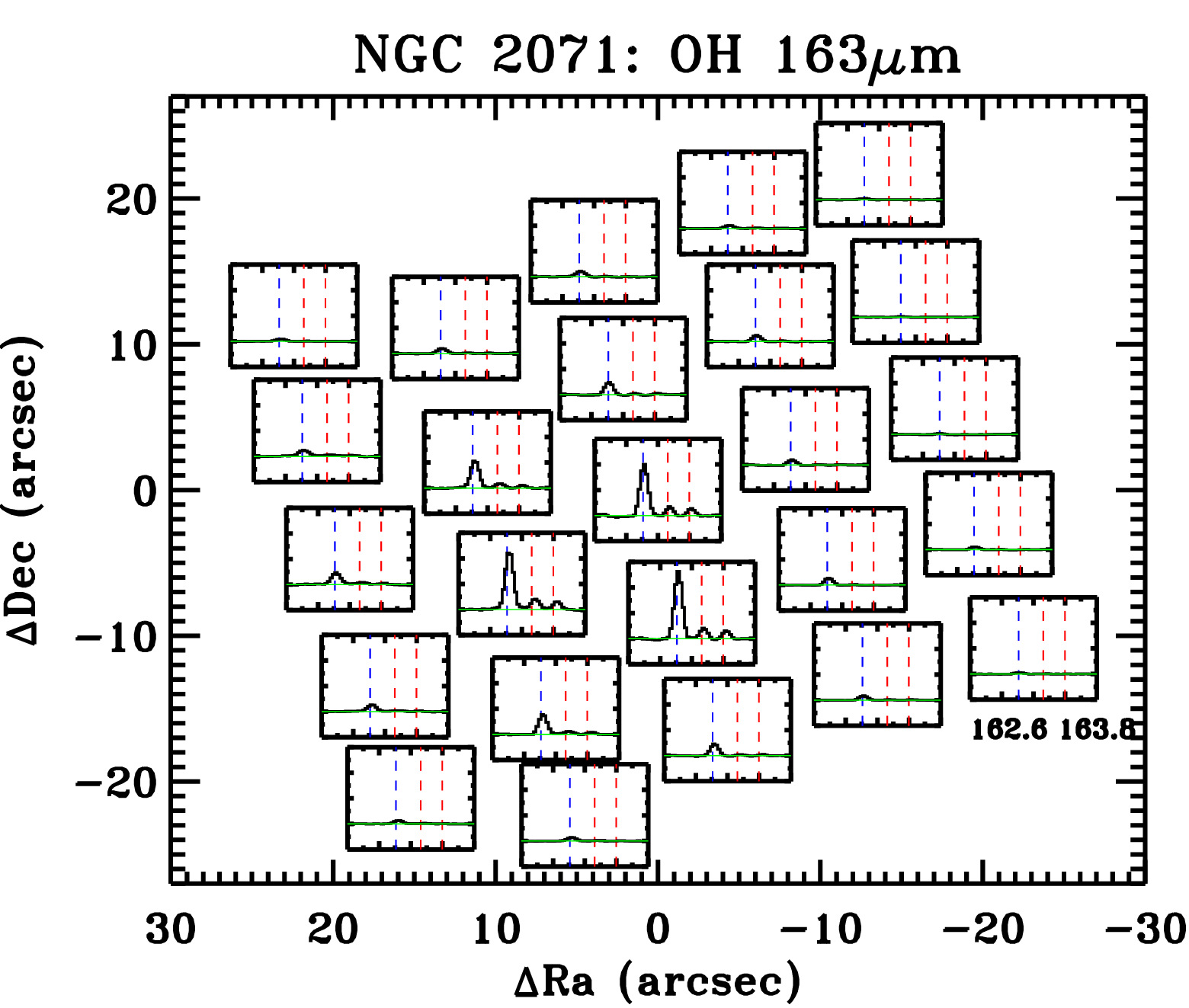}} \\
  \end{tabular}
  \caption{Maps of the OH emission from NGC~2071. The red and blue dashed lines indicate the rest wavelengths of the OH and CO transitions, respectively.}
 \label{fig:ngc2071maps}
\end{figure}
\end{landscape}
}

\def\placefigureVela17maps{
\begin{landscape}
\begin{figure}
 \centering
  \begin{tabular}{c@{\extracolsep{40pt}}c}
   \resizebox{0.35\hsize}{!}{\includegraphics[angle=0,bb=28 714 651 1280]{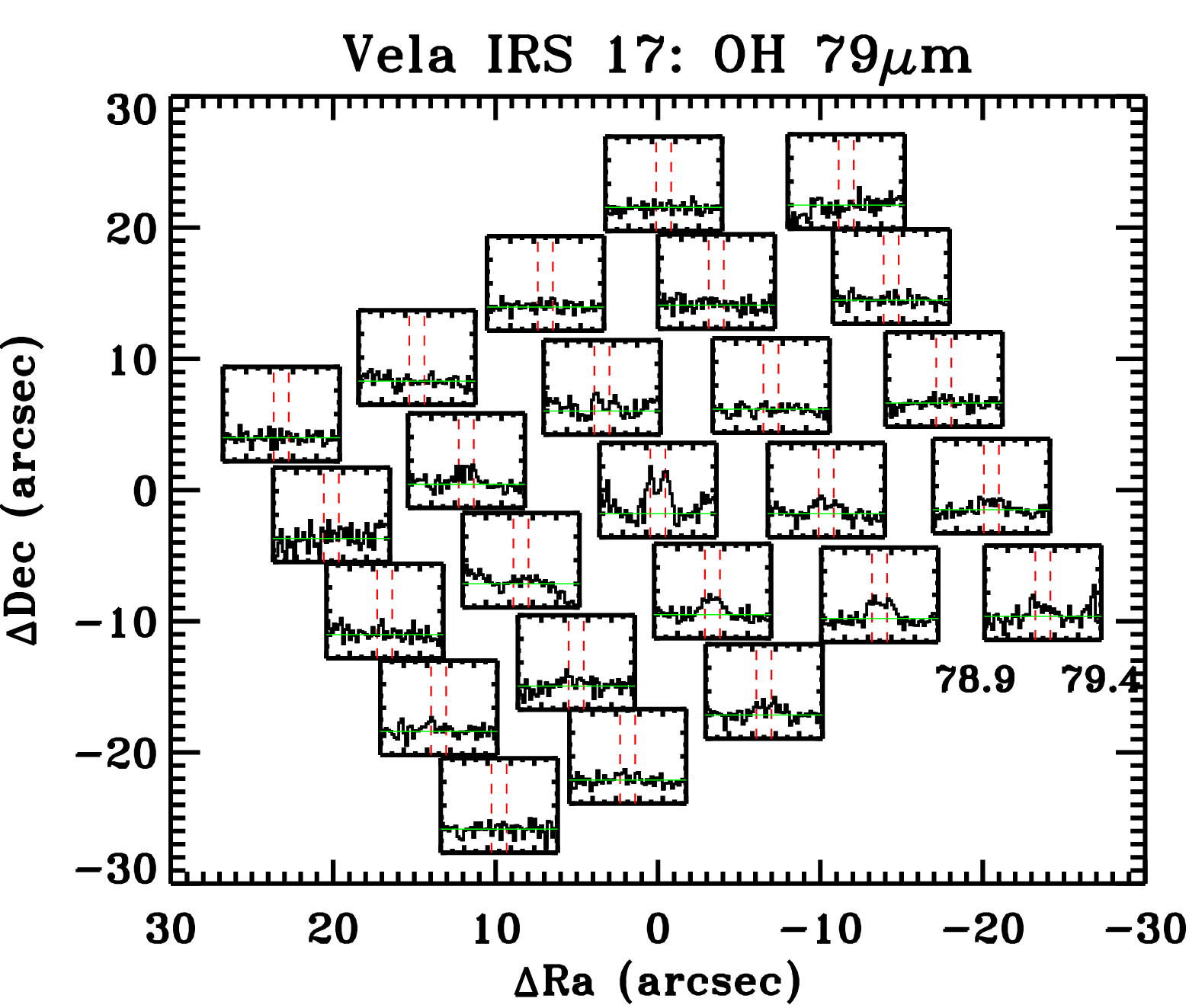}} &
   \resizebox{0.35\hsize}{!}{\includegraphics[angle=0,bb=28 714 651 1280]{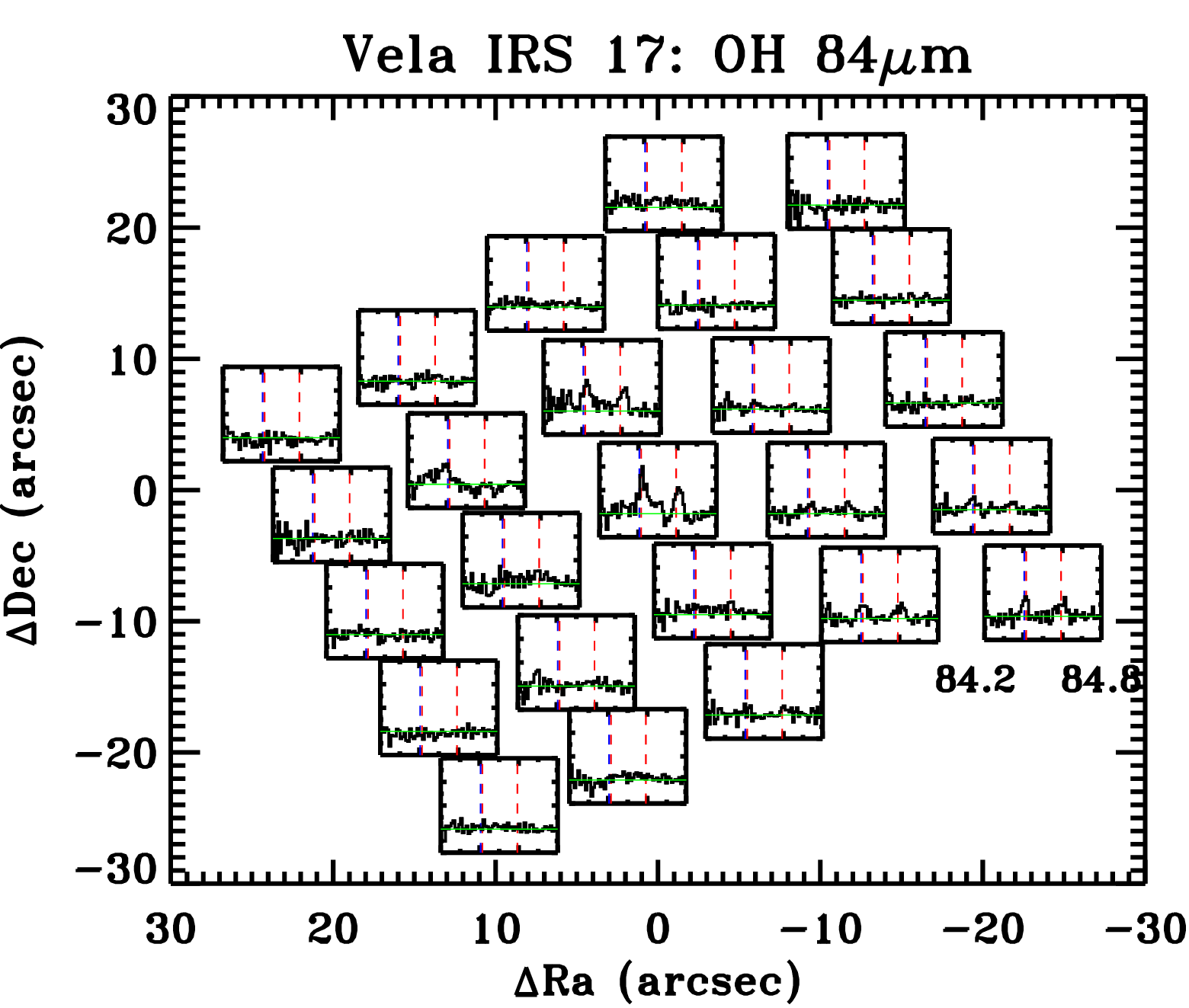}} \\
   & \\
   \resizebox{0.35\hsize}{!}{\includegraphics[angle=0,bb=28 714 651 1280]{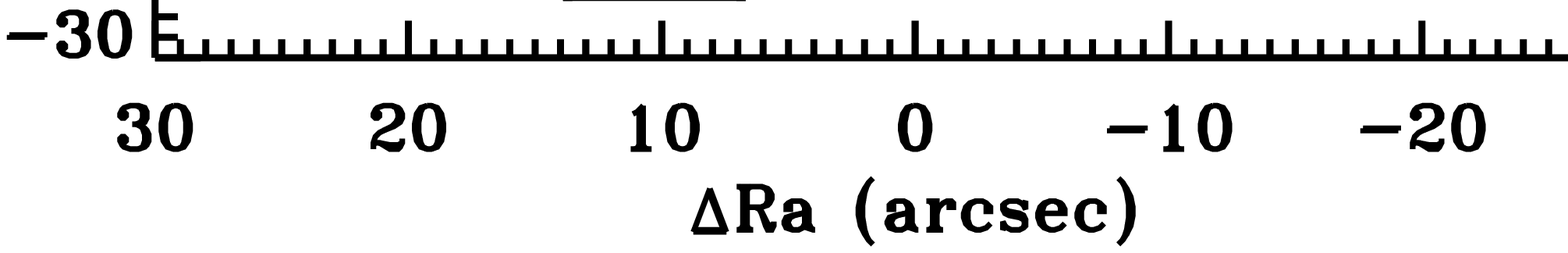}} &
   \resizebox{0.35\hsize}{!}{\includegraphics[angle=0,bb=28 714 651 1280]{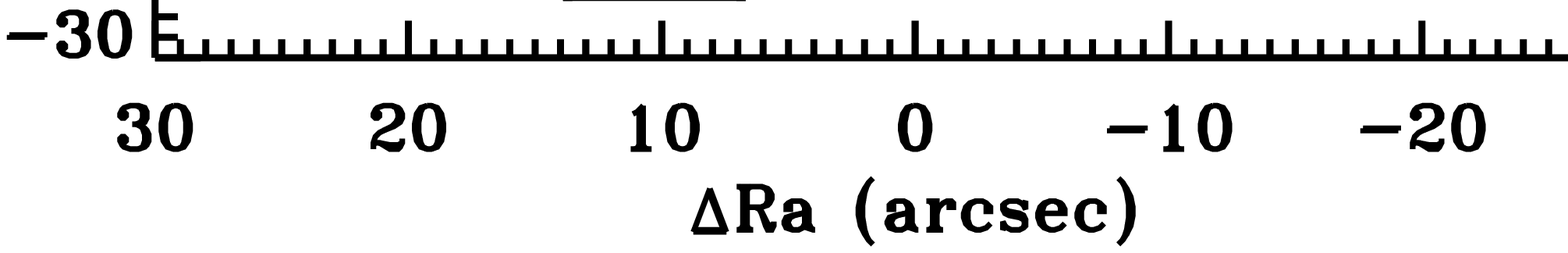}} \\
  \end{tabular}
  \caption{Maps of the OH emission from Vela~IRS~17. The red and blue dashed lines indicate the rest wavelengths of the OH and CO transitions, respectively.}
 \label{fig:vela17maps}
\end{figure}
\end{landscape}
}

\def\placefigurevela19maps{
\begin{landscape}
\begin{figure}
 \centering
  \begin{tabular}{c@{\extracolsep{40pt}}c}
   \resizebox{0.35\hsize}{!}{\includegraphics[angle=0,bb=28 714 651 1280]{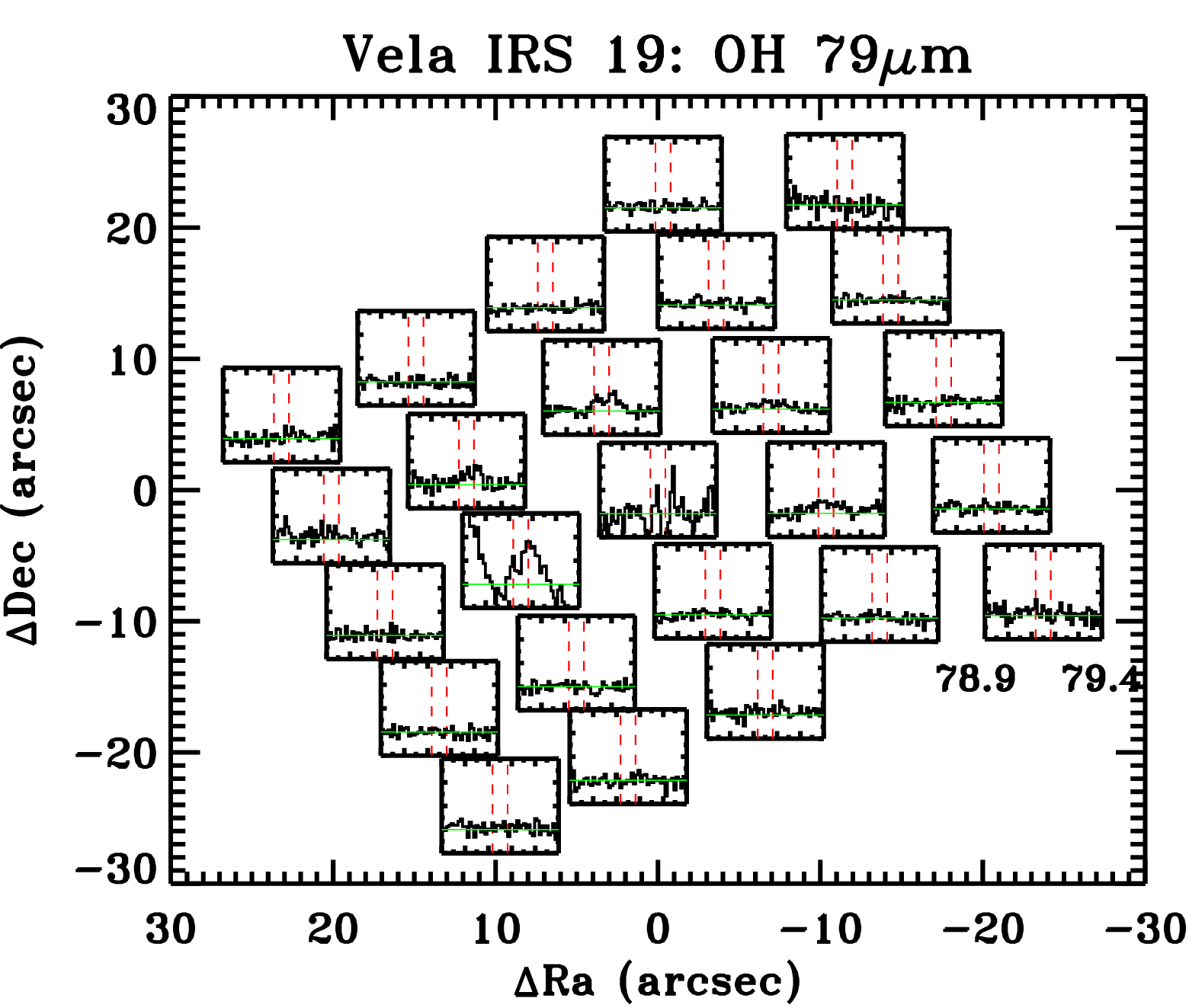}} &
   \resizebox{0.35\hsize}{!}{\includegraphics[angle=0,bb=28 714 651 1280]{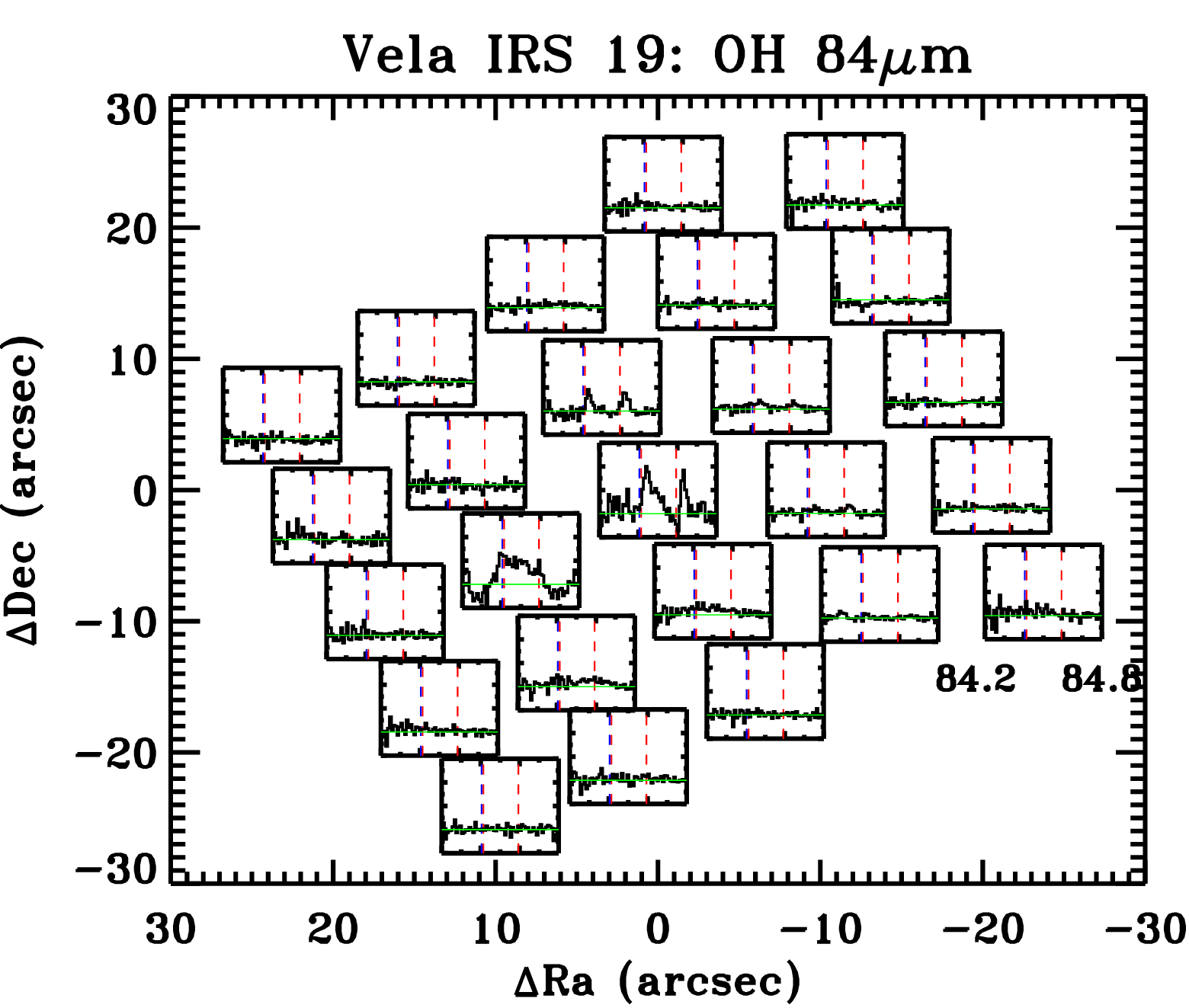}} \\
   & \\
   \resizebox{0.35\hsize}{!}{\includegraphics[angle=0,bb=28 714 651 1280]{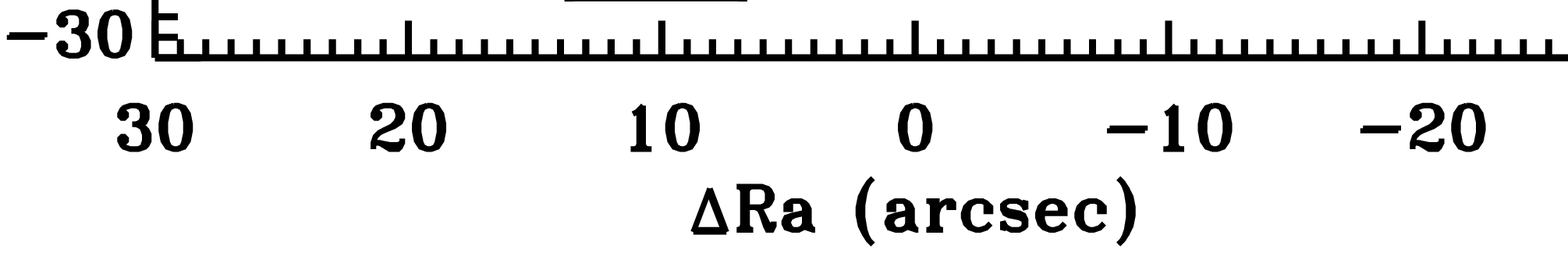}} &
   \resizebox{0.35\hsize}{!}{\includegraphics[angle=0,bb=28 714 651 1280]{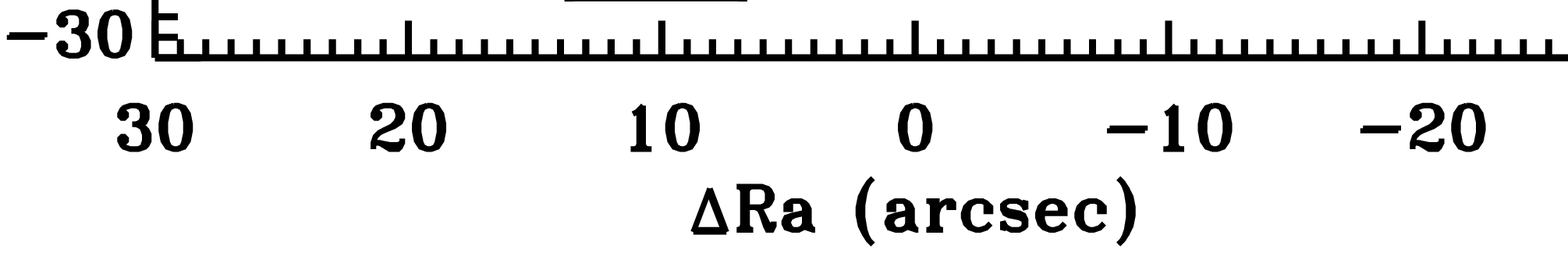}} \\
  \end{tabular}
  \caption{Maps of the OH emission from Vela~IRS~19. The red and blue dashed lines indicate the rest wavelengths of the OH and CO transitions, respectively.}
 \label{fig:vela19maps}
\end{figure}
\end{landscape}
}

\def\placefiguren7129maps{
\begin{landscape}
\begin{figure}
 \centering
  \begin{tabular}{c@{\extracolsep{40pt}}c}
   \resizebox{0.35\hsize}{!}{\includegraphics[angle=0,bb=28 714 651 1280]{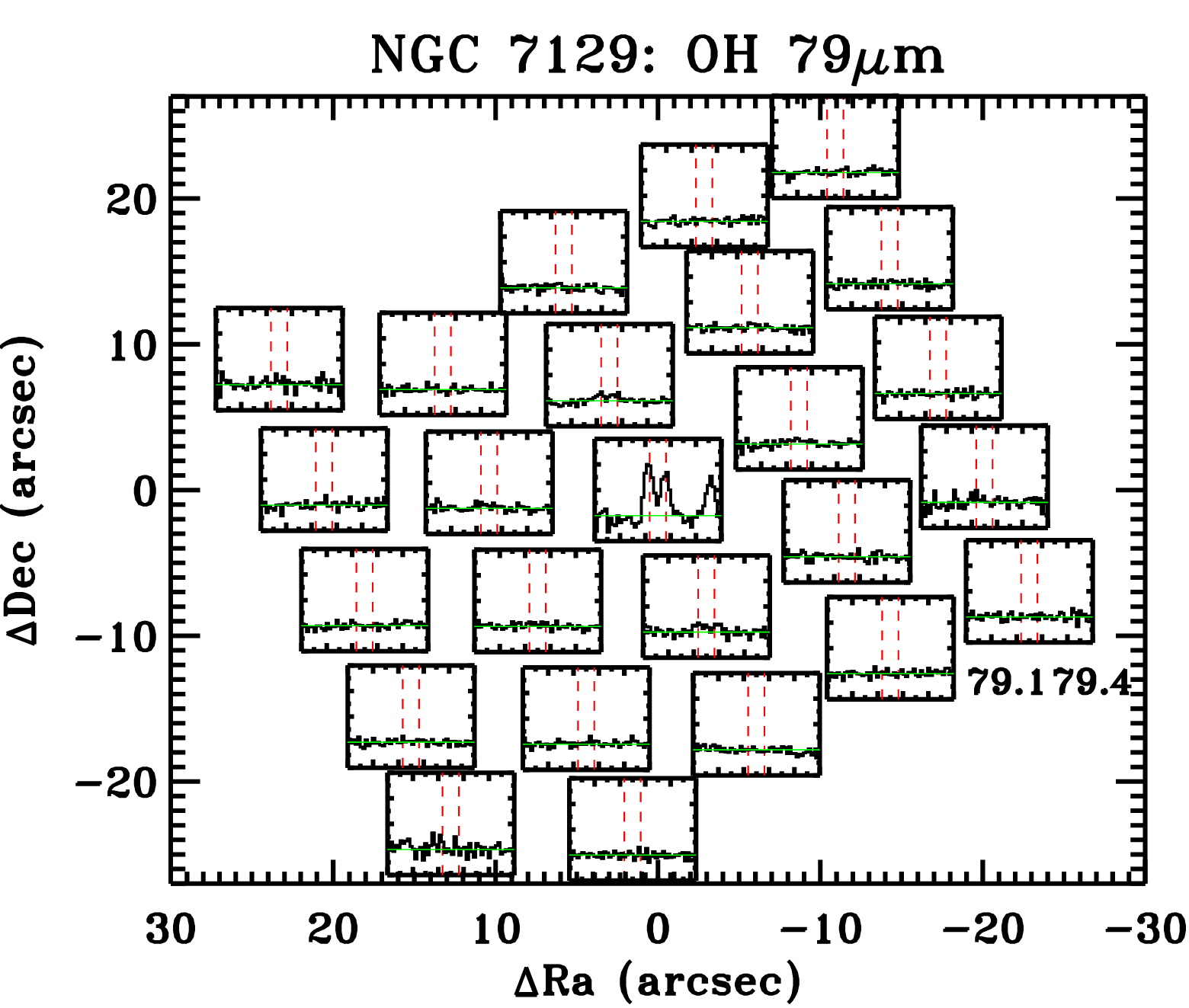}} &
   \resizebox{0.35\hsize}{!}{\includegraphics[angle=0,bb=28 714 651 1280]{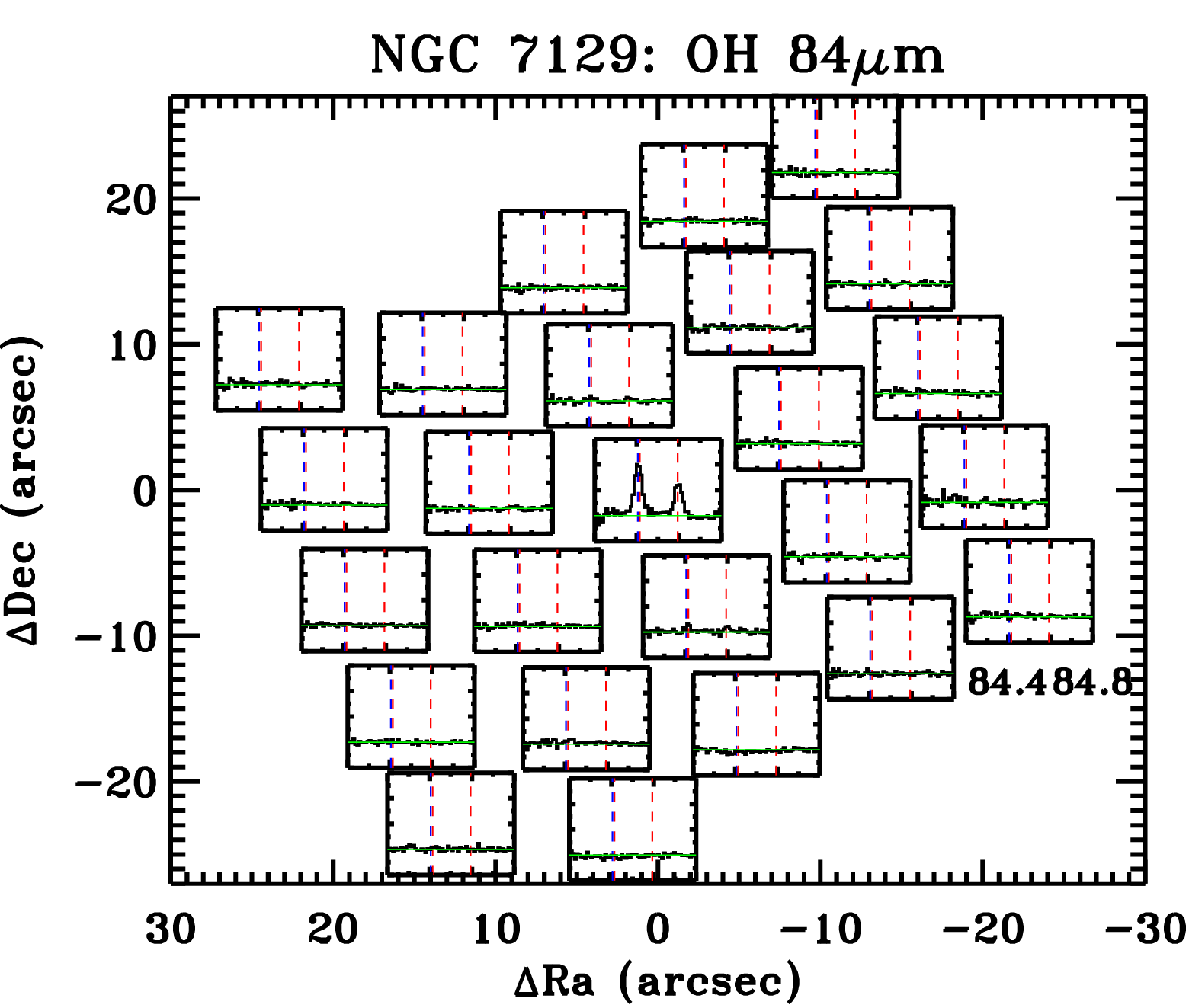}} \\
   & \\
   \resizebox{0.35\hsize}{!}{\includegraphics[angle=0,bb=28 714 651 1280]{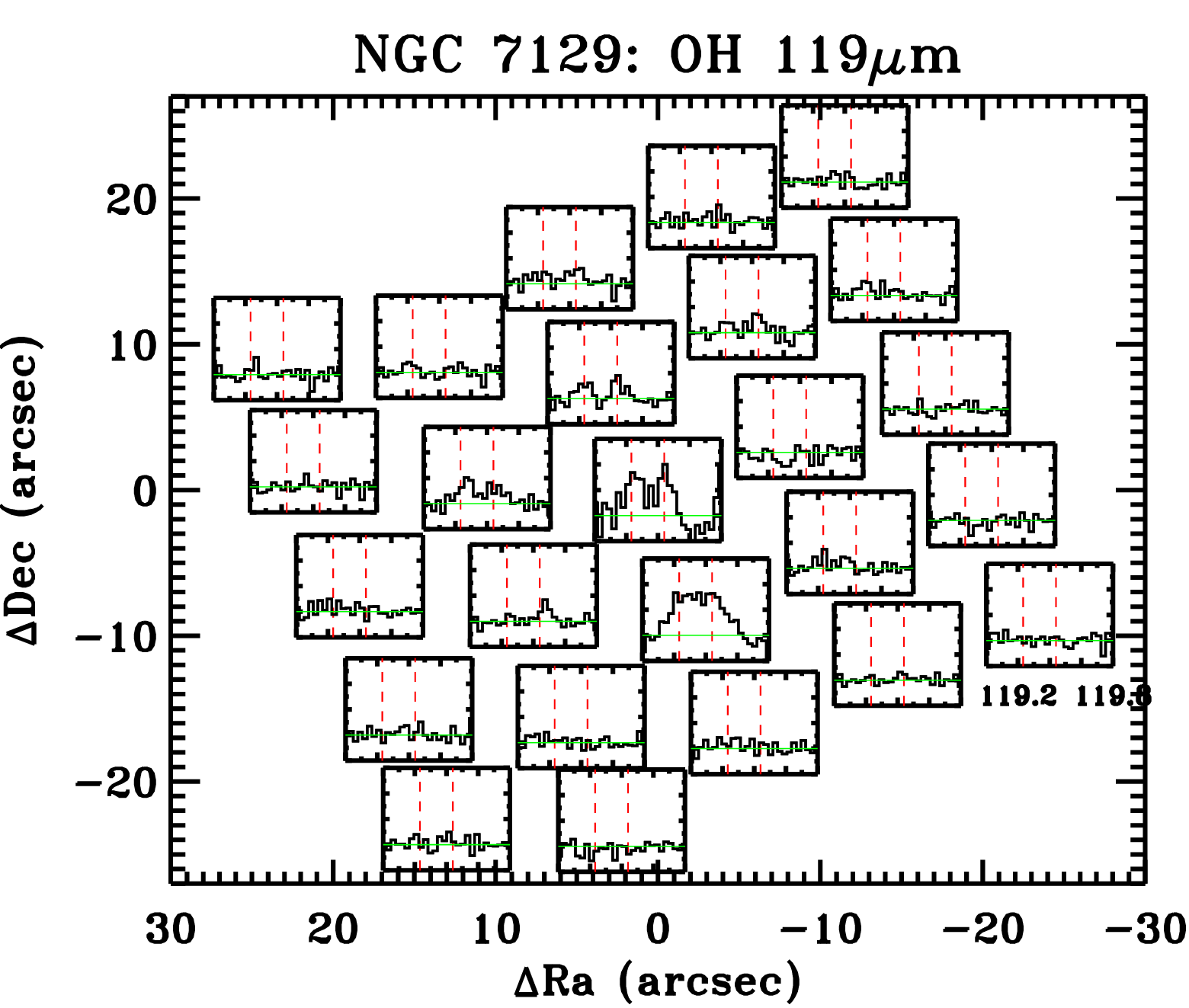}} &
   \resizebox{0.35\hsize}{!}{\includegraphics[angle=0,bb=28 714 651 1280]{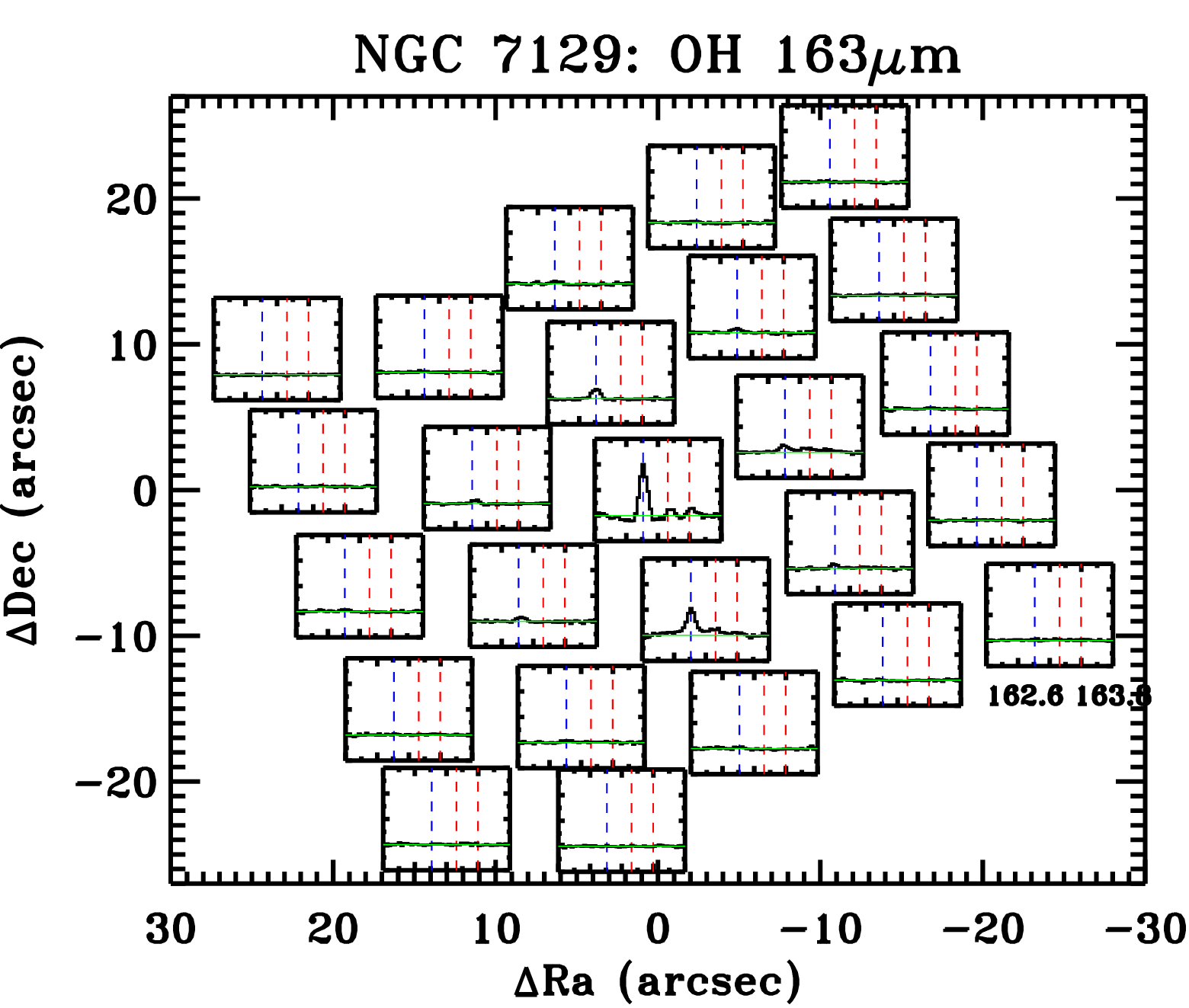}} \\
  \end{tabular}
  \caption{Maps of the OH emission from NGC~7129. The red and blue dashed lines indicate the rest wavelengths of the OH and CO transitions, respectively.}
 \label{fig:ngc7129maps}
\end{figure}
\end{landscape}
}

\def\placefigurl1641maps{
\begin{landscape}
\begin{figure}
 \centering
  \begin{tabular}{c@{\extracolsep{40pt}}c}
   \resizebox{0.35\hsize}{!}{\includegraphics[angle=0,bb=28 714 651 1280]{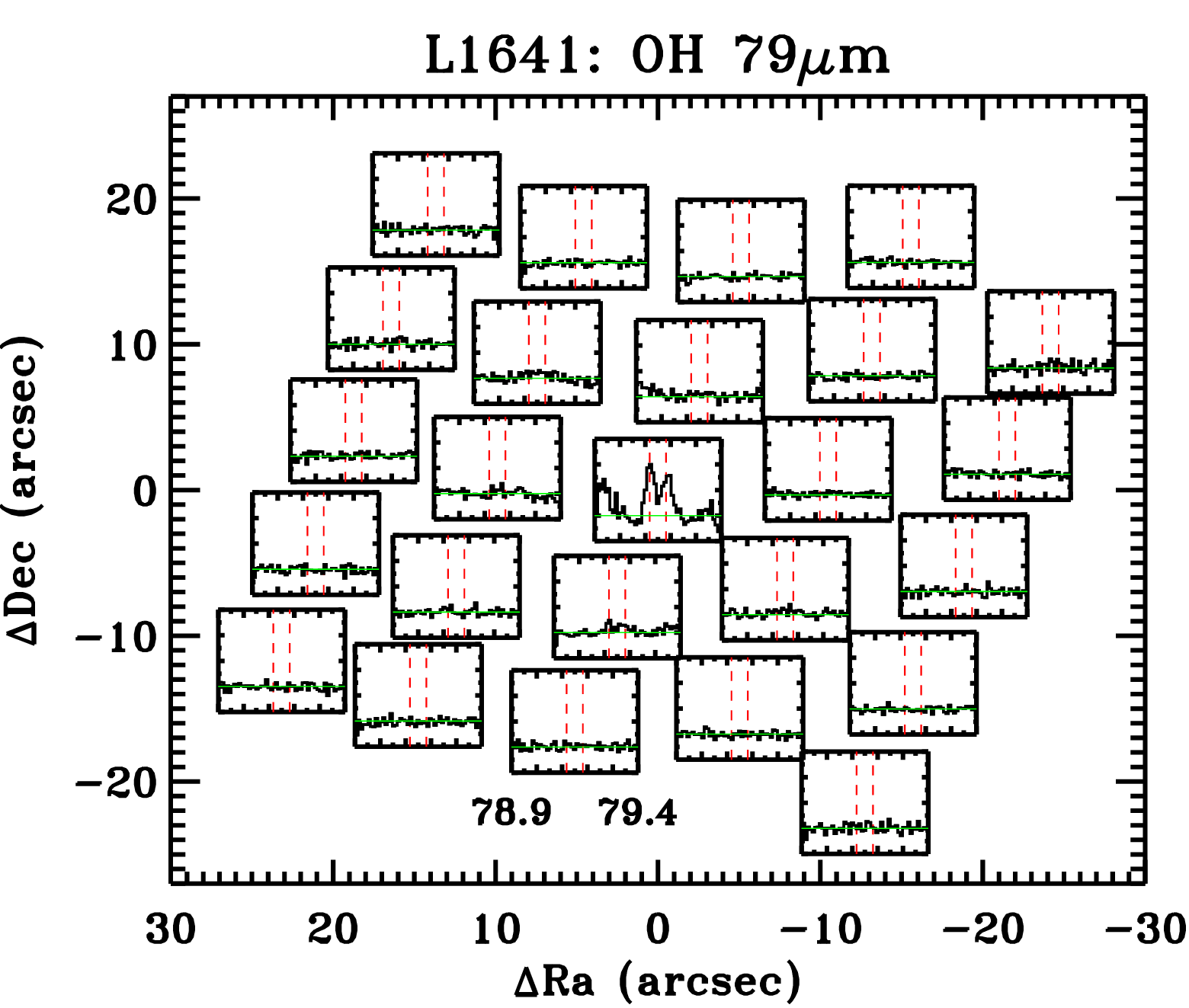}} &
   \resizebox{0.35\hsize}{!}{\includegraphics[angle=0,bb=28 714 651 1280]{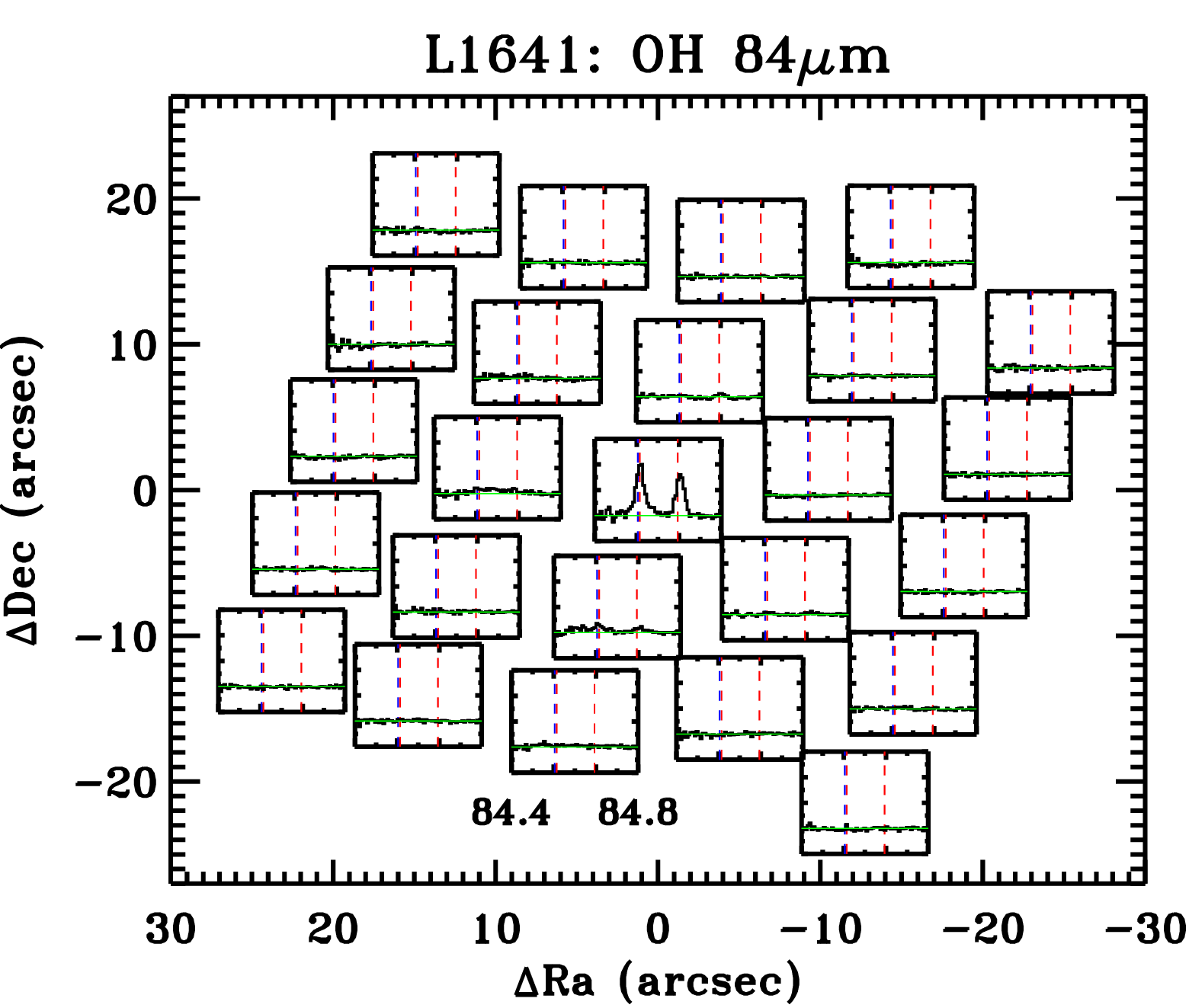}} \\
   & \\
   \resizebox{0.35\hsize}{!}{\includegraphics[angle=0,bb=28 714 651 1280]{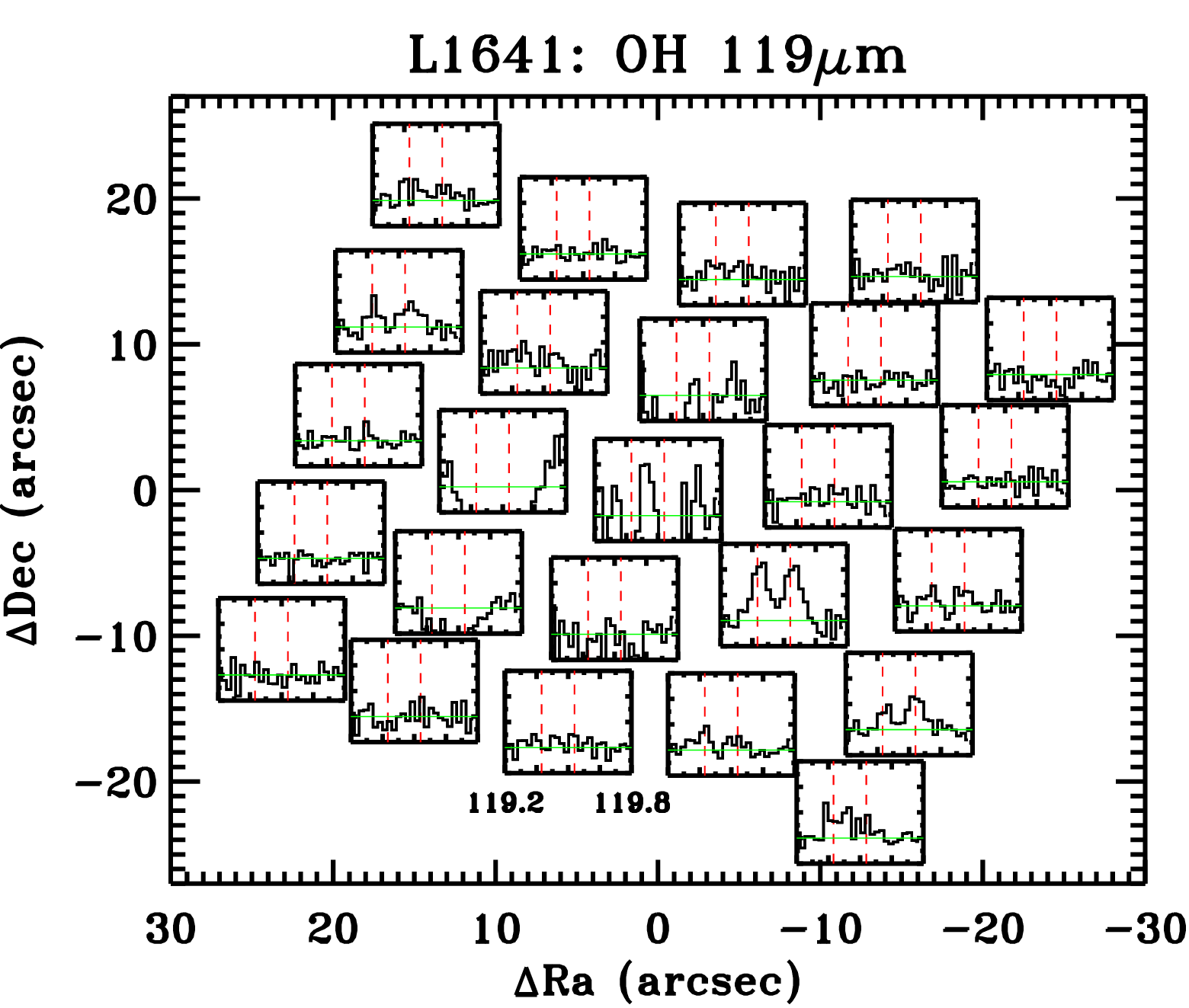}} &
   \resizebox{0.35\hsize}{!}{\includegraphics[angle=0,bb=28 714 651 1280]{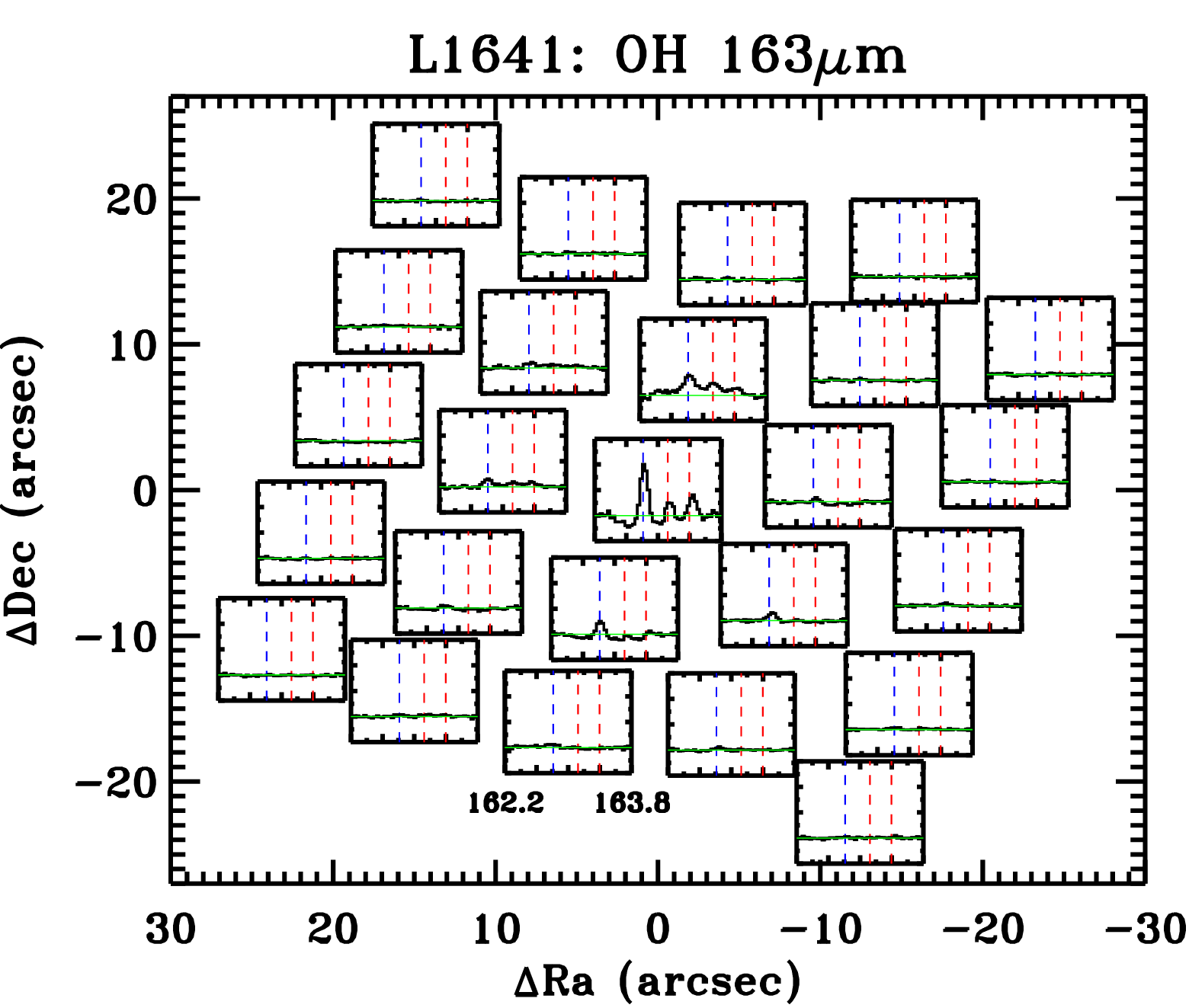}} \\
  \end{tabular}
  \caption{Maps of the OH emission from L~1641. The red and blue dashed lines indicate the rest wavelengths of the OH and CO transitions, respectively.}
 \label{fig:l1641maps}
\end{figure}
\end{landscape}
}

\def\placetableMoldata{
\begin{table*}
\caption{Molecular data from the LAMDA database \citep{Schoier05} for the OH transitions detected with PACS. Wavelengths marked with a star were only observed in the full range scans. Frequencies are rest frequencies. $A(B) \equiv A \times 10^B$.}
\begin{center}
\begin{tabular}{c c c c c c c}
\hline 
\hline
Transition      & Wavelength        & Frequency & $E_\mathrm{up}$ &$A_{\mathrm{ul}}$ & $g_\mathrm{u}$ & $g_\mathrm{l}$  \\
$\Omega$, J, P  & [$\mu\mathrm{m}$] & [GHz]     & [K]             &[s$^{-1}$]        &       &        \\ 
\hline
3/2-3/2, 9/2-7/2, $-$ - + & \phantom{1}65.13$^\star$ & 4602.9 & 512.1 & 1.276(+0)   &           10 & 8 \\
3/2-3/2, 9/2-7/2, + - $-$ & \phantom{1}65.28$^\star$ & 4592.5 & 510.9 & 1.267(+0)   &           10 & 8 \\
1/2-1/2, 7/2-5/2, $-$ - + & \phantom{1}71.17$^\star$ & 4212.3 & 617.6 & 1.014(+0)   & \phantom{1}8 & 6 \\ 
1/2-1/2, 7/2-5/2, + - $-$ & \phantom{1}71.22$^\star$ & 4209.7 & 617.9 & 1.012(+0)   & \phantom{1}8 & 6 \\ 
1/2-3/2, 3/2-5/2, + - $-$ & \phantom{1}96.31$^\star$ & 3114.0 & 270.2 & 9.270($-$3) & \phantom{1}4 & 6 \\
1/2-3/2, 3/2-5/2, $-$ - + & \phantom{1}96.37$^\star$ & 3111.1 & 269.8 & 9.250($-$3) & \phantom{1}4 & 6 \\
1/2-1/2, 5/2-3/2, + - $-$ & \phantom{1}98.74$^\star$ & 3036.3 & 415.5 & 3.530($-$1) & \phantom{1}6 & 4 \\
1/2-1/2, 5/2-3/2, $-$ - + & \phantom{1}98.76$^\star$ & 3035.4 & 415.9 & 3.531($-$1) & \phantom{1}6 & 4 \\
1/2-3/2, 1/2-3/2, $-$ - + & \phantom{1}79.12         & 3789.3 & 181.9 & 3.606($-$2) & \phantom{1}2 & 4 \\
1/2-3/2, 1/2-3/2, + - $-$ & \phantom{1}79.18         & 3786.3 & 181.7 & 3.598($-$2) & \phantom{1}2 & 4 \\
3/2-3/2, 7/2-5/2, + - $-$ & \phantom{1}84.42         & 3551.2 & 291.2 & 5.235($-$1) & \phantom{1}8 & 6 \\
3/2-3/2, 7/2-5/2, $-$ - + & \phantom{1}84.60         & 3543.8 & 290.5 & 5.202($-$1) & \phantom{1}8 & 6 \\
3/2-3/2, 5/2-3/2, $-$ - + &           119.23         & 2514.3 & 120.7 & 1.388($-$1) & \phantom{1}6 & 4 \\
3/2-3/2, 5/2-3/2, + - $-$ &           119.44         & 2510.0 & 120.5 & 1.380($-$1) & \phantom{1}6 & 4 \\
1/2-1/2, 3/2-1/2, + - $-$ &           163.12         & 1837.8 & 270.1 & 6.483($-$2) & \phantom{1}4 & 2 \\
1/2-1/2, 3/2-1/2, $-$ - + &           163.40         & 1834.7 & 269.8 & 6.450($-$2) & \phantom{1}4 & 2 \\
\hline
\end{tabular}
\end{center}
\label{tab:moldata}
\end{table*}
}

\def\placetablesourceprops{
\begin{table*}
\caption{Coordinates, distance, bolometric luminosity, and envelope masses of the low-mass class 0, class I, and intermediate-mass protostars in the sample.}
\centering
\begin{tabular}{l@{\extracolsep{2pt}}cccccccc}
\hline \hline
Source           & RA         & Dec                                & Class & $d$    & $L_\mathrm{bol}$ & $M_\mathrm{env}$  \\ 
                 & [h m s]    & [${}^{\circ}~{\arcmin}~{\arcsec}$] &       & [pc]   & [L$_{\odot}$]  & [M$_{\odot}$]          \\
\hline
NGC~1333~IRAS~2A & 03:28:55.6 &   +31:14:37.1                      & 0     & \phantom{0}235${}^{a}$ &\phantom{00}37.0${}^{b}$  & \phantom{0}5.1${}^{c}$      \\
NGC~1333~IRAS~4A & 03:29:10.5 &   +31:13:30.9                      & 0     & \phantom{0}235${}^{a}$ &\phantom{000}8.5${}^{b}$  & \phantom{0}5.6${}^{c}$ \\
NGC~1333~IRAS~4B & 03:29:12.0 &   +31:13:08.1                      & 0     & \phantom{0}235${}^{a}$ &\phantom{000}4.9${}^{b}$  & \phantom{0}3.0${}^{c}$ \\
L~1527           & 04:39:53.9 &   +26:03:09.8                      & 0     & \phantom{0}140${}^{d}$ &\phantom{000}1.8${}^{b}$  & \phantom{0}0.9${}^{c}$ \\
Ced110~IRS4      & 11:06:47.0 & $-$77:22:32.4                      & 0     & \phantom{0}125${}^{e}$ &\phantom{000}1.0${}^{b}$  & \phantom{0}0.2${}^{c}$ \\
IRAS~15398/B228  & 15:43:01.3 & $-$34:09:15.0                      & 0     & \phantom{0}130${}^{e}$ &\phantom{000}1.2${}^{b}$  & \phantom{0}0.5${}^{c}$ \\
L~483~mm         & 18:17:29.9 & $-$04:39:39.5                      & 0     & \phantom{0}200${}^{d}$ &\phantom{00}10.8${}^{b}$  & \phantom{0}4.4${}^{c}$ \\
Ser~SMM1         & 18:29:49.8 &   +01:15:20.5                      & 0     & \phantom{0}230${}^{f}$ &\phantom{00}31.7${}^{b}$  & 
16.1${}^{c}$     \\
Ser~SMM3         & 18:29:59.2 &   +01:14:00.3                      & 0     & \phantom{0}230${}^{f}$ &\phantom{00}11.0${}^{b}$  & \phantom{0}3.2${}^{c}$ \\
L~723~mm         & 19:17:53.7 &   +19:12:20.0                      & 0     & \phantom{0}300${}^{d}$ &\phantom{000}3.9${}^{b}$  & \phantom{0}1.3${}^{c}$ \\
\hline
L~1489           & 04:04:43.0 &   +26:18:57.0                      & I     & \phantom{0}140${}^{g}$ &\phantom{000}3.7${}^{b}$  & \phantom{0}0.2${}^{c}$      \\
TMR~1            & 04:39:13.7 &   +25:53:21.0                      & I     & \phantom{0}140${}^{g}$ &\phantom{000}3.2${}^{b}$  & \phantom{0}0.2${}^{c}$      \\
TMC~1A           & 04:39:34.9 &   +25:41:45.0                      & I     & \phantom{0}140${}^{g}$ &\phantom{000}2.7${}^{b}$  & \phantom{0}0.2${}^{c}$      \\
TMC~1            & 04:41:12.4 &   +25:46:36.0                      & I     & \phantom{0}140${}^{g}$ &\phantom{000}0.8${}^{b}$  & \phantom{0}0.2${}^{c}$      \\
HH~46            & 08:25:43.9 & $-$51:00:36.0                      & I     & \phantom{0}450${}^{h}$ &\phantom{00}25.8${}^{b}$  & \phantom{0}4.4${}^{c}$      \\
IRAS~12496/DK~Cha& 12:53:17.2 & $-$77:07:10.6                      & I     & \phantom{0}178${}^{i}$ &\phantom{00}35.3${}^{b}$  & \phantom{0}0.8${}^{c}$      \\
RNO~91           & 16:34:29.3 & $-$15:47:01.4                      & I     & \phantom{0}125${}^{j}$ &\phantom{000}2.4${}^{b}$  & \phantom{0}0.5${}^{c}$      \\
\hline
AFGL~490         & 03:27:38.4 &   +58:47:08.0                      & IM    &           1000${}^{k}$ &          2000\phantom{.0}${}^{l}$ & 45\phantom{.0}${}^{m}$               \\
NGC~2071         & 05:47:04.4 &   +00:21:49.0                      & IM    & \phantom{0}422${}^{n}$ &\phantom{0}520\phantom{.0}${}^{o}$ & 30\phantom{.0}${}^{p}$               \\
Vela~IRS~17      & 08:46:34.7 & $-$43:54:30.5                      & IM    & \phantom{0}700${}^{q}$ &\phantom{0}715\phantom{.0}${}^{r}$ & 
\phantom{0}6.4 ${}^{s}$                \\
Vela~IRS~19      & 08:48:48.5 & $-$45:32:29.0                      & IM    & \phantom{0}700${}^{q}$ &\phantom{0}776\phantom{.0}${}^{q}$ & \phantom{0}3.5 ${}^{s}$                 \\
NGC~7129~FIRS~2  & 21:43:01.7 &   +66:03:23.6                      & IM    & 1250${}^{t}$           &\phantom{0}430\phantom{.0}${}^{u}$ & 
50\phantom{.0} ${}^{v}$                 \\
L1641~S3MMS1     & 05:39:55.9 & $-$07:30:28.0                      & IM    & \phantom{0}465${}^{n}$ &\phantom{00}70\phantom{.0}${}^{w}$ & 
20.9 ${}^{x}$        \\
\hline
\end{tabular}
\begin{flushleft}
\footnotesize{References: 
${}^{a}$ \citet{Hirota08}, 
${}^{b}$ Values from \citet{Kristensen12} using PACS data from the WISH and DIGIT key programs (Karska et al. in prep, Green et al. in prep.),
${}^{c}$ Derived from DUSTY modeling of the sources as described in \citet{Kristensen12}, 
${}^{d}$ \citet{Andre00}, 
${}^{e}$~\citet{Knude98}, 
${}^{f}$ \citet{Eiroa08}. Based on VLBA measurements of a star thought to be associated with the Serpens cluster, \citet{Dzib10} inferred a distance of 415~pc. 
${}^{g}$ \citet{Kenyon08}, 
${}^{h}$ \citet{Heathcote96}, 
${}^{i}$ \citet{Whittet97}, 
${}^{j}$ \citet{deGeus89}, 
${}^{k}$ \citet{Snell84}, 
${}^{l}$ \citet{Mozurkewich86}, 
${}^{m}$ \citet{Schreyer02} derived an envelope mass of 40-50~M$_{\odot}$, 
${}^{n}$ \citet{Wilson05}, 
${}^{o}$ \citet{Butner90}, 
${}^{p}$ \citet{Johnstone01}, 
${}^{q}$ \citet{Liseau92}, 
${}^{r}$ \citet{Giannini05}, 
${}^{s}$ \citet{Massi99}, 
${}^{t}$ \citet{Shevchenko89},
${}^{u}$~\citet{Johnstone10}, 
${}^{v}$ \citet{Crimier10}, 
${}^{w}$ \citet{Stanke00}}
${}^{x}$ \citet{vanKempen12}
\end{flushleft}
\label{tab:source_props}
\end{table*} 
}

\def\placetablesourceobsids{
\begin{table*}[!h]
\caption{Observing modes, identities, and pipeline versions for the sources sample.}
\centering
\begin{tabular}{l@{\extracolsep{2pt}}lcccc}
\hline \hline \noalign{\smallskip} 
Source           & Mode  & Setting 1  & Pipeline          & Setting 2  & Pipeline         \\
\hline
NGC~1333~IRAS~2A & Range & 1342190686 & v6  & 1342191149 & v6 \\
NGC~1333~IRAS~4A & Line  & 1342216082 & v8  & 1342216083 & v8  \\
NGC~1333~IRAS~4A & Range & 1342216084 & v5  & 1342216085 & v5 \\
NGC~1333~IRAS~4B & Line  & 1342216180 & v8  & 1342216179 & v8 \\
NGC~1333~IRAS~4B & Range & 1342216178 & v6  & 1342216177 & v6 \\
L~1527           & Line  & 1342225836 & v8  & 1342192983 & v8 \\
Ced110~IRS4      & Line  & 1342210188 & v8  & 1342210187 & v8 \\
IRAS~15398/B228  & Line  & 1342191302 & v8  & 1342191301 & v8 \\
L~483~mm         & Line  & 1342217806 & v8  & 1342192156 & v8 \\
Ser~SMM1         & Range & 1342207781 & v6  & 1342207780 & v6 \\
Ser~SMM3         & Line  & 1342207778 & v8  & 1342207779 & v8 \\
L~723~mm         & Line  & 1342208919 & v8  & 1342208918 & v8 \\
\hline
L~1489           & Line  & 1342214361 & v8  & 1342214360 & v5 \\ 
TMR~1            & Line  & 1342225835 & v8  & 1342225834 & v8 \\
TMC~1A           & Line  & 1342225833 & v8  & 1342225832 & v8 \\
TMC~1            & Line  & 1342225831 & v8  & 1342225830 & v8 \\
HH~46            & Line  & 1342186315 & v8  & 1342186316 & v8 \\
DK~Cha (DIGIT)   & Range & 1342188039 & v6  & 1342188040 & v6 \\
RNO~91           & Line  & 1342215640 & v8  & 1342215639 & v8 \\
\hline
AFGL~490         & Line  & 1342202582 & v8  & 1342191353 & v8 \\
NGC~2071         & Line  & 1342218760 & v8  & 1342218761 & v8 \\
Vela~IRS~17      & Line  & 1342209407 & v8  & 1342211844 & v8 \\
Vela~IRS~19      & Line  & 1342210189 & v8  & 1342210190 & v8 \\
NGC~7129~FIRS~2  & Line  & 1342186321 & v8  & 1342186322 & v8 \\
L1641 S3MMS1     & Line  & 1342226194 & v8  & 1342226195 & v8 \\
\hline
\end{tabular}
\label{tab:source_obsids}
\end{table*}
}

\def\placetablecorrfc{
\begin{landscape}
\begin{table}
\caption{Resulting fluxes in units of $10^{-16}~\mathrm{W}~\mathrm{m}^{-2}$ in the spaxel with the strongest continuum (assumed to contain the central source position), corrected by the spillover factor (0.70 at $79~\mu\mathrm{m}$, 0.69 at $84~\mu\mathrm{m}$, 0.62 at $119~\mu\mathrm{m}$, and 0.49 at $163~\mu\mathrm{m}$).}
\centering
\begin{tabular}{l@{\extracolsep{2pt}}llllllll}
\hline \hline
Source           & \multicolumn{8}{c}{OH transition $[\mu\mathrm{m}]$}\\
                 & 79.12 & 79.18 & 84.42 & 84.60 & 119.23 & 119.44 & 163.12 & 163.40\\ 
\hline
NGC~1333~IRAS~2A (scan) & - & - & - & - & abs. & abs. & - & -\\
NGC~1333~IRAS~4A        & - & - & $0.60\pm0.08$ & $0.35\pm0.07$ & [$0.26\pm0.10$] & ($0.37\pm0.10$) & $0.26\pm0.05$ & $0.56\pm0.05$ \\
NGC~1333~IRAS~4B & $1.03\pm0.08$ & $0.77\pm0.07$ & $1.96\pm0.08$ & $1.18\pm0.07$ & $1.45\pm0.07$ & $1.57\pm0.06$ & $0.33\pm0.04$ & $0.35\pm0.04$ \\
L~1527           & $0.33\pm0.05$ & $0.27\pm0.05$ & $0.50\pm0.05$ & $0.49\pm0.06$ & $0.60\pm0.03$ & $0.71\pm0.03$ & [$0.07\pm0.03$] & ($0.12\pm0.03$)\\
Ced110~IRS4      & $0.31\pm0.06$ & ($0.23\pm0.06$) & $0.53\pm0.06$ & $0.47\pm0.06$ & ($0.16\pm0.04$) & $0.25\pm0.06$ & $0.16\pm0.03$ & - \\
IRAS~15398/B228  & $0.36\pm0.06$ & $0.46\pm0.06$ & $0.46\pm0.06$ & $0.43\pm0.06$ & $0.50\pm0.04$ & $0.59\pm0.04$ & - & [$0.13\pm0.05$] \\ 
L~483~mm         & $1.05\pm0.09$ & $1.07\pm0.09$ & $1.17\pm0.09$ & $1.04\pm0.09$ & $0.78\pm0.04$ & $0.95\pm0.05$ & $0.25\pm0.03$ & $0.22\pm0.03$\\
Ser~SMM1 (scan)  & $5.71\pm0.69$ & blend & $8.57\pm0.44$ & $6.25\pm0.44$ & $3.70\pm0.52$ & $3.29\pm0.51$ & $2.39\pm0.34$ & $1.67\pm0.36$ \\
Ser~SMM3         & ($0.44\pm0.11$) & ($0.36\pm0.11$) & $0.63\pm0.06$ & $0.42\pm0.05$ & - & - & - & - \\
L~723~mm         & -             & [$0.17\pm0.06$]  & $0.61\pm0.06$ & $0.41\pm0.05$ & ($0.12\pm0.03$) & ($0.13\pm0.04$) & - & ($0.14\pm0.03$) \\
\hline
L~1489           & $0.79\pm0.07$ & $0.77\pm0.06$ & $1.03\pm0.07$ & $0.94\pm0.08$ & $0.52\pm0.03$ & $0.67\pm0.02$ & ($0.14\pm0.04$) & ($0.15\pm0.04$) \\
TMR~1            & $0.41\pm0.08$ & $0.73\pm0.09$ & $0.80\pm0.10$ & $0.88\pm0.10$ & ($0.13\pm0.04$) & ($0.19\pm0.04$) & $0.34\pm0.03$ & $0.27\pm0.04$ \\
TMC~1A           & $0.18\pm0.03$ & ($0.14\pm0.03$) & $0.72\pm0.09$ & $0.71\pm0.09$ & (abs.) & (abs.) & ($0.10\pm0.03$) & - \\ 
TMC~1            & $0.61\pm0.05$ & $0.65\pm0.05$ & $0.89\pm0.07$ & $1.01\pm0.07$ & $0.52\pm0.04$ & $0.60\pm0.04$ & $0.22\pm0.04$ & ($0.19\pm0.04$)\\
HH~46            & $0.63\pm0.09$ & $0.49\pm0.09$ & $0.87\pm0.08$ & $0.91\pm0.09$ & $0.22\pm0.04$ & $0.34\pm0.04$ & ($0.19\pm0.06$) & - \\
IRAS~12496/DK~Cha (scan) & $1.18\pm0.55$ & blend & $2.85\pm0.34$ & $2.44\pm0.30$ & - & - & - & - \\
RNO~91           & $0.61\pm0.11$ & blend & ($0.36\pm0.07$) & ($0.40\pm0.08$) & - & [$0.08\pm0.03$] & - & - \\
\hline
AFGL~490         & (abs.) & (abs.) & - & (abs.) & abs. & abs. & ($0.31\pm0.08$) & ($0.36\pm0.08$) \\ 
NGC~2071         & $10.65\pm0.29$ & $8.83\pm0.32$ & $16.57\pm0.56$ & $10.97\pm0.52$ & $6.40\pm0.24$ & $6.96\pm0.22$ & $4.09\pm0.17$ & $3.22\pm0.16$ \\
Vela~IRS~17      & $1.15\pm0.07$ & $1.15\pm0.07$ & $1.18\pm0.11$ & $0.61\pm0.10$ & abs. & abs. & - & - \\
Vela~IRS~19      & [$0.71\pm0.31$] & ($1.26\pm0.26$) & $1.75\pm0.40$ & $1.01\pm0.36$ & abs. & (abs.) & $0.78\pm0.06$ & $0.85\pm0.06$ \\
NGC~7129~FIRS~2  & $1.91\pm0.13$ & $1.56\pm0.14$ & $3.19\pm0.15$ & $1.70\pm0.13$ & - & - & $0.55\pm0.07$ & $0.67\pm0.08$ \\
L1641 S3MMS1     & $1.78\pm0.18$ & $1.94\pm0.19$ & $4.72\pm0.15$ & $3.38\pm0.15$ & (abs.) & (abs.) & $0.79\pm0.10$ & $1.17\pm0.10$ \\
\hline
\end{tabular}
\label{tab:fluxes1x1corr}
\end{table} 
\end{landscape}
}

\def\placetablef33{
\begin{landscape}
\begin{table}
\caption{Resulting fluxes in units of $10^{-16}~\mathrm{W}~\mathrm{m}^{-2}$ in the 3x3 central spaxels.}
\centering
\begin{tabular}{l@{\extracolsep{2pt}}llllllll}
\hline \hline
Source           & \multicolumn{8}{c}{OH transition $[\mu\mathrm{m}]$}\\
                 & 79.12 & 79.18 & 84.42 & 84.60 & 119.23 & 119.44 & 163.12 & 163.40\\ 
\hline
NGC~1333~IRAS~2A (scan) & - & - & - & - & abs. & abs. & - & -\\
NGC~1333~IRAS~4A & $0.73\pm0.13$ & ($0.64\pm0.13$) & $1.72\pm0.15$ & ($0.48\pm0.14$) & - & $0.94\pm0.13$ & $0.62\pm0.04$ & $0.52\pm0.04$ \\ NGC~1333~IRAS~4B & $2.17\pm0.14$ & $1.64\pm0.11$ & $4.72\pm0.12$ & $2.47\pm0.11$ & $2.24\pm0.08$ & $2.78\pm0.07$ & 0.82$\pm0.07$ &0.50$\pm0.07$ \\
L~1527           & $0.40\pm0.08$ & $0.42\pm0.08$ & ($0.41\pm0.12$) & - & $0.40\pm0.07$ & $0.52\pm0.07$ & - & ($0.12\pm0.04$)\\
Ced110~IRS4      & - & - & $0.77\pm0.14$ & ($0.59\pm0.14$) & - & - & - & - \\
IRAS~15398/B228  & $0.63\pm0.14$ & $0.65\pm0.14$ & $1.18\pm0.13$ & $0.58\pm0.12$ & $0.73\pm0.09$ & $0.69\pm0.08$ & - & - \\
L~483~mm         & $1.17\pm0.18$ & $1.15\pm0.18$ & $1.46\pm0.14$ & $0.88\pm0.14$ & $0.36\pm0.06$ & $0.65\pm0.07$ & $0.31\pm0.04$ & ($0.17\pm0.04$) \\
Ser~SMM1 (scan)  & $9.97\pm0.82$ & blend & $12.5\pm0.43$ & $9.29\pm0.43$ & $8.75\pm0.71$ & $8.18\pm0.71$ & $2.76\pm0.30$ & $1.92\pm0.33$ \\
Ser~SMM3         & $1.01\pm0.18$ & $1.03\pm0.17$ & $1.34\pm0.18$ & $1.01\pm0.17$ & (abs.) & (abs.) & - & - \\ 
L~723~mm         & -             & -             & ($0.45\pm0.12$) & ($0.44\pm0.10$) & -             & -             & - & - \\
\hline
L~1489           & $0.82\pm0.15$ & ($0.52\pm0.14$) & $0.88\pm0.12$ & $0.74\pm0.12$ & $0.41\pm0.07$ & $0.52\pm0.07$ & ($0.17\pm0.04$) & ($0.10\pm0.03$) \\
TMR~1            & ($0.72\pm0.16$) & $1.17\pm0.16$ & $1.35\pm0.16$ & $1.12\pm0.16$ & ($0.35\pm0.08$) & $0.47\pm0.08$ & $0.30\pm0.05$ & $0.28\pm0.05$ \\
TMC~1A           & - & - & $0.77\pm0.13$ & $0.66\pm0.13$ & (abs.) & (abs.) & - & -\\ 
TMC~1            & $1.00\pm0.08$ & $0.89\pm0.08$ & $1.40\pm0.14$ & $1.11\pm0.14$ & $0.45\pm0.07$ & $0.67\pm0.06$ & ($0.18\pm0.06$) & - \\
HH~46            & $0.69\pm0.17$ & $0.73\pm0.18$ & $0.84\pm0.17$ & $1.00\pm0.18$ & $0.37\pm0.06$ & $0.58\pm0.06$ & - & - \\
IRAS~12496/DK~Cha (scan) & - & blend & $4.00\pm0.39$ & $3.22\pm0.35$ & - & - & - & - \\
RNO~91           & ($0.52\pm0.17$) & blend & - & ($0.70\pm0.17$) & (abs.) & - & - & - \\ 
\hline
AFGL~490         & abs. & abs. & - & - & abs. & abs. & $1.12\pm0.08$ & $0.91\pm0.08$ \\
NGC~2071         & $24.36\pm0.66$ & $24.26\pm0.72$ & $35.72\pm0.64$ & $27.21\pm0.59$ & mix. & mix. & $10.11\pm0.30$ & $7.04\pm0.28$ \\
Vela~IRS~17      & $1.67\pm0.15$ & $1.70\pm0.16$ & $1.81\pm0.22$ & $1.43\pm0.19$ & mix. & mix. & $0.69\pm0.08$ & $0.35\pm0.07$ \\
Vela~IRS~19      & $2.50\pm0.25$ & $2.46\pm0.21$ & $4.71\pm0.29$ & $1.88\pm0.26$ & mix. & mix. & $1.13\pm0.06$ & $0.84\pm0.06$ \\
NGC~7129~FIRS~2  & $2.05\pm0.18$ & $1.71\pm0.20$ & $2.85\pm0.19$ & $1.57\pm0.16$ & $0.73\pm0.10$ & $0.91\pm0.11$ & $0.71\pm0.07$ & $0.51\pm0.07$ \\
L1641 S3MMS1     & $1.95\pm0.24$ & $2.09\pm0.25$ & $4.52\pm0.19$ & $3.13\pm0.19$ & mix. & mix. & $0.79\pm0.06$ & $0.93\pm0.06$\\
\hline
\end{tabular}
\label{tab:fluxes3x3}
\end{table} 
\end{landscape}
}

\def\placetableft{
\begin{landscape}
\begin{table}
\caption{Resulting fluxes in units of $10^{-16}~\mathrm{W}~\mathrm{m}^{-2}$ on the total 5 by 5 detector array.}
\centering
\begin{tabular}{l@{\extracolsep{2pt}}llllllll}
\hline \hline
Source           & \multicolumn{8}{c}{OH transition $[\mu\mathrm{m}]$}\\
                 & 79.12 & 79.18 & 84.42 & 84.60 & 119.23 & 119.44 & 163.12 & 163.40\\ 
\hline
NGC~1333~IRAS~2A (scan) & - & - & - & - & abs. & abs. & - & -\\
NGC~1333~IRAS~4A & ($0.97\pm0.23$) & $1.22\pm0.23$ & $3.11\pm0.24$ & $1.17\pm0.22$ & ($0.75\pm0.16$) & $1.86\pm0.16$ & $0.75\pm0.08$ & $0.63\pm0.08$ \\
NGC~1333~IRAS~4B & $2.36\pm0.30$ & $1.43\pm0.25$ & $5.91\pm0.35$ & $2.87\pm0.31$ & $3.17\pm0.14$ & $4.09\pm0.12$ & $0.91\pm0.10$ & ($0.53\pm0.11$) \\
L~1527           & - & - & - & - & ($0.49\pm0.10$) & $0.70\pm0.11$ & - & -\\
Ced110~IRS4      & - & - & ($0.86\pm0.27$) & - & (abs.) & - & - & - \\
IRAS~15398/B228  & ($1.48\pm0.34$) & ($1.04\pm0.31$) & $1.63\pm0.29$ & - & $1.12\pm0.11$ & $1.25\pm0.10$ & - & - \\
L~483~mm         & - & ($1.22\pm0.38$) & ($0.97\pm0.28$) & - & ($0.43\pm0.11$) & $0.82\pm0.13$ & ($0.32\pm0.07$) & ($0.22\pm0.07$) \\
Ser~SMM1 (scan)  & $12.20\pm1.22$ & blend & $12.89\pm0.95$ & $1.09\pm0.95$ & $10.79\pm0.87$ & $10.66\pm0.87$ & $3.23\pm0.40$ & $2.37\pm0.45$ \\
Ser~SMM3         & ($1.66\pm0.40$) & $2.22\pm0.38$ & $2.34\pm0.38$ & $1.73\pm0.36$ & abs. & abs. & - & - \\ 
L~723~mm         & -               & -             & -             & -             & -             & - & (abs.) & - \\ 
\hline
L~1489           & ($1.06\pm0.26$) & - & $1.21\pm0.20$ & ($0.72\pm0.20$) & ($0.36\pm0.10$) & $0.54\pm0.09$ & - & - \\
TMR~1            & -               & -                 & ($1.02\pm0.33$) & ($1.11\pm0.33$) & ($0.49\pm0.10$) & ($0.45\pm0.10$) & ($0.36\pm0.08$) & ($0.35\pm0.08$)\\
TMC~1A           & ($0.47\pm0.11$) & ($0.40\pm0.12$) & ($1.38\pm0.35$) & - & - & - & - & - \\
TMC~1            & $1.32\pm0.24$ & - & $1.44\pm0.27$ & ($1.10\pm0.27$) & $0.51\pm0.09$ & $0.69\pm0.09$ & - & ($0.29\pm0.09$)\\
HH~46            & ($0.85\pm0.28$) & - & ($1.02\pm0.28$) & - & $0.60\pm0.10$ & $0.85\pm0.10$ & - & - \\
IRAS~12496/DK~Cha (scan) & [$2.81\pm1.14]$ & blend & ($5.07\pm1.19$) & [$2.95\pm1.06$] & - & - & - & - \\
RNO~91           & - & - & - & ($1.03\pm0.31$) & (abs.) & - & - & - \\ 
\hline
AFGL~490         & mix. & mix. & - & - & abs. & abs. & $1.19\pm0.08$ & $0.90\pm0.08$ \\
NGC~2071         & $30.96\pm0.73$ & $30.47\pm0.79$ & $42.54\pm0.81$ & $33.68\pm0.75$ & mix. & mix. & $12.62\pm0.39$ & $8.33\pm0.37$ \\
Vela~IRS~17      & $2.60\pm0.23$ & $2.50\pm0.24$ & $3.12\pm0.27$ & $2.92\pm0.24$ & mix. & mix. & $0.95\pm0.13$ & ($0.51\pm0.12$) \\
Vela~IRS~19      & $3.08\pm0.44$ & $2.62\pm0.37$ & $4.99\pm0.49$ & $2.20\pm0.44$ & mix. & mix. & $1.27\pm0.07$ & $1.01\pm0.08$ \\
NGC~7129~FIRS~2  & $2.60\pm0.33$ & ($1.41\pm0.36$) & $1.86\pm0.30$ & ($1.30\pm0.26$) & $1.05\pm0.11$ & $1.20\pm0.13$ & $0.63\pm0.08$ & ($0.42\pm0.08$) \\
L1641 S3MMS1     & $2.49\pm0.43$ & $2.93\pm0.45$ & $4.28\pm0.31$ & $2.87\pm0.30$ & mix. & mix. & $0.87\pm0.08$ & $0.93\pm0.08$ \\
\hline
\end{tabular}
\label{tab:fluxes5x5}
\end{table} 
\end{landscape}
}

\def\placetableusedfluxes{
\begin{landscape}
\begin{table}
\caption{OH fluxes used throughout the paper in units of $10^{-16}~\mathrm{W}~\mathrm{m}^{-2}$ with $1\sigma$ errors from the flux extraction. Calibration errors are not included. Values in round brackets are detected below the $5\sigma$ but above the $3\sigma$ level, those in square brackets are above $2\sigma$ but below $3\sigma$.}
\centering
\begin{tabular}{l@{\extracolsep{2pt}}llllllll}
\hline \hline
Source           & \multicolumn{8}{c}{OH transition $[\mu\mathrm{m}]$}\\
                 & 79.12 & 79.18 & 84.42 & 84.60 & 119.23 & 119.44 & 163.12 & 163.40\\ 
\hline
NGC~1333~IRAS~2A (scan) & - & - & - & - & abs. & abs. & - & -\\
NGC~1333~IRAS~4A & ($0.98\pm0.23$) & $0.99\pm0.23$ & $2.40\pm0.16$ & $0.84\pm0.14$ & [$0.41\pm0.15$] & $1.51\pm0.15$ & $0.64\pm0.07$ & $0.53\pm0.07$ \\
NGC~1333~IRAS~4B & $2.08\pm0.16$ & $1.30\pm0.14$ & $4.88\pm0.19$ & $2.50\pm0.17$ & $2.23\pm0.09$ & $2.77\pm0.08$ & $0.82\pm0.07$ & ($0.58\pm0.08$) \\
L~1527           & $0.33\pm0.05$ & $0.27\pm0.05$ & $0.50\pm0.05$ & $0.49\pm0.06$ & $0.60\pm0.03$ & $0.71\pm0.03$ & [$0.07\pm0.03$] & ($0.12\pm0.03$)\\
Ced110~IRS4      & $0.31\pm0.06$ & ($0.23\pm0.06$) & $0.53\pm0.06$ & $0.47\pm0.06$ & ($0.16\pm0.04$) & $0.25\pm0.06$ & $0.16\pm0.03$ & - \\
IRAS~15398/B228  & $0.63\pm0.14$ & $0.65\pm0.14$ & $1.18\pm0.13$ & $0.58\pm0.12$ & $0.73\pm0.09$ & $0.69\pm0.08$ & - & - \\
L~483~mm         & $1.17\pm0.18$ & $1.15\pm0.18$ & $1.46\pm0.14$ & $0.88\pm0.14$ & $0.36\pm0.06$ & $0.65\pm0.07$ & $0.31\pm0.04$ & ($0.17\pm0.04$) \\
Ser~SMM1 (scan)  & $9.97\pm0.82$ & blend & $12.5\pm0.43$ & $9.29\pm0.43$ & $8.75\pm0.71$ & $8.18\pm0.71$ & $2.76\pm0.30$ & $1.92\pm0.33$ \\
Ser~SMM3         & ($1.46\pm0.32$) & $1.57\pm0.31$ & $2.01\pm0.30$ & $1.38\pm0.29$ & (abs.) &  & - & - \\
L~723~mm         & -             & [$0.17\pm0.06$]  & $0.61\pm0.06$ & $0.41\pm0.05$ & ($0.12\pm0.03$) & ($0.13\pm0.04$) & - & ($0.14\pm0.03$) \\
\hline
L~1489           & $0.79\pm0.07$ & $0.77\pm0.06$ & $1.03\pm0.07$ & $0.94\pm0.08$ & $0.52\pm0.03$ & $0.67\pm0.02$ & ($0.14\pm0.04$) & ($0.15\pm0.04$) \\
TMR~1            & ($0.72\pm0.16$) & $1.17\pm0.16$ & $1.35\pm0.16$ & $1.12\pm0.16$ & ($0.35\pm0.08$) & $0.47\pm0.08$ & $0.30\pm0.05$ & $0.28\pm0.05$ \\
TMC~1A           & - & - & $0.77\pm0.13$ & $0.66\pm0.13$ & (abs.) & (abs.) & - & -\\
TMC~1            & $1.00\pm0.08$ & $0.89\pm0.08$ & $1.40\pm0.14$ & $1.11\pm0.14$ & $0.45\pm0.07$ & $0.67\pm0.06$ & ($0.18\pm0.06$) & - \\
HH~46            & $0.69\pm0.17$ & $0.73\pm0.18$ & $0.84\pm0.17$ & $1.00\pm0.18$ & $0.37\pm0.06$ & $0.58\pm0.06$ & - & - \\
DK CHA           & ($2.29\pm0.68$) & blend & $4.82\pm0.44$ & $2.88\pm0.40$ & - & - & - & - \\ 
RNO~91           & $0.61\pm0.11$ & blend & ($0.36\pm0.07$) & ($0.40\pm0.08$) & - & [$0.08\pm0.03$] & - & - \\
\hline
AFGL~490         & mix. & mix. & - & - & abs. & abs. & $1.19\pm0.08$ & $0.90\pm0.08$ \\
NGC~2071         & $30.96\pm0.73$ & $30.47\pm0.79$ & $42.54\pm0.81$ & $33.68\pm0.75$ & mix. & mix. & $12.62\pm0.39$ & $8.33\pm0.37$ \\
Vela~IRS~17      & $2.60\pm0.23$ & $2.50\pm0.24$ & $3.12\pm0.27$ & $2.92\pm0.24$ & mix. & mix. & $0.95\pm0.13$ & ($0.51\pm0.12$) \\
Vela~IRS~19      & $3.08\pm0.44$ & $2.62\pm0.37$ & $4.99\pm0.49$ & $2.20\pm0.44$ & mix. & mix. & $1.27\pm0.07$ & $1.01\pm0.08$ \\
NGC~7129~FIRS~2  & $2.60\pm0.33$ & ($1.41\pm0.36$) & $1.86\pm0.30$ & ($1.30\pm0.26$) & $1.05\pm0.11$ & $1.20\pm0.13$ & $0.63\pm0.08$ & ($0.42\pm0.08$) \\
L1641 S3MMS1     & $2.49\pm0.43$ & $2.93\pm0.45$ & $4.28\pm0.31$ & $2.87\pm0.30$ & mix. & mix. & $0.87\pm0.08$ & $0.93\pm0.08$ \\
\hline
\end{tabular}
\label{tab:usedfluxes}
\end{table} 
\end{landscape}
}

\def\placetableIMH2OandOI{
\begin{table*}
\caption{H$_2$O and [OI] fluxes of the intermediate-mass sources in units of $10^{-16}~\mathrm{W}~\mathrm{m}^{-2}$ measured on the full 5x5 array.}
\centering
\begin{tabular}{l@{\extracolsep{2pt}}cccc}
\hline \hline\noalign{\smallskip} 
Source           & \multicolumn{2}{c}{H$_2$O transitions} &\multicolumn{2}{c}{[OI] transitions}\\
                 & $3_{2,2}-2_{1,1}$ & $2_{1,2}-1_{0,1}$ & ${}^{3}$P$_{1}-{}^{3}$P$_{2}$ & ${}^{3}$P$_{0}-{}^{3}$P$_{1}$ \\
                 & $89.99~\mu\mathrm{m}$ & $179.52~\mu\mathrm{m}$ & $63.18~\mu\mathrm{m}$ & $145.53~\mu\mathrm{m}$\\ 

\hline
AFGL~490         & -                        & (abs.)                   & $121.1\phantom{0}\pm1.4\phantom{0}$ & $\phantom{0}11.88\pm0.26$ \\
NGC~2071         & $12.80\pm1.03$           & $43.70\pm0.65$           & $552.2\phantom{0}\pm1.2\phantom{0}$ & $\phantom{0}42.42\pm0.41$ \\
Vela~IRS~17      & $\phantom{0}2.46\pm0.25$ & $12.22\pm0.18$           & $237.6\phantom{0}\pm0.79$ & $\phantom{0}56.85\pm0.33$ \\
Vela~IRS~19      & -                        & $\phantom{0}3.28\pm0.13$ & $\phantom{0}95.48\pm1.08$ & $\phantom{0}11.04\pm0.14$ \\
NGC~7129~FIRS~2  & $\phantom{0}1.84\pm0.38$ & $\phantom{0}4.27\pm0.10$ & $\phantom{0}15.72\pm0.61$ & $\phantom{00}0.90\pm0.13$ \\
L1641 S3MMS1     & $\phantom{0}2.71\pm0.24$ & $\phantom{0}1.84\pm0.17$ & $\phantom{0}22.63\pm0.54$ & $\phantom{00}3.03\pm0.12$ \\ 
\hline
\end{tabular}
\label{tab:im_h2o_oi}
\end{table*} 
}

\def\placetableratios{
\begin{table}
\caption{Observed line flux ratios (see text for details). 
}
\centering
\begin{tabular}{ccccc}
\hline \hline
Ratio & Minimum & Maximum & Mean & Median \\
\hline
\phantom{1}79/\phantom{1}84 & 0.40 & 1.54 & 0.90 & 0.85 \\ 
\phantom{1}79/119           & 0.46 & 2.30 & 1.32 & 1.31 \\
\phantom{1}79/163           & 1.68 & 5.38 & 3.22 & 3.01 \\
\phantom{1}84/119           & 0.75 & 3.28 & 1.65 & 1.58 \\
\phantom{1}84/163           & 1.44 & 6.48 & 3.44 & 3.57 \\
          119/163           & 1.41 & 4.10 & 2.66 & 2.14 \\ 
\hline
\end{tabular}
\label{tab:ratios}
\end{table} 
}

\def\placetablespatialextent{
\begin{table*}
\caption{Description whether the flux measured in the on-source spaxel and the surrounding three by three spaxels are consistent with the pure spillover factor (0.70 at $79~\mu\mathrm{m}$, 0.69 at $84~\mu\mathrm{m}$, 0.62 at $119~\mu\mathrm{m}$, and 0.49 at $163~\mu\mathrm{m}$). If the fraction of flux measured in the on-source spaxel compared to the three by three central spaxels is smaller than the spillover value, it is an indication that the OH line is spatially extended, as summarized in the last column.}
\centering
\begin{tabular}{l@{\extracolsep{2pt}}ccccc}
\hline \hline
Source           & $79~\mu\mathrm{m}$  & $84~\mu\mathrm{m}$ & $119~\mu\mathrm{m}$ & $163~\mu\mathrm{m}$ & spatially extended \\
\hline
\hline
NGC~1333~IRAS~2A & - & - & no (abs.) & - & - \\
NGC~1333~IRAS~4A & - & no & no & - & yes\\
NGC~1333~IRAS~4B & no & no & no & no & yes\\
L~1527           & no & yes & yes & yes & (no)\\
Ced110~IRS4      & yes  & (yes) & yes & yes & no \\
IRAS~15398/B228  & no & no & no & no  & yes\\
L~483~mm         & (no) & yes & yes & -  & (no) \\
Ser~SMM1         & no & no & no & no & yes\\
Ser~SMM3         & no & no & - & - & yes\\
L~723~mm         & - & yes & yes  & - & no \\
\hline
L~1489           & yes & yes & yes & yes & no \\
TMR~1            & no & no & no & yes & (mispointed)\\
TMC~1A           & no & yes & - & yes & (no)\\
TMC~1            & no & no & - & yes & (yes) \\ 
HH~46            & no & yes & no & - & no \\
DK~Cha           & - & no & - & - & (yes)\\
RNO~91           & yes & - & - & - & no \\
\hline
AFGL~490         & no (abs.) & no (abs.) & no (abs.) & no & yes \\
NGC~2071         & no & no & no (abs./em.) & no & yes\\
NGC~7129~FIRS~2  & no & yes & no & -  & yes\\
Vela~IRS~17      & no & no & no (abs./em.) & no & yes \\
Vela~IRS~19      & no & no & no (abs./em.) & no & yes\\
L1641 S3MMS1     & no & yes & no & yes & yes\\
\hline
\end{tabular}
\label{tab:spatial_extent}
\end{table*}
}

\def\placetablecorrcoeffs{
\begin{table}
\caption{Pearson's correlation coefficients $\rho$, number of sources $N$, corresponding significance levels $p$, and number of standard deviations $\sigma$ for the OH $84.6~\mu\mathrm{m}$ line luminosity or flux and various envelope parameters.}
\centering
\begin{tabular}{lcccc}
\hline \hline
Parameter & $\rho$ & $N$ & $p$ & $\sigma$\\
\hline
$L_\mathrm{bol}$                       & \phantom{$-$}0.93 & 21 & $< 10^{-3}$ & 4.17 \\
$M_\mathrm{env}$                       & \phantom{$-$}0.84 & 21 & $< 10^{-3}$ & 3.77 \\
$T_\mathrm{bol}$                       & $-0.23$         & 16 & 0.20 & 0.87 \\
$F_\mathrm{CO}$                        & \phantom{$-$}0.46 & 15 & 0.04 & 1.71 \\ 
$L_\mathrm{bol}^{0.6}/M_\mathrm{env}$  & $-0.21$         & 21 & 0.18 & 0.95 \\
100~K model radius                     & \phantom{$-$}0.86 & 16 & $< 10^{-3}$ & 3.33 \\
$n_\mathrm{1000~AU}$                   & \phantom{$-$}0.47 & 16 & 0.03 & 1.82 \\
H$_2$ column density                   & \phantom{$-$}0.30 & 16 & 0.26 & 1.16 \\
\hline
$[$OI$]$ $63~\mu\mathrm{m}$            & \phantom{$-$}0.71 & 21 & $< 10^{-3}$ & 3.16 \\
$[$OI$]$ $145~\mu\mathrm{m}$           & \phantom{$-$}0.79 & 18 & $< 10^{-3}$ & 3.27 \\
H$_2$O $89.99~\mu\mathrm{m}$           & \phantom{$-$}0.83 & 17 & $< 10^{-3}$ & 3.33 \\
\hline
\end{tabular}
\label{tab:corrcoeffs}
\end{table} 
}

 
\abstract
   {The OH radical is a key species in the water chemistry network of star-forming regions, because its presence is tightly related to the formation and destruction of water. Previous studies of the OH far-infrared emission from low- and intermediate-mass protostars suggest that the OH emission mainly originates from shocked gas and not from the quiescent protostellar envelopes.}
   {We aim to study the excitation of OH in embedded low- and intermediate-mass protostars, determine the influence of source parameters on the strength of the emission, investigate the spatial extent of the OH emission, and further constrain its origin.}
   {This paper presents OH observations from 23 low- and intermediate-mass young stellar objects obtained with the PACS integral field spectrometer on-board \textit{Herschel} in the context of the ``Water In Star-forming Regions with Herschel (WISH)'' key program. Radiative transfer codes are used to model the OH excitation.}
   {Most low-mass sources have compact OH emission ($\lesssim 5000~\mathrm{AU}$ scale), whereas the OH lines in most intermediate-mass sources are extended over the whole 47\farcs0 x 47\farcs0 PACS detector field-of-view ($\gtrsim 20000~\mathrm{AU}$). The strength of the OH emission is correlated with various source properties such as the bolometric luminosity and the envelope mass, but also with the [OI] and H$_2$O emission. Rotational diagrams for sources with many OH lines show that the level populations of OH can be approximated by a Boltzmann distribution with an excitation temperature at around 70~K. Radiative transfer models of spherically symmetric envelopes cannot reproduce the OH emission fluxes nor their broad line widths, strongly suggesting an outflow origin. Slab excitation models indicate that the observed excitation temperature can either be reached if the OH molecules are exposed to a strong far-infrared continuum radiation field or if the gas temperature and density are sufficiently high. Using realistic source parameters and radiation fields, it is shown for the case of Ser~SMM1 that radiative pumping plays an important role in transitions arising from upper level energies higher than 300~K. The compact emission in the low-mass sources and the required presence of a strong radiation field and/or a high density to excite the OH molecules points towards an origin in shocks in the inner envelope close to the protostar.}
   {}


\maketitle

\section{Introduction} 
Oxygen is the most abundant element in the interstellar medium apart from hydrogen and helium. Many oxygen-bearing species, most importantly water and its precursors, have a very limited observability from the ground because of atmospheric constraints. The \textit{Herschel Space Observatory} \citep{Pilbratt10} outperforms previous space-borne facilities in sensitivity as well as spatial and spectral resolution. It is thus well suited to study the water chemistry in young stellar objects in more detail. Because water undergoes large gas-phase abundance variations with varying temperature or radiation field, it is an excellent probe of the physical conditions and dynamics in star-forming regions \citep[][]{Kristensen12,vanDishoeck11}. 

A key connecting piece between atomic oxygen and water is the hydroxyl radical (OH). In the high-temperature gas-phase chemistry regime ($T > 230~\mathrm{K}$), all available gas-phase oxygen is first driven into OH and then into water by the $\mathrm{O} + \mathrm{H}_{2} \rightarrow \mathrm{OH} + \mathrm{H}$ and subsequent $\mathrm{OH} + \mathrm{H}_{2} \rightarrow  \mathrm{H}_2\mathrm{O} + \mathrm{H}$ reactions. The importance of the backward reaction $\mathrm{H}_2\mathrm{O} + \mathrm{H} \rightarrow \mathrm{OH} + \mathrm{H}_{2}$ depends on the atomic to molecular hydrogen ratio and therefore on the local UV field or shock velocity. H$_2$ is dissociated in J-type shocks when the velocity exceeds $\sim25~\mathrm{km}~\mathrm{s}^{-1}$ \citep[e.g.][]{Hollenbach80}. OH is also a byproduct of the H$_2$O photo-dissociation process. Thus, OH is most abundant in regions with physical conditions different from those favoring the formation and existence of water and it therefore provides a complementary view of oxygen in the gas phase. In addition to its importance in the oxygen and water chemistry, OH also contributes to the FIR cooling budget of warm gas in embedded YSOs \citep[e.g.,][Karska et al.\ subm.]{Neufeld89b,Kaufman96,Nisini02}. The goal of the ``Water In Star-Forming regions with Herschel'' \citep[WISH,][]{vanDishoeck11} Herschel key program is to study the H$_2$O and associated OH chemistry for a comprehensive picture of H$_2$O during protostellar evolution.  

Detections of OH far-infrared (FIR) transitions from star-forming regions were made previously with the Kuiper Airborne Observatory \citep[e.g.][]{Melnick87,Betz89}, the Infrared Space Observatory \citep[e.g.][]{Ceccarelli98,Giannini01,Larsson02,Goicoechea02,Goicoechea04,Goicoechea06}, \textit{Herschel} \citep[e.g.][]{vanKempen10a,Wampfler10,Wampfler11,Goicoechea11}, and recently SOFIA \citep[][]{Csengeri12,Wiesemeyer12}. Masers of OH hyperfine transitions are also commonly observed towards high-mass star-forming regions at cm wavelengths, but are not detected from low- and intermediate-mass protostars. Because this work focuses on low- and intermediate-mass young stellar objects (YSOs), we will not discuss OH maser emission \citep[a detailed overview can be found in e.g.][]{Elitzur92}.

From first \textit{Herschel} results using the Photodetector Array Camera and Spectrometer \citep[PACS,][]{Poglitsch10}, \citet{vanKempen10b} found that the OH emission from the low-mass class I young stellar object (YSO) HH~46 is not spatially extended, in contrast to the H$_2$O and high$-J$ CO emission from the same source. They speculated that at least parts of the OH emission could stem from a dissociative shock caused by the impact of the wind or jet on the dense inner envelope. Spectrally resolved \textit{Herschel} observations of OH with the Heterodyne Instrument for the Far-Infrared \citep[HIFI,][]{deGraauw10} support an outflow scenario based on the inferred broad line widths of more than $10~\mathrm{km}~ \mathrm{s}^{-1}$ \citep{Wampfler10,Wampfler11}.
Furthermore, analysis of OH PACS lines from a set of six low- and intermediate-mass protostars yielded similar excitation conditions of OH in all these sources \citep{Wampfler10}. A tentative correlation of the OH line luminosities with the [OI] luminosities as well as the bolometric luminosities of the protostars were found, indicating that the observed OH emission might originate from a dissociative shock.

In this paper we present \textit{Herschel}-PACS observations of OH in an extended sample of 23 low-and intermediate-mass YSOs. Our first goal is to test whether the tentative correlations from earlier work can be confirmed and to determine the influence of different source properties on the strength of the emission. The second goal is to study the OH excitation and the spatial extent of the OH emission in the target sources. A detailed analysis of the OH excitation and spatial extent is important to determine the origin of the OH emission and whether 
this is consistent with the picture from H$_2$O observations.

The paper is organized as follows: Section \ref{sec:observations} describes the source sample, the observations, and the data reduction methods. In Sect.~\ref{sec:results} we present the observational results. Section \ref{sec:models} contains a discussion of spherically symmetric envelope models for OH (Sect.~\ref{sec:ratran_models}), the description and results of slab radiative transfer models to study the OH excitation in the outflow (Sect.~\ref{sec:slab_models}), as well as the discussion and data interpretation (Sect.~\ref{sec:discussion}).  Finally, the conclusions are summarized in Sect.~\ref{sec:conclusions}.

\section{Observations and data reduction} \label{sec:observations}

\placetablesourceprops

Observations of 23 low- and intermediate-mass young stellar objects were carried out with the Photodetector Array Camera and Spectrometer \citep[PACS,][]{Poglitsch10} on the \textit{Herschel Space Observatory}. All observations were obtained within the ``Water in Star-Forming Regions with Herschel'' key program \citep[WISH,][]{vanDishoeck11} except for \mbox{IRAS~12496-7650} \citep[DK Cha,][]{vanKempen10a}, which was observed in the ``Dust, Ice and Gas In Time (DIGIT)'' key program (PI N.~Evans). The coordinates and properties of the targets can be found in Table~\ref{tab:source_props} and the data identity numbers (obsids), observing modes, and the pipeline versions are given in Table~\ref{tab:source_obsids} in the appendix. 

\placefigureOHLevels

Two different observing modes are available for the PACS spectrometer, line and range spectroscopy. The ``range spectroscopy'' mode provides a full scan of the $50-220~\mu\mathrm{m}$ wavelength regime. The ``line spectroscopy'' mode covers only small windows around selected target lines, but generally at higher spectral sampling and sensitivity than range spectroscopy. The majority of the sources in our sample have been observed with the PACS line spectroscopy mode, targeting four main OH rotational doublets: ${}^2\Pi_{1/2} (J = 1/2) \rightarrow {}^2\Pi_{3/2} (J = 3/2)$ at $79~\mu\mathrm{m}$, ${}^2\Pi_{3/2} (J = 7/2 \rightarrow 5/2)$ at $84~\mu\mathrm{m}$, ${}^2\Pi_{3/2} (J = 5/2 \rightarrow 3/2)$ at $119~\mu\mathrm{m}$, and ${}^2\Pi_{1/2} (J = 3/2 \rightarrow 1/2)$ at $163~\mu\mathrm{m}$. The integration time and the noise level for all sources observed in line spectroscopy mode is similar. NGC~1333~IRAS~4A and NGC~1333~IRAS~4B have been observed in both line and full range spectroscopy. Range spectroscopy only was used for NGC~1333~IRAS~2A, Ser SMM~1, and DK~Cha. An illustration of the OH pure rotational transitions that are accessible with PACS on-board \textit{Herschel} is provided in Fig.~\ref{fig:OH_levels}. An overview on the molecular data used in this work can be found in Table~\ref{tab:moldata}.

\placetableMoldata

The full spectral scan of DK~Cha was presented previously in \citet{vanKempen10a} and the line spectroscopy of HH~46 and NGC~7129~FIRS~2 in \citet{vanKempen10b} and \citet{Fich10}, respectively. These three spectra plus IRAS~15398, TMR~1, and NGC~1333~IRAS~2A were part of the sample analyzed in our previous work \citep{Wampfler10}. We have now re-reduced all spectra with a newer version of the PACS pipeline and calibration files. The full spectral scans of NGC~1333~IRAS~4B and Ser SMM1 are discussed in great detail in \citet{Herczeg12} and \citet{Goicoechea12}, respectively.

The PACS spectrometer is an integral field spectrometer and operates simultaneously in a blue and a red channel. The detector consist of 5 by 5 square spatial pixels (``spaxels'') with a pixel size of 9\farcs4. The Herschel half power beam width is smaller than a spaxel in the blue wavelength regime, but exceeds the spaxel size on the sky in the red part of the spectrum. The spatial resolution is therefore limited by the pixel size for short wavelengths and by the diffraction beam pattern at longer wavelengths. 
The spectral resolution depends on the grating order and varies from $R = 3000 - 4000$ at wavelengths $\lambda < 100~\mu\mathrm{m}$ to  $R = 1000 - 2000$ for $\lambda > 100~\mu\mathrm{m}$. All individual OH lines are therefore unresolved and at the lower resolutions, even blending of the OH doublets occurs, mostly at $79~\mu\mathrm{m}$ and $119~\mu\mathrm{m}$. When doublet components are blended, they were assumed to be of equal strength in the analysis.
 
The spectra were reduced with the Herschel interactive processing environment \citep[HIPE,][]{Ott10}, versions 8 (for line scans) and 6 (for range scans). The wavelength grid was rebinned to four pixels per resolution element for line scans and two pixels per resolution element for range scans. The spectra were flat-fielded to improve the signal-to-noise ratio. The fluxes were normalized to the telescopic background and then calibrated using measurements of Neptune as a reference. The relative calibration uncertainty on the fluxes is currently estimated to be below 20\%.
The PACS spectrometer suffers from spectral leakage in the wavelength ranges $70-73~\mu\mathrm{m}$, $98-105~\mu\mathrm{m}$, and $190-220~\mu\mathrm{m}$ where the next higher grating order ranges $52.5-54.5~\mu\mathrm{m}$, $65-70~\mu\mathrm{m}$, and $95-110~\mu\mathrm{m}$ are superimposed on the spectrum. The fluxes from lines in these wavelength bands might therefore be less reliable than in parts of the spectra that are not affected by spectral leakage. This applies to the OH $71~\mu\mathrm{m}$ and $98~\mu\mathrm{m}$ doublets.

The lines were then subsequently analyzed in IDL using first or, if required, second order polynomials as baselines and the flux was measured by integrating over the line. The fraction of the point-spread function (PSF) seen by the central spaxel is wavelength-dependent, reaching from around 0.7 for a perfectly centered point source at $60-80~\mu\mathrm{m}$ down to about 0.4 at $200~\mu\mathrm{m}$. A significant fraction of the flux might therefore fall onto neighboring spaxels and even more so if the source is off-centered on the central spaxel. It is therefore important to investigate the flux distribution on the detector, which can be caused by spatially extended emission, the telescope PSF, or a combination of both. Extracting the flux from all 25 spaxels is not an optimal solution, because of contamination from nearby sources and because the signal-to-noise ratio drops if many spaxels without line emission but extra noise are added. This is particularly problematic for weak lines or lines with little spatial extent. For a subset of our targets, where contamination from nearby sources occurs, we chose a set of spaxels that excludes the contribution from close-by sources. This applies to NGC~1333~IRAS~4A, NGC~1333~IRAS~4B, Ser~SMM3, and DK~Cha. Spaxels excluded in the flux measurements of the WISH sources are marked in gray in the maps in the online appendix \ref{sec:oh_maps}. For DK~Cha, a 3 by 4 set around the on-source spaxel was used. 
For all other sources, we used either the on-source spaxel, corrected for the spillover (L~1527, Ced~110~IRS~4, L~723, L~1489, RNO~91), 3 by 3 spaxels centered on the on-source spaxel (IRAS~15398, L~483, TMR~1, TMC~1A, TMC~1, HH~46, and Ser~SMM1), or the full array (all intermediate-mass sources). 

\section{Results and Analysis} \label{sec:results}
\subsection{Detected lines and spatial extent}
We detected at least one OH doublet in all 23 sources. The fluxes integrated over the emitting area, obtained as described in Sect.~\ref{sec:observations}, can be found in Table~\ref{tab:usedfluxes}. The doublet that is most often detected, in 21 out of 23 sources, is the ${}^2\Pi_{3/2} (J = 7/2 \rightarrow 5/2)$ doublet at $84~\mu\mathrm{m}$, thanks to a combination of intrinsic strength and higher spectral resolution of the instrument at the shorter wavelengths. The component at $84.42~\mu\mathrm{m}$ is however blended with CO(31-30) at $84.41~\mu\mathrm{m}$ because the spectral resolution is around $0.037~\mu\mathrm{m}$. The sources in which the $84~\mu\mathrm{m}$ lines were not detected are NGC~1333~IRAS~2A and AFGL~490. Figures~\ref{fig:class0_spectra} and \ref{fig:classI_spectra} present the OH spectra of the low-mass class~0 and I sources, Fig.~\ref{fig:im_spectra} the spectra of the intermediate-mass sources. In 22 out of 23 sources at least one line was in emission, with the exception being NGC~1333~IRAS~2A, where the only detection is the $119~\mu\mathrm{m}$ doublet feature in absorption \citep{Wampfler10}. 
We mainly detected emission lines, but absorption is also found towards higher envelope masses. In the subsequent analysis, we only considered the fluxes of lines that are purely in emission. Absorption in the $119~\mu\mathrm{m}$ OH ${}^2\Pi_{3/2}$ intra-ladder doublet transitions is observed from NGC~1333~IRAS~2A and in some spaxels of AFGL~490, Vela~IRS~17, and Vela~IRS~19, but there are also spaxels with emission from these targets. For even higher envelope masses like AFGL~490, the $79~\mu$m OH cross-ladder transitions are also in absorption. This behavior can be explained by the fact that both lines are directly coupled to the ground rotational state of OH: because the first excited state is at $E_\mathrm{up} \approx 120~\mathrm{K}$, almost only the ground state is populated at temperatures below $\sim100~\mathrm{K}$. The $119~\mu\mathrm{m}$ transitions couple the ground state to the first excited state and have a large Einstein A coefficient ($\sim 1.4 \times 10^{-1}~\mathrm{s}^{-1}$). The $79~\mu$m lines are cross-ladder transitions between the ground states of both rotational ladders. Cross-ladder transitions generally have much smaller Einstein A coefficients ($\sim 3.6 \times 10^{-2}~\mathrm{s}^{-1}$ for the $79~\mu$m lines) than the intra-ladder transitions. Therefore, absorption at $79~\mu$m does not occur as easily as for the $119~\mu\mathrm{m}$ doublet.

\placetableusedfluxes

\placefigureClass0Spectra
\placefigureClassISpectra
\placefigureIMSpectra

\clearpage

Several sources show OH emission that is extended beyond what would be expected from a point source folded with the point-spread function of the telescope or leakage onto neighboring spaxels. The line emission is usually extended along the outflow direction (see Karska et al. subm.) and strongly correlated with the spatial extent of the atomic oxygen transitions, as illustrated in Fig.~\ref{fig:map_oh_oi_n2071} and further discussed in Karska et al. (subm).  

Maps of the OH $79, 84, 119$, and $163~\mu\mathrm{m}$ transitions for all WISH sources can be found in the appendix. An example of compact OH emission at $84~\mu\mathrm{m}$ from the low-mass class I YSO L~1489 is shown in Fig.~\ref{fig:map_l1489}. 
Despite their location at further distances than the low-mass objects, all six intermediate-mass sources show signatures of extended OH emission. The largest spatial extent is seen in the $119~\mu\mathrm{m}$ ground state lines, which can be excited most easily. Figure~\ref{fig:map_afgl490} presents the spaxel map of the $119~\mu\mathrm{m}$ line from AFGL~490. The absorption is strongest towards the YSO and extended over an area of $\sim 25\arcsec$ (25000~AU) around the central position in the direction perpendicular to the outflow. The spatial extent of the emission is comparable to the 20000~AU by 6000~AU envelope structure discussed in \citet{Schreyer06}. The OH absorption changes into weak emission along the outflow direction. 
For the full sample we provide an overview whether the flux observed outside the on-source spaxel is consistent or inconsistent with the spillover factor in Table~\ref{tab:spatial_extent}.

\placefigureSpatOHOI

\placefigureMapL1489

\placetablespatialextent

\placefigureMapAFGL490

\subsection{OH emission line flux ratios} \label{sec:lineratios}
Comparison of the OH $84~\mu$m line luminosities (flux corrected for the source distance) between the different source types shows that the intermediate-mass sources in our sample have higher OH luminosities than the low-mass sources by about two orders of magnitude on average. Among the low-mass sources, the class 0 sources are on average about a factor of two more luminous in OH than the class I sources. 

In earlier work \citep{Wampfler10}, we found that the line flux ratios among the sources in the sample were relatively constant. We have therefore also calculated the $79~\mu\mathrm{m}/84~\mu\mathrm{m}$, $79~\mu\mathrm{m}/119~\mu\mathrm{m}$, $79~\mu\mathrm{m}/163~\mu\mathrm{m}$,  $84~\mu\mathrm{m}/119~\mu\mathrm{m}$, $84~\mu\mathrm{m}/163~\mu\mathrm{m}$, and $119~\mu\mathrm{m}/163~\mu\mathrm{m}$ line flux ratios for the
extended sample, i.e. the sources in Tab. \ref{tab:source_props} from which emission was detected. The fluxes of doublets are added except for the $84~\mu\mathrm{m}$ doublet, because the $84.42~\mu\mathrm{m}$ is blended with CO(31-30). Instead we use twice the flux of the $84.60~\mu\mathrm{m}$ component assuming that both components are of similar strength. Cases where only one doublet component was clearly detected were not considered. The results are listed in Table~\ref{tab:ratios}.

\placetableratios

Again we find that the line ratios remain relatively constant within less than a factor of four around their mean values over the whole luminosity, mass, and age range of several orders of magnitude spanned by the sample.
The excitation of OH is therefore likely to be similar in all sources, indicating that either the OH emission stems from gas at similar physical conditions or that the ratios remain stable over a spread of parameter values. The latter possibility is supported by the models (cf. Sect.~\ref{sec:models}), but does not exclude the first option. 

\subsection{OH rotational temperature} \label{sec:rottemp}
The full spectral scans cover the OH transitions from about $55-200~\mu\mathrm{m}$ ($E_\mathrm{up} \approx 120 - 875~\mathrm{K}$) and therefore allow us to study the excitation conditions by means of rotational diagrams. Figure~\ref{fig:smm1_rotdiag} presents the diagram for Ser~SMM1. 
The derived rotational temperature is $T_\mathrm{rot} \approx 72 \pm 8~\mathrm{K}$ if all the transitions shown on the plot except the $119~\mu\mathrm{m}$ (optically thick), $84.42~\mu\mathrm{m}$ (blended with CO), and $98~\mu\mathrm{m}$ (in leaking region) are included.
The rotational diagram for NGC~1333~IRAS~4B can be found in \citet{Herczeg12}, giving $T_\mathrm{rot} \approx 60 \pm 15~\mathrm{K}$.
No emission lines were detected in the full spectral scan of NGC~1333~IRAS~2A and only very few line detections are available for NGC~1333~IRAS~4A and DK~Cha \citep[see also][]{vanKempen10a}, so that the rotational diagram for these sources is very sparsely populated and a fit of the rotational temperature is therefore not feasible. 
For Ser~SMM1, the corresponding OH column density would be $N_{\mathrm{OH}} = 1.0 \times 10^{14}~\mathrm{cm}^{-2}$ assuming a source size of $20\arcsec$ and using an interpolated value for the partition function $Q$ from the JPL catalog \footnote{http://spec.jpl.nasa.gov}
\citep{Pickett98}. However, the ${}^{2}\Pi_{3/2}$ points fall below the fit, suggesting that these transitions are either optically thick, which is well possible because the ${}^{2}\Pi_{3/2}$ ladder contains the ground rotational state, or that the two rotational ladders might have a different rotational temperature.

\placefigureRotDiag

\subsection{Dependence of OH luminosity on source properties}\label{sec:corrsourceprops}
Understanding which source parameters influence or even determine the strength of the OH emission and the excitation of the different transitions is important to constrain the origin of the emission. We therefore test whether the OH line luminosity correlates with various envelope parameters and the emission from other molecular and atomic species. Because the different OH transitions are fairly well correlated (Sect.~\ref{sec:lineratios}), we restrict the analysis to the $84~\mu\mathrm{m}$ luminosity, where we have most detections.

\placefigureLOHvsLbol

\placefigureLOHvsMenv

\placefigureLbolvsMenv

In Fig.~\ref{fig:LOH_Lbol} and Fig.~\ref{fig:LOH_Menv} the dependence of the OH line luminosity on the bolometric source luminosity and the envelope mass are shown, respectively. We use new values for $L_\mathrm{bol}$ that were derived by \citet{Kristensen12} based on additional continuum values from PACS observations, which are presented in Karska et al. (subm.).
The envelope masses for the low-mass sources are calculated from spherical models based on continuum radiative transfer and include all material at temperatures above $10~\mathrm{K}$. For the intermediate-mass sources, we use literature values and the method that was used to determine the envelope mass may therefore be different.

\placetablecorrcoeffs

The Pearson correlation coefficient, defined as \mbox{$\rho_{X,Y} = cov(X,Y)/[\sigma(X) \cdot \sigma(Y)]$}, i.e. the covariance of two random variables X and Y divided by their standard deviations, for $L_\mathrm{OH}$ with $L_\mathrm{bol}$ is 0.93, including all 21 sources where the OH $84~\mu\mathrm{m}$ line was detected. An overview on all obtained correlation coefficients and their corresponding significance levels is given in Table~\ref{tab:corrcoeffs}. Values of $\rho$ close to 1 and $-1$ describe a tight correlation or anticorrelation of the parameters $X$ and $Y$, respectively, while values close to 0 indicate that $X$ and $Y$ are uncorrelated. At what level of $\rho$ a correlation is considered to be significant, depends on the number of data points. The significance of $\rho$ can be expressed as a probability value, the significance level $p$, and tells how likely the actual observed or a more extreme value of $\rho$ is found under the assumption that the null hypothesis is true, i.e. that there is no relationship between the parameters. Another option is to express the significance in terms of number of standard deviations $\sigma$, calculated from $\rho \sqrt{N-1}$, where $N$ is the number of sources. 

The correlation coefficient for $L_\mathrm{OH}$ with $M_\mathrm{env}$ is smaller ($\rho = 0.84$). As illustrated by Fig.~\ref{fig:Lbol_Menv}, the bolometric luminosity and the envelope mass are not independent properties of the sources. This is consistent with \citet[][Fig.~1]{Bontemps96} and \citet[][Fig.~6b]{Andre00}, who found that $L_\mathrm{bol}$ and $M_\mathrm{env}$ are strongly correlated for the class 0 sources and that class I sources move down in the diagram with time along the evolutionary tracks. From an $L_\mathrm{OH} - M_\mathrm{env}$ correlation one would also expect $L_\mathrm{OH}$ to be correlated with $T_\mathrm{bol}$ \citep[Fig.~8 of][]{Jorgensen02}. Such a correlation is however not found for our low-mass source sample, as illustrated by Fig.~\ref{fig:LOH_Tbol}, indicating that the underlying $L_\mathrm{bol} - M_\mathrm{env}$ correlation might create the observed $L_\mathrm{OH} - M_\mathrm{env}$ trend.
A correlation between two variables can be caused by an underlying correlation of the two variables with a third one. A method to measure such effects is the concept of the partial correlation coefficient, defined as $\rho_{(X,Y)/Z} = (\rho_{X,Y} - \rho_{X,Z} \cdot \rho_{Y,Z}) /\sqrt{(1- \rho_{X,Z}^2) \cdot (1- \rho_{Y,Z}^2)}$ \citep[see also appendix A from][]{Marseille10}. The resulting value for the partial correlation coefficient of $L_\mathrm{OH}$ with $M_\mathrm{env}$ under the influence of $L_\mathrm{bol}$ is 0.56, i.e. significantly reduced, thus indicating that the trend might indeed be based on an underlying relation between mass and luminosity. 
Class I YSOs that lie above the $L_\mathrm{OH} - M_\mathrm{env}$ correlation in Fig.~\ref{fig:LOH_Menv} are also the ones that do not follow the $L_\mathrm{bol}$-$M_\mathrm{env}$ relation (Fig.~\ref{fig:LOH_Menv}), in particular DK~Cha. These sources have a higher $L_\mathrm{OH}$ than what would be expected from their mass and a lower value than what would be expected from their bolometric luminosity. They are found in the region of more evolved class I sources \citep[see Fig.~6 in][]{Andre00}, i.e. they have a significantly lower envelope mass at a given bolometric luminosity than less evolved sources like e.g. HH~46, and are in the transitional stage to class II where disk emission may contribute as well. 

\placefigureLOHvsTbol

We find that $L_\mathrm{OH}$ (the luminosity of the OH $84~\mu\mathrm{m}$ transitions) seems to be well correlated with luminosity for both evolutionary stages and that class 0 and I sources lie on the same straight line. In contrast, \citet{Nisini02} found from ISO data that class~I sources fall systematically below class~0 sources in their plot of the total far-infrared luminosity \mbox{$L_\mathrm{FIR} = L_\mathrm{[OI]} + L_\mathrm{CO} + L_{\mathrm{H}_2\mathrm{O}} + L_\mathrm{OH}$} versus $L_\mathrm{bol}$. Furthermore, they concluded that class 0 and I sources fall onto the same straight line in the $L_\mathrm{FIR} - M_\mathrm{env}$ plot, but in Fig.~6 of \citet{Nisini02}, class I sources with $M_\mathrm{env} \lesssim 10^{-1}~\mathrm{L}_{\odot}$ tend to branch off as well.

A correlation of $L_\mathrm{OH}$ with the outflow momentum flux \citep[``outflow force''][]{Bontemps96} $F_\mathrm{CO}$ is not significant from our Fig.~\ref{fig:LOH_fco}. The values for $F_\mathrm{CO}$ are taken from the literature and were not derived in a consistent way. New estimates of $F_\mathrm{CO}$ are currently work in progress (Y{\i}ld{\i}z et al. in prep.). 

\placefigureLOHvsFCO

Furthermore, $L_\mathrm{OH}$ seems to decrease with evolution for class~0 sources, but not for class I, as can be seen from Fig.~\ref{fig:LOH_Evo} showing $L_\mathrm{OH}$ plotted against $L_\mathrm{bol}^{0.6}/M_\mathrm{env}$, a quantity that was proposed as an evolutionary tracer by \citet{Bontemps96}. 

\placefigureLOHvsEvo

\placefigureLOHrd100K

\placefigureLOHn1000au

The different behavior of the class 0 and I sources could be caused by their different geometry. If the OH was radiatively pumped, then the location of the warm dust would be an important parameter. Warm dust at $100~\mathrm{K}$ (cf. Sect.~\ref{sec:rottemp}) is expected from two physical components in young stellar objects, the inner envelope and the outflow walls. Class 0 and class I sources differ in the geometrical alignment of their warm dust components: In class 0 sources, high-energetic radiation from the protostar and accretion processes is reprocessed to a far-infrared field by the dust in the inner envelope, which represents a small solid angle. Thus the OH luminosity for sources where this component dominates should be mass dependent. In class I sources, the warm dust from the outflow walls can directly irradiate the local OH molecules and therefore the OH luminosity for these sources should depend on the bolometric luminosity. We extracted the radius at which the temperature reaches $100~\mathrm{K}$ from the spherically symmetric envelope models to test this hypothesis. Figure~\ref{fig:LOH_rd100K} shows the OH luminosity plotted against the $100~\mathrm{K}$ radius from the spherically symmetric envelope models \citep{Kristensen12} and there seems to be a trend of increasing OH luminosity when the size of the inner hot envelope $T \ge 100~\mathrm{K}$ is larger. The OH luminosity does not seem to depend significantly on the density at 1000~AU as shown in Fig.~\ref{fig:LOH_n1000au}. A correlation with density would not necessarily imply that collisions are the dominant excitation mechanism anyway as the dust mass also increases with density.

\subsection{Correlation of the OH flux with other species} \label{sec:corrotherspecies}

\placefigureLOHLOI

OH is tightly related to the formation and destruction processes of H$_2$O and a connecting piece between H$_2$O and atomic oxygen in the chemistry. We therefore test whether the OH emission is correlated with the [OI] and H$_2$O emission. 
The [OI] and H$_2$O fluxes for the low-mass protostars are tabulated in Karska et al. (subm.) and those of the intermediate-mass protostars in Table~\ref{tab:im_h2o_oi} in the appendix. 
Fig.~\ref{fig:LOH_LOI} shows the OH $84~\mu\mathrm{m}$ flux plotted versus the [OI] ${}^{3}$P$_1 -{}^{3}$P$_2$ and ${}^{3}$P$_0 -{}^{3}$P$_1$ fluxes at $63~\mu\mathrm{m}$ (left panel) and $145~\mu\mathrm{m}$ (right panel), respectively. 
The Pearson correlation coefficients for OH $84~\mu\mathrm{m}$ vs. [OI] $63~\mu\mathrm{m}$ and $145~\mu\mathrm{m}$ are 0.71 and 0.79, respectively, corresponding to a three sigma result. The [OI] $63~\mu\mathrm{m}$ could be optically thick or more affected by contamination than the $145~\mu\mathrm{m}$, which would explain the slightly lower correlation coefficient for OH $84~\mu\mathrm{m}$ vs. [OI] $63~\mu\mathrm{m}$ compared to [OI] $145~\mu\mathrm{m}$. The OH and [OI] emission thus seem to be correlated both spatially and in total flux, hinting at a common origin,  most likely from gas associated with the outflow, because in the cases of non-compact emission it is usually found to be extended along the outflow direction.

The correlation of the OH $84~\mu\mathrm{m}$ flux with \mbox{p-H$_2$O $3_{2,2}-2_{1,1}$} ($\lambda \approx 89.99~\mu\mathrm{m}$, $E_\mathrm{up} = 296.8~\mathrm{K}$) is shown in Fig.~\ref{fig:LOH_H2O}. The \mbox{p-H$_2$O $3_{2,2}-2_{1,1}$} was chosen because it has a very similar upper level energy to the OH $84~\mu\mathrm{m}$ transition ($E_\mathrm{up} = 290.5~\mathrm{K}$) and a similar wavelength, which strongly reduces the influence of instrumental effects, as the PSF is very similar. The fluxes were measured on the same spaxel sets. The correlation of the OH $84~\mu\mathrm{m}$ flux with p-H$_2$O $3_{2,2}-2_{1,1}$ is significant at more than 3$\sigma$ with $\rho = 0.83$. The same plot for o-H$_2$O $2_{1,2}-1_{0,1}$ ($\lambda \approx 179.53~\mu\mathrm{m}$, $E_\mathrm{up} = 114.4~\mathrm{K}$) can be found in Karska et al. (subm.). The o-H$_2$O $2_{1,2}-1_{0,1}$ line has a lower upper level energy and was observed in a larger beam, but was detected in more sources. Again, there is a statistical correlation between the fluxes of the OH and H$_2$O lines at similar energy, but the H$_2$O and OH spatial extents can differ (Karska et al. subm.). 

\placefigureLOHH2O

\placefigureLbolvsOHH2O89ratio

\placefigureMenvvsOHH2O89ratio

\placefigureTbolvsOHH2O89ratio

\placefigureEvovsOHH2O89ratio

The ratio of the two fluxes versus the bolometric luminosity is shown in Fig.~\ref{fig:Lbol_OHH2O89ratio}. The low-mass class I sources lie in the upper part of the plot as they have OH $84~\mu\mathrm{m}$ to H$_2$O $89~\mu\mathrm{m}$ ratios above $2-3$, whereas the class 0 sources have a slightly lower ratio on average. On the other hand, there is no significant trend with bolometric luminosity ($\rho = -0.34$). The ratio seems to decrease with envelope mass, as illustrated by Fig.~\ref{fig:Menv_OHH2O89ratio}, and to increase with bolometric temperature (Fig.~\ref{fig:Tbol_OHH2O89ratio}), but there is a large spread. Fig.~\ref{fig:Evo_OHH2O89ratio} shows that an increase in the OH $84~\mu\mathrm{m}$ to H$_2$O $89~\mu\mathrm{m}$ ratio could be an evolutionary effect. The increase in OH compared to H$_2$O could for instance be caused by the envelope becoming more tenuous in the more evolved stages, allowing high-energy photons from the protostar to penetrate further and dissociate the H$_2$O. The intermediate-mass sources are missing in Fig.~\ref{fig:Tbol_OHH2O89ratio}, because a consistent set of bolometric temperatures is not available.

\citet{Goicoechea11} find that the OH $84~\mu\mathrm{m}$ emission is not spatially correlated with o-H$_2$O $3_{0,3}-2_{1,2}$ in the Orion Bar PDR, but with high-$J$ CO and CH$^+$. We cannot test such a spatial correlation here because the fluxes drop rapidly outside the on-source spaxels in our low-mass source sample. 

\section{Excitation and origin of OH} \label{sec:models} 
Where does the OH emission arise? The environment of low-mass protostars is known to consist of different physical components, including the collapsing envelope, shocks, outflow cavities heated by UV radiation and entrained outflow gas, all of which are contained in the \textit{Herschel} beams \citep[see][for an overview]{Visser12}. 
To derive physical properties like the density and temperature of the emitting gas, the observed fluxes and line ratios are compared to the results from radiative transfer models. The obtained physical conditions can help in constraining the origin of the emission in combination with information about the spatial distribution of the emission. Furthermore, radiative transfer models permit to study what mechanism dominates the excitation under given physical conditions.

The lowest rotational transitions of OH lie in the far-infrared and therefore significantly interact with the dust continuum field, which peaks in the same wavelength regime for dust temperatures typical of embedded YSOs. Moreover, the properties of OH such as the large dipole moment and large rotational constant make radiative pumping an important excitation process. Therefore, the excitation of OH should be modeled using a code that includes radiative effects. 

Before describing the models, we recall the clues on the origin of the OH emission that come from spectrally resolved OH line profiles. \citet{Wampfler10} present upper limits on the OH $163~\mu$m (1837 GHz) line intensities using HIFI for two low-mass sources which, when combined with measured PACS fluxes for the same lines, imply FWHM line widths of at least $10~\mathrm{km}~\mathrm{s}^{-1}$. For a third low-mass source, Ser~SMM1, a recent unpublished HIFI spectrum shows a detection of the broad ($\mathrm{FWHM} \approx 20~\mathrm{km}~\mathrm{s}^{-1}$) component, but no sign of a narrow component (Wampfler et al., in prep.). In contrast, \citet{Wampfler11} found a narrow component ($\mathrm{FWHM} \approx 4-5~\mathrm{km}~\mathrm{s}^{-1}$) in the OH $1837~$GHz spectrum from the high-mass protostar W3~IRS~5 which can be attributed to the quiescent envelope, on top of a broad outflow component with $\mathrm{FWHM} \approx 20~\mathrm{km}~\mathrm{s}^{-1}$. Therefore, the OH line profiles from low-mass YSOs seem to be dominated by the outflow contribution, whereas those of high-mass sources can contain both an outflow and envelope component.

\subsection{Spherical envelope models} \label{sec:ratran_models}
To further quantify any contribution from the protostellar envelope, spherically symmetric source models such as derived by \citet{Kristensen12} are commonly used in combination with a non-LTE line radiative transfer code like RATRAN \citep{Hogerheijde00} or LIME \citep{Brinch10} to model the line emission.

We ran RATRAN calculations in spherical symmetry for a few low-mass sources, including Ser~SMM1, using the same RATRAN code as for W3~IRS~5 (see Sect.~\ref{sec:model_description} for molecular data input). The OH fractional abundances considered range from $x_\mathrm{OH} = 10^{-9}$ to $10^{-7}$. Note that the source structure on small scales is often not well constrained by the single-dish continuum observations and that the spherical symmetry starts to break down inside about 500~AU \citep[e.g.][]{Jorgensen05}. Thus this type of model is not ideally suited for lines with a high critical density, where the bulk of the emission usually arises from the innermost parts of low-mass protostellar envelopes. The inner envelope is also the area where the excited states of OH are mainly populated, and thus the model results should only be
regarded as a rough order of magnitude estimate.

It is clear that these spherically symmetric source models cannot reproduce the observed OH emission in low-mass sources for a variety of reasons. First, the synthetic spectra from the model fail to reproduce the observed broad lines (FWHM of $\sim 20~\mathrm{km}~\mathrm{s}^{-1}$). The model lines are typically much narrower (FWHM of a few $\mathrm{km}~\mathrm{s}^{-1}$) for any reasonable Doppler$-b$ parameter that fits emission from other molecules. Therefore, the line width measured from the HIFI spectra is inconsistent with an envelope-only or envelope-dominated scenario for the low-mass sources observed with HIFI. Moreover, the 79, 84, and $119~\mu\mathrm{m}$ lines from the RATRAN models are always in absorption, while the detected PACS lines from the low-mass sources are in emission, providing another indication that the bulk of the emission does not arise from the envelope. Although a minor envelope contribution to the $163~\mu$m lines cannot be excluded, the bulk of the emission is likely associated with the outflow. For the intermediate-mass sources, absorption is observed at $119~\mu\mathrm{m}$ in at least one spaxel (except for NGC~7129~FIRS~2), suggesting that a potential envelope contribution to the spectra of intermediate-mass sources cannot be excluded, in particular in the lowest rotational transitions.

\subsection{Outflow model} \label{sec:slab_models}
\subsubsection{Slab model Description} \label{sec:model_description}
To model OH emission from outflows or shocks, a simple slab geometry is appropriate. We use a radiative transfer code based on the escape probability formalism as described in \citet{Takahashi83}, which includes radiative pumping. 
The physical conditions such as density and temperature and the excitation of the molecule are assumed to be constant throughout the region. The free parameters of the model are the kinetic temperature of the gas $T_\mathrm{gas}$, the temperature of the dust $T_\mathrm{dust}$, the density of the collision partner $n$, the molecular column density of OH $N_\mathrm{OH}$, and the dust column density $N_\mathrm{dust}$. The line profile function has a rectangular shape with a width of $\Delta V = \sqrt{\pi}/(2 \sqrt{\ln 2}) \Delta v$ where $\Delta v$ is the full width at half maximum (FWHM) of a Gaussian line. Here, $\Delta v$ was assumed to be $10~\mathrm{km}~\mathrm{s}^{-1}$ for all models. As discussed above, this FWHM is motivated by the non-detection of OH from low-mass sources with HIFI \citep{Wampfler10} that lead to the conclusion that the line width must be broader than $10~\mathrm{km}~\mathrm{s}^{-1}$. 

Our slab code is similar to the ``RADEX'' code \citep{vanderTak07}, but includes an active dust continuum, which is capable of both absorbing and emitting radiation. Results from our code in the limiting case where no dust is present agree very well with the RADEX results. An earlier OH excitation study by \citet{Offer92} included a dust continuum radiation field for the radiative excitation but not the absorption of line photons by dust grains. Furthermore, different dust properties were used in their calculations. Therefore, the comparability of the two models is limited.

We use the molecular data file from the Leiden atomic and molecular database \citep[LAMDA,][]{Schoier05} for OH without hyperfine structure, because the resolution of PACS does not allow to resolve the hyperfine components. The file contains frequencies, energy levels, and Einstein A coefficients from the JPL catalog \citep{Pickett98} and collision rates with ortho- and para-H$_2$ for temperatures in the range of $15-300~\mathrm{K}$ from \citet{Offer94}. The highest excited state contained in the file has an energy of $875~\mathrm{K}$.  
The ortho-to-para-H$_2$ ratio in the model is temperature dependent according to
\begin{equation}
 r = \min \left[3.0, 9.0 \times \exp \left( \frac{-170.6}{T}\right) \right].
\end{equation}
The dust opacities for dust grains with thin ice mantles for the wavelength range of $1-1300~\mu\mathrm{m}$ are taken from \citet[][ Table~1, Col.~5]{Ossenkopf94}. 

The parameter space explored in the slab models covers the full range of physical conditions expected in the protostellar environment, i.e., temperature range $T = 50-800~\mathrm{K}$, densities of $n = 10^4-10^{12}~\mathrm{cm}^{-3}$, dust column densities of $N_\mathrm{OH} = 10^{18}-10^{23}~\mathrm{cm}^{-2}$, and OH molecular column densities of $10^{14}-10^{18}~\mathrm{cm}^{-2}$. Temperatures above $800~\mathrm{K}$ are problematic because the highest rotational level included in the molecular data file is $875~\mathrm{K}$ and the collision rates are only available up to $300~\mathrm{K}$. The collisional de-excitation rates are kept constant above $300~\mathrm{K}$, and the excitation rates are calculated from the de-excitation rates from the detailed balance relations and thus depend on the kinetic temperature through a factor $\exp(-\Delta E/(k_\mathrm{B}T_\mathrm{kin}))$.

\subsubsection{Slab model results} \label{sec:model_results}

\placefigureModelTexTgas200KTdust100K
\placefigureModelRatiosTgas200KTdust100K

A useful quantity to describe the excitation of a molecule is the excitation temperature, defined as
\begin{equation}
 T_\mathrm{ex} = - \frac{\Delta E}{k_\mathrm{B} \ln \left( \frac{n_\mathrm{u}}{n_\mathrm{l}} \frac{g_\mathrm{l}}{g_\mathrm{u}}\right)}
\end{equation}
where $g_\mathrm{l}$ and $g_\mathrm{u}$ are the statistical weights of the lower and upper level, respectively, and $n_\mathrm{l}$ and $n_\mathrm{u}$ their normalized populations. The energy of the transition is $\Delta E = h \nu$ and $k_\mathrm{B}$ denotes Boltzmann's constant.

Figure~\ref{fig:model_tex_Tgas200K_Tdust100K} shows the excitation temperature of the OH transitions at $79.18~\mu\mathrm{m}$, $84.60~\mu\mathrm{m}$, $119.44~\mu\mathrm{m}$, and $163.12~\mu\mathrm{m}$ as a function of $n_{\mathrm{H}_2}$ and $N_\mathrm{dust}$ for a model of fixed gas and dust temperatures ($T_\mathrm{gas} = 200~\mathrm{K}, T_\mathrm{dust} = 100~\mathrm{K}$) and three values of $N_\mathrm{OH}$. The $N_\mathrm{OH} = 10^{14}~\mathrm{cm}^{-2}$ model largely represents the optically thin case, the $N_\mathrm{OH} = 10^{18}~\mathrm{cm}^{-2}$ model the optically thick regime, and $N_\mathrm{OH} = 10^{16}~\mathrm{cm}^{-2}$ is an intermediate case. The adopted parameters are 
representative for the OH emitting region.

We notice from Fig.~\ref{fig:model_tex_Tgas200K_Tdust100K} that all transitions behave relatively similar with respect to their excitation. Furthermore, this behavior is also comparable for all considered gas temperatures ($50, 100, 200, 500$, and $800~\mathrm{K}$).  For densities above the critical density ($\sim 10^8~\mathrm{cm}^{-3}$), the excitation temperature approaches the kinetic temperature of the gas, as expected, irrespective of the dust column density. The populations are thermalized at the kinetic temperature. Collisions must therefore be the dominant excitation mechanism in this regime. 
For densities below the critical density, dust pumping dominates the excitation for dust column densities above $\sim 10^{21}~\mathrm{cm}^{-2}$. This is illustrated by the fact that $T_\mathrm{ex}$ approaches $T_\mathrm{dust}$, i.e. the populations are thermalized at the temperature of the dust radiation field. 
Below $N_\mathrm{dust} \approx 10^{21}~\mathrm{cm}^{-2}$, both mechanisms contribute to the excitation, whereby one order of magnitude increase in $n_{\mathrm{H}_2}$ and $N_\mathrm{dust}$ have similar effects on $T_\mathrm{ex}$. 
If the OH column density is enhanced, the gas becomes optically thick and line photons are trapped. Then, the level populations thermalize at the gas temperature already at lower densities because the critical density is reduced with an effective critical density of $n_\mathrm{cr} (\tau_\mathrm{line}) \approx {n_\mathrm{cr} (0)}/\tau_\mathrm{line} $. 

In models of equal dust and gas temperature, we find that the $79~\mu\mathrm{m}$ cross-ladder transition even experiences supra-thermal excitation for low densities ($n_{\mathrm{H}_2} \lesssim 10^8~\mathrm{cm}^{-3}$), temperatures above $50~\mathrm{K}$, OH column densities $10^{14}-10^{16}~\mathrm{cm}^{-2}$, and dust column densities in the range of $2 \times 10^{20}-1 \times 10^{22}~\mathrm{cm}^{-2}$. 

As for line optical depths, the models show that the $119~\mu\mathrm{m}$ transition reaches the highest optical depth as expected from the fact that this transition is connected to the ground state. It is followed by the  $84~\mu\mathrm{m}$ transition, originating from the same rotational ladder, just above the $119~\mu\mathrm{m}$. The $79~\mu\mathrm{m}$ line is a cross-ladder transition with a much smaller Einstein A coefficient and does therefore not become optically thick as easily as the two afore mentioned ${}^{2}\Pi_{3/2}$ intra ladder transitions. 

Figure~\ref{fig:model_ratios_Tgas200K_Tdust100K} presents the line flux ratios for the six possible combinations of the observed four transitions. We decided to use line ratios instead of fluxes because the ratios are less dependent on the source size. The black contours represent the minimum and maximum value of the observed ratios and the shaded area indicates where the models match the observed values, given in Table~\ref{tab:ratios}. White areas at high dust column densities indicate regions where at least one line is in absorption. 
The observed line ratios can be modeled with a wide range of parameters, but the observed line ratios still allow us to exclude parts of the parameter space. The very optically thin model ($N_\mathrm{OH} = 10^{14}~\mathrm{cm}^{-2}$) fails to reproduce the observed $79~\mu\mathrm{m}/119~\mu\mathrm{m}$ line flux ratio for temperatures below $200 - 500~\mathrm{K}$ and can also hardly match the $119~\mu\mathrm{m}/163~\mu\mathrm{m}$ ratio. Similar difficulties occur for the very optically thick case ($N_\mathrm{OH} = 10^{18}~\mathrm{cm}^{-2}$), where the observed $79~\mu\mathrm{m}/163~\mu\mathrm{m}$ ratio can only be found in a very narrow range of dust column densities. 
A direct comparison of the column density with observations is however difficult because the model does not include any geometrical effects. 

There are basically two main regimes in the intermediate case ($N_\mathrm{OH} = 10^{16}~\mathrm{cm}^{-2}$) that are able to reproduce to the observed values. First, there is the radiatively dominated regime at low densities ($10^4-10^8~\mathrm{cm}^{-3}$) for dust column densities in the range of $N_\mathrm{dust} \approx 1 \times 10^{19} - 3 \times 10^{21}~\mathrm{cm}^{-2}$, slightly depending on temperature (lower $N_\mathrm{dust}$ for higher temperature and vice versa). In the regime where collisions start to become important ($n \gtrsim 10^{6}~\mathrm{cm}^{-3}$) and at low dust column densities ($N_\mathrm{dust} \lesssim 10^{20}~\mathrm{cm}^{-2}$), it is harder to find a set of parameters that is able to reproduce all line ratios simultaneously, in particular the $79~\mu\mathrm{m}/84~\mu\mathrm{m}$, $79~\mu\mathrm{m}/163~\mu\mathrm{m}$, and $84~\mu\mathrm{m}/163~\mu\mathrm{m}$ ratios. For densities of $10^{6}-10^{8}~\mathrm{cm}^{-3}$, small ranges can however be found, especially given the uncertainties of our line ratios. At temperatures below $\sim100~\mathrm{K}$, a collision dominated solution exists towards higher densities ($n \gtrsim 10^{8}~\mathrm{cm}^{-3}$) and dust column densities above $\sim 3 \times 10^{20}~\mathrm{cm}^{-2}$. So given the degeneracies in the models, we cannot well distinguish between collisional excitation and radiative pumping unless we have additional constraints on the density of the emitting gas and the dust column density. 

\placefigureRotDiagModel

To constrain which of the two scenarios is more likely, we ran additional RADEX slab models \citep{vanderTak07} that allow us to include the actual continuum field of a source at a given distance. The line width is chosen to be $10~\mathrm{km}~\mathrm{s}^{-1}$ and the density and temperature at a given distance are taken from the continuum model of the source presented in \citet{Kristensen12}. The OH column density is a free parameter. The continuum field is only included in the excitation of the OH molecules, but not in the line formation, thus assuming a geometry in which the OH is not right in front of the line of sight to the continuum. Models which include the source continuum in both the excitation and the line formation yield several lines in absorption and cannot reproduce the observations. To study the importance of the continuum, models without the continuum field were calculated as well. The model fluxes were scaled such that the $84.60~\mu\mathrm{m}$ flux is equal to the observed value. 

Figure~\ref{fig:smm1_rotdiag_model} shows the results for Ser~SMM1 ($n = 8 \times 10^{7}~\mathrm{cm}^{-3}$ and $T = 90.3~\mathrm{K}$ at a distance of 100~AU) and $N_\mathrm{OH} = 10^{16}~\mathrm{cm}^{-2}$. While the modeled fluxes are fairly similar for the transitions with an upper level energy less than $300~\mathrm{K}$, the model including the continuum field matches the observed values better than that without a source continuum for the higher excited lines ($98~\mu\mathrm{m}$, $65~\mu\mathrm{m}$, and $71~\mu\mathrm{m}$). 

Another indication that the continuum may be important is illustrated by Fig.~\ref{fig:LOH_63continuum} showing the correlation of the OH $84~\mu\mathrm{m}$ flux with the continuum at $63~\mu\mathrm{m}$ and $84~\mu\mathrm{m}$ as measured from the PACS data. A continuum field with a temperature similar to the observed rotational temperature of OH, i.e. $T_\mathrm{rad} \approx 100~\mathrm{K}$, would peak at around $30~\mu\mathrm{m}$, which cannot be observed with Herschel. Therefore, we use the continuum at $63~\mu\mathrm{m}$ and $84~\mu\mathrm{m}$ instead.

\placefigureLOH63continuum

For the Orion bar, \citet{Goicoechea11} found line ratios that differ from the values derived for our source sample, except for the $79~\mu\mathrm{m}/119~\mu\mathrm{m}$ and $79~\mu\mathrm{m}/163~\mu\mathrm{m}$ ratios. The $79~\mu\mathrm{m}/84~\mu\mathrm{m}$ ratio of almost four is higher than in our sample, as is the $119~\mu\mathrm{m}/163~\mu\mathrm{m}$ ratio. On the other hand, the $84~\mu\mathrm{m}/119~\mu\mathrm{m}$ and $84~\mu\mathrm{m}/163~\mu\mathrm{m}$ ratios are lower. The temperature, density, and the column density of the dust grains are lower in the Orion bar, therefore both collisional excitation and pumping by the FIR field are likely lower.

\subsection{Discussion} \label{sec:discussion}
From the PACS observations we find that the excitation of the OH emission in our sample of 17 low- and 6 intermediate-mass protostars is similar in all sources, as indicated by the relatively constant line ratios among the target sample. This could either mean that the OH emission from all sources arises from gas at very similar physical conditions or that the observed line ratios can be realized by a broad range of gas properties. Comparison of the ratios to the modeling results (cf. Sect.~\ref{sec:model_results}) shows that the observed values can indeed be reproduced by a wide parameter range, which may explain why similar ratios are observed in sources that span a large range in luminosity, envelope mass, and evolutionary stage. 

From the full range scans of the two class 0 YSOs Ser~SMM1 and NGC~1333~IRAS~4B we find that the rotational temperature of OH in these sources is around $70~\mathrm{K}$. From the fact that the points can be relatively well fitted with a straight line in the rotational diagram, we conclude that the level population of OH can be represented by a Boltzmann distribution at $T \approx 70~\mathrm{K}$. The constant line ratios found in the sample may be an indication that other sources could have similar rotational temperatures. However, the models demonstrate that different physical conditions are able to reproduce the observed line ratios and thus full range scan data, which includes more than four OH transitions, of additional sources is needed to test how the OH rotational temperatures vary from source to source and whether there is any difference between evolutionary stages. If the rotational temperature were equal to the ambient kinetic temperature, the population would be in LTE. However, because of the very high critical densities of most OH transitions (around $10^8~\mathrm{cm}^{-3}$), this scenario seems to be rather unlikely, as a significant amount of the emission stems from the outflow \citep{Wampfler10,Wampfler11}. 
Compression factors of a hundred are needed to obtain a post-shock density of $10^8~\mathrm{cm}^{-3}$ for a pre-shock density of $10^6~\mathrm{cm}^{-3}$ \citep{Neufeld89}. Alternatively, the OH excitation may be in equilibrium with the temperature of the far-infrared dust continuum field, that can strongly pump OH.

The models show that the degeneracy in the parameter space is large, allowing different combinations of parameters for the realization of the observational results. Therefore, it is only possible to distinguish between the collision-dominated and radiative pumping controlled scenarios described in Sect.~\ref{sec:model_results} when additional constraints on the density and dust column density of gas emitting or absorbing in OH far-infrared lines are available. The similar line ratios could be a hint that the excitation is not controlled by a combination of both collisions and radiative pumping but rather by one dominant mechanism, because it would be much harder to match the same or similar combination in all targets. The RADEX results point towards radiative pumping as the main excitation process, at least in Ser~SMM1, as models including a far-infrared continuum field fit the observations better, in particular the high excited lines ($65~\mu\mathrm{m}$ and $71~\mu\mathrm{m}$). Another indication that the radiation field is important in the excitation is provided by the strong correlation of the OH $84~\mu\mathrm{m}$ luminosity with the bolometric luminosity of the sources regardless of their evolutionary stage and type, and a tentative trend with the continuum measurements.

An envelope-only or strongly envelope-dominated scenario can be excluded for the low-mass sources from the fact that spherically symmetric radiative transfer models yield the $79$, $84$, and $119~\mu\mathrm{m}$ transitions in absorption, in conflict with the observed emission lines. Furthermore, the FWHM of the modeled $163~\mu\mathrm{m}$ line is at least a factor of two narrower than the observed value, pointing also at an outflow origin of the emission. Several physical regimes have been identified in the outflow from Herschel H$_2$O and CO observations \citep[e.g.][]{Nisini10,Yildiz10,vanKempen10a,Lefloch10,Kristensen12,Herczeg12,Vasta12}: the cold entrained outflow material, with typical densities around $10^{5}~\mathrm{cm}^{-1}$ and temperatures around $100~$K, an intermediate warm region ($T \sim 300~\mathrm{K}$, $n \sim 10^{6}-10^{7}~\mathrm{cm}^{-1}$), and the shock front, where temperatures reach $T \sim 800~\mathrm{K}$ and densities are found up to $n \sim 10^{8}~\mathrm{cm}^{-1}$. The OH rotational temperature of $T \sim 70~\mathrm{K}$ is low compared to the values found for H$_2$O and CO and favors the intermediate warm regime over a shock front origin \citep{Goicoechea12}. The entrained material however has a density that is significantly lower than the critical density, which would require the presence of a strong radiation field to excite the OH molecules, for instance close to the protostar.

\section{Conclusions} \label{sec:conclusions}

We analyze \textit{Herschel}/PACS spectroscopy of the OH lines at $79, 84, 119$ and $163~\mu\mathrm{m}$ in a set of 23 low- and intermediate-mass protostars. In addition, we use slab radiative transfer models to study the OH excitation. Our main results are:
\begin{enumerate}
 \item At least one OH line is detected in all 23 sources. For NGC~1333~IRAS~2A, only the $119~\mu\mathrm{m}$ doublet is detected in absorption against the continuum. Most sources show only emission lines. Absorption occurs towards higher envelope masses.
 \item The derived rotational temperatures for Ser~SMM1 and NGC~1333~IRAS~4B, where additional OH lines are available from a full spectral scan, are similar and around $70~\mathrm{K}$. 
 \item The ratios between the fluxes of different OH emission lines are relatively constant among the source sample, i.e. the fluxes in the different transitions are well correlated.
 \item There is a strong correlation (Pearson correlation coefficient of $\rho = 0.93$) between the OH line luminosity and the bolometric luminosity of the sources. The correlation with envelope mass is less tight ($\rho = 0.84$) and could even be introduced by an underlying correlation of the bolometric luminosity and the envelope mass. We do not find evidence for an evolutionary effect, i.e. a strong correlation with the bolometric temperature or the evolutionary tracer $L_\mathrm{bol}^{0.6}/M_\mathrm{env}$. 
 \item There are also trends of an increasing OH flux with increasing [OI] and H$_2$O fluxes. 
 \item The OH 84$~\mu\mathrm{m}$/H$_2$O 89$~\mu\mathrm{m}$ ratio is slightly higher in the low-mass class I YSOs than in the class 0 sources in our sample. The ratio seems to increase with increasing bolometric temperature and to decrease towards higher envelope masses, indicating that a change in the ratio could be an evolutionary effect. 
 \item Spherically symmetric protostellar envelope models do not fit the observed OH line widths and produce 79, 84 and $119~\mu$m lines in absorption rather than emission. 
 \item Slab radiative transfer models representative of an outflow show that the observed line ratios can be reproduced by a range of parameters. The degeneracy in the parameter space of the models and the currently available data make it challenging to clearly distinguish whether the excitation of OH is dominated by collisions or by radiative pumping or even a combination of both. Fluxes of transitions with $E_\mathrm{up} \gtrsim 300~\mathrm{K}$ are clearly better reproduced by RADEX models with radiative pumping compared to those without a source continuum. The OH cannot be right along the line of sight to the continuum, however, since this would yield several lines in absorption.
 \item Both the excitation and spatial extent observed by \textit{Herschel}, combined with independent information on broad line widths, provide strong evidence that the OH emission from low-mass protostars is dominated by shocks impacting on the dense envelope close to the protostar.
\end{enumerate}

\bibliographystyle{aa}
\bibliography{mybib}

\begin{acknowledgements}
We thank the referee for constructive comments that helped to improve the paper. 
The work on star formation at ETH Zurich is partially funded by the Swiss National Science Foundation (grant nr. 200020-113556). This program is made possible thanks to the Swiss HIFI guaranteed time program. WISH research in Leiden is supported by the Netherlands Research School for Astronomy (NOVA), by grant 614.001.008 from the Netherlands Organization for Scientific Research (NWO) and by EU-FP7 grant 238258 (LASSIE).
HIFI has been designed and built by a consortium of 
institutes and university departments from across Europe, Canada and the 
United States under the leadership of SRON Netherlands Institute for Space
Research, Groningen, The Netherlands and with major contributions from 
Germany, France and the US. Consortium members are: Canada: CSA, 
U.Waterloo; France: CESR, LAB, LERMA, IRAM; Germany: KOSMA, 
MPIfR, MPS; Ireland, NUI Maynooth; Italy: ASI, IFSI-INAF, Osservatorio 
Astrofisico di Arcetri- INAF; Netherlands: SRON, TUD; Poland: CAMK, CBK; 
Spain: Observatorio Astron{\'o}mico Nacional (IGN), Centro de Astrobiolog{\'i}a 
(CSIC-INTA). Sweden: Chalmers University of Technology - MC2, RSS $\&$ 
GARD; Onsala Space Observatory; Swedish National Space Board, Stockholm 
University - Stockholm Observatory; Switzerland: ETH Zurich, FHNW; USA: 
Caltech, JPL, NHSC.

\end{acknowledgements}

\Online

\begin{appendix}

\section{Observational Details}
\placetablesourceobsids

\placetableIMH2OandOI

\placetablecorrfc

\placetablef33

\placetableft

\section{OH maps} \label{sec:oh_maps}
\placefigu2IRASAmaps
\placefigIRAS4Amaps
\placefigBIRAS4maps
\placefigL1527maps
\placefigureced110irs4maps
\placefigureiras15398maps
\placefigurel483maps
\placefiguresMM1maps
\placefigureSMM3maps
\placefigure723Lmaps
\placefigureL1489maps
\placefigureTMR1maps
\placefigureATMC1maps
\placefigureTMC1maps
\placefigureHH46maps
\placefigureRNO91maps
\placefigureAFGL490maps
\placefigureN2071maps
\placefigureVela17maps
\placefigurevela19maps
\placefiguren7129maps
\placefigurl1641maps

\end{appendix}

\end{document}